%% file: main.tex
\documentclass[12pt]{article}

\usepackage{amsmath, amssymb, amsthm, amsfonts, float, comment}
\usepackage{afterpage}

\usepackage[linesnumbered, ruled, vlined]{algorithm2e}

\usepackage{graphicx, subfigure, color, colortbl}

\usepackage{multirow}

\usepackage{enumitem}

\usepackage{fancyhdr}

\usepackage[section]{placeins}

\usepackage{url}

\usepackage{etoolbox}

\usepackage{indentfirst}

\usepackage[explicit]{titlesec}

\usepackage[T1]{fontenc}

\usepackage{charter}

\usepackage[subfigure]{tocloft}

\usepackage[doublespacing]{setspace}

\usepackage{pgfplots}
\pgfplotsset{compat=1.5}

\titleformat{\section}[block]{\centering\singlespace\large\bf}{CHAPTER \thesection \hspace{0.4em} \MakeUppercase{#1}}{0em}{}{}

\titleformat{name=\section,numberless}[block]{\centering\singlespace\large\bf}{\hspace{0.4em} \MakeUppercase{#1}}{0em}{}{}

\titlespacing{\section}{0pt}{0pt}{0pt}
\titlespacing{\subsection}{0pt}{0pt}{0pt}
\titlespacing{\subsubsection}{0pt}{0pt}{0pt}

\cftsetindents{section}{0em}{6em}

\makeatletter
\def\thm@space@setup{\thm@preskip=0pt
\thm@postskip=0pt}
\makeatother
\newtheoremstyle{newstyle}      
{} 
{} 
{\mdseries} 
{} 
{\bfseries}  
{.} 
{ } 
{} 

\makeatletter
\renewenvironment{proof}[1][\proofname]{\par
  \pushQED{\qed}%
  \normalfont \topsep0\p@\relax
  \trivlist
  \item[\hskip\labelsep\itshape
  #1\@addpunct{.}]\ignorespaces 
}{
  \popQED\endtrivlist\@endpefalse
}
\makeatother

\patchcmd{\thebibliography}{\section*{\refname}}{}{}{}

\let\OLDthebibliography\thebibliography
\renewcommand\thebibliography[1]{
    \OLDthebibliography{#1}
    \setlength{\parskip}{0pt}
    \setlength{\itemsep}{0pt plus 0.3ex}
}

\usepackage[left=1in,right=1in, top=1in, bottom=1in]{geometry}

\newcommand{\Mdef}[2]{\newcommand{#1}{\relax \ifmmode #2 \else $#2$\fi}}

\newcommand{\Vspc}{\vspace*{0.1in}}

\makeatletter \@addtoreset{equation}{section} \makeatother

\renewcommand{\thesection}{\arabic{section}}

\newcommand*{\revise}{\textcolor{black}}
\definecolor{LightCyan}{rgb}{0.0,1,1}
\definecolor{LightGray}{rgb}{0.8,0.8,0.8}

\newcommand{\code}[1]{\texttt{#1}}

\newtheorem{theorem}{Theorem}

\begin{document}

\input{title}
\newpage

\pagestyle{fancy} \chead{} \rhead{} \lhead{}
\pagenumbering{roman} \lfoot{}\cfoot{\thepage}\rfoot{}
\setcounter{page}{2}

\input{dedication}
\newpage

\input{acknowledgements}
\newpage


\begin{singlespace}

\renewcommand{\contentsname}{\hfill\large TABLE OF CONTENTS \hfill}
\tableofcontents
\newpage

\addcontentsline{toc}{section}{List of Tables}
\renewcommand{\listtablename}{\hfill\large LIST OF TABLES \hfill}
\listoftables
\newpage

\addcontentsline{toc}{section}{List of Figures}
\renewcommand{\listfigurename}{\hfill\large LIST OF FIGURES \hfill}
\listoffigures

\end{singlespace}

\clearpage

\pagestyle{fancy} \chead{\thepage} \rhead{} \lhead{}
\pagenumbering{arabic} \lfoot{}\cfoot{}\rfoot{}
\setcounter{equation}{0}

\input{abstract}
\clearpage

\input{chapter1}
\clearpage

\input{chapter2}
\clearpage

\input{chapter3}
\clearpage

\input{chapter4}

\clearpage

\input{chapter5}

\clearpage


\input{chapter7}

\clearpage

\input{chapter8}

\clearpage





\input{bibliography}

\end{document}

%% file: title.tex
\centerline{\bf LOW-POWER WIDE-AREA NETWORK DESIGN}

\vskip-0.4cm

\thispagestyle{empty}

\begin{center}
    \vspace{-0.4cm}
    by \\
    {\bf MD MAHBUBUR RAHMAN}\\ 
    {\bf DISSERTATION}\\  
    Submitted to the Graduate School,\\
    of Wayne State University,\\
    Detroit, Michigan\\
    in partial fulfillment of the requirements\\
    for the degree of\\
    {\bf DOCTOR OF PHILOSOPHY} 
\end{center}

\begin{flushleft}
    \vspace*{-0.20in}
    \hspace*{3.09in}2020 
    \hspace*{3.09in}MAJOR: COMPUTER SCIENCE\\ 
\end{flushleft}

%% file: dedication.tex
\section*{DEDICATION}
\addcontentsline{toc}{section}{Dedication}
\begin{center}
Dedicated to my parents and Moni.
\end{center}

%% file: acknowledgements.tex
\section*{ACKNOWLEDGEMENTS}
\addcontentsline{toc}{section}{Acknowledgements}



I am immensely grateful to my advisor, Prof. Abusayeed Saifullah, for his invaluable guidance during my PhD. Without his constant care and patience, I doubt that I would  ever survive this journey. He has always been there for me during good times or difficult times. I consider myself lucky to have him as my advisor.

I am thankful to all of my fellow labmates for their constructive feedback on my research and support in many other ways. Thanks to Dali Ismail for running with me into the woods and helping me collect experimental data. Being a member of the CRI Lab, I have grown more, both personally  and professionally.

I am grateful to my parents for their dedication and sacrifices for me. I would like to thank my brother, Md Mamunur Rahman, for his support during my higher secondary and undergraduate studies, without which I would have never made it this far. A special thanks to my wife, Mirza Ishrat Noor (Moni), for her kind support and sacrifices during my PhD.

Finally, I sincerely acknowledge the financial support from my advisor, NSF, and the Graduate School and Department of Computer Science at Wayne State University.

%% file: abstract.tex
\section*{ABSTRACT}

\addcontentsline{toc}{section}{Abstract}
\centerline{\bf LOW-POWER WIDE-AREA NETWORK DESIGN}

{\setlength\baselineskip{0.3in}
\begin{center}
by\\
\medskip
{\bf MD MAHBUBUR RAHMAN}\\
\medskip
{\bf August 2020}\\
\end{center}
\Vspc
\begin{tabular}{ll}
	{\bf Advisor:} & Dr. Abusayeed Saifullah \\
	{\bf Major:} & Computer Science \\
	{\bf Degree:} & Doctor of Philosophy
\end{tabular}
}

\bigskip \bigskip

Low-Power Wide-Area Network (LPWAN) is an enabling technology for long-range, low-power, and low-cost Internet of Things (IoT) and Cyber-Physical Systems (CPS) applications. Due to their escalating demand in the IoT/CPS applications, recently, multiple LPWAN technologies have been developed that operate in the cellular/licensed (e.g., 5G, LTE Cat M1, and NB-IoT) and unlicensed/ISM (e.g., LoRa and SigFox) bands. To avoid the crowd in the limited ISM band (where most LPWANs operate) and the cost of the licensed band, we propose a {\em novel} LPWAN technology called {\em Sensor Network Over White Spaces (SNOW)} by utilizing the TV white spaces. {\em White spaces} refer to the allocated but locally unused TV channels (54--698MHz in the US) and can be used by the unlicensed devices as secondary users. White spaces offer less crowded and much wider spectrum in both urban and rural areas, boasting an abundance in rural and suburbs and have excellent propagation and obstacle penetration characteristics that enable long-range communication.

In this thesis, we design, develop, and experiment SNOW that is highly scalable, energy-efficient, and can connect thousands of sensors over a single-hop distance of several kilometers. SNOW achieves scalability and energy efficiency by enabling concurrent packets reception (Rx) at a base station (BS) using a single radio from numerous sensors and concurrent packets transmission (Tx) to numerous sensors from the BS using a single radio, simultaneously, which we achieve by proposing a {\em distributed} implementation of {\em Orthogonal Frequency Division Multiplexing}. We also demonstrate the feasibility of SNOW design by implementing on a prototype hardware called Universal Software Radio Peripheral(USRP).

To enable the low-cost and scalable SNOW deployment in practical IoT/CPS applications, we then implement SNOW using the low-cost and small form-factored commercial off-the-shelf (COTS) devices, where we address multiple practical challenges including the peak-to-average power ratio (PARP) problem handling, channel state information estimation, and carrier frequency offset estimation. Additionally, we propose an adaptive transmission power protocol for the SNOW nodes to handle the near-far power problem in SNOW. To demonstrate the feasibility of COTS SNOW implementation, we use TI CC1310 and TI CC1350 devices as SNOW nodes and deploy in the city of Detroit, Michigan.

To enable connecting tens of thousands of nodes over hundreds of kilometers, we further propose a network architecture called {\em SNOW-tree} through a seamless integration of multiple SNOWs where they form a tree structure and are under the same management/control at the tree root. We address the intra-SNOW and inter-SNOW interferences in SNOW-tree by formulating a constrained optimization problem called the {\em scalability optimization problem (SOP)} whose objective is to maximize scalability by managing the spectrum sharing across the SNOWs. By proving the NP-hardness of SOP, we then propose two polynomial-time methods to solve it: a greedy heuristic algorithm and a $\frac{1}{2}$-approximation algorithm. Our deployment covering approximately (25x15)km$^2$ in the Detroit metropolitan area demonstrates that both of our algorithms are highly efficient in practice. 

%% file: chapter1.tex
\section{Introduction}   \label{chap:Introduction}

Low-Power Wide-Area Network (LPWAN) is an emerging communication technology that supports low-power and low-cost connectivity to numerous devices (e.g., sensors) over long distances. It is considered as a key enabling technology for many Internet of Things (IoT) and Cyber-Physical Systems (CPS) applications, including urban sensing~\cite{citysense}, connected vehicles~\cite{whitespaceSurvey}, oil field monitoring~\cite{oilffield}, and agriculture and smart farming~\cite{vasisht17farmbeats}. These applications often need to connect tens of thousands of sensors over several kilometers. The existing Wireless Sensor Network (WSN) technologies operating in the ISM bands such as IEEE 802.15.4~\cite{ieee154}, IEEE 802.11~\cite{wifi}, and Bluetooth Low Energy (BLE)~\cite{Bluetooth} have short range (e.g., 30--50m for IEEE 802.15.4-based networks operating in the 2.4GHz band) and are impractical to be adopted in these IoT/CPS applications. For example, to cover a wide area with numerous sensors, WSNs form multi-hop mesh networks at the expense of energy, cost, and complexity, limiting scalability and lifetime of the IoT/CPS applications~\cite{snow_ton}.

Due to their escalating demand in the IoT/CPS applications, multiple LPWAN technologies have been developed recently that operate in the licensed/cellular (e.g., 5G~\cite{akpakwu2017survey}, LTE Cat M1~\cite{cat}, NB-IoT~\cite{nbiot}, and EC-GSM-IoT~\cite{ecgsmiot}) and unlicensed/non-cellular (e.g., LoRa~\cite{lorawan}, SigFox~\cite{sigfox}, RPMA~\cite{rpma},  IQRF~\cite{iqrf}, Telnesa~\cite{telensa}, DASH7~\cite{dash7}, WEIGHTLESS-N/P~\cite{weightless, weightless-p}, IEEE 802.11ah~\cite{IEE80211_ah}, IEEE 802.15.4k/g~\cite{IEE802154_k, IEE802154_g}) bands. LPWAN technologies that operate on the licensed/cellular bands require high network maintenance costs due to high service fees and costly infrastructure. On the other hand, LPWANs operating on the unlicensed/non-cellular (e.g., sub-1GHz band) bands face severe inter-technology, intra-technology, and/or other interferences due to the proliferation of LPWANs as well as other wireless technologies in these bands.

To avoid the high cost of the licensed band and the crowd of the ISM band, in this thesis research, we design, develop, and experiment a novel LPWAN technology by exploiting the TV white spaces. {\em White spaces} refer to the allocated but locally unused TV channels (54--698MHz in the US) and can be used by unlicensed devices as the secondary users~\cite{FCC_first_order, fcc_second_order}. Compared to the crowded ISM band, white spaces offer less crowded and much wider spectrum in both urban and rural areas, boasting an abundance in rural and suburbs~\cite{whitespaceSurvey}. Due to their low frequency, white spaces have excellent propagation and obstacle penetration characteristics that enable long-range communication~\cite{snow}. White spaces thus hold the potentials for LPWAN to support various IoT/CPS applications. Specifically, we make the following novel research contributions in this thesis.

First, we propose a highly scalable and energy efficient LPWAN technology called {\em Sensor Network Over White Spaces (SNOW)}. SNOW achieves scalability and energy-efficiency by enabling concurrent packets reception (Rx) at a base station (BS) using a single radio (Rx-radio) from numerous sensors and concurrent packets transmission (Tx) to numerous sensors from the BS using a single radio (Tx-radio), simultaneously. The Rx-radio and Tx-radio at the BS use the same set of TV channels split into narrowband subcarriers, i.e., subchannels optimized for scalability, energy efficiency, and communication reliability. Every sensor transmits/receives a spectral component of a Distributed Orthogonal Frequency-Division Multiplexing (D-OFDM) symbol asynchronously using an assigned subcarrier to encode/decode its packets. The BS uses a Fast Fourier Transformation (FFT)/Inverse-FFT to decode (using Rx-radio)/encode (using Tx-radio) different data from/to different sensors, which can scale to thousands of nodes at the same complexity over several kilometers. We also demonstrate the feasibility of SNOW by implementing it on a prototype
hardware called Universal Software Radio Peripheral (USRP). 

Second, to enable the low-cost and scalable SNOW deployment in practical IoT/CPS applications, we implement SNOW on the low-cost and small form-factored commercial off-the-shelf (COTS) IoT devices. In this implementation, we address the following practical challenges that are related to low-cost COTS devices with cheap radios. 
(1) The D-OFDM-based SNOW physical (PHY) layer degrades the reliability in the nodes by introducing severe inter-(sub)carrier interference (ICI) due to the signal power
asymmetries in different subcarriers. 
(2) Due to the bandwidth asymmetries in the BS and nodes, subcarrier noise estimation is extremely difficult, thereby degrading the reliability and communication range. 
(3) Due to radio imperfections (e.g., frequency mismatch) in the BS and nodes, the orthogonality of the D-OFDM subcarriers breaks, introducing severe ICI and degrading reliability at the BS and nodes.
Through this implementation, we address the above challenges in SNOW. Specifically, we handle the subcarrier power asymmetries by lowering their peak-to-average power ratio (PAPR). We also propose a preamble-based (known sequence of bits) subcarrier noise estimation technique for the asymmetric subcarrier bandwidths. Additionally, we proposed a preamble-based subcarrier frequency offset estimation technique to handle the radio imperfections. Through this implementation, we also address the classic near-far power problem for wide-area SNOW deployment by formulating a predictive control-based adaptive transmission power mechanism at the nodes. We demonstrate COTS SNOW implementation on the TI CC1310 and CC1350 devices, reducing the cost and form-factor of a SNOW node by 30x and 10x, respectively, compared to the USRP-based SNOW implementation.

Finally, as the LPWANs are evolving rapidly, they still face limitations in meeting the scalability and coverage
demand of very wide-area IoT/CPS deployments (e.g., (74x8)km$^2$ East Texas oilfield monitoring with tens of thousands of sensors, especially in the infrastructure-limited rural areas. To enable this, we propose a network architecture called {\em SNOW-tree} through a seamless integration of multiple SNOWs where they form a tree structure and are under the same management/control at the tree root. Such integration, however, requires simultaneous intra-SNOW and inter-SNOW communications while avoiding scalability-limiting inter-SNOW interference. We address this by formulating a constrained optimization problem whose objective is to maximize scalability by managing the spectrum sharing across the SNOWs, where each pair of neighboring BSs (i.e., SNOWs) communicate using a distinct special subcarrier. By proving the NP-hardness nature of the problem, we then propose two polynomial-time methods to solve it: a greedy heuristic algorithm and a $\frac{1}{2}$-approximation algorithm. We demonstrate the feasibility of this SNOW-tree by deploying 15 SNOWs, covering approximately (25x15)km$^2$. These experimentations demonstrate that both greedy heuristic and approximation algorithms are highly effective in practice, the latter providing a performance bound as well.

This thesis is organized as follows. Chapters~\ref{chap:snow1} and~\ref{chap:snow2} concentrate on designing the SNOW architecture and enabling reliable, asynchronous, and bidirectional communications in SNOW. Chapter~\ref{chap:snow3} focuses on the practical implementation of SNOW using COTS IoT devices. Chapter~\ref{chap:snow4} concentrates on the SNOW-tree architecture for handling the inter-SNOW and intra-SNOW communications and interferences. Chapter~\ref{chap:future} presents the future research directions. Finally, Chapter~\ref{chap:conclusion} concludes this thesis.


%% file: chapter2.tex
\section{SNOW: Sensor Network over White Spaces}   \label{chap:snow1}

Wireless sensor networks (WSNs)  face significant scalability challenges due to the proliferation of wide-area wireless monitoring and control systems that require thousands of  sensors to be connected over long distances. Due to their short communication range, existing WSN technologies such as those based on IEEE 802.15.4 form many-hop mesh networks complicating the protocol design and network deployment. To address this limitation, we  propose a scalable sensor network architecture - called  Sensor Network Over White Spaces (SNOW) -  by exploiting the TV  white spaces. Many WSN applications need low data rate, low power operation, and scalability in terms of geographic areas and the number of nodes. The long communication range of white space radios significantly increases the chances of packet collision at the base station. 
We achieve scalability and energy efficiency by  splitting channels into narrowband orthogonal subcarriers and enabling packet receptions on the subcarriers in parallel with a single radio. The physical layer of SNOW   is designed through a distributed implementation of OFDM that enables distinct orthogonal signals from distributed nodes. 
Its MAC protocol handles subcarrier allocation among the nodes and transmission scheduling. We implement SNOW in  GNU radio using USRP devices. Experiments demonstrate that it can correctly decode in less than 0.1ms  multiple packets received in parallel at different subcarriers, thus drastically enhancing the scalability of WSN.

\input{chapter2/introduction}

\input{chapter2/background}

\input{chapter2/architecture}

\input{chapter2/phy}

\input{chapter2/mac}

\input{chapter2/implementation}

\input{chapter2/experiment}

\input{chapter2/simulation}
\input{chapter2/comparison}
\input{chapter2/related_work}

\subsection{Summary}\label{chap2:conclusion}
We have designed and implemented SNOW, a scalable and energy-efficient WSN architecture over white spaces.  It achieves scalability and energy efficiency through a PHY design that splits channels into narrow band orthogonal subcarriers and enables simultaneous packet receptions with a single radio. SNOW is implemented in  GNU radio using USRP devices. Experiments demonstrate that it can decode correctly all simultaneously received packets, thus enabling the scalability for thousands of nodes. In the future, SNOW will be designed based on O-QPSK  modulation which is used in IEEE 802.15.4 at 2.4GHz.   


%% file: chapter2/introduction.tex
\subsection{Introduction}\label{chap2:introduction}

Despite the advancement in wireless sensor network (WSN) technology, we still face significant challenges in supporting large-scale and wide-area applications (e.g.,   urban sensing~\cite{citysense},  civil infrastructure monitoring~\cite{GGB, Bo}, oil field management~\cite{oilffield}, and precision agriculture~\cite{agriculture}). These applications often need thousands of  sensors to be connected over long distances.   Existing WSN technologies operating in ISM bands such as IEEE 802.15.4~\cite{ieee154}, Bluetooth~\cite{Bluetooth}, and IEEE 802.11~\cite{wifi}   have short range (e.g., 30-40m for IEEE 802.15.4 in 2.4GHz) that poses    a significant limitation in meeting this impending demand. To cover a large area with numerous devices, they form many-hop mesh networks at the expense of  energy cost and complexity.   To address this limitation,  we  propose a scalable sensor network architecture - called {\bf\em Sensor Network Over White Spaces (SNOW)} -  by designing sensor networks to operate over the TV {\em   white spaces}, which refer to the allocated but unused TV channels.


In a historic ruling in 2008, the Federal Communications Commission (FCC) in the US allowed unlicensed devices to operate on TV white spaces~\cite{FCC_first_order}. To learn about unoccupied TV channels at a location,  a device needs to either (i) sense the medium before transmitting, or (ii) consult with a cloud-hosted geo-location database, either periodically or every time it moves 100 meters~\cite{fcc_second_order}. Similar regulations are being adopted in many countries including Canada, Singapore, and UK. Since TV transmissions are in lower frequencies -- VHF and lower UHF (470 to 698MHz) --  white spaces have excellent propagation characteristics over long distance. They can easily penetrate obstacles, and hence hold enormous potential for WSN applications that need long transmission range. Compared to the ISM bands used by traditional WSNs, white spaces are less crowded and have wider availability in both rural and urban areas, with rural areas tending to have more~\cite{ws_sigcomm09, guangzhou, singapore, europe, uk, chicago}. Many wide-area WSNs such as those for monitoring habitat~\cite{GreatDuckIsland}, environment~\cite{Murphy}, volcano~\cite{Deployment_volcano} are in rural areas, making them perfect users of white spaces.   However, to date, the potential of white spaces is mostly being tapped into for wireless broadband access by industry leaders such as Microsoft~\cite{4Africa, MSRAfrica} and Google~\cite{GoogleAfrica}. Various standards bodies such as IEEE 802.11af~\cite{IEE802_af},  IEEE 802.22~\cite{IEEE802_22},  and  IEEE 802.19~\cite{IEE802_19} are modifying existing standards to exploit white spaces for broadband access.


The objective of our proposed SNOW architecture is to exploit white spaces for long range, large-scale WSNs. Long range will reduce many WSNs to a single-hop topology that has potential to avoid the complexity, overhead, and latency associated with multi-hop mesh networks.   Many WSN applications need low data rate, low cost nodes, scalability, and energy efficiency. Meeting these requirements  in  SNOW introduces significant challenges. Besides, long communication range increases the chances of packet collision at the base station as many nodes may simultaneously transmit to it. SNOW achieves scalability and energy efficiency through channel splitting and enabling simultaneous packet receptions at a base station with a single radio. The base station has a single transceiver that uses available wide spectrum from white spaces. The spectrum is split into narrow orthogonal  subcarriers whose bandwidth is optimized for scalability, energy efficiency, and reliability. Narrower bands have lower throughput but longer range, and consume less power~\cite{channelwidth}. Every sensor node transmits on an assigned subcarrier and the nodes can transmit asynchronously.  The base station is able to receive at any number of subcarriers simultaneously. The availability of wide white space spectrum will thus allow massive parallel receptions at the base station. Today, all communication paradigms in WSN are point to point, even though convergecast is the most common scenario. Simultaneous packet receptions at low cost and low energy in SNOW represents a key enabling technology for highly scalable WSN. Enabling such  simultaneous receptions at a node is  challenging as it requires a novel decoding technique which, to our knowledge, has not been studied before.

In SNOW, we implement concurrent transmissions  through a Distributed implementation of  Orthogonal Frequency Division Multiplexing (OFDM), called {\bf\em D-OFDM},  
to enable distinct orthogonal signals from distributed nodes. To extract spectral components from an aggregate OFDM signal, we exploit the Fast Fourier Transformation (FFT) that runs on the entire spectrum of the receiver's radio.  A traditional decoding technique would require a strict synchronization among the transmissions if it attempts to extract  the symbols from multiple subcarriers using FFT.   We address this challenge by designing SNOW as an asynchronous network, where no synchronization among the transmitters is needed. The decoder at the base station extracts information from all subcarriers irrespective of their packets' arrival time offsets. Thus, the nodes transmit on their subcarriers whenever they want.  The specific contributions of this paper are:  
\begin{itemize}
\vspace{-1mm}
\item The Physical layer (PHY)  of  SNOW  that includes white space spectrum splitting into narrowband orthogonal subcarriers and a demodulator design for  simultaneous packet receptions;  It can decode packets from any number of subcarriers in parallel without increasing the demodulation time complexity. 
The demodulator also allows to exploit fragmented spectrum.
\vspace{-1.1mm}
\item The Media Access Control (MAC) protocol for SNOW that handles subcarrier allocation among the nodes and their transmission scheduling. 
\vspace{-1.1mm}
\item  Implementation of SNOW  in  GNU radio using Universal Software Radio Peripheral (USRP)  devices; Our experiments show that it can decode in less than 0.1ms  all packets simultaneously received at different subcarriers, thus drastically enhancing WSN scalability. 
\end{itemize}


In the rest of this chapter, Section~\ref{chap2:background} outlines the background.  Section~\ref{chap2:architecture} describes the SNOW architecture. Section~\ref{chap2:phy} presents the  PHY of SNOW. Section~\ref{chap2:mac} presents the MAC protocol.  Sections~\ref{chap2:implement},~\ref{chap2:experiment}, and ~\ref{chap2:simulation} present the implementation,  experiments, and simulations, respectively. 
 Section~\ref{sec:comparison} compares SNOW against the upcoming Low-Power Wide-Area Network (LPWAN) technologies. 
Section~\ref{chap2:related} overviews related work.  Section~\ref{chap2:conclusion} concludes the paper.

%% file: chapter2/background.tex
\subsection{Background and Motivation}\label{chap2:background}
%
%

A WSN is a network of sensors that deliver their data to a base station. It has myriads of applications such as process management~\cite{RTSS2015, pieee}, data center management~\cite{capnet}, and monitoring of habitat~\cite{GreatDuckIsland}, environment~\cite{Murphy},   volcano~\cite{Deployment_volcano}, and civil infrastructure~\cite{GGB}. Many WSNs are characterized by a dense and large number of nodes, small packets, low data rate, low power, and low cost. The nodes are typically battery powered. Thus, scalability and energy are the key concerns in WSN design. Currently, IEEE 802.15.4 is a prominent standard for WSN that operates at 2.4GHz  with a   bit rate of 250kbps, a communication range of 30-40m at 0dBm, and a maximum packet size of 128 bytes (maximum 104 bytes payload). In this section, we explain the advantages and challenges of adopting white space in WSN.

\subsubsection{White Spaces Characteristics for WSN}\label{sec:advantages}

%

{\bf Long transmission range.}   
Due to lower frequency, white space radios have very long
communication range. Previous~\cite{ws_sigcomm09} as well as our study in this paper have shown their communication range to be of several kilometers. 
Time synchronization,  a critical requirement in many WSN applications, incurs considerable overhead in 
large-scale and multi-hop deployments which can be avoided in a single-hop structure. Single hop in turn results in 
shorter end-to-end communication latency by avoiding multi-hop routing.

{\bf Obstacle penetration.}  
Wireless communication in 5/2.4GHz band is more susceptible to obstacles. Hence, for example, WirelessHART networks in process monitoring adopt high redundancy where a packet is transmitted multiple times  through multiple paths, hindering their scalability~\cite{WirelessHART}. In contrast, lower frequencies of white space allow propagation with negligible signal decay through obstacles. 


%

Many WSN applications need to collect data from sensors spread over a large geographic area. 
For example, ZebraNet tracks zebras in 200,000$m^2$~\cite{ZebraNet}. It lacks continuous connectivity due to the short  communication range, and is managed through a delay-tolerant network which cannot deliver information in real time.  Also,  
with the growing applications,  industrial process management networks such as WirelessHART networks 
  need to scale up to tens of thousands of nodes~\cite{WH10000}. A WirelessHART  network  relies on global time synchronization  and central management  
 that limits network scalability~\cite{TECS}. Having long communication range, white spaces can greatly simplify such wide-area 
 applications.

\subsubsection{Challenge and Approach}


WSN characteristics and  requirements for scalability and energy efficiency pose unique challenges to adopt white spaces. To achieve  energy efficiency, many WSNs try to reduce the idle listening time, employing techniques like low power listening~\cite{LPL} or receiver initiated MAC~\cite{RIMAC}. However, both cases require one side of the link to send extremely long preambles. Blindly applying existing WSN MAC designs in long communication range will cause most nodes to wake up unintentionally. Besides, long communication range significantly increases the chances of packet collision.


SNOW achieves scalability and energy efficiency through  splitting channels into narrowband orthogonal subcarriers and enabling multiple sensors to transmit simultaneously to the base station with a single radio. Today, all communication paradigms in WSN (and at large) are point to point, even though convergecast is the most common  scenario. An $n$-to-1 convergecast is achieved through $n$ 1-to-1 links. Simultaneous packet receptions at low cost and low energy in SNOW represents a key and novel enabling technology for highly scalable WSN. Such  simultaneous receptions at a node is  challenging as it requires a novel decoding technique. Our design is based on a distributed implementation of OFDM and we exploit FFT to extract information from all subcarriers.  A traditional decoding technique would require that the $i$-th symbols from all subcarriers be in the same FFT window, requiring  strict time synchronization among the transmitting nodes which is difficult  for commercially available hardware.  We design SNOW as an asynchronous network, where no time synchronization is needed. The decoder can extract information from any number of subcarriers carrying packets irrespective of their packets' arrival time offsets.


%% file: chapter2/architecture.tex
\subsection{SNOW Architecture}\label{chap2:architecture}

Our proposed SNOW architecture is a WSN with a single base station (BS)  and a set of sensor nodes, each equipped with a single half-duplex white space radio. Due to long communication range, all sensor nodes are within a single hop of the BS, and vice versa. We observed in experiment that a node's communication range can be over 1.5km at low transmission power (e.g., 0 dBm). The BS is line-powered, Internet-connected, and  powerful. The sensor nodes  are power constrained and not directly connected to the Internet. 

\begin{figure}[!htb]
\includegraphics[width=0.9\textwidth]{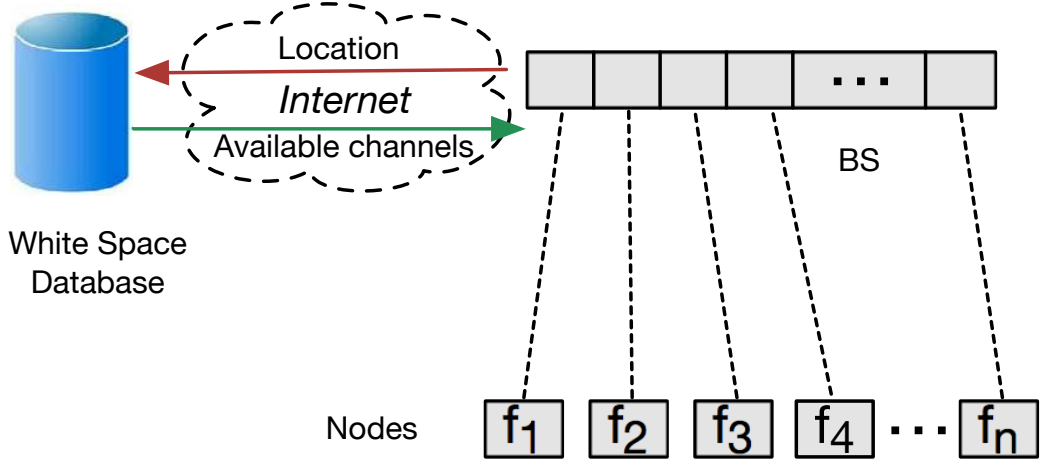}
\caption{System architecture}
\label{fig:chap2-arch}
\end{figure}

The BS uses a wide channel for reception which is split into subcarriers, each of equal spectrum width (bandwidth). Each node is assigned one subcarrier on which it transmits to the BS. Subcarrier allocation to nodes is handled in the MAC protocol.  We use the IEEE 802.15.4~\cite{ieee154} packet structure. For integrity check, the senders add cyclic redundancy check (CRC) at the end of each packet.  For better energy efficiency, the network does not employ any carrier sensing, RTS/CTS, frame acknowledgment (ACK), or  time synchronization protocol. We leave most complexities at the BS and keep the other nodes very simple and energy-efficient. For simplicity, sensors  do not do spectrum sensing or cloud access.  The BS determines white spaces by accessing a cloud-hosted database  through the Internet.   We assume that it knows the locations of the nodes  either through manual configuration or through some  existing WSN localization technique~\cite{wsnlocalizationsurvey}.  The BS thus selects white space channels that are available at its own location and at the locations of all other nodes.  Figure~\ref{fig:chap2-arch} shows the system architecture of SNOW.



%% file: chapter2/phy.tex
\subsection{SNOW PHY Design}\label{chap2:phy}
For scalability and energy efficiency, we design the PHY based on channel splitting and by enabling simultaneous packet receptions on different subcarriers at the BS with a single radio. This is done through D-OFDM which is a distributed implementation of OFDM to enable distinct orthogonal signals from distributed sources. We first explain how D-OFDM is realized in SNOW. Then we explain how each subcarrier is modulated for data encoding and how the BS demodulates from multiple subcarriers simultaneously.

\subsubsection{Adopting D-OFDM in SNOW}\label{subsec:chap2-spacing}
{\em OFDM} is a frequency-division multiplexing (FDM) scheme for digital multi-carrier modulation that uses a large number of closely spaced orthogonal subcarrier signals to carry data on multiple parallel data streams. The key aspect in OFDM is maintaining carrier orthogonality. If the integral of the product of two signals is zero over a time period,  they are {\bf\em orthogonal} to each other. Two sinusoids with frequencies that are integer multiples of a common one satisfy this criterion. Therefore, two subcarriers at center frequencies $f_i$  and $f_j$, $f_i\not=f_j$, are orthogonal when over time $T$~\cite{ofdmbook}:
\begin{equation*}
\int_0^T  \cos (2\pi f_i t) \cos (2\pi f_j t) dt=0.
\end{equation*}

The orthogonal subcarriers can be {\bf overlapping}, thus increasing the spectral efficiency. The guardbands that were necessary to allow individual demodulation of subcarriers in an FDM system would no longer be necessary. As long as orthogonality is maintained, it is still possible to recover the individual subcarriers' signals despite their overlapping spectrums.  
In OFDM modulation, the subcarrier frequency  $f_i$, $i=1,2, \cdots, $  is defined as $f_i= i \Delta f$, where $\Delta f$ is the {\em subcarrier spacing}, $T$ is one symbol period and  $\Delta f$ is set to $\frac{1}{T}$ for optimal effectiveness. When there are $n'$ subcarrier center frequencies, 
 $\Delta f= \frac{W}{n'} = \frac{1}{n'T}$  with $W$ being the entire bandwidth. The number of usable subcarriers may be less than $n'$ due to the unavailability of side band at the first/last subcarrier.  For example, using one TV channel (6MHz) between 547 - 553MHz, if we want each subcarrier of 400kHz bandwidth, we have $n'=30$, $\Delta f = 200$kHz. The relative subcarrier frequencies become $200$, $400$, $600$, $\cdots$, $1000$kHz. Thus, there will be 29 orthogonal subcarriers with center frequencies 547.2, 547.4, $\cdots$, 552.8MHz from this one TV channel.   

While the traditional OFDM is used between a single sender and a single receiver for increased data rate or to increase the symbol duration for enhanced reliability,  we adopt D-OFDM in SNOW by assigning the orthogonal subcarriers to different nodes. Each node transmits on the assigned subcarrier. Thus the nodes that are assigned different subcarriers can transmit simultaneously. These component sinusoids form an aggregate time domain signal as follows.  
\begin{equation}\label{eqn:chap2-fft}
X(t)=  \sum_{i=0}^{n'-1} x(k)\sin(\frac{2\pi kt}{n'})  -   j  \sum_{i=0}^{n'-1}   x(k)\cos(\frac{2\pi kt}{n'}).
\end{equation}
where $X(t)$ is the value of the signal at time $t$ which is composed of frequencies denoted by $(2\pi kt/n')$, $k$ is the index of frequency over $n'$ spectral components that divides the available bandwidth with equal spacing and $x(k)$ gives the value of the spectrum at $k$-th frequency. As seen in Equation (\ref{eqn:chap2-fft}), any part of the spectrum can be recovered by suitably selecting the spectral coefficients $x(k)$. This is the key principle we adopt in decoding parallel receptions at the BS. We design the demodulator for the receiver of this signal in a way so that no synchronization among these transmitters is needed.

\subsubsection{Modulation Technique}\label{subsec:chap2-modulation}
The method for extracting information from multiple subcarriers from an aggregate D-OFDM  signal depends on the modulation technique used for encoding the baseband  in the carrier signal. We design the PHY of SNOW  based on amplitude-shift-keying (ASK) modulation that was adopted in the IEEE 802.15.4 (2006) standard at 868/915MHz~\cite{ieee154}. 



  {\em ASK} is a form of Amplitude Modulation (AM) that represents digital data as variations in the amplitude of a carrier wave. In an ASK system, the binary symbol 1 is represented by transmitting a fixed-amplitude carrier wave and fixed frequency for a  duration of $T$ seconds, where $T$ is the symbol duration. If the signal value is 1 then the carrier signal will be transmitted; otherwise, a signal value of 0 will be transmitted. Every symbol thus carriers one bit. We use the simplest and most common form of ASK, called on-off keying (OOK), in which  the presence of a carrier wave indicates a binary one and its absence  indicates a binary zero. While AM is not as noise-immune as Frequency Modulation (FM) or Phase Modulation  (PM) because the amplitude of the signal can be affected by many factors (interference, noise, distortion) resulting in bit errors, this limitation can be mitigated through bit spreading techniques~\cite{cc2420}.

The simplicity of AM receiver design is a key advantage of AM over FM and PM~\cite{wirelessbok}. Compared to AM, PM needs more complex receiving hardware. Low bandwidth efficiency is another limitation of PM. The easiest method for  AM  receiver is to use a simple diode detector. AM transmitter also is simple and cheap as no specialized components are needed. Such a simple circuitry consumes less energy.  FM needs comparatively wider bandwidth to handle frequency leakage while AM needs narrower bandwidth as it can be implemented by just making the carrier signal present or absent. Narrower bandwidth in turn consumes much less energy  as transmission (Tx) energy is consumed by every Hz of bandwidth. At the same Tx power, the transmitter with narrower bandwidth has longer range. As AM needs narrower bandwidth, the available white space spectrum can be split into a larger number of subcarriers, enhancing SNOW scalability. Thus, there are trade-offs between AM and FM or PM as a modulation technique which is not the focus of this paper. 
 
For robustness in decoding, the modulation maps each bit to a $r$-bit sequence that simply repeats the bit $r$ times using bit spreading technique. We discuss the choice of parameter $r$ in the following section. At the transmitter, bits are mapped to symbols, and then a complex signal is generated. There are only two types of symbols, each consisting of one bit, the signal level above a threshold representing `1' and `0'  otherwise. Our work can easily be extended to Quadrature Amplitude Modulation (QAM) that encodes data on both $I$-signal and $Q$-signal, thereby doubling the bit rate.

\subsubsection{Demodulator Design}\label{subsec:chap2-demodulator}

The BS receives an analog D-OFDM signal in time domain and converts it to a digital signal and feeds the digital samples into the SNOW demodulator. We now detail the  technique for decoding data from multiple subcarriers.

The transmitters transmit on subcarriers whenever they want without coordinating among themselves. The idea for handling such an asynchronous scenario is to allow the BS to receive anytime. Since the BS  is line-powered and has no energy constraints, this is always possible.  The BS keeps running an FFT algorithm. The key idea in our demodulator design is to apply an FFT as a {\bf\em global FFT Algorithm} on the entire range of the spectrum of the BS, instead of running a separate FFT for each subcarrier. The demodulator starts processing by storing time domain sequential samples of the received aggregate signal into a vector $v$ of size equal to the number of FFT bins. The global FFT (called FFT for simplicity throughout the paper) is performed on vector $v$. This  repeats at every cycle of the baseband signal.

\begin{figure}[!htb]
 \centering
\includegraphics[width=0.75\textwidth]{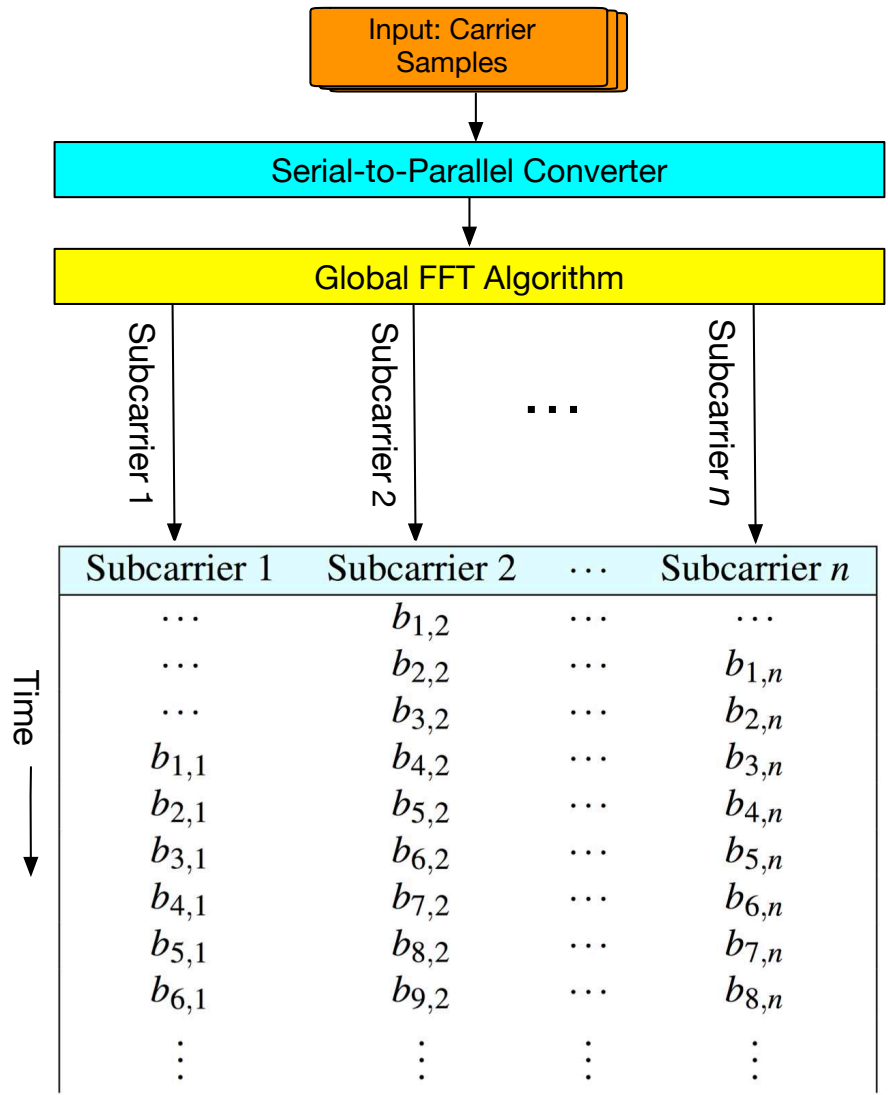}
\caption{Steps of packet decoding}
\label{fig:chap2-demodWorkFlow}
\end{figure}

A workflow showing the various steps for decoding packets from multiple subcarriers in our demodulator is given in Figure~\ref{fig:chap2-demodWorkFlow}.  A Fourier transform decomposes a time domain signal into a frequency domain representation. The frequency domain represents energy level at each frequency (frequency bins) of that time domain signal. To handle $n$ subcarriers, we apply  an $m$ point FFT algorithm, where $m\ge n$,  which is a variation of discrete Fourier transform at $m$ frequency  bins. Note that the number of subcarriers $n$ depends on the available spectrum, subcarrier spacing, desired bit rate and subcarrier bandwidth which are theoretically explained  in Sections~\ref{subsec:chap2-spacing} and~\ref{subsec:chap2-widthbitrate},  and are experimentally evaluated in Section~\ref{chap2:experiment}. Each subcarrier corresponds to $\frac{m}{n}$ bins with one middle bin representing its center frequency. The frequency bins are ordered from left to right with the left most $\frac{m}{n}$ bins representing the first subcarrier.  Each FFT output gives us a set of $m$ values. Each index in that set represents a single energy level at the corresponding frequency at a time instant. Since our FFT size is fixed no matter how many nodes transmit concurrently, it can decode packets from any number of subcarriers in parallel without increasing the demodulation time complexity. However, the more the number of bins per subcarrier, the cleaner  the signal on it.


{\bf Handling Spectrum Leakage.} FFT algorithm works on a finite set of time domain samples that represent one period of the signal. However, in practice, the captured signal may not be an integer multiple of periods. In that case, finiteness of measured signal results in a truncated waveform. Thus, the end-points become discontinuous and FFT outputs some spectral components that are not in the original signal,  letting the energy at one spectral component leak into others. To mitigate the effects of such {\em spectral leakage} on the neighboring subcarriers, we adopt the {\em Blackman-Harris windowing}~\cite{blackman}. {\em Windowing} multiplies  a discontinuous time domain records by a finite length window. This window has amplitudes that vary  smoothly and gradually towards zero at the edges, minimizing the effects of  leakage. Blackman-Harris windowing works for random or mixed signals and gives the best resolution in terms of minimizing  spectral leakage.


{\bf Packet Decoding.} To detect the start of a packet at any subcarrier, the  demodulator keeps track of FFT outputs. Since the FFT outputs energy level at each subcarrier, the demodulator applies a threshold to decide whether there is data in the signal.  
It uses the same threshold to detect preamble bits and the data bits. Once a preamble is detected on a subcarrier, the receiver immediately gets ready to receive subsequent bits of the packet. If the modulation technique spreads one bit into $r$ bits, the demodulator collects samples from $r$ FFT outputs for that subcarrier and then decides whether  the actual bit was zero or one. First the packet header is decoded and payload and CRC length is calculated. Then it knows how many data bits it has to receive to decode the packet. Since any node can transmit any time without any synchronization, the correct decoding of all packets is handled by maintaining a 2D matrix where each column represents a subcarrier or its center frequency bin that stores the bits decoded at that subcarrier. The last step in Figure~\ref{fig:chap2-demodWorkFlow} shows the 2D matrix where entry $b_{i,j}$ represents $i$-th bit of $j$-th subcarrier. The demodulator starts storing in a column only if a preamble is detected in the corresponding subcarrier. Hence, it stores data and CRC bits for every transmitter when needed. On each subcarrier, when the number of bits stored in the corresponding column of the 2D matrix equals the length of data and CRC bits, we check the CRC and test the validity of reception, and then continue the same process.

{\bf Handling Fragmented Spectrum.} An added advantage of our  design is that it allows to use fragmented spectrum. Namely, if we cannot find consecutive white space channels when we need more spectrum, we may use non-consecutive white spaces. The global FFT is run on the entire spectrum (as a single wide channel) that includes all fragments (including the occupied TV channels between the fragments). The occupied spectrum will not be assigned to any node and the corresponding bins will be ignored in decoding.

\subsubsection{Design Parameters}\label{subsec:chap2-widthbitrate}
We now discuss some design parameters that play key roles in SNOW operation. We perform signal processing at digitized baseband samples. Those samples are fixed-point precision once converted from the analog domain. For baseband processing, the true measured values in units of current or voltage are not important because those values depend on number representation in the design and the dynamic range of the ADC and prior analog components. Thus, the units of all our parameters are to be interpreted as absolute values.

    \begin{figure*}[!htb]
      \centering
      \subfigure[CDF of RSS magnitudes for `0' transmission\label{fig:chap2-thold_zeros}]{
        \includegraphics[width=.5\textwidth]{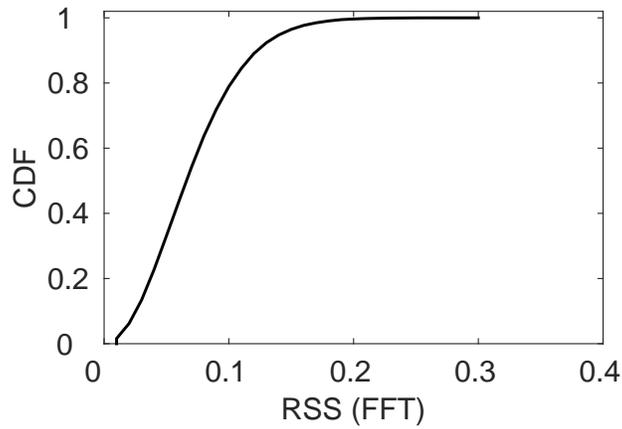}
      }
      \hfill
      \subfigure[CDF of RSS magnitudes for `1' transmission\label{fig:chap2-thold_ones}]{
        \includegraphics[width=.5\textwidth]{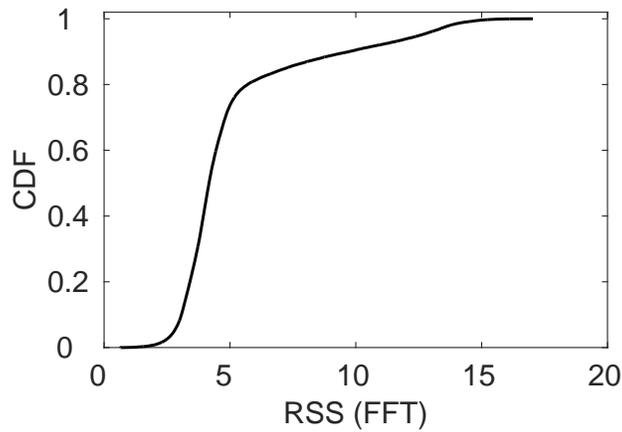}
      }
      \hfill
      \subfigure[Distribution of RSS magnitudes over distances\label{fig:chap2-thold_distance}]{
        \includegraphics[width=.5\textwidth]{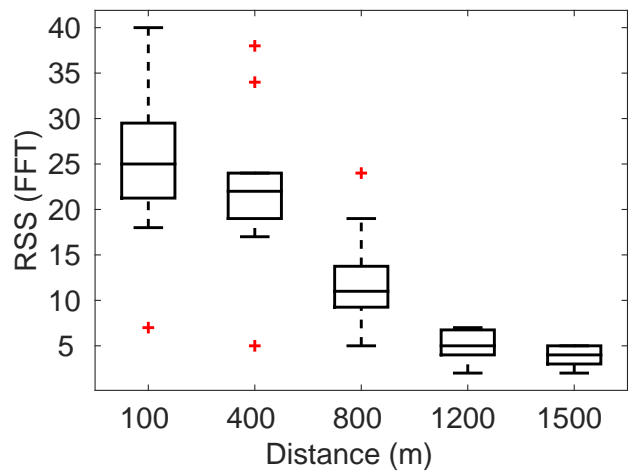}
      }
      \caption{Threshold behavior} 
      \label{fig:chap2-threshold}
    \end{figure*}

{\bf Threshold Selection.}
In our decoding, threshold selection on signal strength is a key design parameter to extract information from the received signal. Specifically, the received signal value above the threshold will be considered  bit `1', and `0' otherwise. 
We consider the average signal power to decide the threshold. The average Received Signal Strength (RSS) is estimated using the formula  $\sum_{i=1}^{M} \sqrt{I^{2}+Q^{2}}$, where the $I$ and $Q$ are the in-phase and quadrature components, respectively, of the signal, and $M$ is the averaging number of samples. 

For selecting the threshold, we observe the variability of the spectrum over a period of time and the effect on the RSS at the receiver. We analyzed the spectrum and collected the spectrum data  from radio front-ends for a period of 3 weeks. In the receiver, we gathered the RSS values for over 50000 samples for the whole duration of the experiment in indoor and outdoor environment that showed us that we can select a steady threshold for packet decoding.  Figure~\ref{fig:chap2-thold_zeros}  shows the cumulative distribution function (CDF) of the magnitudes of 50,000 samples for `0' transmission. As it shows, all 100\% samples have magnitudes below 0.4 FFT magnitudes. Figure~\ref{fig:chap2-thold_ones}  shows the CDF of the RSS values for 50000 samples at the same receiver for `1' transmission. In more than 80\% cases, 
the magnitude is above 4.5 while in more than 98.5\% cases, it is above 3, implying that we can set a  threshold of 3. Figure~\ref{fig:chap2-thold_distance} shows the distribution in boxplot for `1' transmission over various distances. At each distance, the boxplot shows the distribution of 5000 samples. All RSS magnitudes including the outliers in all cases are above 5 FFT magnitudes. The results show that a threshold between 0.4 and 5 can distinguish between 1 and 0.

{\bf Bit Spreading.}
{\em Bit spreading} is a well-known technique for reducing bit errors in noisy environments by robustly discerning the expected signal and the noise in many wireless technologies such as IEEE 802.15.4~\cite{ieee154} and IEEE 802.11b~\cite{wifi}. In IEEE 802.15.4 based hardware, the Direct Sequence Spread Spectrum (DSSS) technique maps the actual data bits to a different set of bits called {\em chip-sequence} whose number of bits is 8 times the number of actual data bits~\cite{cc2420}.  Similarly, in our design using ASK modulation, we  adopt bit spreading where every data bit is spread over 8 bits. Our experimental results (Section~\ref{chap2:experiment}) confirm that this bit spreading helps decode packets  correctly even in various noisy conditions. 

{\bf Packet size, Subcarrier Width, and Bit Rate.}
We use 28 bytes payload along with 12 bytes header totaling 40-byte as our default packet size in our experiment. TelosB mote~\cite{telosb}, a representative WSN mote based on  IEEE 802.15.4, uses a default payload of 28 bytes in TinyOS~\cite{tinyos}. All results shown in the paper are based on 40-byte packets. The subcarrier bandwidth is another important parameter to decide. The maximum transmission bit rate $C$ of an AWGN channel of bandwidth $B$ based on Shannon-Hartley Theorem is given by  $C=B\log_2 (1+S/N)$, where $S$ is the signal power  and $N$ is the noise power. The ratio $S/N$ is called {\em Signal to Noise Ratio (SNR)}. The 802.15.4 specification for lower frequency band, e.g., 430-434MHz band (IEEE 802.15.4c~\cite{802154c}), has a bit rate of 50kbps. We also aim to achieve a bit rate of 50kbps. We consider a minimum value of $3$dB for SNR in decoding. Taking into account the bit spreading, we need to have $50*8$kbps bit rate in the medium. Thus, a subcarrier of bandwidth 200kHz can have a bit rate up to  $50*8$kbps  in the medium. Based on Nyquist Theorem, $C=2B\log_2 2^k$ where $2^k$ is the number of signal levels needed to support  bit rate $C$ for a noiseless channel, a modulation technique that uses $2$ signal levels can support  $50*8$kbps bit rate for a noiseless channel of bandwidth 200kHz. Since ASK modulation uses 2 signal levels, it is theoretically sufficient for this bit rate and bandwidth under no noise. However, to support this bit rate under noise in practical scenarios we determine a required bandwidth of 400kHz through exhaustive experiments in Section~\ref{subsec:chap2-widthbitrateexp}.

%% file: chapter2/mac.tex
\subsection{MAC Protocol for SNOW}\label{chap2:mac}


The MAC protocol operates in two phases - one phase for upward communication (i.e., the nodes transmit to the BS) of duration $t_u$  and the other  for downward communication (i.e., the BS transmits to the nodes) of duration $t_d$, where $t_u\gg t_d$.  


The BS first adopts a greedy approach to select the widest free spectrum in available white spaces. If it needs even wider spectrum it can also use the neighboring white spaces in addition to this widest one, thus using fragmented spectrum. For simplicity of presentation, we consider a single (widest) fragment of spectrum. This spectrum is split into  $n$ overlapping orthogonal subcarriers, each of equal width.  Each node is then assigned one subcarrier. We first explain the case where the number of nodes $N'\le n)$, thus allowing each node to be assigned a unique subcarrier. We denote the subcarrier assigned to node $i$, $1\le i\le N'$, by $f_i$. The BS also chooses a control subcarrier denoted by $f_c$. This channel is used for control operations during the downward communications. Initially and in the downward phase all nodes switch to $f_c$.  The network starts with a downward control command where the BS assigns the subcarriers to the nodes. 



The upward communication phase starts right after the BS notifies all the nodes their assigned subcarriers. The BS informs the nodes that the next $t_u$ seconds will be for upward communication. In this way, the nodes do not need to have synchronized absolute times. The BS switches to the entire spectrum and remains in receive mode. In this phase, all nodes asynchronously transmit their data to the BS on the respective subcarriers.  After   $t_u$  seconds, each node switches to control subcarrier $f_c$  and remains in receive mode for the downward  phase, and remains so until it receives a control command from the BS. The BS now switches to $f_c$ and broadcasts control command. This same process repeats.


When the number of nodes $N'>n$,  the nodes are grouped, each group having $n$ nodes except the last group that gets $(N' \mod n)$ nodes when $(N' \mod n)\not=0$. 
Every node in a group is assigned a unique subcarrier so that all nodes in the group can transmit together. The BS, in a downward phase,  asks a group to transmit their messages in the next upward phase. The next group can be selected in round robin. Thus, the nodes can safety sleep and duty cycle. In upward phase, a node can transmit its own packets and then immediately go to sleep till the end of the upward phase if it has no more data. In downward phase, the node must stay awake to receive any packets from the BS. We can reduce  energy consumption further by having the BS notify the nodes in the first downward packet whether it will send more packets in the same phase.




Spectrum and network dynamics are handled through the downward phase.  If the spectrum availability changes, then the new channel assignment is informed in the downward phase. The network uses redundant control channels so that if one control channel becomes noisy or unavailable, it can switch to another. If a new node joins the network, it can use  the control channel to communicate with the BS. When it detects signals in the control channel, it waits until the channel becomes idle and transmits its ID and location (assumed to be known) to the BS. The BS then checks the available white space and assigns it an available subcarrier. Similarly, any node from which the BS has not received any packet for a certain time window can be excluded from the network.

Since we do not use per packet ACK, a node can proactively repeat a transmission $\gamma$  times for enhanced reliability. The BS can send to the nodes  an aggregate ACK
to the nodes, e.g., by sending total  received packets from a node in the last cycle based on which a node can decide a value of $\gamma$.

%% file: chapter2/implementation.tex

\subsection{SNOW Implementation}\label{chap2:implement}
We have implemented SNOW on USRP devices using  GNU Radio.  GNU Radio is a  toolkit for implementing software-defined radios and signal processing~\cite{gnuradio}. USRP is a software-defined radio platform with RF front-ends to transmit and receive in a specified frequency~\cite{usrp}. We have 6 sets of USRP B210 devices for experiment, 5 of which are used as SNOW nodes and one as the BS. On the transmitter (Tx) side,  packets are generated in IEEE 802.15.4 structure. We represent the preamble and the packet (data, CRC)  using a default GNU radio vector.  The vector is then sent to the GNU radio repeat block, which performs bit spreading by repeating each bit 8 times. This baseband signal is then modulated with the carrier frequency.   For the BS to receive on multiple subcarriers, we implement the decoder using a 64-point FFT. The decoder incorporates    serial-to-parallel converter, FFT, parallel-to-serial converter, and signal processing. We do not need FFT size larger than 64-point because of the limited number of devices we have (as every subcarrier already corresponds to multiple FFT bins). Large-scale implementation is done through simulations in QualNet~\cite{qualnet}. 


%% file: chapter2/experiment.tex
\begin{table}[!htb]
\centering
\begin{tabular}{|l|l|}
\hline
\textbf{Parameter}     & \textbf{Value}                                                                         \\ \hline
Frequency  Band        & 547 -- 553MHz                                                                    \\ \hline
Orthogonal Frequencies & \begin{tabular}[c]{@{}l@{}}549.6, 549.8, 550.0,\\ 550.2, 550.4, 550.6MHz\end{tabular} \\ \hline
Tx Power               & 0dBm                                                                                  \\ \hline
Receive Sensitivity    & -94dBm                                                                     			      \\ \hline
Tx Bandwidth           & 400kHz                                                                                \\ \hline
Rx Bandwidth           & 6MHz                                                                                  \\ \hline
Packet Size            & 40 bytes                                                                               \\ \hline
SNR                    & 6dB                                                                                   \\ \hline
Distance               & \begin{tabular}[c]{@{}l@{}}Indoor: 100m\\ Outdoor: 1.5km\end{tabular}                \\ \hline
\end{tabular}
\vspace{.1 in}
\caption{Default parameter settings}
\label{tab:chap2-default_param}
\end{table}

\subsection{Experiments}\label{chap2:experiment}
In this section,  we perform experiments on different aspects of the SNOW architecture.
\subsubsection{Setup}
\begin{figure}[t!]
    \centering
      \subfigure[Outdoor nodes location in the City of Rolla\label{fig:chap2-outdoor}]{
      \includegraphics[width=.75\textwidth]{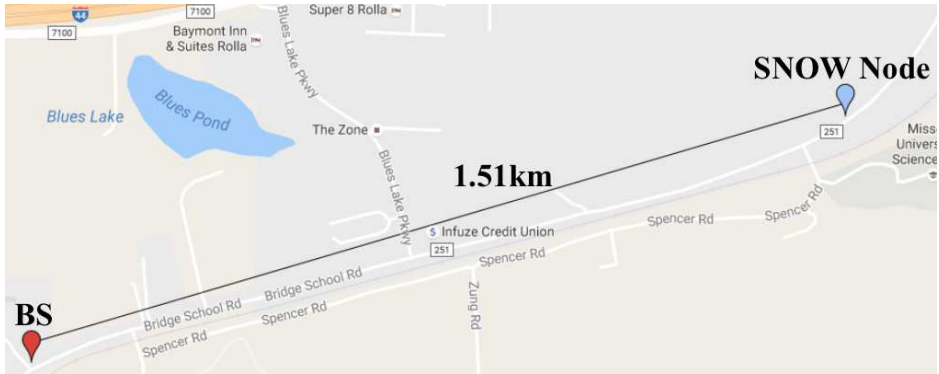}      }
      \hfill
    \subfigure[Node positions shown on the CS building floor plan\label{fig:chap2-floorplan}]{
      \includegraphics[width=.75\textwidth]{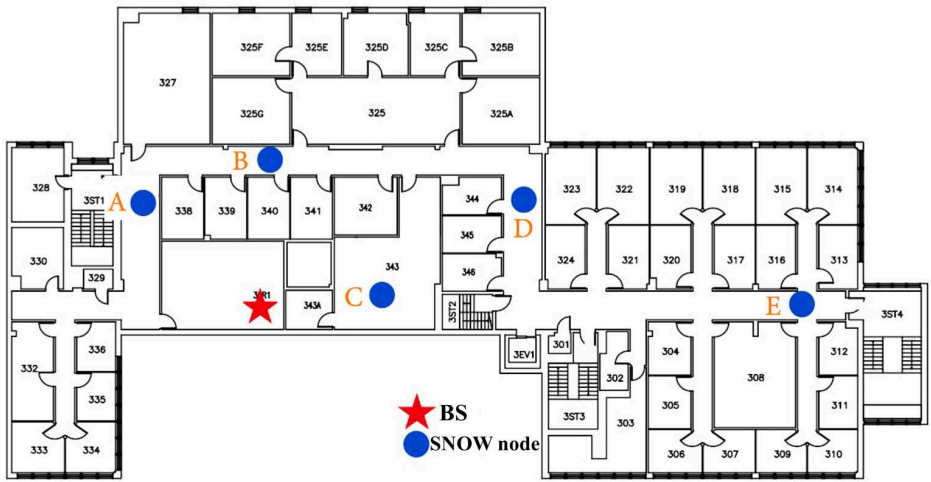}
      }
    \caption{Node positions in experiments}
    \label{fig:chap2-positions}
\end{figure}

We perform experiments using the SNOW implementation on USRP devices in both indoor and outdoor environments. 
Figure~\ref{fig:chap2-outdoor} shows outdoor node positions for the longest distance we have tested in the City of Rolla.  Figure~\ref{fig:chap2-floorplan} shows the positions of the nodes and the BS in the Computer Science building at Missouri University of Science \& Technology. It shows 5 different positions (only the positions and not the actual number of nodes) where the nodes were placed in various experiments. We fixed the antenna height approximately 5 ft above the ground. We experimented in the band between 547MHz and 553MHz that was a fraction of white spaces in the experimental locale. 
We define {\bf\em Correctly Decoding Rate (CDR)} as the percentage of packets that are correctly decoded at a receiver (Rx) among the transmitted ones. CDR is used to measure the decoding performance of SNOW. We first present the results on determining the subcarriers. Then we present the results running the MAC protocol. Unless stated otherwise, Table~\ref{tab:chap2-default_param} shows the default parameter settings for all the experiments.

\subsubsection{Subcarrier Determination}\label{subsec:chap2-widthbitrateexp}
We perform experiments to determine how to split a wide spectrum into narrowband subcarriers. Narrower bands have lower throughput but they have longer range, are more resilient to multipath effects, and consume less power~\cite{channelwidth}. Therefore, we  first determine through experiments a feasible bandwidth that is narrow but is sufficient to provide the desired bit rate  and to carry WSN packets. In practice,   the devices such as TelosB~\cite{telosb} based on IEEE 802.15.4 standard has a default payload size of 28 bytes in TinyOS~\cite{tinyos} which is sufficient to carry WSN data. Therefore, first we set a packet size of 40 bytes of which 28 bytes is payload and 12 bytes is header. We also aim to achieve at least  50kbps bit rate as discussed before. These experiments are performed between two nodes: one node as Tx and the BS as Rx.

{\bf Feasibility of Different Bandwidths over Distances and Obstacles.} 
\begin{figure}[t!]
      \centering
 	\includegraphics[width=.5\textwidth]{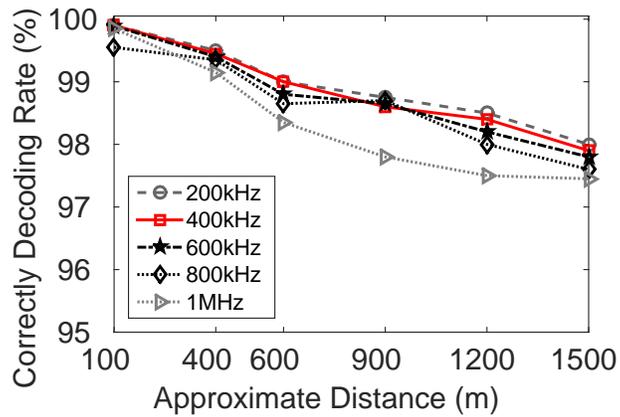}
        \caption{Reliability over long distances (outdoor)}
      \label{fig:chap2-distance_vs_reliability}
\end{figure}
We tested in outdoor environments with subcarriers of  bandwidths 200kHz, 400kHz, 600kHz, 800kHz, and 1MHz in the band 550 - 551MHz using 0dBm Tx power (which is our default Tx power). Considering 10,000 consecutive packet transmissions, Figure~\ref{fig:chap2-distance_vs_reliability} shows we have CDR over 97\% for each bandwidth when the receiver is up to 1.5km from the transmitter. As expected, at the same Tx power,  the narrower bandwidth has better performance over long distances.  While we achieve reliability using 200kHz bandwidth (that was the required theoretical bandwidth as we analyzed in Section~\ref{subsec:chap2-demodulator}), the bit rate becomes much less than 50kbps.  In contrast, when we use  400kHz, we can achieve an effective bit rate of at least 50kbps (8*50kbps in the medium considering spread bits) making 400kHz as our desired subcarrier bandwidth. These  results also verify that 40 bytes is a feasible packet size for this bandwidth.
\begin{figure}[t!]
    \centering
      \subfigure[Reliability at various SNR\label{fig:chap2-snr_vs_reliability}]{
        \includegraphics[width=.5\textwidth]{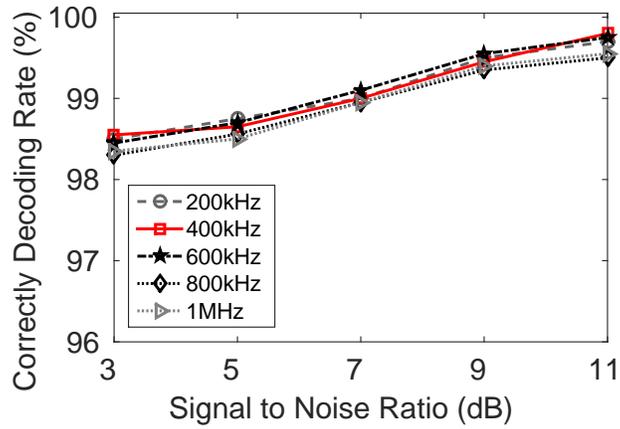}
      }
    \hfill
      \subfigure[Propagation through walls \label{fig:chap2-wall_vs_reliability}]{
        \includegraphics[width=.5\textwidth]{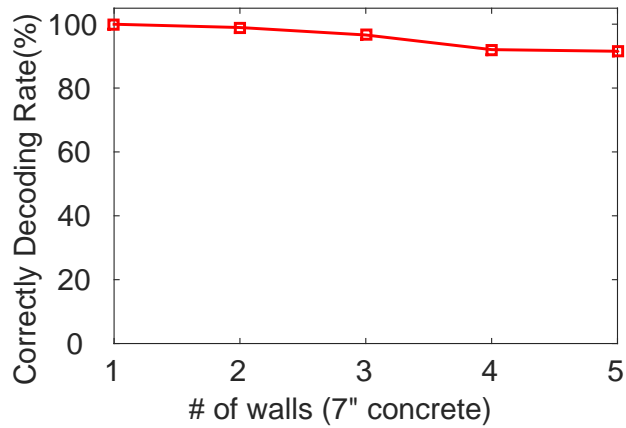}
      }
    \caption{Link level experiment over obstacles (indoor)}
    \label{fig:chap2-distance}
\end{figure}

We also perform experiments in indoor environments. Figure~\ref{fig:chap2-floorplan} shows different positions of the transmitter while the receiver is placed in a fixed position. Considering 10,000 consecutive packet transmissions, Figure~\ref{fig:chap2-snr_vs_reliability} shows the CDR over various SNR conditions for different subcarrier bandwidth. An SNR of 3dB gives a CDR around 98.5\% for all subcarrier bandwidths. As we increase the distances between the BS and the nodes, the SNR changes due to noise, multipath effect, and obstacles. The higher the SNR, the better the CDR. We observe at least 98\% CDR on all bandwidths and achieve the desired bit rate  when the bandwidth is 400kHz. Based on an experiment using 400kHz bandwidth across obstacles in the same building,  Figure~\ref{fig:chap2-wall_vs_reliability}  shows that there is at least 90\% CDR when the line of sight is obstructed by up to 5 walls (each $7''$  concrete). This shows feasibility of this bandwidth in terms of propagation through obstacles.

{\bf Feasibility under  Different Transmission Power.} 
\begin{figure}[t!]
    \centering
    \includegraphics[width=0.5\textwidth]{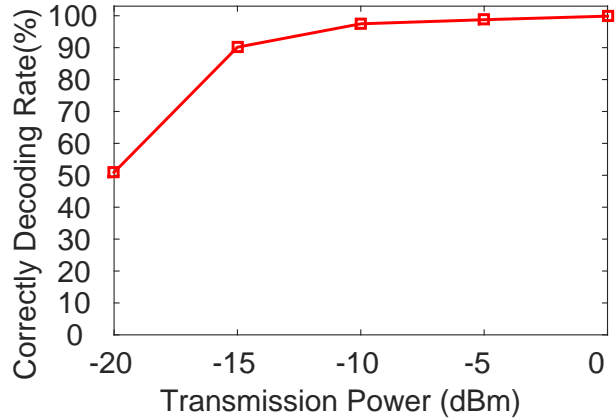}
    \caption{Reliability vs Tx power}
    \label{fig:chap2-txpower}
\end{figure}
We now test the feasibility of 400kHz subcarrier bandwidth under different Tx power. Since USRP devices do not provide any direct mechanism to control Tx power, we perform this experiment by varying the Tx gains at the transmitter to emulate the effect of varying Tx power. Setting a Tx gain of 65dB outputs a Tx power of 0dBm~\cite{usrp}. For 10,000 consecutive packet transmissions in outdoor (Tx and Rx are 1.5km apart), Figure~\ref{fig:chap2-txpower} shows the CDR at the receiver under different Tx powers. For Tx power between -15dBm and -10dBm the CDR is at least 97.4\%, while for that at 0dBm the CDR is at least 98.1\%. The results thus show that when Tx power is not extremely low,  400kHz is a feasible bandwidth.

\subsubsection{Experimenting the SNOW Architecture}
We now perform experiments using the complete SNOW architecture under the scenario when multiple nodes transmit to the BS. 
All of these experiments were done in indoor environments. The node locations are shown in Figure~\ref{fig:chap2-floorplan}.

{\bf Overlaps between Orthogonal Subcarriers.}
\begin{figure}[t!]
    \centering
    \includegraphics[width=0.5\textwidth]{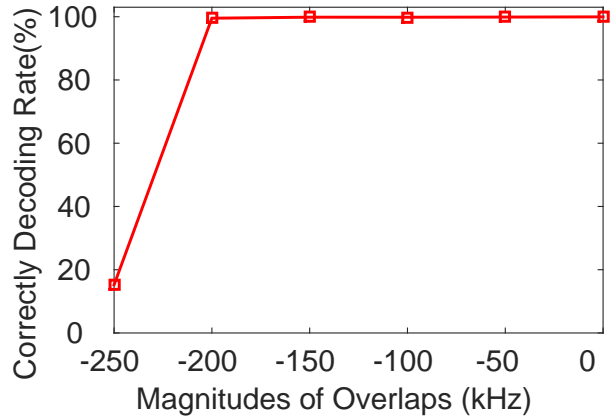}   
   \caption{Reliability vs magnitudes of subcarrier overlap}
    \label{fig:chap2-guardband}   
\end{figure}
In splitting a wideband radio among multiple orthogonal subcarriers, now we need to analyze the magnitudes of overlaps between the subcarriers. Note that OFDM technology does not require guardband between subcarriers; instead it allows them to be overlapping. We used two subcarriers each of 400kHz bandwidth. Starting with 0 guardband (start of the second subcarrier - end of the first subcarrier), we keep decreasing  the value up to the point when the two subcarriers overlap by 50\% (representing a guardband of -200kHz). 

To evaluate the feasibility of simultaneous reception on overlapping subcarriers, we start transmitting at the two transmitters at the same time. Considering 5,000 consecutive packet transmissions from both of the transmitters,  Figure~\ref{fig:chap2-guardband} shows a CDR of at least 99.5\% at the BS when there is an overlap of 50\% or less between these two neighboring subcarriers. While orthogonality allows these overlaps, such a high reliability is achieved not only for orthogonality but also for bit spreading. We observed that there are frequency leakages interfering nearby subcarrier bins, but those were not enough to cause decoding error due to bit spreading. In addition, using multiple bins per subcarrier also helped us reduce the impact of leakage. If we try to move two subcarriers even closer, they affect each other and CDR sharply drops to 5-10\%. The experiment shows that the orthogonal subcarriers, each of 400kHz bandwidth, can safely overlap up to 50\% with the neighboring ones, thereby yielding high spectrum efficiency (a key purpose of OFDM).

{\bf Network Performance.} 
We evaluate some key features of SNOW. First,  its {\em achievable throughput} (total bits received per second at the BS) can be at least $n$ times that of any traditional wireless networks,  both having the same link capacity (bit rate on the link) where $n$ is the number of subcarriers. This is  because SNOW can receive from $n$ nodes simultaneously. Second, as SNOW runs a single FFT with the same number of bins irrespective of the number of simultaneous transmitters,  the time required to demodulate $n$ simultaneous packets is  equal to the time needed for decoding a single packet. Now we test these features in experiments. We also evaluate SNOW in terms of energy consumption and network latency.

  \begin{figure}[t!]
    \centering
    \includegraphics[width=0.5\textwidth]{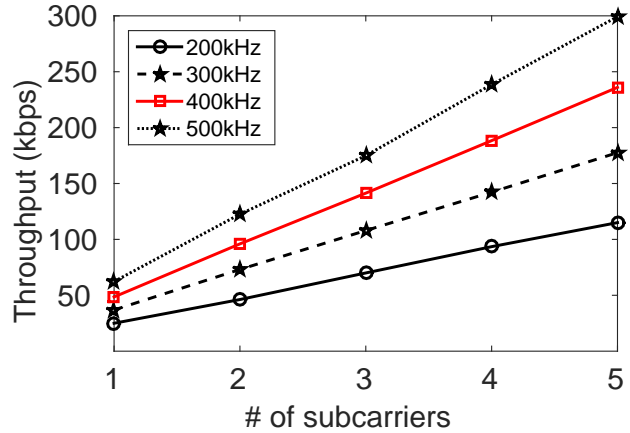}
    \caption{Throughput vs \# of subcarriers in SNOW}
    \label{fig:chap2-throughput}
    \end{figure}


{\bf Throughput.} 
First we observe the throughput under various number of subcarriers up to 5. The positions of the BS and 5 nodes (indexed as A, B, C, D, E) are shown in Figure~\ref{fig:chap2-floorplan}. Each node transmits 40-byte packets consecutively at their maximum bit rate. Thus the throughput measured at the BS indicates the maximum achievable throughput under this setting. The subcarriers are chosen with 50\% overlapping with the neighbor/s. In addition to our chosen 400kHz bandwidth, we also experiment with various bandwidths (200kHz, 300kHz, 500kHz) to see the throughput change. Figure~\ref{fig:chap2-throughput} shows the throughput averaged over a time interval of  1 hour. When each subcarrier has a bandwidth of 400kHz, the throughput using  one transmitter is at least 50kbps. This throughput at the BS  increases linearly as we increase the number of transmitters. This increase happens due to parallel receptions on multiple subcarriers at the BS. Note that under similar settings, a traditional WSN will not observe such increased throughput as its radio can receive only if one transmitter transmits at a time. At wider bandwidth, the throughput in SNOW becomes even higher. Thus when we have small number of nodes (compared to the number of subcarriers) and need high throughput, we can choose wider subcarriers.

\begin{figure}[t!]
    \centering
    \includegraphics[width=0.5\textwidth]{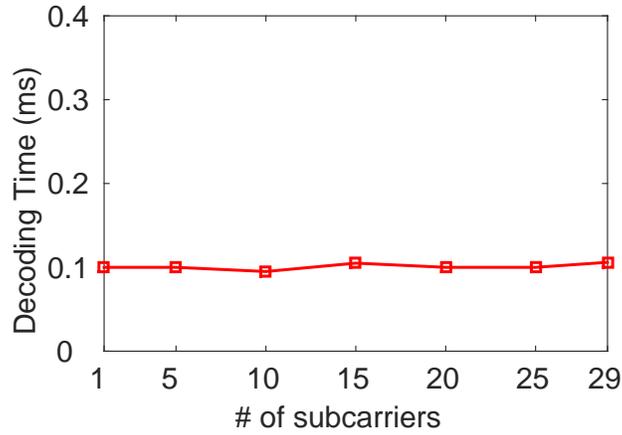}   
	 \caption{Decoding time vs \# of subcarriers}
    \label{fig:chap2-latency}   
    \end{figure}

{\bf Decoding Time.} 
Since the BS in SNOW can receive $n$ packets concurrently, we measure how much time its demodulator takes to handle multiple transmitters.  Within a 6MHz channel, we can accommodate 29 orthogonal subcarriers each of width 400kHz and each overlapping 50\% with the neighbor/s. Even though we have only 5 USRP transmitters, we can calculate the decoding time for all 29 subcarriers. To do this, we simply assume other 24 transmitters are sending packets containing all zero bits. Theoretically, decoding time for any number of subcarriers should be constant as  the FFT algorithm runs with the same number of bins every time. However, assuming 1 to 29 transmitters, we run separate experiments for each number of transmitters (subcarriers) for 7 minutes, and record the worst case time needed for decoding packets. For all cases, Figure~\ref{fig:chap2-latency} shows that the decoding time remains no greater than 0.1ms. This demonstrates the high scalability of SNOW decoding scheme.

{\bf Energy Consumption.} 
We measure  energy consumption in SNOW  and compare with that in A-MAC~\cite{dutta2010design} protocol which, to our knowledge, is the state-of-art energy efficient MAC protocol for IEEE 802.15.4 (2.4GHz) based WSN. A-MAC uses receiver initiated probe to inform the sender to send the packets. Upon receiving the probe the sender sends a hardware generated ACK, followed by the data packet. After receiving the data packet successfully, receiver sends another probe with the ACK bit set. If there are multiple senders, the data packets collide. In that case, the receiver sends a probe containing back-off interval period and backcast channel information.
 
 \begin{table}[!htb]
\centering 
\begin{tabular}{|l|l|}
\hline
\textbf{Device mode} & \textbf{\begin{tabular}[c]{@{}l@{}}Current Consumption\\  (Supply voltage 3 v)\end{tabular}} \\ \hline
Tx                   & 17.5mA                                                                                      \\ \hline
Rx                   & 18.8mA                                                                                      \\ \hline
Idle                 & 0.5mA                                                                                        \\ \hline
Sleep                & 0.2$\mu$A                                                                                     \\ \hline
\end{tabular}
\vspace{.1in}
\caption{Energy profile of CC1070}
\label{tab:chap2-cc1070}
\end{table}
 
\begin{figure}[t!]
    \centering
      \subfigure[Energy Consumption vs \# of nodes\label{fig:chap2-energy_experiment}]{
        \includegraphics[width=.5\textwidth]{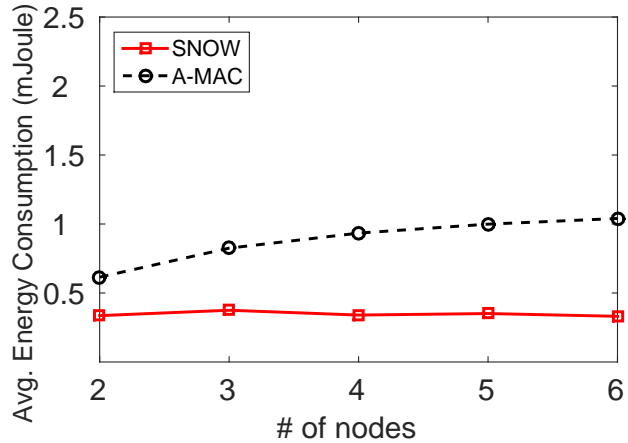}
      }
    \hfill
      \subfigure[Latency vs \# of nodes \label{fig:chap2-latency_experiment}]{
        \includegraphics[width=.5\textwidth]{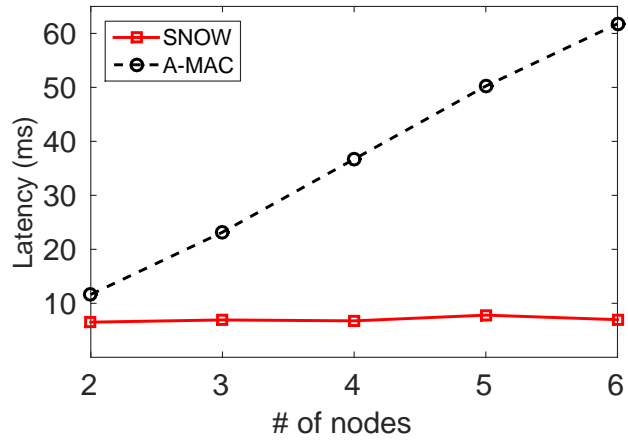}
      }
    \caption{Energy consumption and latency}
  \label{fig:chap2-amac-comparison}
  \end{figure}

To estimate the energy consumption in SNOW nodes, we place 5 SNOW transmitters each 200m apart from the BS. To make a fair comparison with A-MAC, we place A-MAC nodes 40m apart from each other making a linear multi-hop network. In both of the networks, each node (except the BS) sends one 40-byte packet every 60 seconds. Since USRP platform does not provide any energy consumption information, we use CC1070 RF transmitter energy model by Texas Instruments~\cite{cc1070} to determine approximate energy consumptions in SNOW. This off-the-shelf radio chip has the PHY configuration close to SNOW as it operates in low frequency (402-470 and 804-940MHz) and adopts ASK as one of its modulation techniques. CC1070 energy model is shown in Table~\ref{tab:chap2-cc1070}. In this setup, the BS is always connected to a power source and is not considered in energy calculation. We run many rounds of convergecast for one hour. Figure~\ref{fig:chap2-energy_experiment} shows the average energy consumption in each node per convergecast. Regardless of the number of nodes, a SNOW node consumes 0.34mJoule energy. In contrast,  a node under  A-MAC consumes on average 0.62mJoule energy when we consider 2 nodes. Average energy consumption on each node in A-MAC  increases with the number of nodes. This happens as we increase the number of hops (in the linear topology). Figure~\ref{fig:chap2-energy_experiment} shows that average energy consumption is 1.04mJoule for 6 nodes in A-MAC while it is almost constant in SNOW. Due to single-hop topology (thanks to long range) and parallel reception at the BS, each node in SNOW consumes less energy on average. This demonstrates the energy efficiency of SNOW over traditional WSN.

{\bf Network Latency.} 
Figure~\ref{fig:chap2-latency_experiment} shows the comparison of convergecast latency between SNOW and A-MAC with the previous experimental settings. Considering each node has a packet, we measure the latency required to collect all of those packets at the BS. SNOW takes approximately 7ms while A-MAC takes nearly 62ms to collect all the packets in convergecast. It is also noticeable that SNOW needs almost constant time to collect all the packets regardless of the number of nodes as the number of nodes does not exceed the number of subcarriers. Owing to a small network in this experiment (6 nodes), the difference between the latency in A-MAC and that in SNOW cannot be very high. However, for larger networks we will show in simulation that this difference can be very high, demonstrating the scalability of SNOW.

\begin{figure}[!htb]
    \centering
    \includegraphics[width=0.5\textwidth]{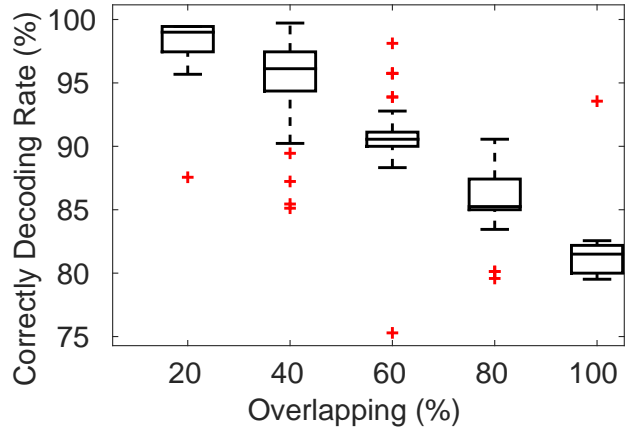}
    \caption{Performance of SNOW under interference}
    \label{fig:chap2-interference}
    \end{figure}

    \begin{figure}[!htb]
    \centering 
    \includegraphics[width=0.5\textwidth]{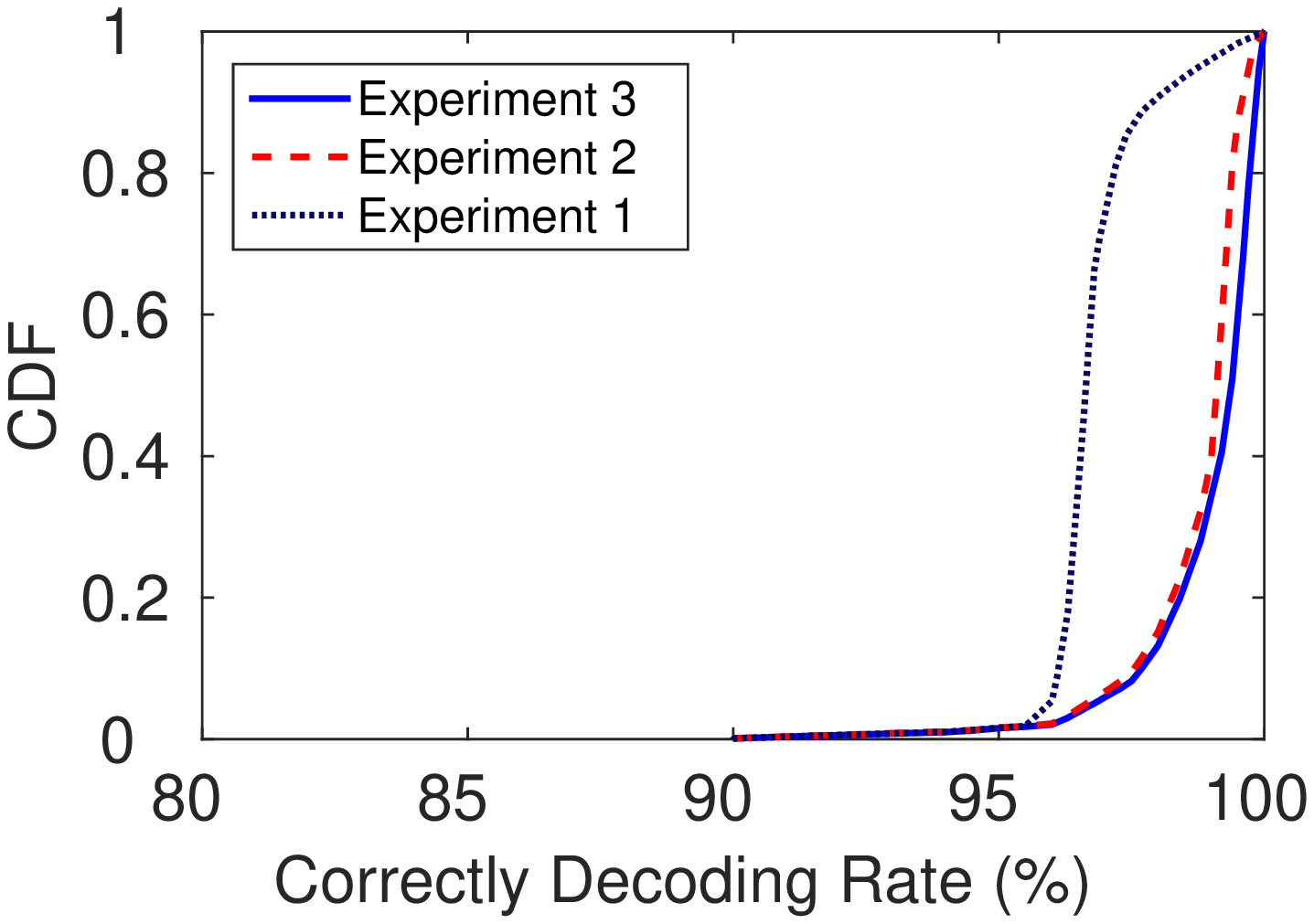}
    \caption{Using fragmented spectrum in SNOW}
    \label{fig:chap2-fragment}
    \end{figure}

{\bf Performance in the Presence of Interference.}
We create interference to see its impact on SNOW performance. We run the upward phase of the MAC protocol where 4 transmitters send packets to the BS concurrently and incessantly. We take another transmitter to act as an interferer. We use the same Tx gain at each transmitter, and place the interferer close (at place A while the legitimate transmitters at places B, C, D, and E in Figure~\ref{fig:chap2-floorplan}) to the BS to have its signal strong at the BS. The interferer operates on different parts of the spectrum of one (of the four) legitimate transmitter, and in every case it uses a timer that fires after every 200ms. At some random time in each of this 200ms window, it transmits a 40-byte packet to the BS. For every magnitude  of subcarrier overlapping, we run the experiments for 2 minutes,  and measure the CDR at the BS. We do 50 runs of this experiment and plot the distribution of CDR values in Figure~\ref{fig:chap2-interference}.  As it shows, with 80\% overlap between the subcarriers of a legitimate Tx and the interferer we can decode at least 79\% of packets from legitimate Tx in all runs.  For 100\% overlap, we can decode at least 77\% of packets in all runs. This result shows how external interferences can affect SNOW performance. As the figure shows, this impact is less severe or negligible when the interferer's spectrum partially overlaps with the transmitter's subcarrier.

{\bf Performance under Fragmented Spectrum.}
An added feature of SNOW is its capability in exploiting fragmented white space spectrum. As primary users may use channels that are far away from each other, white spaces can be largely fragmented. To test the performance of SNOW in fragmented spectrum, we choose different local TV channels such that there are white spaces available on both sides. In this experiment,  the BS uses a bandwidth of 8MHz  where 6MHz in the middle is occupied by some TV channel. We use two transmitters that act as SNOW nodes and consider three different channels to do three experiments under different fragmented spectrum.  Both of the transmitters send 100 consecutive packets and then randomly sleep between 500 to 1000ms. We run this experiment for 2 hours around each channel. In all cases, we run FFT over the entire 8MHz channel and collect data from SNOW nodes only. Under different fragmented spectrum, the SNIR (Signal-to-Noise and Interference Ratio) is different as the TV channels change. Figure~\ref{fig:chap2-fragment} shows three sets of experiments on fragmented spectrum, each having different ranges of SNIR condition. In experiment 1, the SNIR varies from 3 to 5dB and SNOW achieves  at least 95\% CDR in at least 96\% cases. In experiment 2, the SNIR varies from 6 to 8dB that results in at least 99\% CDR in 90\% cases. Experiment 3 with varying SNIR from 9 to 11dB or more shows even better CDR. The results  show that SNOW can exploit fragmented spectrum.

%% file: chapter2/simulation.tex
\subsection{Simulations}\label{chap2:simulation}

We evaluate the performance of SNOW for large-scale networks through simulations in QualNet~\cite{qualnet}. We evaluate in terms of latency and energy consumption. 


\subsubsection{Setup}
For SNOW, we consider 11MHz spectrum from white space and split into 50 (400kHz each) orthogonal subcarriers each overlapping 50\% with the neighbor/s. Putting the BS at the center, we create a star network placing the nodes within 1.5km radius. We generate various numbers of nodes in the network, each  in the direct communication with the BS. Since  A-MAC is designed for short range WSN (e.g., approx. 40m at 0dBm Tx power), for simulations with A-MAC we place nodes to cover 1.5km radius, making a 38-hop network. In both networks, we perform convergecast. Every node has 100 packets to deliver to the BS. A sleep interval of 100ms is used after a node transmits all of its 100 packets. Each packet is of 40 bytes and is transmitted at 0dBm.

Starting with 50 nodes, we test up to 2000 nodes. We  calculate the total latency and the average energy consumption at each node (i.e., the ratio of total energy consumed by all nodes to the number of nodes) to collect all of these 100 packets from all of these nodes at the BS. For SNOW, we assign energy model of CC1070 radio as given in Table~\ref{tab:chap2-cc1070} to each node. For A-MAC, we assign energy model of CC2420 radio which is roughly similar to that of CC1070 radio.



For A-MAC, we run the default TinyOS~\cite{tinyos} Collection Tree Protocol~\cite{CTP} with proper configuration wiring~\cite{dutta2010design}. As the network is multi-hop, many nodes also forward packets received from other nodes.  All the transmitters keep retrying a packet until they receive a probe with ACK bit set. When we receive at least 90\% of all the packets at the BS, we stop data collection for both of the networks.

\subsubsection{Result}
\begin{figure}[t!]
    \centering 
          \subfigure[Latency for convergecast\label{fig:chap2-qualnet_latency}]{
         \includegraphics[width=.5\textwidth]{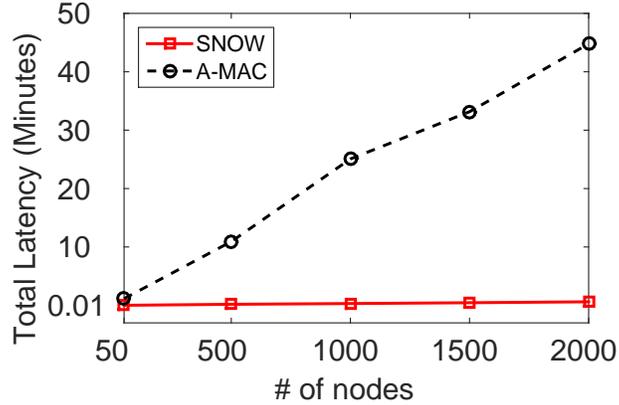}
      }
      \hfill
      \subfigure[Avg. energy consumption per node\label{fig:chap2-average_energy_consumption_convergecast}]{
      \includegraphics[width=.5\textwidth]{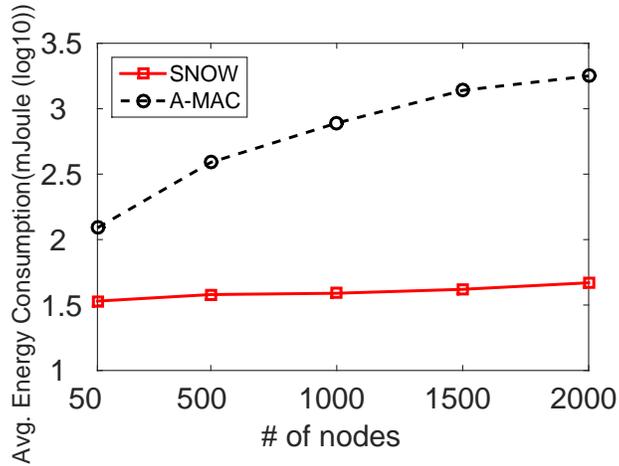}
      }
    \caption{Latency and energy consumption in simulation}
    \label{fig:chap2-energy_consumption}
\end{figure}

Figure~\ref{fig:chap2-qualnet_latency} shows the overall latency for both SNOW and A-MAC for collecting 100 packets from each node at the BS. The latency in A-MAC  increases sharply as the number of nodes increases. Up to 50 nodes, SNOW has a total latency of 0.013 minutes as opposed to 1.15 minutes in A-MAC. For 1000 nodes, the A-MAC latency is 25 minutes (vs 0.31 minutes in SNOW) which increases to 45 minutes (vs 0.67 minutes in SNOW) for 2000 nodes. The latency in A-MAC is very high due to collisions, back-off,  and probably retransmissions as well. As already acknowledged in~\cite{dutta2010design}, A-MAC  tends to perform worse in dense neighborhood and high packet delivery scenarios. On the other hand, latency in SNOW is negligible compared to A-MAC. In SNOW,  increasing the number of nodes above 50  increases the overall latency because only 50 nodes can transmit simultaneously.


Figure~\ref{fig:chap2-average_energy_consumption_convergecast} shows average energy consumption on each node when there are various numbers of nodes. We represent the energy information in $\log_{10}$ scale to give a better visibility. 
For 50-node network, an A-MAC  node consumes on average 123.27mJoules for delivering 100 packets compared to 35.2mJoules in SNOW node.
For 1000 nodes, these values are 780.12 and 38.33, respectively. For 2000 nodes, these values are 1765.89 and 45.05, respectively. In A-MAC, average energy consumption per node increases sharply as the total number of nodes increases because  of higher chances of  collisions,  back-offs, and retransmissions. As SNOW does not experience collision, its average energy consumption per node increases negligibly with the number of nodes. This justifies the low energy consumption behavior in SNOW. 


%
%
%
%
%
%
%
%
%

%% file: chapter2/comparison.tex
\subsection{SNOW vs Existing LPWAN Technologies} \label{sec:comparison}


While still in their infancy, LPWAN technologies are gaining momentum in recent years, with multiple competing technologies being offered or under development. The newly certified NB-IoT standard~\cite{nbiot}  operates over existing cellular networks. NB-IoT and 5G~\cite{ngmn} are designed for expensive licensed bands. SIGFOX~\cite{sigfox} and LoRa~\cite{lorawan} operate in unlicensed ISM band. Their field devices usually need to subscribe to the gateway towers. The radio emitters are required to adopt duty cycled transmission of 1\% or 0.1\%, depending on the sub-band. Thus they are less suitable for many WSN applications that need real-time requirements or frequent sampling. SIGFOX supports a data rate of 10 to 1,000bps. A message is of 12 bytes, and a device can send at most 140 messages per day. 
Each message transmission typically takes 3 seconds~\cite{linklab} while SNOW can transmit such a 12-byte message in  less than 2ms.  LoRa data rates range from 0.3 to 50kbps depending on the bit spreading factor (SF), and allows a user-defined packet size that impacts on Tx range. A high SF enhances reliability but reduces the effective data rate. For example, using 125kHz bandwidth, SFs of 11 and 10 give bit rates of 440bps and 980bps, respectively. Using 125kHz bandwidth and SF of 10, a 10-byte payload packet has an air time of 264.2ms typically~\cite{semtech}, which is very large compared to SNOW. SIGFOX and LoRa achieve long ranges using a Tx power up to 20dBm (27dBm for LoRa in USA). SNOW was tested up to 1.5km for which the devices needed a very low Tx power  (0dBm or less) which is similar to that achievable in LoRa~\cite{lorapub1}.




For SIGFOX, there exists no publicly available specification or implementation. Hence, an experimental comparison between SNOW and this  proprietary technology is beyond our reach at this time. The LoRa specification, designed and patented by Semtech Corporation, has recently been made openly available. Version 1.0 of the LoRaWAN specification was released in June 2015, and is still going through several amendments. While an open source MAC implementation for it was recently released by IBM, it is still going through multiple major updates to be compatible with Semtech modules~\cite{compatibility}. It has just been updated to LoRaWAN Specification v1.0.1 in July of 2016~\cite{updated}. Thus, even though this standard is promising, the devices and protocols are still under active development.  Hence, we leave the experimental comparison with LoRa as a future work. However, we provide some numerical comparison in terms of scalability as follows.


Scalability of  SIGFOX/LoRa is achieved assuming extremely low traffic. For example, if a device sends one packet per hour,  a LoRaWAN SX1301 gateway using 8 separate radios to exploit 8 channels can handle about 62,500 devices~\cite{lorawan}. With its 12-byte message and 140 messages per device per day, one SIGFOX gateway can support 1 million devices~\cite{sigfox}. We now estimate the scalability of SNOW for this communication scenario. Using one TV channel (6MHz width), we can get 29 OFDM subcarriers (each 400kHz).   The total time for a 12-byte message transaction between a SNOW node and the BS is less than 2ms (including Tx-Rx turnaround time). A group of 29 nodes can transmit simultaneously, each on a distinct subcarrier. We first consider only upward communication. If every device sends 140 messages per day (like SIGFOX),  every subcarrier can be shared by $\frac{24*3600*1000}{140*2}> 308,571$ devices. Thus 29 subcarriers can be shared by $308,571*29 > 8.9$ million devices. If we consider a downward message after every group of simultaneous transmissions by 29 nodes to schedule the next group of transmissions, SNOW with one white space channel can support at least $8.9/2\approx 4.45 $ million devices. Using $m$ channels, it can support $4.45m$ million devices. This back-of-envelop calculation indicates SNOW may support significantly more devices than SIGFOX and LoRa.  This advantage stems from SNOW's capability to support simultaneous transmissions on multiple subcarriers within a single TV channel.

Another important advantage of SNOW is that it is  designed to exploit white spaces which have widely available free spectrum (between 54 and 698MHz in US). In contrast, SIGFOX/LoRa has much less and limited spectrum to utilize (863--870MHz in EU, 902--928MHz in US). The upcoming IEEE 802.15.4m~\cite{802154m} standard aims to exploit white spaces as an extension to the IEEE 802.15.4 standard. Our results can therefore help shape and evolve such standards.

%

%% file: chapter2/related_work.tex
\subsection{Related Work}\label{chap2:related}



Several measurement and analytical studies have  shown that there exist abundant
white spaces in  outdoor~\cite{uk, europe, chicago, singapore, guangzhou} and indoor~\cite{shortindoor, indoor1, indooroutdoor, testbed, ws_mobicom13, winet} environments.  Prior work focused on opportunistically forming a single link~\cite{ws_dyspan08},  spectrum sensing~\cite{ws_dyspan08_kim, ws_mobicom08},  and identification of  primary users. Later, white spaces were exploited for establishing Wi-Fi like network~\cite{ws_sigcomm09, WATCH}, video-streaming~\cite{videostreaming}, mobile devices~\cite{linkasymmetry, vehiclebased}, dynamic spectrum access~\cite{ws_mobicom09, ws_nsdi10}, and designing a prototype system for spectrum sensing~\cite{ws_dyspan11, FIWEX}. As spectrum sensing is no longer compulsory, the FCC has recently mandated the use of a geo-location service~\cite{dbreq}. The geo-location approach has been widely studied using databases to store  white space information for clients query~\cite{database3, database1, database2, vehiclebased, hysim}.  All of these works consider using white spaces for wireless broadband service. In contrast, we have proposed WSN over white spaces.

Our work is most related to SMACK~\cite{smack} and WiFi-NC~\cite{wifinc}. SMACK~\cite{smack} was designed for allowing ACK of single-hop broadcast made by an access point.  This was done by assigning a subcarrier to each client node that sends an ACK by sending or not sending a tone which is sensed by the access point  through energy detection. All such ACKs need to arrive (almost) at the same time - within a window of few microseconds. SMACK is not capable of decoding data from subcarriers and is not designed for handling  simultaneous packet reception on multiple subcarriers. WiFi-NC uses a wideband radio  as a compound radio that is split into multiple narrowband channels called {\em radiolets}. Each radiolet is entirely implemented as a separate digital circuit allowing for independent carrier sensing, decoding logic, transmission, and reception of packets in its own narrow channel. Specifically, the transmitter circuit of each radiolet  consists of a baseband transmitter, an upsampler, a  low pass filter, and a  mixer.  The receiver circuit of each radiolet consists of a mixer, a low pass filter, a down sampler, and  a baseband receiver. Thus the architecture of a WiFi-NC compound radio with $m'$ radiolets is close to that of $m'$ transceivers with low form factor benefits. In contrast, SNOW needs no extra circuitry for any subcarrier. The BS uses a single radio that can receive simultaneously on multiple subcarriers using a single decoding algorithm with no extra hardware or circuit.  


%% file: chapter3.tex
\section{Enabling Reliable, Asynchronous, and Bidirectional Communication in Sensor Networks over White Spaces}   \label{chap:snow2}

Low-Power Wide-Area Network (LPWAN) heralds a promising class of technology to overcome the range limits 
and scalability challenges in traditional wireless sensor networks. Recently proposed Sensor Network over White Spaces (SNOW) 
technology is particularly attractive due to the availability and advantages of TV spectrum in long-range communication. This paper proposes a new design of SNOW that is asynchronous, 
reliable, and robust. It represents the first highly scalable LPWAN over TV 
white spaces to support reliable, asynchronous, bi-directional, and concurrent communication between numerous sensors 
and a base station. This is achieved through a set of novel techniques.  This new design of SNOW has an OFDM based physical layer 
that adopts robust modulation scheme and allows the base station using a single antenna-radio  (1) to send different data 
to different nodes concurrently and (2) to receive concurrent transmissions made by the sensor nodes asynchronously.  
It has a lightweight MAC protocol that (1)  efficiently implements per-transmission acknowledgments of the asynchronous  
transmissions by exploiting the adopted OFDM design; (2)  combines CSMA/CA and location-aware spectrum allocation 
for mitigating hidden terminal effects, thus enhancing the flexibility of the nodes in transmitting asynchronously. Hardware 
experiments through deployments in three radio environments - in a large metropolitan city, in a rural area, and in an indoor 
environment -  as well as large-scale simulations demonstrated that the new SNOW design drastically outperforms other LPWAN 
technologies in terms of scalability, energy, and latency.

\input{chapter3/introduction}

\input{chapter3/relatedwork}

\input{chapter3/model}

\input{chapter3/phy}

\input{chapter3/mac}

\input{chapter3/implementation}

\input{chapter3/experiment}

\input{chapter3/simulation}

\subsection{Summary}\label{sec:chap3-conclusion}
In this paper, we have proposed the design of an asynchronous, reliable, and robust Sensor Network over White spaces (SNOW).
This new design of SNOW represents the first low power and long range sensor network over TV 
white spaces to support reliable, asynchronous, bi-directional, and concurrent communication between numerous sensors 
and a base station. Hardware experiments through deployments  in multiple geographical areas as well as simulations demonstrated that it significantly outperforms the state-of-the-art designs in terms of scalability, energy, and latency.


%% file: chapter3/introduction.tex
\subsection{Introduction}\label{sec:chap3-introduction}
Sensor networking over TV white spaces has gained interest  recently~\cite{snow, 802154m, vasisht17farmbeats}. Wireless sensor network (WSN) in large-scale and wide-area applications (e.g.,   urban sensing~\cite{citysense},  civil infrastructure monitoring~\cite{GGB}, oil field management~\cite{oilffield},  precision agriculture~\cite{Murphy}) often needs  to connect thousands of  sensors over long distances.  Due to their short communication range, the existing WSN technologies  in the ISM band such as IEEE 802.15.4~\cite{ieee154},  802.11~\cite{wifi}, and Bluetooth~\cite{Bluetooth} cover a large area with numerous devices as multi-hop mesh networks at the expense of  energy, cost, and complexity. These limitations can be overcome by letting WSNs operate over TV white spaces. Such a network architecture is called {\slshape Sensor Network Over White Spaces ({\bf \slshape SNOW})}.

{\slshape White spaces} refer to the allocated but locally unused TV spectra, and can be used by unlicensed devices~\cite{FCC_first_order, fcc_second_order}. The Federal Communications Commission (FCC) in the US mandates that 
a device needs to either sense the channel before transmitting, or  consult with a cloud-hosted geo-location database~\cite{fcc_second_order} to learn about unoccupied TV channels at a location. Similar regulations are adopted in many countries. Compared to IEEE 802.15.4 or Wi-Fi, they offer a large number of and less crowded channels, each 6MHz, available  in both rural and urban areas~\cite{ws_sigcomm09, europe, chicago, guangzhou, shortindoor, testbed, ws_mobicom13, winet}. Thanks to their lower frequencies (54 -- 862MHz in the US), white spaces have excellent propagation characteristics over long distance and obstacles.  Long range will reduce many WSNs to a star-topology that has potential to avoid the complexity, overhead, and latency associated with many-hop mesh networks. Such a paradigm shift must also deal with the challenges that stem from the long range such as increased chances of packet collision. It must also satisfy the typical requirements of WSNs such as low cost nodes, scalability, reliability, and energy efficiency.


Exploiting white spaces for sensor networking is the goal of the on-going IEEE 802.15.4m standardization effort~\cite{802154m}.  As an early research effort in this space, we proposed 
 the first design of SNOW in~\cite{snow}, referred to as {\bf SNOW 1.0} in this paper, to address some of the above challenges. It was designed based on D-OFDM, a distributed implementation of Orthogonal Frequency Division Multiplexing (OFDM), that allowed its base station (BS) to receive multiple packets in parallel. The BS uses wide white space spectrum which is split into narrowband orthogonal subcarriers. Each sensor node is assigned a subcarrier on which it transmits. Despite its promise, SNOW 1.0 has several important limitations as follows. 

\begin{enumerate}

\item D-OFDM in SNOW 1.0 is not implemented for {\em bi-directional} 
communication over different subcarriers.  Its BS can 
receive packets from multiple nodes in parallel but {\em cannot} concurrently 
transmit different packets to different nodes. 
\item SNOW 1.0 cannot support
per-transmission acknowledgment (ACK) which limits its {\em reliability}.  

\item It does not support fully asynchronous operation as the nodes can transmit asynchronously only if their number is no greater than that of the subcarriers. It schedules transmissions from multiple sensors sharing the same subcarriers based on Time-Division Multiple Access (TDMA), which limits their flexibility in transmitting asynchronously.

\item It uses amplitude-shift-keying (ASK) which provides simplicity but is not a robust modulation scheme. 

\end{enumerate}



In this chapter, we address the above challenges and important limitations of SNOW 1.0, and propose a new design of SNOW, referred to as {\bf SNOW 2.0}, that is asynchronous, reliable, and robust. 
Throughout this paper, with `SNOW' we shall mean SNOW 2.0. The terms `SNOW 2.0' and `SNOW 1.0' will be used when we need to distinguish between this new design and the earlier one. 
SNOW 2.0 is the first design of a highly scalable, low power, and long range WSN over TV white spaces which is fully asynchronous and enables reliable massive parallel and asynchronous receptions with a single antenna-radio and multiple concurrent  data transmissions with a single antenna-radio. This is achieved through a full-fledged physical layer (PHY) design by implementing D-OFDM for multiple access in both directions and through a reliable, light-weight Media Access Control (MAC) protocol. While OFDM has been embraced for multiple access in various wireless broadband and cellular technologies recently (see Section~\ref{subsec:chap3-lpwan}), its adoption in low power, low data rate, narrowband, and WSN design remains quite new. Taking the advantage of low data rate  and short packets, we adopt OFDM in WSN through a much simpler and energy-efficient design. The BS's wide white space spectrum is split into narrowband orthogonal subcarriers that 
 D-OFDM uses to enable parallel data streams to/from the distributed nodes from/to the BS. SNOW 2.0 thus represents a promising platform for many cyber-physical systems and Internet of Things (IoT) applications that depend on bidirectional sensor data (e.g., Microsoft's FarmBeats in IoT for agriculture~\cite{vasisht17farmbeats}). 
 
 The specific contributions of this paper are as follows. 
\begin{itemize}
\item We design a D-OFDM based PHY for SNOW with the following features for enhanced scalability, low power, long range.  
{\bf (1)} It adopts robust modulation scheme such as Binary Phase Shift Keying (BPSK) and Quadrature Phase Shift Keying (QPSK).  
{\bf (2)} Using a single antenna-radio, the BS can receive concurrent transmissions  made by the sensor nodes asynchronously.  
{\bf (3)} Using a single antenna-radio, the BS can send different data to different nodes concurrently.  
Note that the above design is different from MIMO radio adopted in various wireless domains such as LTE, WiMAX, 802.11n~\cite{mimo} as the latter uses multiple antennas to enable multiple transmissions and receptions. 

\item We develop a lightweight  MAC protocol for operating the nodes with greater freedom, low power, and reliability.  The SNOW MAC has the following features.  
{\bf (1)} Considering a single half-duplex radio at each node and two half-duplex radios at the BS, we efficiently implement per-transmission ACK of the asynchronous and concurrent transmissions by taking the advantage of D-OFDM design. {\bf (2)} It combines CSMA/CA and location-aware subcarrier assignment for mitigating hidden terminals effects, thus enhancing the flexibility of the nodes that need to transmit asynchronously. 
{\bf (3)} The other key features include the capability of handling peer-to-peer communication, spectrum dynamics, load balancing, and network dynamics. 
\item We implement SNOW  in  GNU Radio~\cite{gnuradio} using Universal Software Radio Peripheral (USRP)~\cite{usrp}  devices. In our experiments, a single radio of the SNOW BS can encode/decode 29 packets on/from 29 subcarriers within 0.1ms to transmit/receive simultaneously, which is similar to standard encoding/decoding time for just one packet. 

\item We perform experiments through SNOW deployments in three different radio environments -  a large metropolitan city, a rural area, an indoor testbed - as well as simulations. 
All results demonstrate the superiority of SNOW over  several LPWAN technologies in terms of scalability, latency, and energy. Large-scale simulations show a 100\% increase in SNOW throughput while having  both latency and energy consumption half compared to our earlier design. 

\end{itemize}


In the rest of this chapter, Section~\ref{sec:chap3-related} overviews related work.  Section~\ref{sec:chap3-architecture} describes the model. Section~\ref{sec:chap3-phy} presents the  PHY. Section~\ref{sec:chap3-mac} presents the MAC protocol.  Sections~\ref{sec:chap3-implementation},~\ref{sec:chap3-experiment}, and ~\ref{sec:chap3-simulation} present implementation,  experiments, and simulations, resp. Section~\ref{sec:chap3-conclusion} is the conclusion.

%% file: chapter3/relatedwork.tex
\subsection{Related Work}\label{sec:chap3-related}

\subsubsection{White Spaces Network}

To date, the potential of white spaces is mostly being tapped into for broadband access by industry leaders such as Microsoft~\cite{4Africa, MSRAfrica} and Google~\cite{GoogleAfrica}. Various standards bodies such as IEEE 802.11af~\cite{IEE802_af},  IEEE 802.22~\cite{IEEE802_22},  and  IEEE 802.19~\cite{IEE802_19} are modifying existing standards to exploit white spaces for broadband access. In parallel, the research community has been investigating techniques to access white spaces through spectrum sensing~\cite{ws_dyspan08_kim, ws_mobicom08, ws_dyspan11, FIWEX} or geo-location approach~\cite{database1, database2, database3, vehiclebased, hysim} mostly for broadband service. A review of white space networking for broadband access can be found in~\cite{snow, ws_mobicom13}. In contrast, the objective of our work is to exploit white spaces for designing highly scalable, low-power, long range, reliable, and robust SNOW. We proposed SNOW 1.0  in~\cite{snow}. As already pointed out in Section~\ref{sec:chap3-introduction}, SNOW 1.0 does not support bidirectional, reliable, and fully asynchronous communication.  Hence, it is not a suitable platform for applications that need ACK, sensing and control~\cite{RTSS2015, capnet}, or bidirectional sensor data~\cite{vasisht17farmbeats}. Our proposed new SNOW design overcomes all of these limitations  and achieves enhanced scalability, reliability, and robustness.

 \subsubsection{Low-Power Wide-Area Network (LPWAN)}\label{subsec:chap3-lpwan}

{\bf SNOW vs LoRa/SIGFOX.} 
LPWAN technologies are gaining momentum in recent years, with multiple competing technologies being offered or under development.  SIGFOX~\cite{sigfox} and LoRa~\cite{lorawan, bor2016lora, voigt2016mitigating, marcelis2017dare} are two very recent LPWAN technologies that operate in unlicensed ISM band. Their devices require to adopt duty cycled transmission of only 1\% or 0.1\% making them less suitable for many WSNs that involve real-time applications or that need frequent sampling. SIGFOX supports a data rate of 10 to 1,000bps. 
A  message is of 12 bytes, and a device can send at most 140 messages per day. Each message transmission typically takes 3 seconds~\cite{linklab} while SNOW can transmit such a 12-byte message in less than 2ms as we experimented in~\cite{snow}.  

Semtech LoRa modulation employs Orthogonal Variable Spreading Factor (OVSF) which enables multiple spread signals to be transmitted at the same time on the same channel.  
OVSF is an implementation of traditional Code Division Multiple Access (CDMA) where before each signal is transmitted, the signal is spread over a wide spectrum range through the use of a user's code. Using 125kHz bandwidth and LoRa spreading factor (LoRa-SF) of 10, a 10-byte payload packet in LoRa has an air time of 264.2ms typically~\cite{semtech}, which is  at least 100 times that in SNOW for the same size message~\cite{snow}.  The higher the LoRa-SF, the slower the transmission and the lower the bit rate in LoRa. This problem is exacerbated by the fact that
large LoRa-SFs are used more often than the smaller ones. For instance, as studied in~\cite{loralimitation}, considering a scenario with end-devices distributed
uniformly within a round-shaped area centered at the gateway, and a path loss calculated with the Okumura-Hata model~\cite{Molisch} for
urban cells, the probability that an end-device uses a LoRa-SF of 12 would be $0.28$, while that of 8 would be $0.08$.


One important limitation of OVSF is that the users' codes have to be mutually orthogonal to each other, limiting the scalability of the network that adopts this technique. LoRa uses 6 orthogonal LoRa-SFs (12 to 7), thus allowing up to 6 different transmissions on a channel simultaneously. Using one TV channel (6MHz wide), we can get 29 OFDM subcarriers (each 400kHz) for SNOW which enables 29 simultaneous transmissions  on a single TV channel. Using a narrower bandwidth like SIGFOX/LoRa would yield even a higher number of subcarriers per channel in SNOW. Note that white spaces can consist of more than one TV channel. Using $M$ channels, the number of simultaneous transmissions multiplies by $M$ in SNOW. Hence, our back-of-envelop calculation even  for SNOW 1.0 in~\cite{snow} showed its superiority in scalability over SIGFOX/LoRa.  Since there exists no publicly available specification for SIGFOX, we compare SNOW with LoRa in Section~\ref{sec:chap3-simulation} through simulation to demonstrate higher efficiency and scalability of SNOW.


{\bf Comparison with The Other LPWAN Technologies.}
SNOW achieves high scalability by exploiting the existing OFDM technology for multi-access. OFDM is a well-known modulation technique and it has been adopted for multi-access in various forms in various wireless broadband and cellular technologies recently. However, its usage in low-power, low-rate, narrowband and wireless sensor network domain is still new. Our adopted technique, D-OFDM, in SNOW has similarity with several OFDM multiple access techniques such as OFDMA (Orthogonal Frequency Division Multiple Access) and SC-FDMA (Single Carrier Frequency Division Multiple Access)   adopted in WiMAX~\cite{wimax, OFDMAWiMAX} and LTE~\cite{3gpp, lte_whitepaper, scfdma}. For uplink communication  in both OFDMA and SC-FDMA adopted in WiMAX and LTE,  respectively, the BS uses {\bf multiple antennas to receive from multiple nodes}.  In contrast, D-OFDM enables multiple receptions using a single antenna and also enables different data transmissions to different nodes using a single antenna.  Both WiMAX and LTE use OFDMA in downlink direction. WiMAX uses OFDMA in uplink direction also. OFDMA is known to be more sensitive to a null in the channel spectrum and it requires channel coding or power/rate control to overcome this deficiency. Specifically, for its usage in uplink communication, the transmit power of the senders need to be adjusted so that the received signal strengths from different senders are close. In low power network, this becomes difficult. Also, OFDMA has a high peak-to-average power ratio (PAPR) which leads to difficulties in transceiver design~\cite{3gpp, lte_whitepaper, scfdma, OFDMAWiMAX}. This also implies high power consumption and lower battery life for the sending nodes in uplink communication. Therefore, the 3GPP standardization group has decided to use SC-FDMA  instead in LTE for uplink communication~\cite{3gpp, lte_whitepaper, scfdma}.

While SC-FDMA has relatively lower PAPR, to meet the high data rate requirement in LTE (86 Mbps in uplink), its receiver design for allowing multiple simultaneous transmitters is complicated, and is designed by using multiple antennas at the cost of high energy consumption~\cite{3gpp, lte_whitepaper, scfdma}.  Such issues are less severe for low data rate and small packet sizes and we can realize with much simpler design. Therefore, our similar design, D-OFDM, remains much simpler and multiple receptions and multi-carrier transmission can be done using a single antenna of the radio in SNOW. 


\begin{figure}[!htb]
\centering
\includegraphics[width=0.7\textwidth]{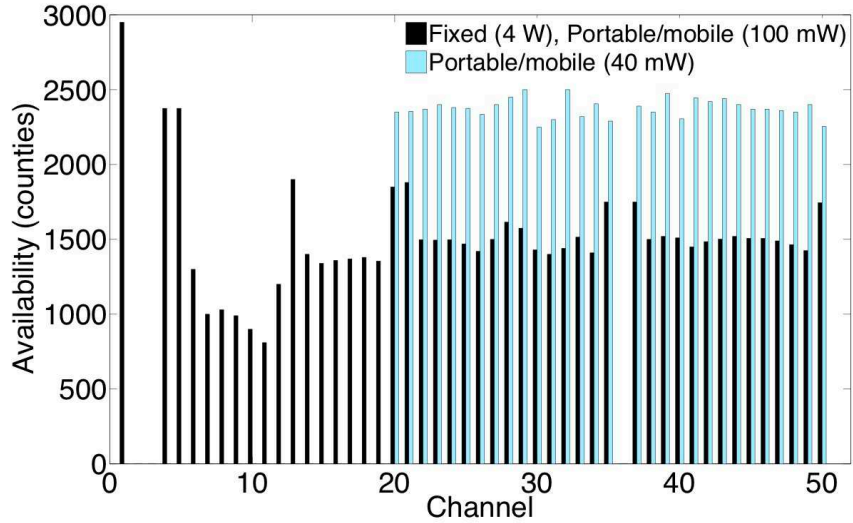}
\caption{White space in the US counties~\cite{spectrumbridge}: showing the number of counties (y-axis) where the channels (x-axis) are white space.}
\label{fig:chap3-availability}
\end{figure}

5G~\cite{ngmn} is envisioned to meet IoT use cases in addition to telecommunications applications using the cellular infrastructure. Currently, the 5G standard is still under development. NB-IoT~\cite{nbiot}  is a narrowband LPWAN technology standard to operate on cellular infrastructure and bands. Its specification was frozen at Release 13 of the 3GPP specification (LTE-Advanced Pro~\cite{LTE_advancedpro}) in June 2016. \revise{These technologies would require devices to periodically wake up to synchronize with the network, giving a burden on battery life. Also, the receiver design to enable multiple packet receptions simultaneously using SC-FDMA requires {\bf multiple antennas}. Note that setting up multiple antennas is {\bf difficult} for lower frequencies as the antenna form factor becomes large due to lower frequency.} The antennas need to be spaced $\lambda/2$ apart, where $\lambda$ is the wavelength. Doing this is difficult as $\lambda$ is {\bf large} for lower frequencies, and even more difficult and expensive to do this for every sector to be served by the base station. Having low data rate and small packet sizes, SNOW PHY design remains much simpler and both the transmitters and the receiver can have a single antenna and the BS can receive multiple packets simultaneously using single antenna radio.  We also design a complete MAC protocol for SNOW which features a location-aware spectrum allocation for mitigating hidden terminal problems, per-transmission ACK for asynchronous transmissions, and the capability of handling peer-to-peer communication, spectrum dynamics, load balancing, and network dynamics. Another important advantage of SNOW is that it is designed to exploit white spaces which have widely available free spectrum (as shown in Figure~\ref{fig:chap3-availability}), while the above standards are designed to use licensed band or limited ISM band.

%% file: chapter3/model.tex
\subsection{System Model}\label{sec:chap3-architecture}

\begin{figure}[!htb]
\centering
\includegraphics[width=0.7\textwidth]{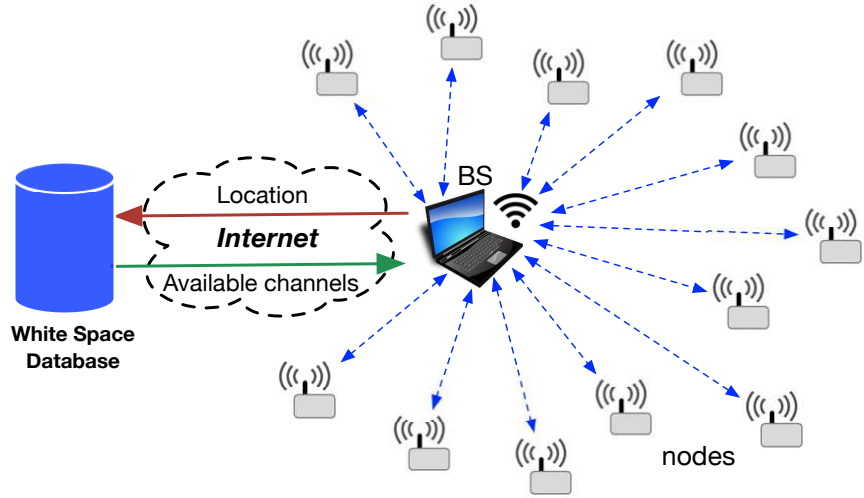}
\caption{The network structure of SNOW.}
\label{fig:chap3-network}
\end{figure}

WSNs are characterized by small packets, low data rate, and low power~\cite{snow}. The nodes are typically battery powered. Thus, scalability and energy-efficiency are the key concerns in WSN design. 
We consider a WSN where a lot of sensor nodes are associated with a BS.  Each sensor node (called {\bf `node'} throughout the paper) is equipped with a single half-duplex narrow-band radio operating in the white space spectrum. Due to long communication range even at low power (e.g., several kilometers at 0 dBm transmission power in our experiment in Section~\ref{sec:chap3-experiment}) of this radio, we consider that the nodes are directly connected (with a single hop) to the BS and vice versa as shown in Figure~\ref{fig:chap3-network}. However, the nodes may or may not be in communication ranges of the other nodes. That is, some nodes can remain as hidden terminal to some other nodes. The BS and its associated nodes thus form a star topology.    The nodes  are power constrained and not directly connected to the Internet. 


The BS uses a wide channel split into subcarriers, each of equal spectrum width (bandwidth). Each node is assigned one subcarrier on which it transmits to and receives from the BS.  For integrity check, the senders add cyclic redundancy check (CRC) at the end of each packet.  We leave most complexities at the BS and keep the other nodes very simple and energy-efficient. The nodes do not do spectrum sensing or cloud access.  The BS determines white spaces by accessing a cloud-hosted database  through the Internet as shown in Figure~\ref{fig:chap3-network}. We assume that it knows the locations of the nodes  either through manual configuration or through some  existing WSN localization techniques such as those based on ultrasonic sensors or other sensing modalities~\cite{wsnlocalizationsurvey}. Localization is out of scope of this paper. The BS  selects white space channels that are available at its own location and at the locations of all other nodes. We use two radios at the BS to support concurrent transmission and reception as described in Section~\ref{sec:chap3-mac}.


%% file: chapter3/phy.tex
\subsection{Physical Layer Design}\label{sec:chap3-phy}
The PHY-layer of SNOW is designed to achieve scalable and robust
bidirectional communication between the BS and numerous nodes.
Specifically, it has three key design goals: (1) to allow the BS to
receive concurrent and asynchronous transmissions from multiple nodes
using a single antenna-radio; (2) to allow the BS to
send different packets to multiple nodes concurrently using a single antenna-radio; (3) support robust modulation such as
BPSK.

%

\subsubsection{Design Rationale}\label{subsec:chap3-spacing}

For scalability and energy efficiency, we design the PHY based on D-OFDM. {\em OFDM} is a frequency-division multiplexing scheme that uses a large number of closely spaced orthogonal subcarrier signals to carry data on multiple parallel data streams between a sender and a receiver. As discussed before, it has been adopted for multi-user in various forms in various wireless broadband and cellular technologies recently.  
D-OFDM is a distributed implementation of OFDM  introduced in~\cite{snow} for multi-user access. Unlike OFDMA and SC-FDMA for multi-access, D-OFDM enables multiple receptions using a single antenna and also enables different data transmissions to different nodes using a single antenna.

In SNOW, the BS's wide white space spectrum is split into narrowband orthogonal subcarriers which carry parallel data streams to/from the distributed nodes from/to the BS as D-OFDM. 
Narrower bands have lower bit rate but longer range, and consume less power~\cite{channelwidth}. Thus, we adopt D-OFDM by assigning the orthogonal subcarriers to different nodes. Each node transmits and receives on the assigned subcarrier. Each subcarrier is modulated using BPSK which is highly robust due to difference of $180^\circ$ between two constellation points, and is widely used (e.g, in WiMAX 16d, 16e;  WLAN 11a, 11b, 11g, 11n).  Since BPSK and QPSK are fundamentally similar with the latter being less robust with higher bit rate, with minor modification QPSK (which is used in IEEE 802.15.4 at 2.4 GHz~\cite{ieee154}) is also adoptable in SNOW.

\begin{figure}[!htb]
\centering
\includegraphics[width=0.7\textwidth]{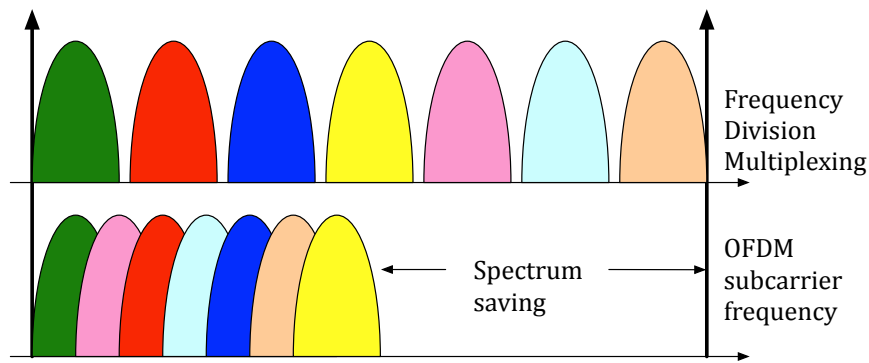}
\caption{Typical frequency-division multiplexing  vs OFDM.}
\label{fig:chap3-ofdm}
\end{figure}

The key feature in OFDM is to maintain subcarrier orthogonality. If the integral of the product of two signals is zero over a time period,  they are {\bf\em orthogonal} to each other. Two sinusoids with frequencies that are integer multiples of a common one satisfy this criterion. The orthogonal subcarriers can be {\bf overlapping}, thus increasing the spectral efficiency (as shown in Figure~\ref{fig:chap3-ofdm}). As long as orthogonality is maintained, it is still possible to recover the individual subcarriers' signals despite their overlapping spectrums.  Specifically, in the {\bf downward communication} in SNOW (i.e. when a single radio of the BS transmits different data to different nodes using  a single transmission),  OFDM encoding happens at a single radio at the BS while the distributed nodes decode their respective data from their respective subcarriers. In the {\bf upward communication} in SNOW (i.e. when many nodes transmit on different subcarriers to a single radio of the BS),  OFDM encoding happens in a distributed fashion on the nodes while a single radio at the BS decodes their data from the respective subcarriers.   

Note that if the BS radio has $n$ subcarriers it can receive from at most $n$ nodes simultaneously. Similarly, it can carry at most $n$ different data at a time. When the number of nodes is larger than $n$, a subcarrier is shared among multiple nodes and their communication is governed by the MAC protocol (Section~\ref{sec:chap3-mac}). To explain the PHY design we ignore subcarrier allocation and consider only the $n$ nodes who have occupied the subcarriers for transmission.

\subsubsection{Upward Communication}
Here we describe how we enable parallel receptions at a single radio at the BS  when each node's data is modulated based on BPSK or QPSK. In our D-OFDM design, we adopt Fast Fourier Transformation (FFT) to extract information from all subcarriers. We allow the nodes to transmit on their respective subcarriers  whenever they want without coordinating among themselves. 

{\bf Decoding upon Distributed Encoding.}
Every node independently encodes based on BPSK (or QPSK) the data on its subcarrier.  To decode a composite OFDM signal generated from orthogonal subcarriers from the distributed nodes, we adopt {\bf\slshape Global FFT Algorithm (G-FFT)} which runs FFT on the entire range of the spectrum of the BS, instead of running a separate FFT for each subcarrier. To receive asynchronous transmissions, the BS keeps running the G-FFT algorithm.  A vector $v$ of size equal to the number of FFT bins stores the received time domain samples. The G-FFT is performed on $v$ at every cycle of the baseband signal. For $n$ subcarriers, we apply  an $m$ point G-FFT algorithm, where $m\ge n$.  Each FFT output gives a set of $m$ values. Each index in that set represents a single energy level and phase of the transmitted sample at the corresponding frequency at a time instant.

\begin{figure}
\begin{center}
\begin{tabular}{|c c c c|}
\hline\rowcolor{LightGray}
{\bf Subcarrier 1} & {\bf Subcarrier 2} & $\cdots$   &   {\bf Subcarrier} $n$\\
\hline
$\cdots$ &  $b_{1,2}$ & $\cdots$ & $\cdots$\\
$\cdots$ & $b_{2,2}$ & $\cdots$ & $b_{1,n}$\\
$\cdots$ & $b_{3,2}$ & $\cdots$ &  $b_{2,n}$\\
$b_{1,1}$ & $b_{4,2}$ & $\cdots$ &  $b_{3,n}$\\
$b_{2,1}$ & $b_{5,2}$ & $\cdots$ &  $b_{4,n}$\\
$b_{3,1}$ & $b_{6,2}$ & $\cdots$ &  $b_{5,n}$\\
$b_{4,1}$ & $b_{7,2}$ & $\cdots$ &  $b_{6,n}$\\
$b_{5,1}$ & $b_{8,2}$ & $\cdots$ &  $b_{7,n}$\\
$b_{6,1}$ & $b_{9,2}$ & $\cdots$ & $b_{8,n}$\\
$\vdots$ & $\vdots$ & $\vdots$ & $\vdots$
\end{tabular}
\end{center}
\caption{2D matrix for decoding in upward communication}
\label{table:chap3-matrix}
\end{figure}

In BPSK, bit 0 and 1 are represented by keeping the phase of the carrier signal at $180^\circ$ and $0^\circ$ degree respectively. We also use a phase threshold that represents maximum allowable phase deviation in the received samples. For BPSK, one symbol is mapped into one bit, where in QPSK one symbol is mapped to a dibit. Since any node can transmit any time without any synchronization, the correct decoding of all packets is handled by maintaining a 2D matrix where each column represents a subcarrier or its center frequency bin that stores the bits decoded at that subcarrier. Figure~\ref{table:chap3-matrix} shows the 2D matrix where entry $b_{i,j}$ represents $i$-th bit (for BPSK) of $j$-th subcarrier.  The same process thus repeats. We handle spectral leakage through the {\em Blackman-Harris windowing}~\cite{blackman}.

%

\subsubsection{Downward Communication}\label{subsec:chap3-downward}

One of our key objectives is to enable transmission from the BS which will encode different data on different subcarriers. A node's data will be encoded on the associated subcarrier. The BS then makes a single transmission and all nodes will decode data from their respective subcarriers. Such a communication goal is challenging due to asymmetric bandwidth between the transmitter (BS in this case) and the receivers (the nodes in this case). In the following, we describe our approach to achieve this in SNOW.

{\bf Encoding for Distributed Decoding.}
Our design approach based on D-OFDM is to enable distributed demodulation at the nodes without any coordination among them. That is, from the received OFDM signal, every node 
will independently decode based on BPSK/QPSK the data from the signal component on its subcarrier only. In our approach, the main design technique lies in the encoding part at the BS. We enable this by adopting IFFT (Inverse FFT) at the transmitter side that encodes different data on  different subcarriers. IFFT is performed after encoding data on the subcarriers. We can encode data on any subset of the subcarriers. The transmission is made after IFFT. If the OFDM transmitter uses $m$ point IFFT algorithm, consecutive $m$ symbols of the original data are encoded in $m$ different frequencies of the time domain signal with each run of the IFFT algorithm. We encode different symbols for different nodes on different subcarriers, thus obviating any synchronization between symbols. We use a vector $v$ of size equal to the number of IFFT bins. Each index of $v$ is a frequency bin. If the BS has any data for node $i$, it maps one unit of the data to a symbol and puts in the $i$-th index. If it has data for multiple nodes, it creates multiple symbols and puts in the respective indices of  $v$. Then the IFFT algorithm is performed on $v$ and a composite time domain signal with data encoded in different frequencies  is generated and transmitted.  This repeats at every cycle of baseband signal. A node listens to its subcarrier center frequency and receives only the signal component in its subcarrier frequency. The node then decodes data from it.

 \begin{figure}[h]
    \centering
      \subfigure[CDR under varying SF and packet sizes\label{fig:chap3-spreading}]{
        \includegraphics[width=.5\textwidth]{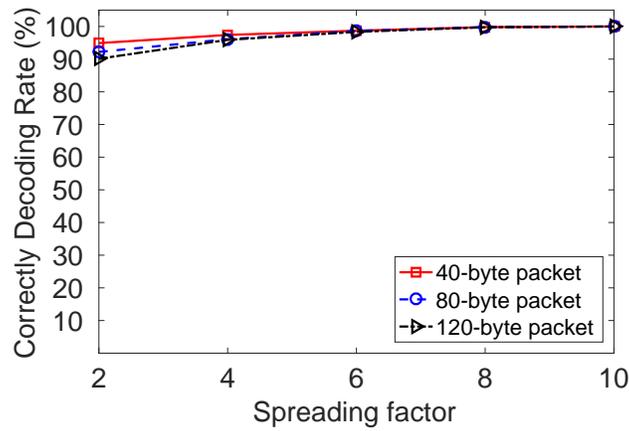}
      }
    \hfill
      \subfigure[BER over distances when SF=8\label{fig:chap3-sf_distance}]{
        \includegraphics[width=.5\textwidth]{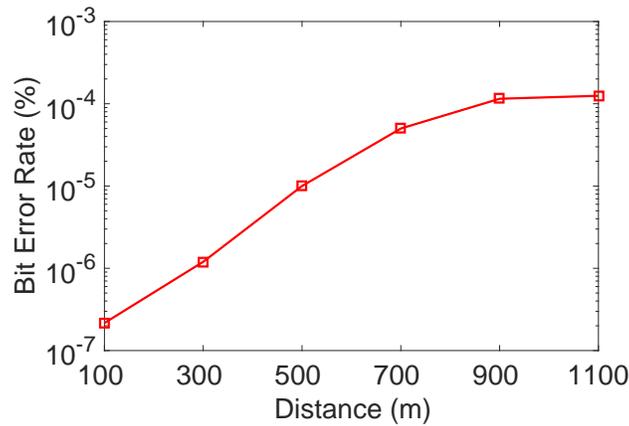}
      }
    \caption{Determining spreading factor}
    \label{fig:chap3-design_params}
  \end{figure}

\subsubsection{Using Fragmented Spectrum} 
White space spectrum may be found fragmented. When we cannot find consecutive white space channels while needing more, we may use non-consecutive white spaces. The G-FFT and IFFT algorithms will be run on the entire spectrum (as a single wide channel) that includes all fragments (including the occupied TV channels between the fragments). The occupied spectrum will not be assigned to any node and the corresponding bins will be ignored in decoding and encoding in G-FFT and IFFT, respectively.

%
%

\subsubsection{Design Considerations}\label{subsubsec:chap3-para}

{\bf Link parameters.}
{\slshape Bit spreading} is a technique to reduce bit errors by transmitting redundant bits for ease of decoding in noisy environments. It is widely used in many wireless technologies such as IEEE 802.15.4~\cite{ieee154} and IEEE 802.11b~\cite{wifi}.  Using USRP devices in TV white spaces and using narrow bandwidth (400kHz) we tested with different packet sizes and bit spreadings factor (SF).  We define \textbf{\slshape Correctly Decoding Rate (CDR)} - as the ratio of the number of correctly decoded packets at the receiver to the total number of packets transmitted.  
A receiver can always decode over 90\% of the packets when the sender is 1.1km away and transmits at 0 dBm (Figure~\ref{fig:chap3-spreading}). Figure~\ref{fig:chap3-sf_distance} shows that bit error rate (BER) remains negligible under varying distances (tested up to 1.1km in this experiment). Note that for wireless communications, a packet is usually dropped if its  BER exceeds $10^{-3}$~\cite{rnr}. 
Thus we will use 8 as default SF. Since the subcarriers can often violate orthogonality in practice, in our low data rate communication using a spreading factor of 8 helps us mitigate its effects and still recover most of the bits.  We have tested the feasibility of different packet sizes (Figure~\ref{fig:chap3-spreading}). WSN packet sizes are usually short. For example, TinyOS~\cite{tinyos} (a platform/OS for WSN motes based on IEEE 802.15.4) has a default payload size of 28 bytes. We use 40-byte (28 bytes payload + 12 bytes header) as our default packet size in our experiment.

\begin{figure}[t]
    \centering 
      \subfigure[Reliability over distance\label{fig:chap3-sub_distance}]{
        \includegraphics[width=.5\textwidth]{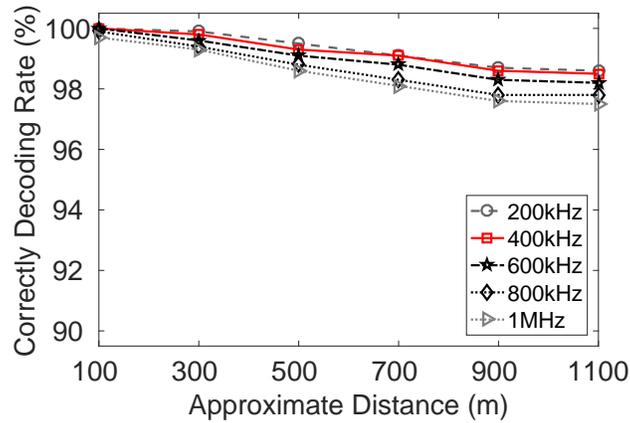}
      }
      \subfigure[Magnitude of overlap between subcarriers\label{fig:chap3-sub_overlap}]{
        \includegraphics[width=.5\textwidth]{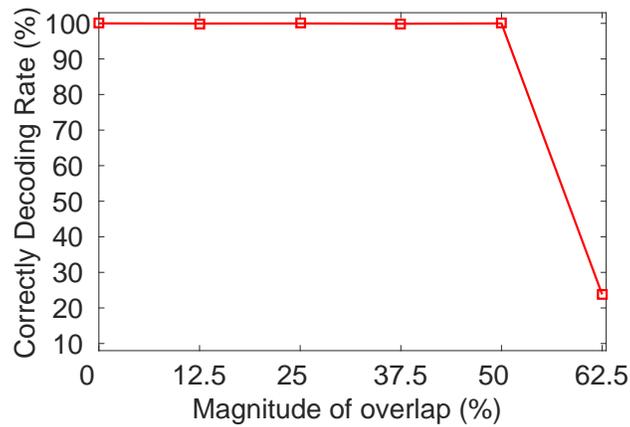}
      }
    \caption{Determining subcarriers}
    \label{fig:chap3-subcarrier}
  \end{figure}

 {\bf Subcarriers.}
 The maximum transmission bit rate $R$ of an AWGN channel of bandwidth $W'$ based on Shannon-Hartley Theorem is given by  $R=W'\log_2 (1+  \textit{SNR})$, where  {\slshape SNR} is the {\slshape Signal to Noise Ratio}. Based on Nyquist Theorem, $R=2W'\log_2 2^k$ where $k$ is the number of bits per symbol ($2^k$ being the number of signal levels) needed to support  bit rate $R$ for a noiseless channel. The 802.15.4 specification for lower frequency band, e.g., 430-434MHz band (IEEE 802.15.4c~\cite{802154c}), has a bit rate of 50kbps. We also aim to achieve this bit rate. We consider a minimum value of $3$dB for SNR in decoding. Taking into account default $SF=8$, we need to have $50*8$kbps bit rate in the medium. Thus, a subcarrier of bandwidth 200kHz can have a bit rate up to  $50*8$kbps  in the medium. Since  BPSK has $k=1$, it is theoretically sufficient for this bit rate and bandwidth under no noise. Using similar setup as the above, Figure~\ref{fig:chap3-sub_distance} shows the feasibility of various bandwidths. In our experiments, 400kHz bandwidth provides our required bit rate  under noise. Hence, we use 400kHz as our default subcarrier bandwidth. We have also experimentally found that our 400kHz subcarriers can safely overlap up to 50\% with the neighboring ones (Figure~\ref{fig:chap3-sub_overlap}).   In our low data rate communication using a spreading factor of 8 helps us mitigate the effects of any orthogonality violation.

%% file: chapter3/mac.tex
\subsection{Reliable MAC Protocol}\label{sec:chap3-mac}

We develop a low overhead MAC protocol for operating the nodes with greater freedom, low power, and reliability. As the nodes transmit asynchronously to the BS, implementing ACK for every transmission is extremely difficult. Considering a single half-duplex radio at each node and two half-duplex radios (both operating on the same spectrum) at the BS, we demonstrate that we can implement ACK immediately after a transmission  in concurrent and asynchronous scenario. Under such a design decision in SNOW, we can exploit the characteristics of our D-OFDM system to enable concurrent transmissions and receptions at the BS. 


%
%

\subsubsection{Location-Aware Spectrum Allocation}\label{subsubsec:chap3-allocation}

This BS spectrum  is split into  $n$ overlapping orthogonal subcarriers, each of equal width. Considering $w$ as the subcarrier bandwidth, $W$ as the total bandwidth at the BS, and $\alpha$ as the magnitude of overlap of the subcarriers (i.e., how much two neighboring subcarriers can overlap), the total number of orthogonal subcarriers $n =  \frac{W}{w \alpha} -1.$ For example, when $\alpha= 50\%$, $W$=6MHz, $w$=400kHz, we can have $n=29$ orthogonal subcarriers. Let us denote the subcarriers by $f_1, f_2, \cdots, f_n$. The BS can use a vector to maintain the status of these subcarriers by keeping their noise level or airtime utilization (considering their usage by surrounding networks), and can dynamically make some subcarrier available or unavailable.  Since our PHY design is capable of handling fragmented spectrum,  such dynamism at the MAC layer is feasible.  


The subcarrier allocation is done at the BS. Each node is assigned one subcarrier. Let $f(u)$ denote the subcarrier assigned to node $u$. When the number of nodes is no greater than the number of subcarriers, i.e. $N\le n$, every node is assigned a unique subcarrier.  Otherwise, a subcarrier is shared by more than one node. The subcarrier allocation will also try to minimize interference as well as contention among the nodes sharing the same subcarrier. Hence, our first goal is to try to assign different subcarriers to a pair of nodes that are {\bf hidden} to each other. That is, if two nodes $u$ and $v$ are hidden to each other, we try to meet the condition $f(u)\not=f(v)$. Our second goal is to ensure there is not excessive contention (among the nodes that are in communication range of each other) on some subcarrier compared to others. Let $H(u)$ denote the estimated set of nodes that are hidden terminal to $u$.  Note that  the BS is assumed to know the node locations  either through manual configuration or through some  existing WSN localization techniques such as those based on ultrasonic sensors or other sensing modalities~\cite{wsnlocalizationsurvey}. Localization is out of the scope of this paper. The BS can estimate $H(u)$ for any node $u$  based on the locations and estimated communication range of the nodes. Let the set of nodes that have been assigned subcarrier  $f_i$ be denoted by $\Omega(f_i)$.   In the beginning,  $\Omega(f_i)=\emptyset, ~\forall i$. For every node $u$ whose subcarrier has not been assigned, we do the following. We assign it a subcarrier such that  $| \Omega(f(u))\cap H(u) |$  is minimum. If there is more than one such subcarrier, then we select the one with minimum $|\Omega(f(u))|$.   This will reduce the impact of {\bf hidden terminal problem}.


%
%

\subsubsection{Transmission Policy}



In SNOW, the nodes transmit to the BS using a CSMA/CA approach. This approach gives us more flexibility and keeps the management more decentralized and energy efficient. Specifically, we do not need to adopt time synchronization, time slot allocation, or to preschedule the nodes. The nodes will sleep by turning off the radios and will turn the radios on (wake up) if they have data to send. After sending the data, a node will go back to sleep again. This will provide high energy-efficiency to the power constrained nodes. We adopt a simple CSMA/CA approach without any RTS/CTS frames. We will adopt a CSMA/CA policy similar to the one implemented in TinyOS~\cite{tinyos} for low power sensor nodes that uses a static interval for random back-off. Specifically, when a node has data to send, it wakes up by turning its radio on. Then it performs a random back-off in a fixed {\slshape initial back-off window}.  When the back-off timer expires, it runs CCA (Clear Channel Assessment) and if the subcarrier is clear, it transmits the data. If the subcarrier is occupied, then the node makes a random back-off in a fixed {\slshape congestion back-off window}. After this back-off expires, if the subcarrier is clean the node transmits immediately. This process is repeated until it makes the transmission. The node then can go to sleep again.


The BS station always remains awake to listen to nodes' requests. The nodes can send whenever they want. There can also be messages from the BS such as management message (e.g., network management, subcarrier reallocation, control message etc.). Hence, we adopt a periodic beacon approach for downward messages. Specifically, the BS periodically sends a beacon containing the needed information for each node through a single message. The nodes are informed of this period. Any node that wants/needs to listen to the BS message can wake up or remain awake (until the next message)  accordingly to listen to the BS. The nodes can wake up and sleep autonomously. Note that the BS can encode different data on different subcarriers, carrying different information on different subcarriers if needed, and send all those as a single OFDM message. As explained in Section~\ref{subsec:chap3-downward}, the message upon reception will be decoded in a distributed fashion at the nodes, each node decoding only the data carried in its subcarrier. 


\begin{figure}[!htbp]
\centering
\includegraphics[width=0.7\textwidth]{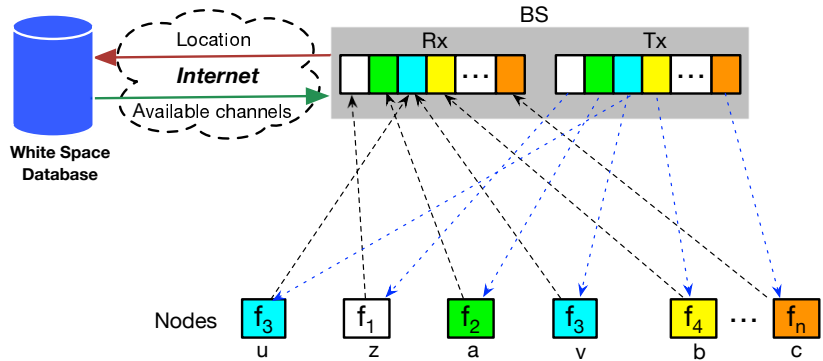}
\caption{SNOW architecture with dual radio BS \& subcarriers}
\label{fig:chap3-dualradio}
\end{figure}

%

\subsubsection{Reliability}

Sending ACK after every transmission is crucial but poses a number of {\bf  challenges}. {\bf First},   since the nodes asynchronously transmit, if the BS sends ACK after every reception, it may lose many packets from other nodes when it switches to Tx mode. {\bf Second}, the BS uses a wide channel while the node needing ACK uses only a narrow subcarrier of the channel. The AP needs to switch to that particular subcarrier which is expensive as such switching is needed after every packet reception. Note that the BS can receive many packets in parallel and asynchronously. Thus when and how these packets can be acknowledged is a {\bf difficult question}. We adopt a dual radio design at the BS of SNOW which is a practical choice as the BS is power-rich. Thus the BS will have two radios - one for only transmission, called {\bf\slshape Tx radio},  and the other for only reception, called {\bf\slshape Rx radio}. The Tx radio will make all transmissions whenever needed and can sleep when there is no Tx needed. The Rx radio will always remain in receive mode to receive packets. As shown in Figure~\ref{fig:chap3-dualradio}, both radios use the same spectrum and have the same subcarriers - the subcarriers in the Rx radio are for receiving while the same in the Tx radio are for transmitting. Such a dual radio BS design will allow us to enable $n$ concurrent transmissions and receptions. Since each node (non BS) has just a single half-duplex radio, it can be either receiving or transmitting, but not doing the both at a time. Thus if $k$ out of $n$ subcarriers are transmitting, the remaining $n-k$ subcarriers can be receiving, thereby making at most $n$ concurrent transmissions/receptions.

Handling ACK and two-way communication using a dual-radio BS still poses the following {\bf challenges}. {\bf First,} while the two radios at the BS are connected in the same module and the Tx radio can send an ACK immediately after a packet is received on the Rx radio, handling ACKs for asynchronous transmissions is a difficult problem in wireless domain. The radio needs to send ACK only to the nodes from which it received packet. Thus some subcarriers will need to have ACK frame while the remaining ones may carry nothing or some data packet. While our PHY design allows to handle this, the challenge is that some ACK/s can be due while the radio is already transmitting some ACK/s.  That is, while sending some ACK/s another packet's reception can be complete making its ACK due immediately. The key question is:  {\slshape ``How can we enable ACK immediately after a packet is received at the BS?"} {\bf Second,} another serious challenge is that the receptions at the Rx radio can be severely interfered by the ongoing transmissions at the Tx radio as both radios operate on the same spectrum and are close to each other.   {\bf Third,}  ACK on a subcarrier can be interfered if a node sharing it starts transmitting before the said ACK is complete.


{\bf Handling the above Challenges in SNOW.}
D-OFDM allows us to encode any data on any subcarrier while the radio is transmitting. Thus the design will allow us to encode any time on any number of subcarriers and enable ACKs to asynchronous transmissions.  When there is nothing to transmit, the Tx radio can sleep. Since a node has a single half-duplex radio, it will either transmit or receive. Let us first consider for a subcarrier which is assigned to only one node such as subcarrier $f_1$ in  Figure~\ref{fig:chap3-dualradio} which is assigned only to node $z$.  Node $z$ will be in receive mode (waiting for ACK) when the Tx radio at the BS sends ACK on $f_1$. Now consider for a subcarrier which is assigned to more than one node such as subcarrier $f_3$ in  Figure~\ref{fig:chap3-dualradio} which is assigned to two nodes,  $u$ and $v$. When $u$ is receiving ACK from the BS, if $v$ attempts to transmit it will sense the subcarrier busy due to BS's ACK on it and make random back-off. Thus any node sharing a subcarrier $f_i$ will not interfere an ACK on $f_i$. Hence, transmitting ACK on a subcarrier $f_i$ from the Tx radio has nothing to interfere at $f_i$ of the Rx radio at the BS. Subcarrier $f_i$ at Rx will be receiving the ACK on it sent by the Tx radio and can be ignored by the decoder at the Rx radio. Thus the subset of the subcarriers which are encoded with ACKs at the Tx radio will have energy. The remaining subcarriers that are not encoded with ACK or anything will have no energy due to OFDM design on the signal coming out from the Tx radio of the BS. During this time, the nodes may be transmitting on those subcarriers. Thus when the Tx radio transmits, its un-encoded subcarriers will have no energy and will not be interfering the same subcarriers at the Rx radio.  Thus receptions on those subcarriers at the Rx radio can continue without interference. The subcarriers carrying ACKs are orthogonal to them and will not interfere either.

%

\subsubsection{Other Features of The MAC Protocol}

{\bf Partially Handling Hidden Terminal.} 
We partially handle hidden terminal problem in subcarrier allocation and MAC protocol. Consider nodes $u$ and $v$ in Figure~\ref{fig:chap3-dualradio}  both of which are assigned subcarrier $f_3$. 
Now consider $u$ and $v$ are hidden to each other. When the TX radio of the BS sends ACK to node $u$ that has just made a transmission to the BS, this ACK signal will have high energy on the subcarrier $f_3$ at the Rx radio of the BS. At this time, if node $v$ makes a transmission to the BS, it will be interfered. Since $v$ will run CCA and sense the energy on $f_3$ it will not transmit. This result is somewhat similar to that of the CTS frame used in WiFi networks to combat hidden terminal problem. Specifically, based on the ACK frame sent by the BS, node $v$ decides not to transmit to avoid interference from the ACK of $u$'s transmission.

{\bf Peer-to-Peer Communication.} 
Two nodes that want to communicate can be hidden to each other or may have different subcarriers. Hence we realize peer-to-peer communication through the BS. For example, in Figure~\ref{fig:chap3-dualradio}, if node $a$ wants to send a packet to $b$, it cannot send directly as they use different subcarriers. Hence, $a$ will first transmit to the BS on subcarrier $f_2$, and then the BS will transmit on subcarrier $f_4$ to node $b$ in its next beacon time. 


{\bf Handling Various Dynamics.} 
{\bf First}, we handle {\bf spectrum dynamics} as follows. 
When the BS's spectrum availability changes due to primary user activity, the BS performs a new spectrum allocation. The nodes whose subcarriers may no more be available may have no way to get the new subcarrier allocation from the BS. We handle this by allocating one or more backup subcarriers (similar to backup whitespace channels adopted in~\cite{ws_sigcomm09}). If a node does not receive any beacon for a certain number of times, it will determine that its subcarrier is no more available and will switch to a backup subcarrier and wait for BS message. The BS will keep sending this rescue information on that  backup subcarrier which will thus be received by that node. For robustness, we maintain multiple such backup subcarriers. Another scenario can be the case when some subcarrier becomes overly noisy. To handle this, we adopt {\bf \slshape subcarrier swapping} among the nodes. The swapping will be done between bad ones only, not between good ones, not between good and bad ones (as some good subcarrier for a node may become bad after swapping). Exchanging between two nodes who are experiencing a high loss can result in good link quality.  

{\bf Second}, we share the loads among the subcarriers by reallocating or swapping. That is, if a subcarrier becomes congested we can un-assign some node from it and assign it a less congested one.  {\bf Third}, we adopt  {\bf node joining and leaving} by allocating some subcarriers for this purpose. When a new node wants to join the network, it uses  this join subcarrier to communicate with the BS. It can transmit its identity and location to the BS. The BS then checks the available white space and assigns it an available subcarrier. Similarly, any node from which the BS has not received any packet for a certain time window can be excluded from the network.

%% file: chapter3/implementation.tex
\subsection{Implementation}\label{sec:chap3-implementation}

We have implemented SNOW in GNU Radio~\cite{gnuradio} using USRP devices~\cite{usrp}. GNU Radio is software-defined radio toolkit~\cite{gnuradio}. USRP is a hardware platform to transmit and receive for software-defined radio~\cite{usrp}. We have used 9 USRP devices (2 at the BS and 7 as SNOW nodes) in our experiment. Two of our devices were USRP B210 while the remaining are USRP B200, each operating on band 70 MHz - 6GHz. The packets are generated in IEEE 802.15.4 structure with random payloads. We implement the decoder at the BS using 64-point G-FFT which is sufficient due to our limited number of devices. In downward communication, multiple parallel packet lines are modulated on the fly and fed into a {\slshape streams-to-vector} block that is fed into  IFFT that generates a composite time domain signal.  



%% file: chapter3/experiment.tex
\begin{table}[!htb]
\centering
\begin{tabular}{ll}
\hline
\textbf{Parameter}     & \textbf{Value}                                                                         \\ \hline
Frequency  Band        & 572-578MHz                                                                             \\ 
Orthogonal Frequencies & \begin{tabular}[c]{@{}l@{}}574.4, 574.6, 574.8, 575.0,\\ 575.2, 575.4, 575.6, 575.8MHz\end{tabular}  \\ 
Subcarrier modulation            & BPSK                                                                               \\ 
Packet Size            & 40 bytes                                                                               \\ 
BS Bandwidth           & 6MHz                                                                                   \\ 
Node Bandwidth         & 400kHz                                                                                 \\ 
Spreading Factor       & 8                                                                                      \\ 
Transmit (Tx) Power         & 0dBm                                                                                   \\ 
Receive Sensitivity    & -94dBm                                                                           \\ 
SNR                    & 6dB                                                                                    \\ 
Distance               & 1.1km                  \\ \hline
\end{tabular}
\vspace{0.1 in}
\caption{Default parameter settings}
\label{tab:chap3-default_param}
\end{table}

\subsection{Experiments}\label{sec:chap3-experiment}
To observe the performance of SNOW in various radio environments, we deployed it  in the Detroit (Michigan) metropolitan area, in an indoor environment, and in a rural area of Rolla (Missouri).  
Here, we describe our experimental results in these deployments. We also compare its performance with existing similar technologies. 

\subsubsection{Deployment in A Metropolitan City Area}\label{subsec:chap3-metropolitan}

\begin{figure}[!htb]
\centering
\includegraphics[width=0.7\textwidth]{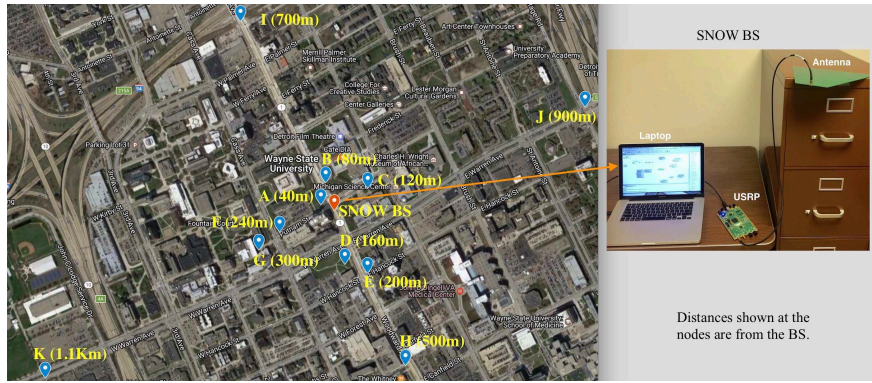}
\caption{Node positions in the Detroit metropolitan area.}
\label{fig:chap3-outdoor}
\end{figure}

{\bf Setup.}
Figure~\ref{fig:chap3-outdoor} shows different nodes and the BS positions in this setting in the Detroit Metropolitan Area. Due to varying distances (max. $\approx 1.1$km) and obstacles between the BS and these nodes, the SNR of received signals varies across these node positions. We keep all of the antenna heights at approximately 5ft above the ground. Unless mentioned otherwise, Table~\ref{tab:chap3-default_param} shows the default parameter settings for all of the experiments. 

{\bf Reliability over Distances and Tx Power.} 
\begin{figure}[!htb]
    \centering
      \subfigure[Uplink reliability\label{fig:chap3-reliability_bs_distance}]{
    \includegraphics[width=0.5\textwidth]{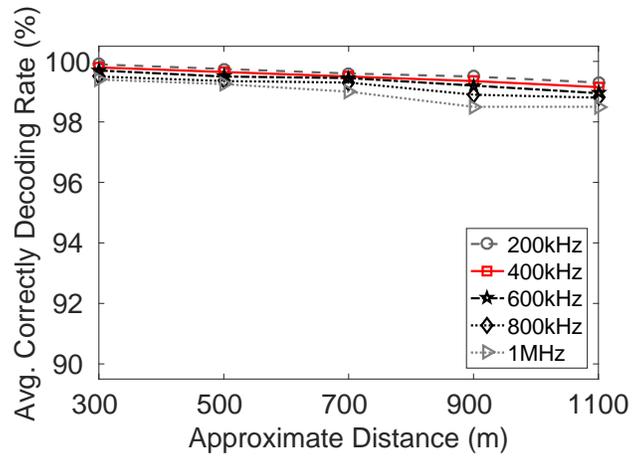}
      }
    \hfill
      \subfigure[Downlink reliability\label{fig:chap3-reliability_node_distance}]{
        \includegraphics[width=.5\textwidth]{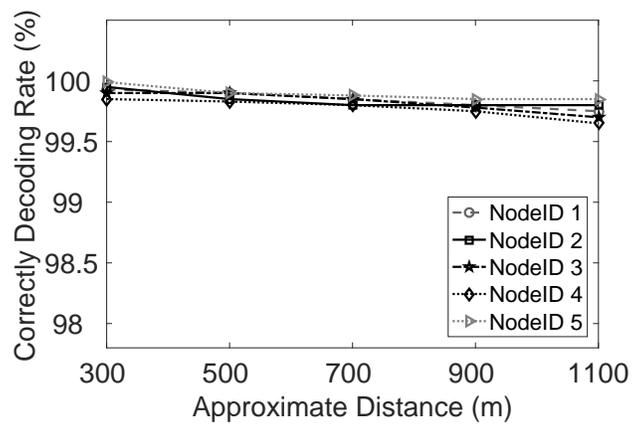}
      }
      \hfill
      \subfigure[Distance with varying Tx powers\label{fig:chap3-distance_txpower}]{
        \includegraphics[width=.5\textwidth]{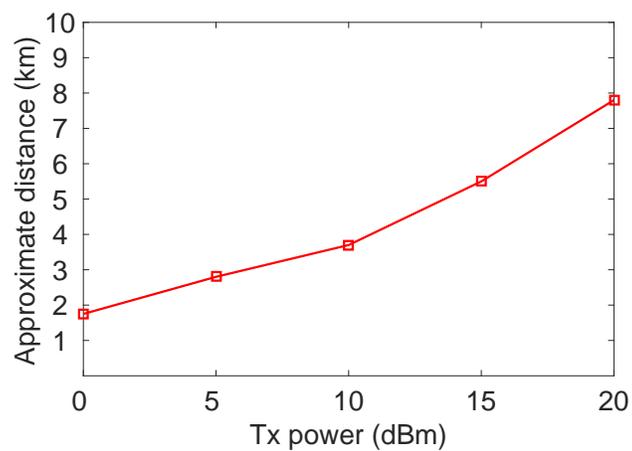}
      }
    \caption{Reliability over distances and varying Tx power.}
    \label{fig:chap3-reliability_bs_node}
 \end{figure}
To demonstrate the reliability at various distances, we place all the nodes at 300m, 500m, 700m, 900m, and 1100m away from the BS, respectively. At each distance, each node transmits 10,000 packets asynchronously to the BS and vice versa. CDR (which indicates the correctly decoding rate as defined in Section~\ref{subsubsec:chap3-para}) is used as a key metric in our evaluation. 
Figure~\ref{fig:chap3-reliability_bs_distance} demonstrates uplink reliability under varying subcarrier bandwidths when the nodes are at different distances from the BS and all transmit at 0dBm. Specifically, with 400kHz of subcarrier bandwidth, the BS can decode on average ~99.15\% of packets from all of the nodes that are 1.1km away. Also, for all other subcarrier bandwidths, the average CDR at the BS stays above 98.5\% at all distances. Similarly, Figure~\ref{fig:chap3-reliability_node_distance} demonstrates high reliability in downlink under varying distances. As shown at five different nodes for subcarrier bandwidth of 400kHz, all the nodes can decode more than 99.5\% of the packets even though they are 1.1km apart from the BS. 


With 0dBm (maximum in WSNs based on IEEE 802.15.4) of Tx power and receiver sensitivity of -94dBm (typical sensor devices), we limited our maximum distance between the BS and a node to 1.1km with high reliability. To demonstrate the feasibility of adopting SNOW in LPWAN, we moved one node much farther away from the BS and vary the Tx power from 0 dBm up to 20 dBm. As shown in Figure~\ref{fig:chap3-distance_txpower}, with 20 dBm of Tx power, SNOW BS can decode from approximately 8km away, hence showing its competences for LPWAN technologies.

\begin{figure}[!htb]
\centering
\includegraphics[width=0.5\textwidth]{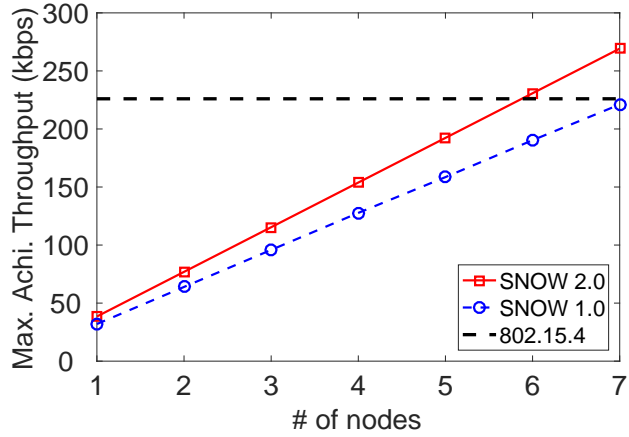}
\caption{ Maximum achievable throughput}
\label{fig:chap3-max_throughput}
\end{figure}

{\bf Maximum Achievable Throughput.}
In this experiment, we compare the {\em maximum achievable throughput} (i.e., maximum total bits that the BS can receive per second)  between the new SNOW design (SNOW 2.0) and the earlier design (SNOW 1.0).  We add ACK capability to SNOW 1.0 in its downward phase where the BS will switch to each node's subcarrier one after another and send an ACK for all transmissions the BS received in the last upward phase from that node. As soon as all ACKs are sent, SNOW 1.0 will switch to upward phase for receiving again from the nodes as we want to measure its maximum achievable throughput by adding ACK. In SNOW 1.0, the upward phase duration was set to 10s. In both of the networks, each node transmits 10,000 40-byte packets. In SNOW 2.0, after each transmission a node waits for its ACK (hence it does not continuously transmit).

 Figure~\ref{fig:chap3-max_throughput} shows that SNOW 2.0 achieves approximately 270kbps compared to 220kbps in SNOW 1.0 when 7 nodes transmit. For better understanding of the maximum achievable throughput, we also draw a baseline, maximum achievable throughput in a typical IEEE 802.15.4 based WSN of 250kbps bit rate. Its maximum achievable throughput is shown considering ACK after each transmission. As expected, the number of nodes does not impact  its maximum achievable throughput as its BS can receive at most one packet at a time. Note that a channel in the IEEE 802.15.4 based network is much wider than a SNOW subcarrier and has a higher bit rate  (250kbps vs 50kbps). Hence, both SNOW 2.0 and SNOW 1.0 surpass the baseline when the number of nodes is 7 or more. But the SNOW throughput keeps increasing linearly with the number of nodes while that in the baseline remains unchanged. Thus, although we have results for up to $7$ nodes, the linear increase in SNOW throughput gives a clear message that it is  superior in throughput and scalability to any protocol used for traditional WSN. Due to a small number of nodes, the throughput improvement of SNOW 2.0  over SNOW 1.0 is not well-visible. Later, in simulation, we show that SNOW 2.0 significantly outperforms SNOW 1.0 in terms of throughput.


{\bf Energy Consumption and Latency.}\label{subsec:chap3-amac_convergecast}
To demonstrate the efficiency in terms of energy and latency,  we compare SNOW 2.0 with a traditional WSN design. Specifically, we consider A-MAC~\cite{dutta2010design} which is an energy efficient MAC protocol for IEEE 802.15.4 based WSN that operates on 2.4GHz band. To estimate the energy consumption and network latency in SNOW 2.0 nodes, we place 7 nodes each 280m apart from the  BS. To compare fairly, we place A-MAC nodes 40m apart from each other making a linear multi-hop network due to their shorter communication ranges. In both of the networks, we start a convergecast after every 60 seconds. That is, each node except the BS generates a packet every 60 seconds that is ready to be transmitted immediately. Our objective is to collect all the packets at the BS.
\begin{figure}[!htb]
    \centering
    \subfigure[Energy consumption\label{fig:chap3-energy_s1}]{
    \includegraphics[width=0.5\textwidth]{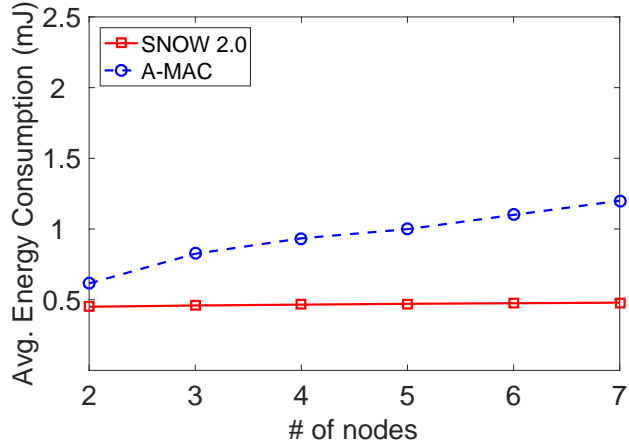}
      }
    \hfill
      \subfigure[Total latency\label{fig:chap3-latency_s1}]{
        \includegraphics[width=.5\textwidth]{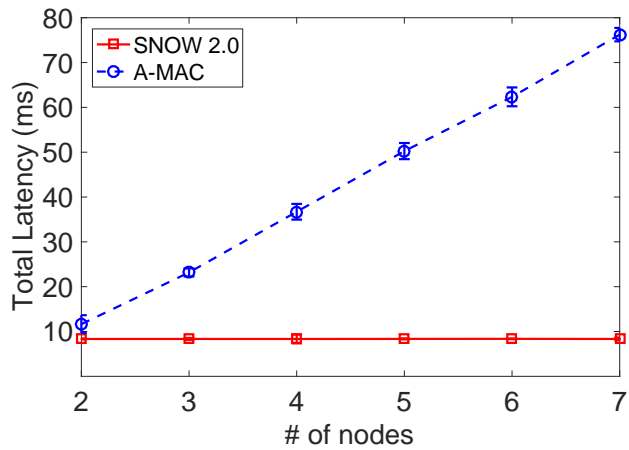}
      }
    \caption{Energy consumption and latency in convergecast}
    \label{fig:chap3-energy_latency_s1}
  \end{figure}
  
  
Since the USRP devices do not provide any energy consumption information, we use the energy model of CC1070 by Texas Instruments~\cite{cc1070}. This off-the-shelf radio chip operates in low frequencies near TV white spaces and also uses BPSK modulation. Table~\ref{tab:chap3-cc1070} shows the energy model of CC1070. Since the BS is line-powered, we keep it out of the energy calculation. We run multiple rounds of convergecast for 2 hours in both of the networks. Figure~\ref{fig:chap3-energy_s1} shows the average energy consumption in each node per convergecast. Regardless of the number of nodes, on average a SNOW 2.0 node consumes nearly 0.46mJ energy. On the other hand in A-MAC, on average each node consumes nearly 1.2mJ when 7 nodes participate in convergecast. In practice, with a large number of nodes, A-MAC node consumes significant amount of energy as we found in~\cite{snow}.  Figure~\ref{fig:chap3-latency_s1} shows the convergecast latency in both SNOW 2.0 and A-MAC. We calculate the total time to collect all the packets at the BS from all the nodes. SNOW 2.0 takes approximately 8.3ms while A-MAC takes nearly 77ms to collect packets from all 7 nodes. Theoretically, SNOW 2.0 should take almost constant amount of time to collect all the packets as long as the number of nodes is no greater than that of available subcarriers. Again, 
due to a small network size, the differences between SNOW 2.0 and A-MAC are not significant in this experiment.   

\begin{table}[!htb]
\centering
\begin{tabular}{ll}
\hline
\textbf{Device mode} & \textbf{\begin{tabular}[c]{@{}l@{}}Current Consumption\end{tabular}} \\ \hline
Tx                   & 17.5 mA                                                                                      \\ 
Rx                   & 18.8 mA                                                                                      \\ 
Idle                 & 0.5 mA                                                                                        \\ 
Sleep                & 0.2 $\mu$A                                                                                     \\ \hline
\end{tabular}\vspace{0.1in}
\caption{Current consumption in CC1070}
\label{tab:chap3-cc1070}
\end{table}

\begin{figure}[!htb]
    \centering
      \subfigure[Energy consumption\label{fig:chap3-energy_s2}]{
    \includegraphics[width=0.5\textwidth]{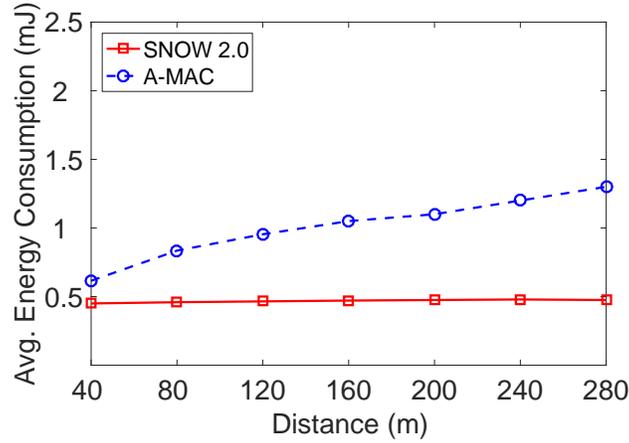}
      }
    \hfill
      \subfigure[Total latency\label{fig:chap3-latency_s2}]{
        \includegraphics[width=.5\textwidth]{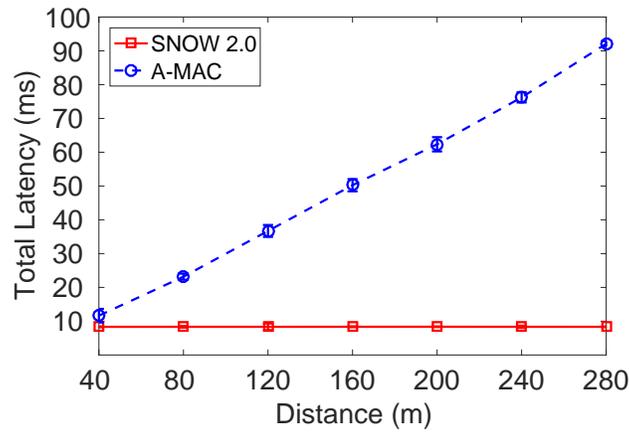}
      }
    \caption{Energy consumption and latency over distance}
    \label{fig:chap3-energy_latency_s2}
  \end{figure}

{\bf Energy Consumption and Latency over Distances.} With the same setups from previous Section~\ref{subsec:chap3-amac_convergecast}, Figure~\ref{fig:chap3-energy_latency_s2} demonstrates the energy and latency comparison between SNOW 2.0 and A-MAC with respect to distances. Figure~\ref{fig:chap3-energy_s2} shows that, a node in SNOW 2.0 consumes on average 0.475mJ of energy to deliver a packet to the BS that is 280m away. On the other hand, an A-MAC node consumes nearly 1.3mJ of energy to deliver one packet to a sink that is 280m away. Also, Figure~\ref{fig:chap3-latency_s2} shows that a SNOW 2.0 and A-MAC node takes 8.33ms and 92.1ms of latency to deliver one packet to the BS, respectively. As the distance increases, the differences become higher, demonstrating SNOW's superiority. 

\begin{figure}[!htb]
\centering
\includegraphics[width=0.5\textwidth]{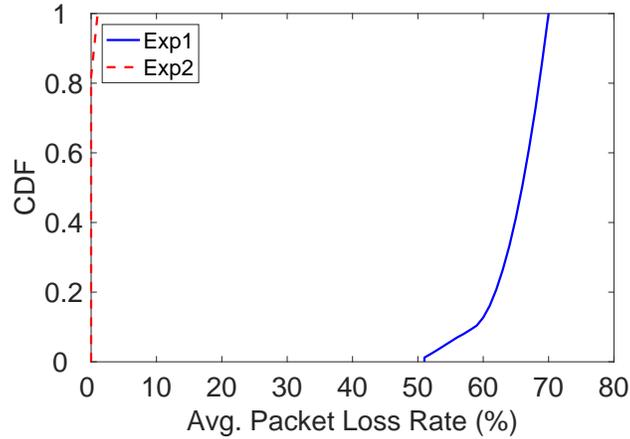}
\caption{Performance under hidden terminals}
\label{fig:chap3-hidden_terminal}
\end{figure}

{\bf Handling Hidden Terminal Problem.}
To test the performance of SNOW 2.0 under hidden terminal, we adjust the Tx powers of the nodes at the positions shown in Figure~\ref{fig:chap3-outdoor} so that (i) nodes A, B and C are hidden to nodes D and E; (ii) D and E are not hidden to each other; (iii) A, B and C are not hidden to each other. We conduct two experiments.  In  experiment 1 (Exp1), the hidden nodes are assigned the same subcarriers. For example, BS assigns one subcarrier to node A and D (hidden to each other), another subcarrier to nodes B, D and E (B is hidden to D and E). In experiment 2 (Exp2), the BS assigns different subcarriers to the nodes hidden to each other. Exp2 reflects the SNOW 2.0 MAC protocol. Each node sends 100 packets to the BS in both experimental setups. After getting the ACK for each packet (or, waiting until ACK reception time), each node sleeps for a random time interval between 0-50ms. After sending 100 packets, each node calculates its \emph{packet loss rate} and we average it. We repeat this experiment for 2 hours. Figure~\ref{fig:chap3-hidden_terminal} shows the CDF of average packet loss in experiments 1 and 2. In Exp1, average packet loss rate is 65\%, for SNOW 2.0 MAC protocol (Exp2) it is 0.9\%, which demonstrates the benefits of combining location-aware subcarrier allocation in SNOW 2.0 MAC. 

\begin{figure}[!htb]
\centering
\includegraphics[width=0.5\textwidth]{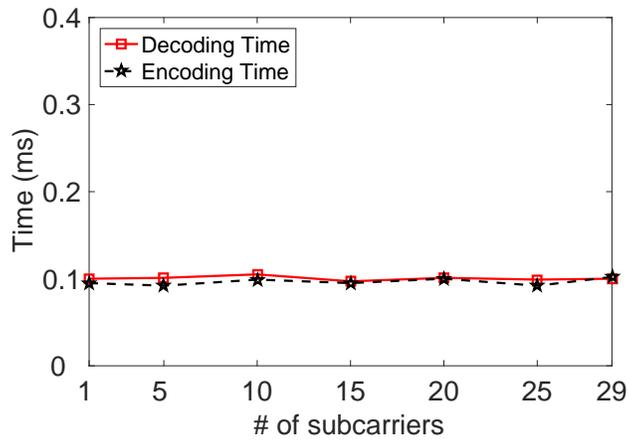}
\caption{Encoding and decoding time at BS}
\label{fig:chap3-encode_decode}
\end{figure}

{\bf BS Encoding Time and Decoding Time.}\label{subsubsec:chap3-time}
While we have seven USRP devices to act as SNOW nodes, we can calculate the data encoding time or decoding time in all 29 subcarriers (in a 6MHz TV channel) at the BS as it depends on the number of bins in the IFFT algorithm. Theoretically, the encoding/decoding time for any number of nodes at the BS should be constant as the IFFT/G-FFT algorithm runs with the same number of bins every time. However, we do separate experiments by encoding/decoding data to/from 1 to 29 nodes. We run each experiment for 10 minutes and record the time needed in the worst case. Figure~\ref{fig:chap3-encode_decode} shows that both encoding time and decoding time are within  0.1ms. This encoding/decoding time is very fast as IFFT/G-FFT runs very fast. Thus our BS encoding/decoding time is almost similar to standard encoding/decoding time for one packet in typical WSN devices.

\begin{figure}[!htb]
\centering
\includegraphics[width=0.5\textwidth]{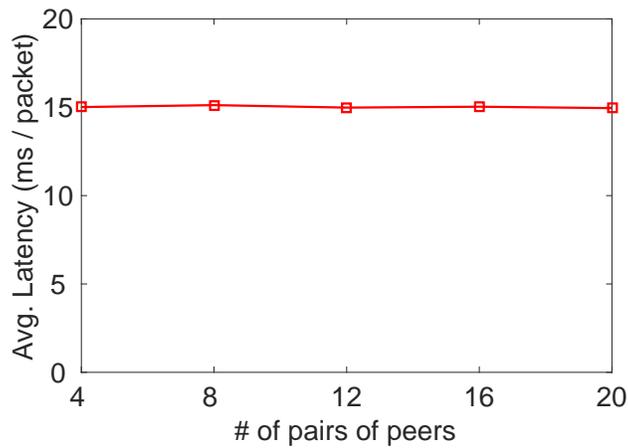}
\caption{Peer-to-peer avg. packet delivery time}
\label{fig:chap3-p2p}
\end{figure}


{\bf Handling Parallel Peer-to-Peer Communication.}
In this experiment, we aim to show the feasibility of parallel peer-to-peer communications in SNOW 2.0. This kind of scenarios are common in wireless control~\cite{WirelessHART}. Having seven SNOW nodes, we generate different numbers of pairs of peers. In each pair of peers, one node delivers 1000 40-byte packets to the other via BS. Figure~\ref{fig:chap3-p2p} shows that the average latency for       one peer-to-peer packet delivery remains within 15ms. While we tested up to 20 pairs, we can expect similar latency  as long as the number of pairs $\le$ the number of subcarriers. Thus, SNOW 2.0 can be a feasible platform even for applications that rely on peer-to-peer communication.

\begin{figure}[!htb]
\centering
\includegraphics[width=0.5\textwidth]{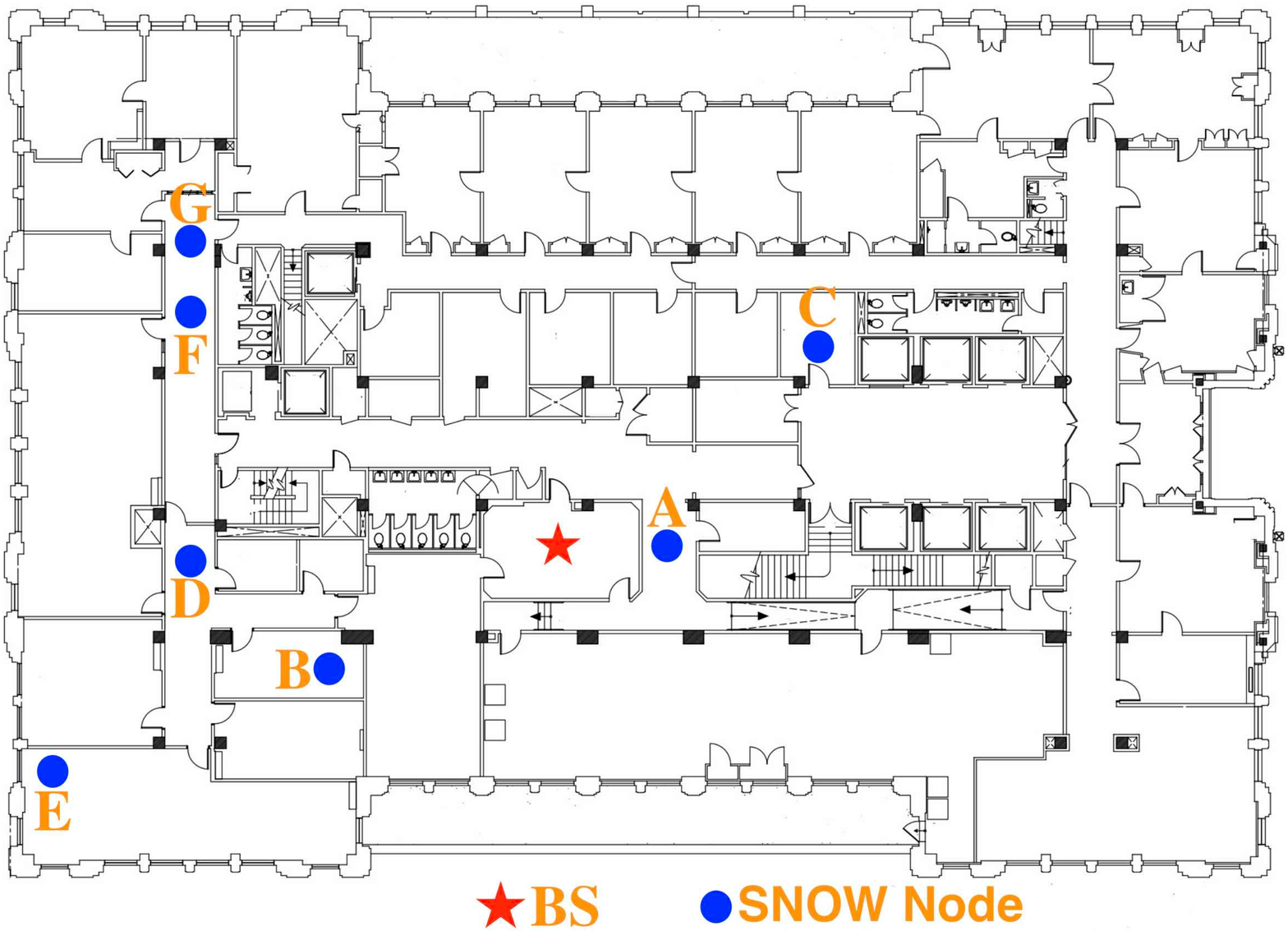}
\caption{Indoor node positions}
\label{fig:chap3-indoor_setup}
\end{figure}
\subsubsection{Indoor Deployment}
{\bf Setup.}
Figure~\ref{fig:chap3-indoor_setup} shows the positions of the SNOW nodes and BS (on floor plan)  all on the same floor (293,000 sq ft) of the Computer Science Building at Wayne State University. We fixed the position of the BS (receiver) while changing the positions of the node. In this experiment a node transmits 10,000 consecutive packets at each position.
\begin{figure}[htb]
    \centering
      \subfigure[Reliability at various SNR\label{fig:chap3-snr_indoor}]{
    \includegraphics[width=0.5\textwidth]{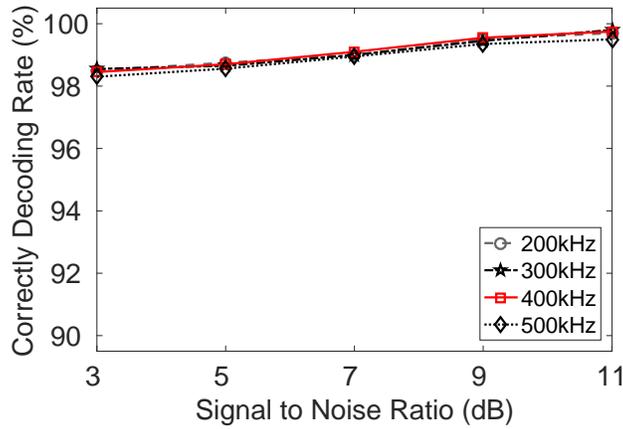}
      }
    \hfill
      \subfigure[Propagation through walls \label{fig:chap3-walls_indoor}]{
        \includegraphics[width=.5\textwidth]{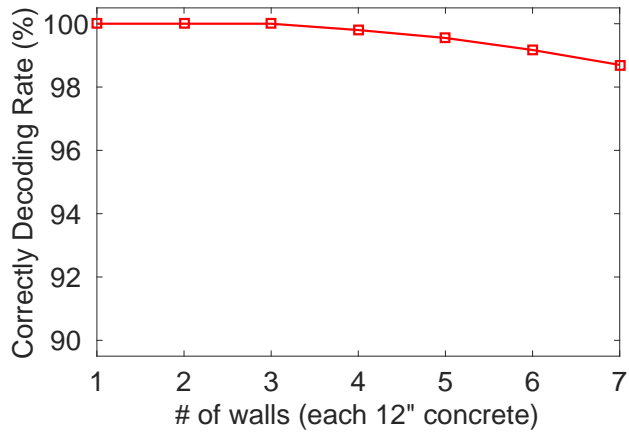}
      }
    \caption{Reliability in indoor environments}
    \label{fig:chap3-indoor}
\end{figure}

{\bf Results.}
Figure~\ref{fig:chap3-snr_indoor} shows the CDR over various SNR conditions under varying subcarrier bandwidths. At SNR of 3dB the CDR is around 98.5\% for all subcarrier bandwidths. We observe that while increasing the SNR, the CDR increases accordingly for all subcarrier bandwidth. This is due to the effect of noise, obstacles, and multipath over SNR.  Figure~\ref{fig:chap3-walls_indoor} shows CDR under varying number of walls between sender and receiver. We achieve at least 98.5\% CDR when the line of sight is obstructed by up to 7 walls (each $12^{\verb+"+}$ concrete). Due to low frequency and narrow bandwidth, SNOW 2.0 can reliably communicate in indoor environments.   

\begin{figure}[!htb]
    \centering
     \subfigure[Reliability vs Tx power\label{fig:chap3-rural_tx_power}]{
    \includegraphics[width=0.5\textwidth]{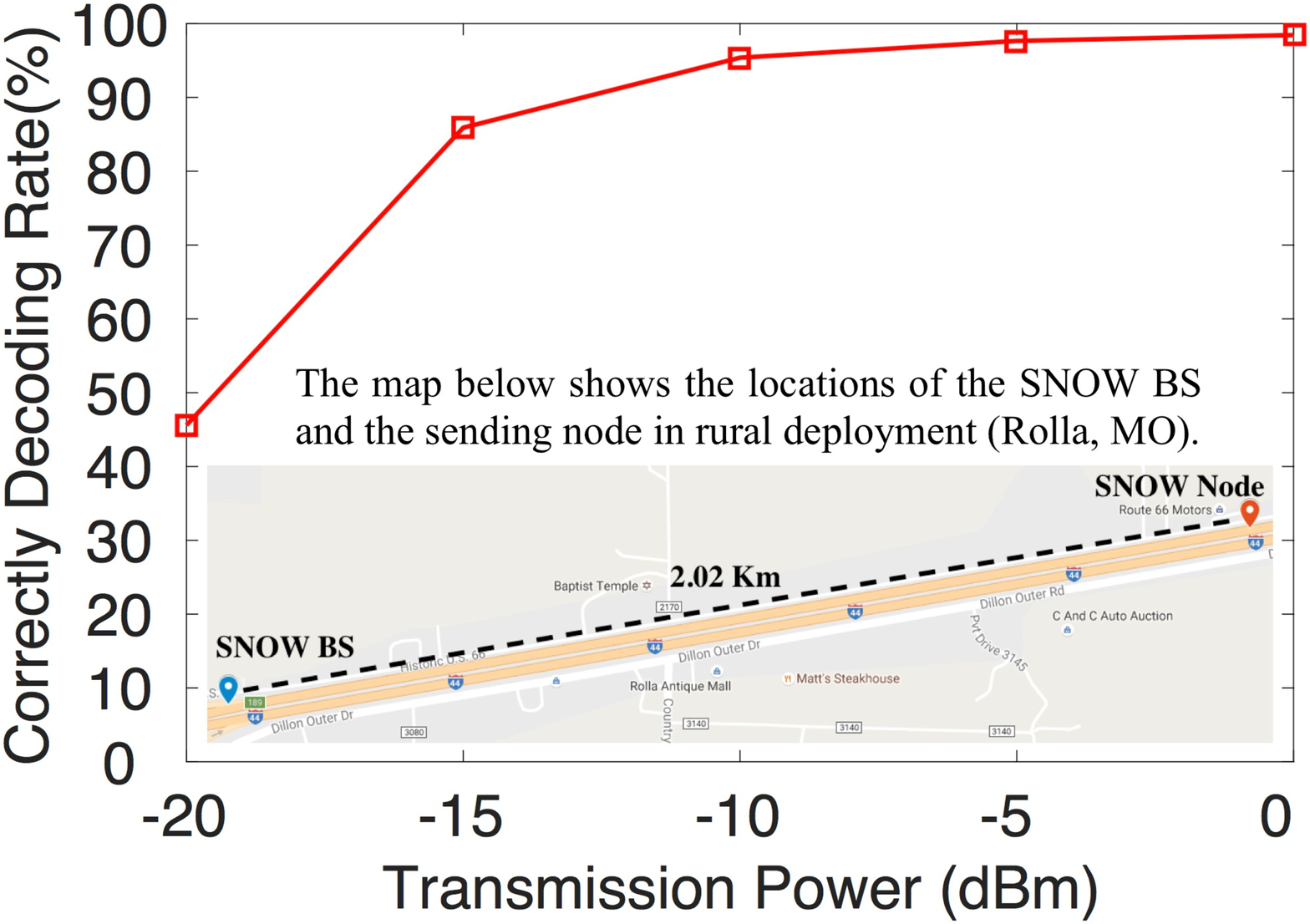}
      }
       \hfill
      \subfigure[BER over distances\label{fig:chap3-sf_distance_rural}]{
        \includegraphics[width=.5\textwidth]{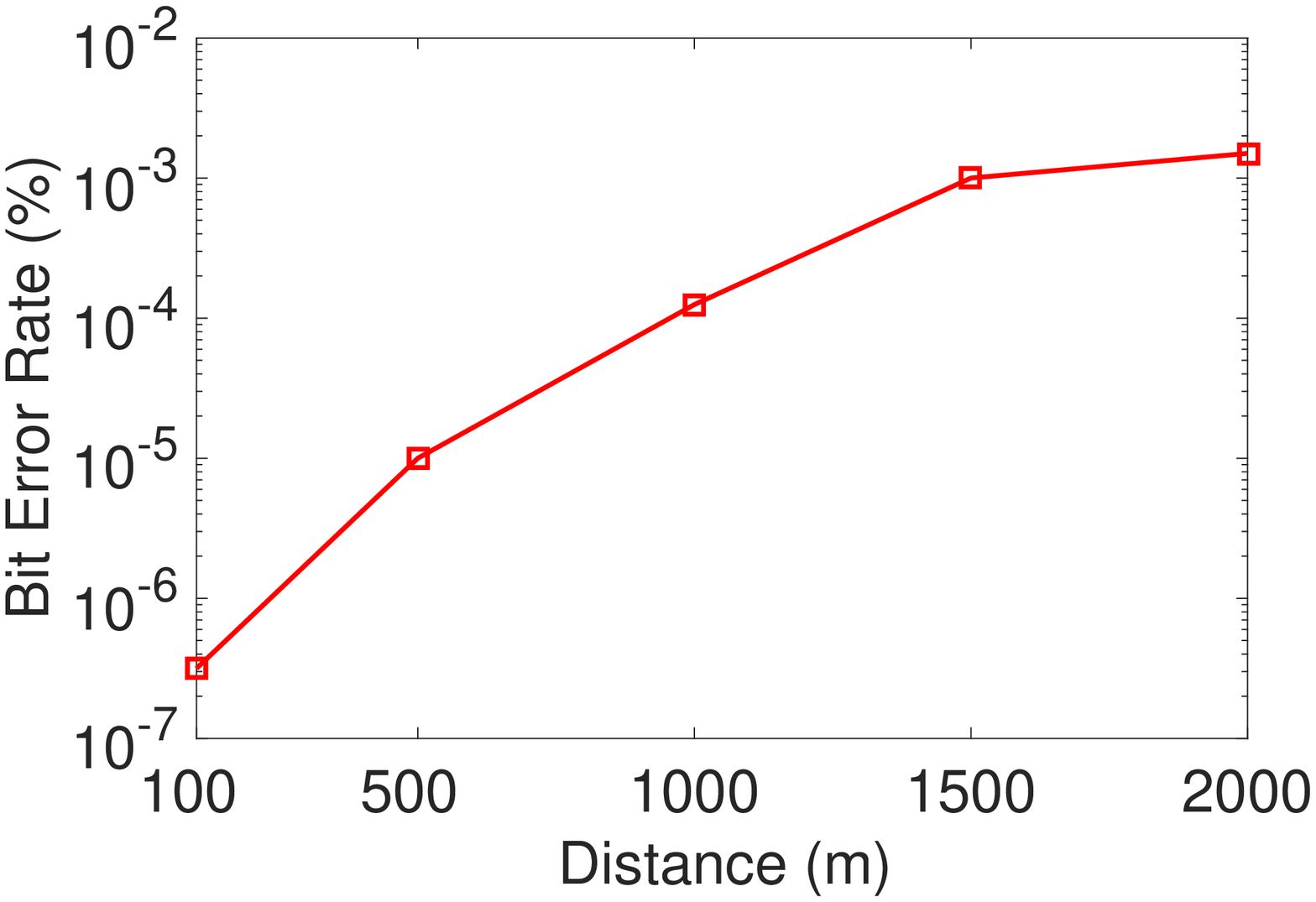}
      }
    \hfill
           \subfigure[Throughput vs bandwidth\label{fig:chap3-bs_throughput_rural}]{
  \includegraphics[width=0.5\textwidth]{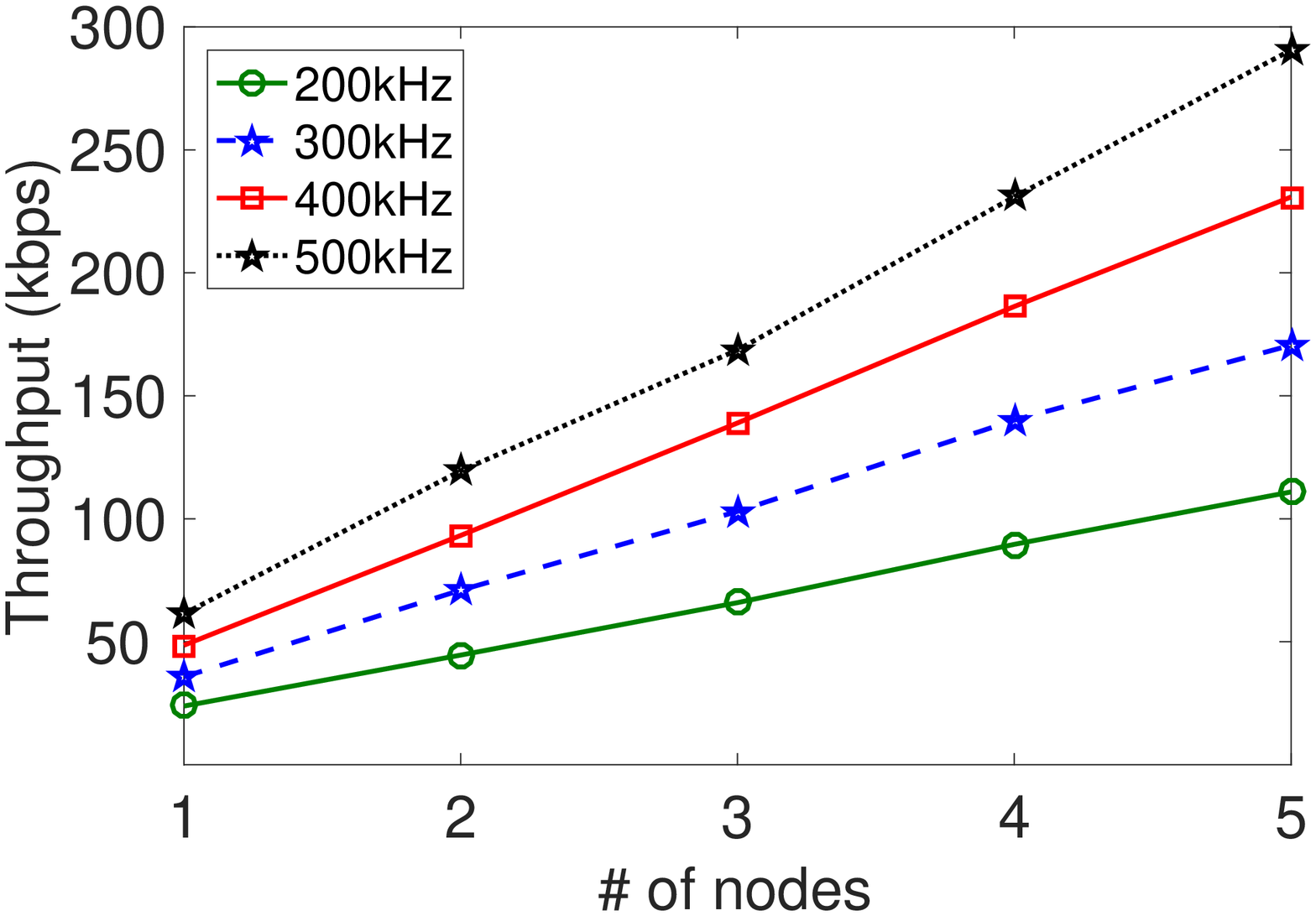}
      }
    \caption{Performance of SNOW 2.0 in rural deployment}
    \label{fig:chap3-rural}
\end{figure}

\subsubsection{Deployment in A Rural Area}


{\bf Setup.} A rural deployment of SNOW is characterized by two key advantages - higher availability of TV white spaces and longer communication range due to lesser absence of obstacles such as buildings. We deployed SNOW 2.0 in a rural area of Rolla, Missouri. We used five USRP devices that acted as SNOW nodes. We follow the similar antenna and default parameter setup as described in Section~\ref{subsec:chap3-metropolitan} and Table~\ref{tab:chap3-default_param}.



{\bf Distance, Reliability, and Throughput.}
The map embedded in Figure~\ref{fig:chap3-rural_tx_power} shows the locations of the BS and a node 2km away from the BS. The node transmits 1000 40-byte packets consecutively. The same figure shows the reliability (in terms of CDR) of the link under varying Tx power. Specifically, SNOW 2.0 achieves 2km+ communication range  at only 0 dBm Tx power which is almost double that we observed in our urban deployment. This happens due to a cleaner light of sight in the former. Similarly, Figure~\ref{fig:chap3-sf_distance_rural} shows the BER at the SNOW BS while decoding packets from various distances. The results show the decodability of the packets transmitted (at 0dBm) from 2km away as BER remains $\le 10^{-3}$.  As expected like in our urban deployment, here also SNOW's maximum achievable throughput linearly increases as we increase the number of nodes (Figure~\ref{fig:chap3-bs_throughput_rural}).


%% file: chapter3/simulation.tex
\begin{figure}[!htb]
    \centering    
    \subfigure[Throughput\label{fig:chap3-throughput_sim_vs_snow}]{
        \includegraphics[width=.5\textwidth]{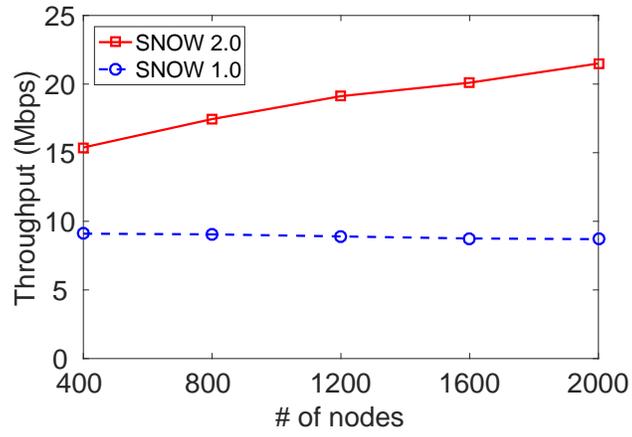}
      }
         \hfill
      \subfigure[Energy Consumption\label{fig:chap3-arrows_vs_snow_energy_sim}]{
        \includegraphics[width=.5\textwidth]{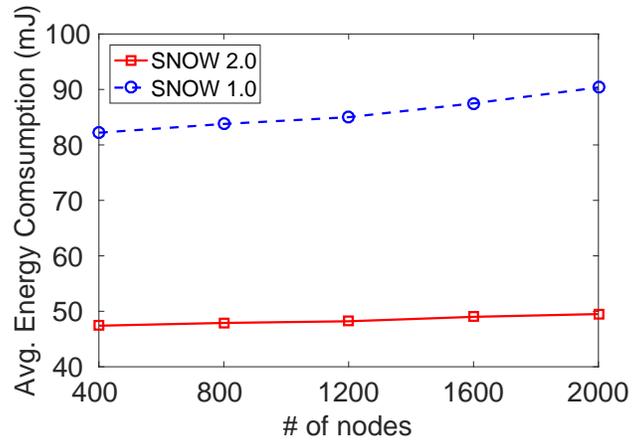}
      }
    \hfill
      \subfigure[Latency\label{fig:chap3-arrows_vs_snow_latency_sim}]{
        \includegraphics[width=.5\textwidth]{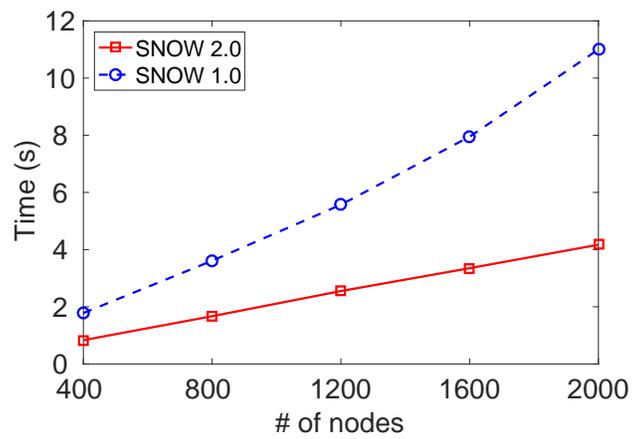}
      }
    \caption{SNOW 2.0 vs SNOW 1.0}    
    \label{fig:chap3-snow-comparison}
\end{figure}

\subsection{Simulation}\label{sec:chap3-simulation}
For large-scale evaluation of SNOW 2.0, we perform simulations in QualNet~\cite{qualnet}. We compare its performance with SNOW 1.0 and LoRa~\cite{lorawan}. 
Note that both SNOW 1.0 and SNOW 2.0 take the advantage of wide white space spectrum while LoRa operates in limited ISM band. Hence, for a fair comparison, we 
compare SNOW 2.0 with SNOW 1.0 and LoRa separately under different setups.


\subsubsection{Comparison with SNOW 1.0}\label{subsec:chap3-snowarrow}

{\bf Setup.} 
For both SNOW 2.0 and SNOW 1.0, we consider 81MHz of BS bandwidth and split it into 400 overlapping (50\%) orthogonal subcarriers each of 400kHz wide. We create a single-hop star network for both. Nodes are distributed within 2km radius of the BS. Then we use a setup similar to Section~\ref{subsec:chap3-metropolitan} for SNOW 2.0 and SNOW 1.0 MAC protocols. Here, the upward phase duration for SNOW 1.0 was set to 1s. In both networks, each node sends 100 40-byte packets and we calculate the throughputs at the BS, average energy consumption per node, and total time needed to collect all packets. As SNOW 1.0 cannot enable per-transmission ACK, we include ACK in SNOW 1.0 after completing upward phase for fair comparison. Thus, when a node sends a packet to its BS the node waits until the end of upward period to receive the ACK.

{\bf Results.} 
Figure~\ref{fig:chap3-throughput_sim_vs_snow} shows that the throughput of SNOW 2.0 is almost double that of SNOW 1.0 under varying number of nodes. Throughput in SNOW 1.0 increases slowly due to the longer downward communication cycles for delivering all the ACKs.  In contrast,  SNOW 2.0 can deliver per transmission ACK to each asynchronous transmission and its throughput increases almost linearly with increase in the number of nodes. (We acknowledge that SNOW 1.0 throughput would be similar to ours without ACK but using ACK for wireless communication is quite critical and for a fair comparison we include ACK in SNOW 1.0.) For the same reason, both the energy consumption and latency in SNOW 1.0 are almost two times that in SNOW 2.0 (Figures~\ref{fig:chap3-arrows_vs_snow_energy_sim} and~\ref{fig:chap3-arrows_vs_snow_latency_sim}). This demonstrates the superiority of SNOW 2.0 over SNOW 1.0 in terms of scalability.

%


\subsubsection{Comparison with LoRa}
\begin{figure}[htb]
    \centering
      \subfigure[Energy Consumption\label{fig:chap3-energy_simulation}]{
        \includegraphics[width=.5\textwidth]{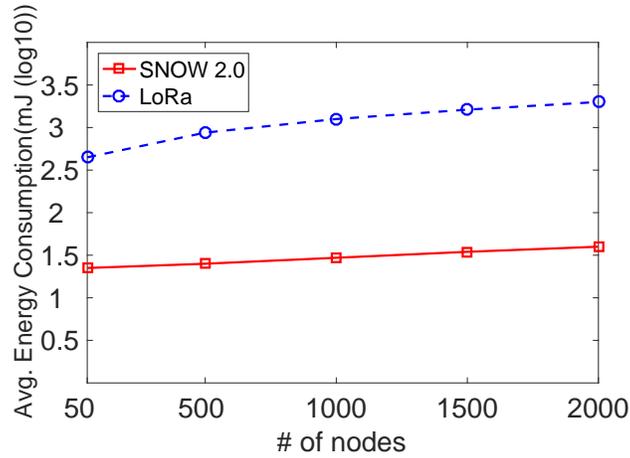}
      }
    \hfill
      \subfigure[Latency\label{fig:chap3-latency_simulation}]{
        \includegraphics[width=.5\textwidth]{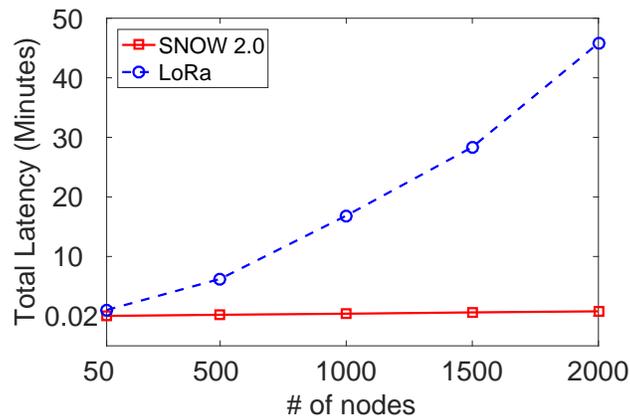}
      }
    \caption{SNOW 2.0 vs LoRa}    
    \label{fig:chap3-lora-comparison}
\end{figure}

{\bf Setup.} 
We consider a LoRa gateway with 8  parallel demodulation paths, each of 500kHz wide (e.g. Semtech SX1301~\cite{sx1301}). For fair comparison, we choose a BS bandwidth of $500kHz * 8$ = 4MHz from white spaces in SNOW 2.0 and split into 19 overlapping (50\%) orthogonal subcarriers, each of 400kHz wide. For each, we create a single-hop star network. All the nodes are within 2km radius of the BS/gateway. We generate various number of nodes in both of the networks. The nodes are distributed evenly in each demodulator path of LoRa gateway. In each demodulator path, LoRa uses the pure ALOHA MAC protocol.  In each network, we perform convergecast. Every node sends 100 40-byte packets with same spreading factor of 8 to the BS/gateway and sleeps for 100ms afterwards. For LoRa, we calculate the airtime of a 40-byte packet (34.94ms) using Lora-calculator~\cite{loracalc} and use it in simulation. For its energy profiling, we consider the LoRa iM880B-L~\cite{im880b} radio chip with its minimum supported Tx power of approximately 5dBm.


{\bf Results.} 
As the superiority of SNOW 1.0 over LoRa in terms of throughput was numerically demonstrated in~\cite{snow} and we have already demonstrated the superiority of SNOW 2.0 over SNOW 1.0 in Section~\ref{subsec:chap3-snowarrow}, here we compare them only in terms of energy consumption and latency.  As shown in Figure~\ref{fig:chap3-energy_simulation} (in $\log_{10}$ scale), for a  2000-node network, the packets are collected at the SNOW BS within  0.79 minutes consuming  22.22mJoule of average energy per node while that  are collected at the LoRa gateway within  45.81 minutes consuming  450.56mJoules of average energy per node. Both energy consumption and latency in SNOW 2.0 are much less since it allows 19 nodes to transmit in parallel, while only 8 nodes can transmit concurrently in LoRa. The MAC protocols in both networks also play role. Our results show that, using the same bandwidth, SNOW 2.0 can support a larger set of nodes.  



%% file: chapter4.tex
\section{LPWAN in the TV White Spaces: A Practical Implementation and Deployment Experiences}   \label{chap:snow3}

Low-Power Wide-Area Network (LPWAN) is an enabling Internet of Things (IoT) technology that supports long-range, low-power, and low-cost connectivity to numerous devices. 
To avoid the crowd in the limited ISM band (where most LPWANs operate) and the cost of licensed band, the recently proposed SNOW (Sensor Network over White Spaces) is a promising LPWAN platform that operates over the TV white spaces. As it is a very recent technology and is still in its infancy, the current SNOW implementation uses USRP devices as LPWAN nodes which have high cost ($\approx$ \$750 USD per device) and large form-factor, hindering its applicability in practical deployment.
In this paper, we implement SNOW using low-cost, low form-factor, low-power, and widely available commercial off-the-shelf (COTS) devices to enable its practical and large-scale deployment. 
Our choice of the COTS device (TI CC13x0: CC1310 or CC1350 LaunchPad) consequently brings down the cost and the form-factor of a SNOW node by 25x and 10x, respectively. 
Such implementation of SNOW on CC13x0 devices faces a number of challenges to enable link reliability and communication range. Our implementation addresses these challenges by handling peak-to-average power ratio problem, channel state information estimation, carrier frequency offset estimation, and near-far power problem.
Our deployment in the city of Detroit, Michigan demonstrates that CC13x0-based SNOW can achieve uplink and downlink throughputs of 11.2kbps and 4.8kbps per node, respectively, over a distance of 1km. 
Also, the overall throughput in the uplink increases linearly with the increase in the number of SNOW nodes.
Our experiments also show that this throughput is several times higher than that in LoRaWAN under typical settings.

\input{chapter4/introduction}

\input{chapter4/snow_overview}

\input{chapter4/technical}

\input{chapter4/deployment}

\input{chapter4/evaluation}

\input{chapter4/related}

\subsection{Summary}\label{sec:chap4-conclusion}
The recently proposed LPWAN technology -- SNOW -- has the potential to enable connectivity to numerous IoT devices over long distances. However, the high cost and the large form-factor of the USRP-based SNOW nodes hinder its practical deployments. In this paper, we have implemented SNOW for practical deployments using the CC13x0 devices as SNOW nodes. Our CC13x0-based SNOW implementation decreases the cost and the form-factor of a single SNOW node by 25x and 10x, respectively. 
We have also addressed several practical deployment challenges that include PAPR reduction, CSI estimation, CFO estimation, and near-far power problem. 
We have deployed our CC13x0-based SNOW in the city of Detroit, Michigan and achieved per node uplink and downlink throughputs of 11.2kbps and 4.8kbps, respectively, over a distance of 1km. Additionally, our overall uplink throughput at the BS have increased linearly with the increase in the number of nodes. Our experiments also show that SNOW can achieve throughput several times higher than LoRaWAN under typical settings.
Finally, our extensive experiments have demonstrated the CC13x0-based SNOW as a feasible LPWAN technology that can be deployed practically at low-cost and in large-scale for future IoT applications.


%% file: chapter4/introduction.tex
\subsection{Introduction}\label{sec:chap4-intro}

Low-Power Wide-Area Network (LPWAN) is an emerging communication technology that supports long-range, low-power, and low-cost connectivity to numerous devices. It is regarded as a key technology to drive the Internet of Things (IoT). Due to their escalating demand, recently multiple LPWAN technologies have been developed that operate in the licensed/cellular (NB-IoT~\cite{nbiot}, LTE-M~\cite{cat}, 5G~\cite{ngmn}) or unlicensed/non-cellular (SNOW~\cite{snow_ton}, LoRa~\cite{lorawan}, SigFox~\cite{sigfox}, etc.) bands. Most of the non-cellular technologies operate in the sub-1GHz ISM band except SNOW (Sensor Network over White Spaces) and WEIGHTLESS-W that operate in the TV white spaces~\cite{whitespaceSurvey, ismail2018low}.

\emph{White spaces} are the allocated but locally unused TV spectrum (54-698MHz in the US) that can be used by unlicensed devices as the secondary users.
Compared to the crowded ISM band, white spaces offer less crowded and much wider spectrum in both urban and rural areas, boasting an abundance in rural and suburbs~\cite{snow2}. Due to their low frequency, white spaces have excellent propagation and obstacle penetration characteristics enabling long-range communication. Thus, they hold the potentials for LPWAN to support various IoT applications. 
To our knowledge, WEIGHTLESS-W (which, to the best of our knowledge, has been decommissioned~\cite{ismail2018low}) and SNOW~\cite{snow_ton} are the only two efforts to exploit the TV white spaces for LPWAN. 
Initially introduced in~\cite{snow}, SNOW is a highly scalable LPWAN technology offering reliable, bi-directional, concurrent, and asynchronous communication between a base station (BS) and numerous nodes~\cite{snow2, snow_ton, snow_cots}.


Despite its promise as a great LPWAN technology, SNOW has not yet received sufficient attention from the research community due to its limited availability for practical deployment.
The current SNOW implementation, which is also available as open-source~\cite{snow_bs}, uses Universal Software Radio Peripheral (USRP) devices as LPWAN nodes, hindering the applicability of this technology in practical and large-scale deployment.
USRP is a hardware platform developed for software-defined radio applications~\cite{usrp}. Using the USRP platform as the SNOW node limits the practical deployment of SNOW in real-world applications due to several factors including its high cost and large form-factor. As of today, a USRP B200 device with a half-duplex radio costs $\approx$ \$750 USD. As such, it inherently becomes costly to deploy a large-scale SNOW network. 
Today, IoT applications including smart city (e.g., waste management, smart lighting, smart grid), transportation and logistics (e.g., connected vehicles), agricultural and smart farming (i.e., Microsoft FarmBeats), process management (e.g.,oil field monitoring) , and healthcare require collection of information from thousands of IoT nodes~\cite{ismail2018low}.

In this paper, we address the above practical limitations of the existing SNOW technology by implementing it on low-cost and low form-factor commercial off-the-shelf (COTS) devices that are deployable as SNOW nodes. Through this implementation, we empirically show that any COTS device with a programmable physical layer (PHY) that operates in the white spaces and supports amplitude-shift-keying (ASK) or binary phase-shift-keying (BPSK), can be practically deployed as SNOW nodes.
Specifically, thanks to its programmable PHY, we use the widely available and low-power TI CC13x0 (CC1310~\cite{cc1310} or CC1350~\cite{cc1350} LaunchPad) IoT device which costs approximately \$30 USD (retail price) and is 10x smaller than a USRP B200 device (including antenna), thereby making SNOW adoptable for practical IoT applications.


The SNOW technology has never been implemented on IoT devices before. The existing USRP-based SNOW implementation does not face the following practical challenges due to the expensive and powerful hardware design of USRP (as reflected by evaluation in~\cite{snow_ton, snow, snow2}), which the implementation on CC13x0 has to address.
{\bf First}, due to its orthogonal frequency division multiplexing (OFDM)-based design, the SNOW BS transmitter is subject to high peak-to-average power ratio (PAPR). Thus, the overall reliability at the CC13x0 device during downlink communication may be degraded severely. 
{\bf Second}, due to the asymmetric bandwidth requirements of the SNOW BS and the nodes, channel state information (CSI) estimation between the BS and a CC13x0 device plays a vital role in both uplink and downlink communications. Without CSI estimation, the overall reliability and the communication range may be decreased. 
{\bf Third}, Carrier frequency offset (CFO) needs to be handled robustly as the effects of CFO are much more pronounced in low-cost CC13x0 devices, leading to severe inter-carrier-interference (ICI). ICI decreases the overall bitrate in both uplink and downlink communications of SNOW. 
Along with addressing these challenges, through this new implementation, we also make SNOW resilient to the classic near-far power problem. Due to the near-far power problem, where a far node's transmission gets buried under a near node's transmission radiation, the reliability in the uplink communication may be degraded. Thus, we address the above challenge as well.
Specifically, we make the following key technical contributions.
\begin{itemize}
	\item We implement SNOW for practical deployment by programming the CC13x0's PHY to work as SNOW nodes. Compared to the current USRP-based SNOW implementation, the cost and the form-factor of a single SNOW node are decreased approximately 25x and 10x, respectively.
	\item In our implementation, we address several practical challenges including the PAPR problem, CSI and CFO estimation, and near-far power problem. Specifically,
	we propose a data-aided CSI estimation technique that allows a CC13x0 device to communicate directly with the SNOW BS from a distance of approximately 1km. Additionally, we propose a pilot-based CFO estimation technique that takes into account the device mobility and increases reliability in both uplink and downlink communications. Finally, we address the near-far power problem in SNOW through an adaptive transmission power control (ATPC) protocol that improves the reliability in uplink communication.
	\item We experiment with the CC13x0-based SNOW implementation through deployment in the city of Detroit, Michigan. Our results demonstrate that we achieve an uplink throughput of 11.2kbps per SNOW node.
	Additionally, our overall uplink throughput increases \emph{linearly} with the increase in the number of SNOW nodes. In downlink, we achieve a throughput of 4.8kbps per SNOW node.
	Compared to a typical LoRa deployment (channel bandwidth: 500kHz, spreading factor: 7, and coding rate: 4/5), our uplink throughput is approximately 3.7x higher when 5 nodes transmit to a gateway that can receive concurrent packets using 3 channels.
\end{itemize}



In the rest of the paper, Section~\ref{sec:chap4-model} provides an overview of SNOW and TI CC13x0 LaunchPads. Section~\ref{sec:chap4-implementation} presents our SNOW implementation detailing how we address several practical challenges. Section~\ref{sec:chap4-near-far} describes the near-far power problem and our ATPC mechanism. Sections~\ref{sec:chap4-deploy} and~\ref{sec:chap4-eval}  analyze the deployment cost and performance of our CC13x0-based SNOW implementation, respectively. Section~\ref{sec:chap4-related} overviews related work. Finally, Section~\ref{sec:chap4-conclusion} concludes our paper.

%% file: chapter4/snow_overview.tex
\subsection{Background and System Model}\label{sec:chap4-model}
\subsubsection{An Overview of SNOW}\label{sec:chap4-snow_overview}
\begin{figure}[!htpb]
\centering
\includegraphics[width=0.7\textwidth]{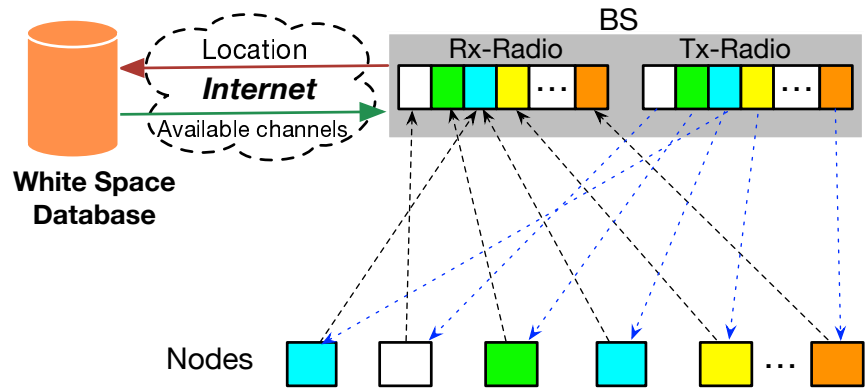}
\caption{Dual-radio BS and subcarriers~\cite{snow_ton}.}
\label{fig:chap4-dualradio}
\end{figure}
In this section, we provide a brief overview of SNOW. Its complete design and description is available in~\cite{snow_ton}. SNOW is a highly scalable LPWAN technology operating in the TV white spaces. It supports asynchronous, reliable, bi-directional, and concurrent communication between a BS and numerous nodes. Due to its long-range, SNOW forms a star topology allowing the BS and the nodes to communicate directly. The BS is powerful, Internet-connected, and line-powered while the nodes are power-constrained and do not have access to the Internet. To determine white space availability in a particular area, the BS queries a cloud-hosted geo-location database via the Internet. A node depends on the BS to learn its white space availability. In SNOW, all the complexities are offloaded to the BS to make the node design simple. 
Each node is equipped with a single half-duplex radio. To support simultaneous uplink and downlink communications, the BS uses a dual-radio architecture for reception (Rx) and transmission (Tx), as shown in Figure~\ref{fig:chap4-dualradio}.

The SNOW PHY uses a distributed implementation of OFDM called {\em D-OFDM}. D-OFDM enables the BS to receive concurrent transmissions from {\em asynchronous} nodes using a single-antenna radio (Rx-radio). Also, using a single-antenna radio (Tx-Radio), the BS can transmit different data to different nodes concurrently~\cite{snow, snow2, snow_ton, snow3, isnow_p2p, snow_cots}. Note that the SNOW PHY is different from MIMO radio design adopted in other wireless domains such as LTE, WiMAX, and 802.11n~\cite{snow2} as the latter use multiple antennas to enable multiple transmissions and receptions.
The BS operates on a wideband channel split into orthogonal narrowband channels/subcarriers (Figure~\ref{fig:chap4-dualradio}). Each node is assigned a single subcarrier. 
For encoding and decoding, the BS runs inverse fast Fourier transform (IFFT) and global fast Fourier transform (G-FFT) over the entire wideband channel, respectively.
When the number of nodes is no greater than the number of subcarriers, every node is assigned a unique subcarrier. Otherwise, a subcarrier is shared by more than one node. 
SNOW supports ASK and BPSK modulation techniques, supporting different bitrates. 


The nodes in SNOW use a lightweight CSMA/CA (carrier sense multiple access with collision avoidance)-based  MAC protocol similar to TinyOS~\cite{tinyos}. Additionally, the nodes can autonomously transmit, remain in receive mode, or sleep. A node runs clear channel assessment (CCA) before transmitting. If its subcarrier is occupied, the node makes a random back-off in a fixed congestion back-off window. After this back-off expires, the node transmits immediately if its subcarrier is free. Then node repeats this operation until it sends the packet and gets the acknowledgment (ACK).

\subsubsection{An Overview of TI CC13x0 LaunchPads}
Texas Instruments introduced TI CC1310 and TI CC1350 LaunchPads as a part of the SimpleLink microcontroller (MCU) platform to support ultra-low-power and long-range communication~\cite{cc1310, cc1350, snow_cots}. 
With a small form-factor (length: 8cm, width: 4cm), both CC1310 and CC1350 are designed to operate in the lower frequency bands (287--351MHz, 359--527MHz, and 718--1054MHz) including the TV band. As an added feature, CC1350 can also operate in the 2.4GHz band. 
The packet structure of the CC13x0 devices includes a \code{preamble}, followed by \code{sync word}, \code{payload length}, \code{payload}, and \code{CRC}, chronologically. They support different data modulation techniques including Frequency Shift Keying (FSK), Gaussian FSK (GFSK), On-Off Keying (OOK), and a proprietary long-range modulation. They are capable of using a Tx/Rx bandwidth that ranges between 39 and 3767kHz. 
Additionally, with a supply voltage in the range of 1.8 to 3.8 volts, their Rx and Tx current consumption is 5.4mA and 13.4mA at +10dBm, respectively, offering ultra-low-power communication.
These devices are commercially available at low cost and support a variety of data modulations techniques. The greatest advantage is that  
they have a programmable and reconfigurable physical layer, offering flexibility and feasibility for customized protocol implementation.

%% file: chapter4/technical.tex
\subsection{SNOW Implementation on TI CC13x0}\label{sec:chap4-implementation}
\begin{figure}[!htb]
\centering
\includegraphics[width=0.9\textwidth]{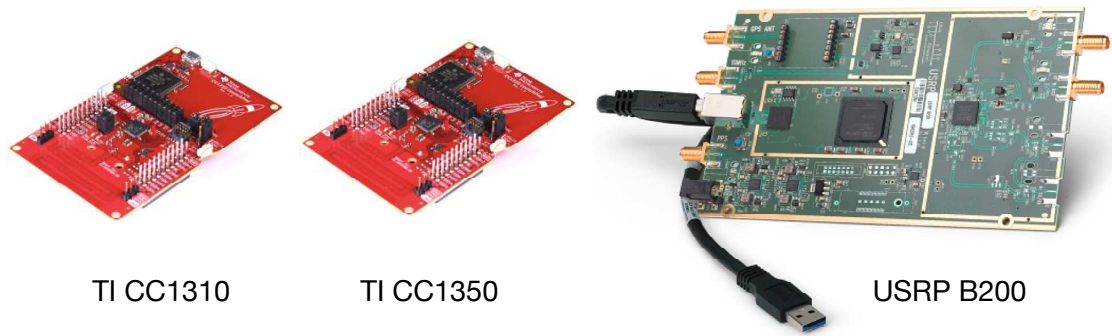}
\caption{Devices used in our SNOW implementation. A node is a CC1310 or CC1350 device (they look alike). The BS has two USRP B200 devices, each having its own antenna. Antenna is not shown in this figure, which is approximately 2x bigger than a USRP B200 device.}
\label{fig:chap4-devices}
\end{figure}
The original SNOW implementation in~\cite{snow_ton} uses the USRP hardware platform for both the BS and the nodes. In our implementation, we use the CC13x0 devices as SNOW nodes and USRP in the BS (Figure~\ref{fig:chap4-devices}).
For BS implementation, we adopt the open-source code provided in~\cite{snow_bs}. The BS uses two half-duplex USRP devices (Rx-Radio and Tx-Radio), each having its own antenna. Also. the BS is implemented on the GNURadio software platform that gives a high magnitude of freedom to perform baseband signal processing~\cite{gnuradio}.
In the following, we explore a number of implementation considerations and feasibility for a CC13x0 device to work as a SNOW node in practical deployments. 
First, we show how to configure a CC13x0 device to make it work as a SNOW node. We then address the practical challenges (e.g., PAPR problem, CSI estimation, and CFO estimation) associated with our CC13x0-based SNOW implementation.

\subsubsection{Configuring TI CC13x0}
To configure the subcarrier center frequency, bandwidth, modulation, and the Tx power we set the appropriate values to the CC13x0 command inputs \code{centerFreq, rxBw, modulation}, and \code{txPower}, respectively, using {\em Code Composer Studio} (CCS) provided by Texas Instruments~\cite{ccs}. A graphical user interface alternative to CCS is {\em SmartRF Studio}, which is also provided by Texas Instruments~\cite{srfs}. The MAC protocol of SNOW in CC13x0 is implemented on top of the example CSMA/CA project that comes with CCS~\cite{ccs}.

\subsubsection{Peak-to-Average Power Ratio Observation}\label{sec:chap4-papr}
By transmitting on a large number of subcarriers simultaneously (in downlink), the BS suffers from a traditional OFDM problem called {\em peak-to-average power ratio (PAPR)}. PAPR of an OFDM signal is defined as the ratio between the maximum instantaneous power and its average power.
In SNOW downlink communication (i.e., BS to nodes), after the IFFT is performed by the BS, the composite signal can be represented as follows.
\begin{equation}
\nonumber x(t) = \frac{1}{\sqrt{N}}\sum_{k=0}^{N-1}X_k~e^{j2 \pi f_k t},~~0 \le t \le NT
\end{equation} 
Here, $X_k$ is the modulated data symbol for node $k = \{0, 1, \cdots, N-1\}$ on subcarrier center frequency $f_k = k\Delta f$, where $\Delta f = \frac{1}{NT}$ and $T$ is the symbol period. Therefore, the PAPR can be calculated as~\cite{jiang2008overview}
\begin{equation}
\nonumber \text{PAPR}[x(t)] = 10\log_{10}\Bigg( \frac{\max\limits_{0~ \le ~t~ \le~ NT} [|x(t)|^2 ]}{P_{\text{avg}}}\Bigg)~~dB.
\end{equation}
Here, the average power $P_{\text{avg}} = E [|x(t)|^2]$.
A node's signal detection on its subcarrier is very sensitive to the nonlinear signal processing components used in the BS, i.e., the digital-to-analog converter (DAC) and high power amplifier (HPA), which may severely impair the bit error rate (BER) in the nodes due to the induced spectral regrowth. If the HPA does not operate in the linear region with a large power back-off due to high PAPR, the out-of-band power will exceed the specified limit and introduce severe ICI~\cite{jiang2008overview}. Moreover, the in-band distortion (constellation tilting and scattering) due to high PAPR may cause severe performance degradation~\cite{kamali2012understanding}. It has been shown that the PAPR reduction results in significant power saving at the transmitters~\cite{baxley2004power}.

\begin{figure}[!htb]
\centering
\includegraphics[width=0.50\textwidth]{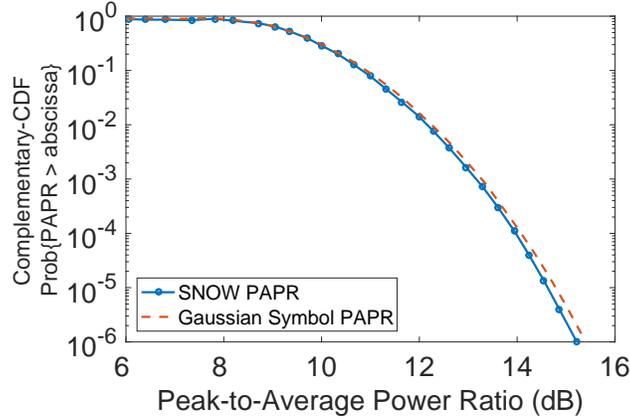}
\caption{PAPR distribution of D-OFDM signal in Tx-Radio.}
\label{fig:chap4-papr}
\end{figure}

\begin{figure}[!htbp]
    \centering
      \subfigure[RSSI under varying distance\label{fig:chap4-csi_rssi}]{
    \includegraphics[width=0.50\textwidth]{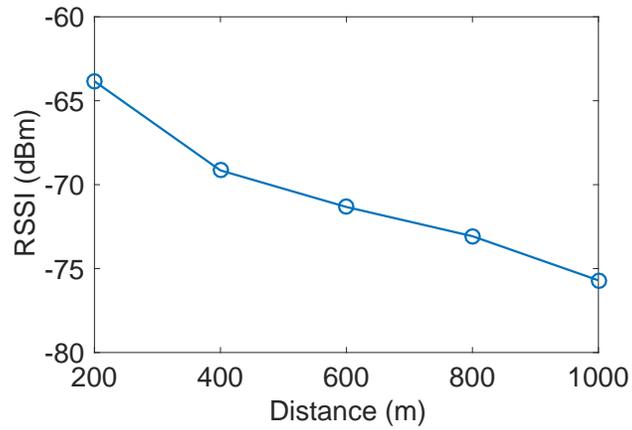}
      }\vfill
      \subfigure[Path Loss under varying distance\label{fig:chap4-csi_pathloss}]{
        \includegraphics[width=.50\textwidth]{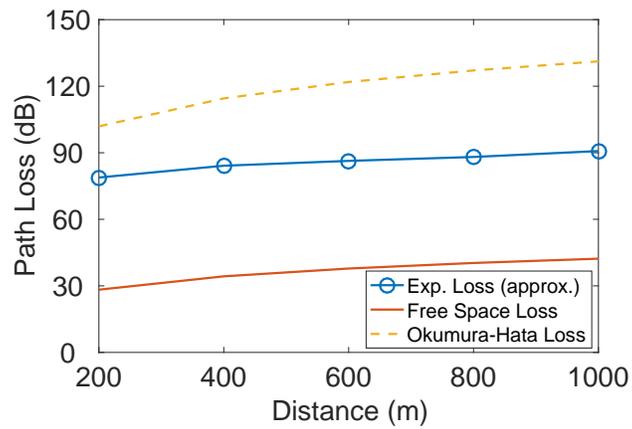}
      }\vfill
      \subfigure[BER under varying distance\label{fig:chap4-csi_ber}]{
        \includegraphics[width=.50\textwidth]{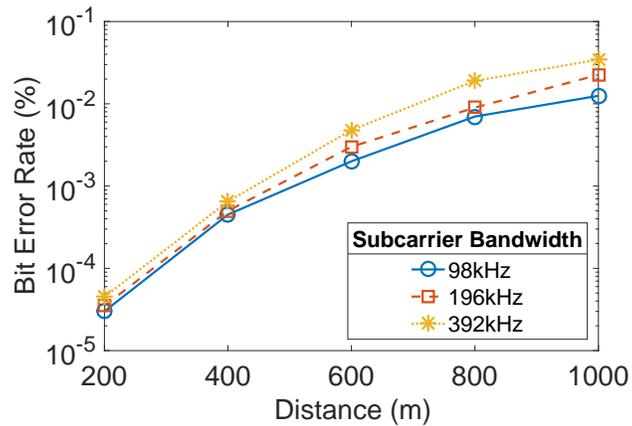}
      }
    \caption{RSSI, path loss, and BER at the SNOW BS for a TI CC1310 node from different distances.}
    \label{fig:chap4-csi}
 \end{figure}
As shown in Figure~\ref{fig:chap4-papr}, the PAPR in SNOW downlink communication (for N = 64) follows the Gaussian distribution. Thus, the peak signal occurs quite rarely and the transmitted D-OFDM signal will cause the HPA to operate in the nonlinear region, resulting in a very inefficient amplification. To illustrate the power efficiency of the HPA for N = 64, let us assume the probability of the clipped D-OFDM frames is less than 0.01\%. We thus need to apply an input back-off (IBO)~\cite{baxley2004power} equivalent to the PAPR at a probability of $10^{-4}$. Here, PAPR $\approx$ 14dB or 25.12. Thus, the efficiency ($\eta = 0.5/\text{PAPR}$) of the HPA~\cite{jiang2008overview} is $\eta = 0.5/25.12 \approx 1.99\%$. Such low efficiency at the HPA motivates us to explore the high PAPR in SNOW for practical deployments.
Several uplink PAPR reduction techniques for single-user OFDM systems have been proposed (see survey~\cite{jiang2008overview}). However, the characteristics of the downlink PAPR in SNOW, where different data are concurrently transmitted to different nodes, are entirely different from the PAPR observed in a single-user OFDM system. To adopt an uplink PAPR reduction technique used in the single-user OFDM systems for the downlink PAPR reduction in SNOW, each node has to process the entire data frame transmitted by the BS and then demodulate its own data. However, a SNOW node has less computational power and does not apply FFT to decode its data~\cite{snow_ton}, or any other node's data. Thus, none of the existing PAPR reduction techniques will work in our implementation.

To this extent, we address the PAPR problem in SNOW by allocating a special subcarrier  called {\em downlink subcarrier} for downlink communication. The BS may send any broadcast message, ACK, or data to the nodes using that downlink subcarrier. A node has to switch to the downlink subcarrier to listen to any broadcast message, ACK, or data.
If the BS requires (downlink subcarrier is being interfered by an external source), it may allocate several redundant downlink subcarriers.
Note that the dual-radio architecture in SNOW BS allows it to receive concurrent packets from a set of nodes (uplink) and transmit broadcast/ACK/data packets to another set of nodes (downlink), simultaneously. The BS can acknowledge several nodes using a single transmission by using a bit-vector of size equals to the number of subcarriers. 
If the BS receives a packet from a node operating on subcarrier $i$, it will set the $i$-th bit in the bit-vector. Upon receiving the bit-vector, that node may get the acknowledgment by looking at the $i$-th bit of the vector.
A node retransmits the packet if that packet is not acknowledged in the first valid ACK received by that node. In the following, we describe our technique to handle a {\bf rare} case in practical SNOW deployments, and hence may be kept optional in implementation.

When a subcarrier (say, $i$) is shared by multiple nodes, the BS may receive a valid second packet (say, from node A) before transmitting the ACK for the valid first packet (say, from node B). In this case, both nodes A and B may be acknowledged by setting the $i$-th bit of the vector. However, if the packet from node A (or, B) is valid and the packet from node B (or, A) is invalid, the BS will reset the $i$-th bit of the vector and transmit the ACK. Thus, none of the packets are acknowledged even if one of them is valid. To compensate for that, the BS (Tx-Radio) will switch to node A's (or, B's) subcarrier and transmit an ACK packet. Thus, in our implementation, if a node finds that its packet is not acknowledged in the first valid ACK it received, before retransmission it listens to its subcarrier for a fixed amount of time. Each node may know this fixed time when it joins the network. Typically, if a subcarrier is shared by $G$ nodes, the fixed amount of time (worst case) may be set to $GD_p$ (ignoring the frequency switching time in the Tx-Radio), where $D_p$ is the time to transmit one packet. Other ways of addressing such issue may include the use of \emph{hash functions}. However, we do not explore that in our implementation for scalability issue due to hash collision.


\subsubsection{Does Channel State Information Estimation Make It More Resilient?}\label{sec:chap4-csi}

Multi-user OFDM communication requires channel estimation and tracking for ensuring high data rate at the BS. One way of avoiding channel estimation is to use the \emph{differential phase-shift keying (DPSK)} modulation technique. However, the use of DPSK results in a lower bitrate at the BS due to a 3dB loss in the singal-to-noise ratio (SNR)~\cite{van1995channel}. Additionally, the current SNOW design does not support DPSK modulation. SNR at the BS for each node is different in SNOW. Also, SNR of each node is affected differently due to channel conditions, deteriorating the overall bitrate in the uplink. Thus, it requires handling of the channel estimation in SNOW.

Figure~\ref{fig:chap4-csi} shows the received signal strength indicator (RSSI), path loss, and BER at the SNOW BS for a CC1310 device that transmits from 200 to 1000m distances with a Tx power of 15dBm, subcarrier center frequency at 500MHz, and a bandwidth of 98kHz. Figure~\ref{fig:chap4-csi_rssi} indicates that the RSSI decreases rapidly with the increase in distance. Also, the path loss in Figure~\ref{fig:chap4-csi_pathloss} shows that it is significantly higher than the theoretical free space loss~\cite{rappaport1996wireless}. We also compare with the Okumura-Hata~\cite{rappaport1996wireless} loss to check if it fits the model, however, it does not. Finally, Figure~\ref{fig:chap4-csi_ber} confirms that the BER goes above $10^{-3}$ (which is not acceptable~\cite{rnr}) beyond 400m due to the unknown channel conditions. Figure~\ref{fig:chap4-csi_ber} also shows that the BER worsens for an increase in the subcarrier bandwidth. Thus, to make our implementation more resilient, we need to incorporate the CSI estimation in SNOW.

We calculate the CSI for each SNOW node independently on its subcarrier. We consider a slow flat-fading model~\cite{tse2005fundamentals}, where the channel conditions vary slowly with respect to the duration of a single node--BS packet duration. Note that joint-CSI estimation~\cite{maniatis2002pilots, jiang2007iterative, ribeiro2008uplink} in SNOW is not our design goal since it would require SNOW nodes to be strongly time-synchronized.  
Similar to IEEE 802.16e~\cite{ieee16e}, we run CSI estimation independently for each node because of their different fading and noise characteristics. In the following, we explain the CSI estimation technique for one node on its subcarrier for each packet. The BS uses the same technique to estimate CSI for other nodes as well. For a node, in a narrowband flat-fading subcarrier, the system is modeled as follows.
\begin{equation}
\nonumber y = Hx + w
\end{equation}
Here, $y$, $x$, and $w$ are the receive vector, transmit vector, and noise vector, respectively. $H$ is the channel matrix. 
We model the noise as additive white Gaussian noise, i.e., a circular symmetric complex normal ($CN$) with $w \sim CN(0, W)$, where the mean is zero and noise covariance matrix $W$ is known.
As the subcarrier conditions vary, we estimate the CSI on a short-term basis based on popular approach called training sequence. We use the known preamble transmitted at the beginning of each packet. $H$ is estimated using the combined knowledge of the received and the transmitted preambles. To make the estimation robust, we divide the preamble into $n$ equal parts (preamble sequence). In our case, n = 4 which yields similar complexity for CSI estimation in IEEE 802.11~\cite{wifi}.

Let, the preamble sequence be $(p_1, p_2, \cdots, p_n)$, where vector $p_i$, for $i = \{1, \cdots, n\}$, is transmitted as
\begin{equation}
\nonumber y_i = Hp_i + w_i.
\end{equation}
Combining the received preamble sequences, we get
\begin{equation}
\nonumber Y = [y_1, \cdots, y_n] = HP + W.
\end{equation}
Here, $P = [p_1, \cdots, p_n]$ and $W = [w_1, \cdots, w_n]$. With combined knowledge of $Y$ and $P$, channel matrix $H$ is estimated. Similar to the CSI estimation in the uplink communication by the BS, each node also calculates the CSI estimation in downlink communication.

\subsubsection{Does Carrier Frequency Offset Estimation Make It More Robust?} \label{sec:chap4-cfo}

Multi-user OFDM systems are very much sensitive to the CFO between the transmitters and the receiver. CFO causes the OFDM systems to lose orthogonality between subcarriers, which results in severe ICI. 
A transmitter and a receiver observe CFO due to (i) the mismatch in their local oscillator frequency as a result of hardware imperfections; (ii) the relative motion that causes a Doppler shift. 
ICI degrades the SNR between an OFDM transmitter and a receiver, which results in significant BER. Thus, we investigate the needs for CFO estimation in our implementation.
\begin{figure}[!htb]
\centering
\includegraphics[width=0.50\textwidth]{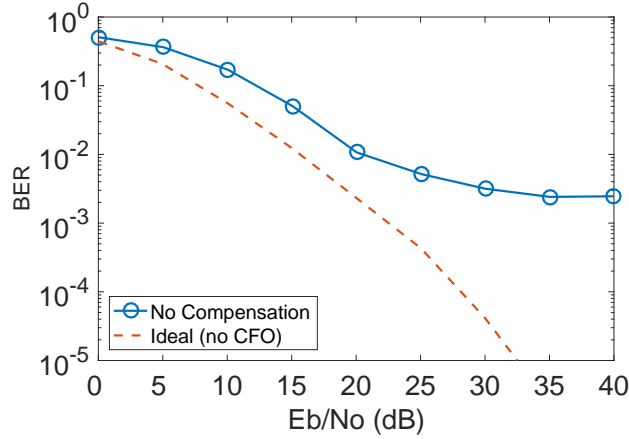}
\caption{BER at different $E_b/N_0$.}
\label{fig:chap4-cfo}
\end{figure}

The loss in SNR due to the CFO between the SNOW BS and a node can be estimated as follows~\cite{nee2000ofdm}.
$$SNR_{loss} = 1 + \frac{1}{3}(\pi \delta f T)^2\frac{E_s}{N_0}$$
Here, $\delta f$ is the frequency offset, $T$ is the symbol duration, $E_s$ is the average received subcarrier energy, and $N_0/2$ is the two-sided spectral density of the noise power. To show the CFO effects, we choose two neighboring orthogonal subcarriers in the BS and send packets from two nodes. Figure~\ref{fig:chap4-cfo} shows the BER at the BS from those two CC1310 nodes at different $E_b/N_0$, where $E_b$ is the average energy per bit in the received signals. This figure shows that BER is nearly $10^{-3}$ even for very high $E_b/N_0$ ($\approx 40$dB), which is also very high compared to the theoretical BER~\cite{choi2000carrier}. Thus, CFO is heavily pronounced in SNOW.
The distributed and asynchronous nature of SNOW does not allow CFO estimation similar to the traditional multi-user OFDM systems.
While the USRP-based SNOW implementation provides a trivial CFO estimation, it is not robust and does not account for mobility of the nodes~\cite{snow_ton}.
We propose a pilot-based CFO estimation technique that is robust and accounts for the node's mobility. We use training symbols for CFO estimation in an ICI free environment for each node independently, while it joins the network by communicating with the BS using a non-overlapping {\em join subcarrier}.

We explain the CFO estimation technique between a node and the BS (uplink) on a join subcarrier $f$ based on time-domain samples. Note that the BS keeps running the G-FFT on the entire BS spectrum. We thus extract the corresponding time-domain samples of the join subcarrier by applying IFFT during a node join. The join subcarrier does not overlap with other subcarriers; hence it is ICI-free. If $f_{\text{node}}$ and $f_{\text{BS}}$ are the frequencies at a node and the BS, respectively, then their frequency offset $\delta f = f_{\text{node}}-  f_{\text{BS}}$.
For transmitted signal $x(t)$ from a node, the received signal  $y(t)$ at the BS that experiences a CFO of $\delta f$ is given by 
$y(t)  = x(t) e^{j2\pi \delta f t}$.
Similar to IEEE 802.11a~\cite{wifi}, we estimate $\delta f$ based on short and long preamble approach. Note that the USRP-based implementation has considered only one preamble to estimate CFO.
In our implementation, the BS first divides a $n$-bit preamble from a node into short and long preambles of lengths $n/4$ and $3n/4$, respectively. Thus for a 32-bit preamble (typically used in SNOW), the lengths of the short and long preambles are  8 and 24, respectively. 
The short preamble and the long preamble are used for coarse and finer CFO estimation, respectively. 
Considering $\delta t_s$ as the short preamble duration and $\delta f_s$ as the coarse CFO estimation, we have
$$y(t-\delta t_s)  = x(t) e^{j2\pi \delta f_s (t-\delta t_s)}.$$
Since $y(t)$ and $y(t-\delta t_s)$ are known at the BS, we have
\begin{align*}
y(t-\delta t_s) y^*(t)  & = x(t) e^{j2\pi \delta f_s (t-\delta t_s)}       x^*(t) e^{-j2\pi  \delta f_s t}\\
                           & = |x(t)|^2  e^{j 2\pi  \delta f_s -\delta t_s }.
\end{align*}
Taking angle of both sides gives us as follows.
\begin{align*}
\sphericalangle  y(t-\delta t_s) y^*(t)   &=  \sphericalangle     |x(t)|^2  e^{j 2\pi  \delta f_s -\delta t_s }\\
                                          &=      - 2\pi  \delta f_s \delta t_s
\end{align*}
By rearranging the above equation, we get
$$\delta f_s   =  - \frac{\sphericalangle  y(t-\delta t_s) y^*(t) }{2\pi\delta t_s}.$$

Now that we have the coarse CFO $\delta f_s$, we correct each time domain sample (say, $P$) received in the long preamble as $ P_a = P_a e^{-ja \delta f_s}$, where $a = \{1, 2, \cdots, A\}$ and $A$ is the number of time-domain samples in the long preamble. Taking into account the corrected samples of the long preamble and considering $\delta t_l$ as the long preamble duration, we estimate the finer CFO as follows. 
\begin{equation} 
\delta f  =  - \frac{\sphericalangle  y(t-\delta t_l) y^*(t) }{2\pi\delta t_l} \label{eqn:chap4-finer_cfo}
\end{equation}
To this extent, considering the join subcarrier $f$, the {\slshape ppm (parts per million)} on the BS's crystal is given by $ \text{ppm}_\text{BS} = 10^6  \big(\frac{\delta f}{f}\big) $. Thus, the BS calculates $ \delta f_i$ on subcarrier $f_i$ (assigned for node $i$) as 
$\delta f_i =  \frac{(f_i * \text{ppm}_\text{BS})}{10^6}.$ The CFO between the Tx-Radio and the Rx-radio can be estimated using a basic SISO CFO estimation technique~\cite{yao2005blind}. Thus, BS also knows the CFO for downlink communication.

We now explain the CFO estimation to compensate for the Doppler shift. Note that if the signal bandwidth is sufficiently narrow at a given carrier frequency and mobile velocity, the Doppler shift can be approximated as a common shift across the entire signal bandwidth~\cite{talbot2007mobility}. Thus, the Doppler shift in the join subcarrier for a node also represents the Doppler shift at its assigned subcarrier, and hence the estimated CFO in Equation (\ref{eqn:chap4-finer_cfo}) is not affected due to the Doppler Shift.
For simplicity, we consider that a node's velocity is constant and the change in Doppler shift is negligible during a single packet transmission in SNOW.
Considering $\delta f_d$ as the CFO due to the Doppler shift, $v$ as the velocity of the node, and $\theta$ as the angle of the arrived signal at the BS from the node, we have~\cite{talbot2007mobility}
\begin{equation}
	\nonumber \delta f_d = f_i\big(\frac{v}{c}\big)\cos(\theta).
\end{equation}
Here, $f_i$ is the subcarrier center frequency and $c$ is the speed of light. The node itself may consider its motion as circular and approximate $\theta = \frac{\delta s}{r}$, where $\delta s$ is the amount of anticipated change in nodes position during a packet transmission and $r$ is the {\em line-of-sight} distance of the node from the BS. Thus, CFO compensation due to the Doppler shift is done at the nodes during uplink communication. In downlink communication, the Tx-Radio of the BS can also compensate for the node's mobility as the node can report its Doppler shift to the BS during uplink communication. 

In summary, as the nodes asynchronously transmit to the BS, doing the joint-CFO estimation for each subcarrier at the BS is quite difficult. Thus, we use a simple feedback approach for proactive CFO correction in uplink communication. 
$\delta f_i$  estimated at the BS for subcarrier $f_i$ is given to the node (during its joining process) that is assigned  subcarrier $f_i$.
The node may then adjust its transmitted signal based on $\delta f_i$ and $\delta f_d$, calculated as $(\delta f_i + \delta f_d)$, which will align its signal so that the BS does not need to compensate for CFO in the uplink communication. Such feedback-based proactive compensation scheme was studied before for multi-user OFDM~\cite{van1999time} and is also used in global system for mobile communication (GSM).

\subsection{Near-Far Power Problem} \label{sec:chap4-near-far}
\begin{figure}[!htb]
\centering
\includegraphics[width=0.7\textwidth]{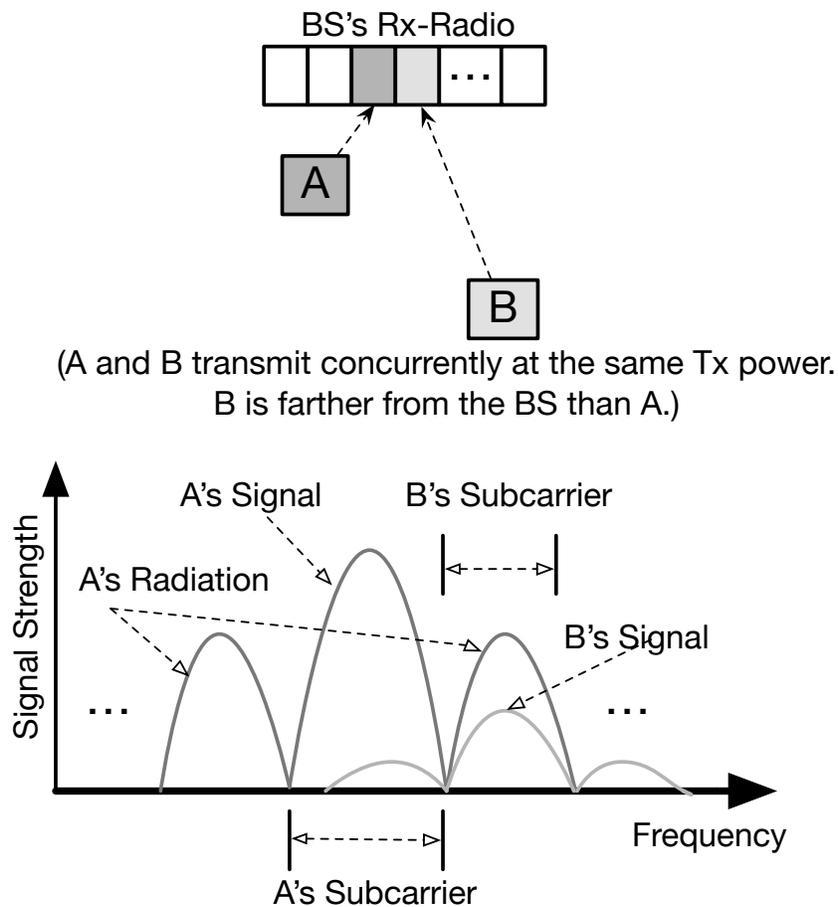}
\caption{An illustration of the near-far power problem.}
\label{fig:chap4-near-far}
\end{figure}
Wireless communication is susceptible to the near-far power problem, especially in CDMA (Code Division Multiple Access)~\cite{muqattash2003cdma}. Multi-user D-OFDM system in SNOW may also suffer from this problem. Figure~\ref{fig:chap4-near-far} illustrates the near-far power problem in SNOW. Suppose, nodes A and B are operating on two adjacent subcarriers. Node A is closer to the BS compared to node B. When both nodes A and B transmit concurrently to the BS, the received frequency domain signals from node A and B may look as shown in the bottom of Figure~\ref{fig:chap4-near-far}. Here, transmission from node B is severely interfered by the strong radiations of node A's transmission. As such, node B's signal may be buried under node A's signal making it difficult for the BS to decode the packet from node B. 
A typical SNOW deployment may have such scenarios if the nodes operating on adjacent subcarriers use the same transmission power and transmit concurrently at the BS from different distances.

\begin{figure}[!htbp]
    \centering
      \subfigure[Avg. PDR at different Tx powers\label{fig:chap4-nf_pdr}]{
    \includegraphics[width=0.50\textwidth]{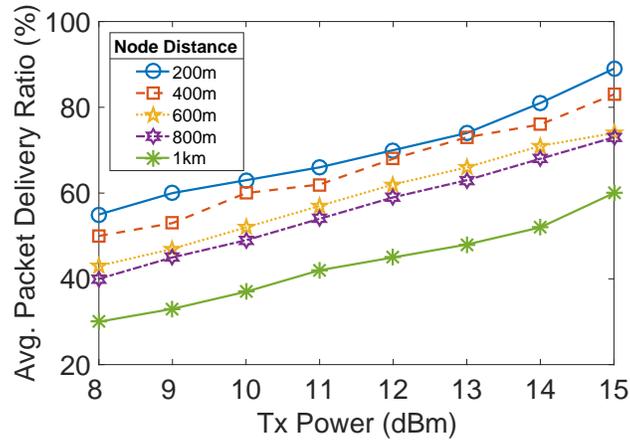}
      }\vfill
      \subfigure[Avg. PDR at different Tx powers at different time.\label{fig:chap4-nf_time}]{
        \includegraphics[width=.50\textwidth]{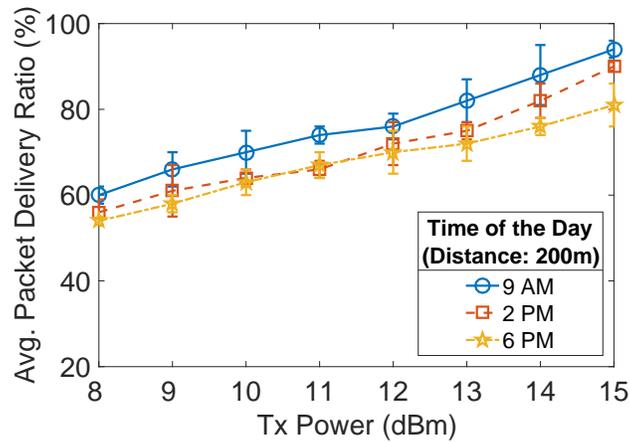}
      }
    \caption{Packet delivery ratio at different Tx powers}
    \label{fig:chap4-nf-effects}
 \end{figure}
To observe the near-far power problem in SNOW, we run experiments by choosing 3 different adjacent subcarriers, where the middle subcarrier observes the near-far power problem introduced by both subcarriers on its left and right. We place two CC1310 nodes within 20m of the BS that use the left and the right subcarrier, respectively. We use another CC1310 node that uses the middle subcarrier and is placed at different distances between 200 and 1000m from the BS. Nodes that are within 20m of the BS transmit packets continuously with a transmission power of 0dBm. At each distance, for each transmission power between 8 and 15dBm, the node that uses the middle subcarrier sends 100 rounds of 1000 consecutive packets (sends one packet then waits for the ACK and then sends another packet, and so on) to the BS and with a random interval of 0-500ms. For each transmission power level, at each distance, that node calculates its average {\em packet delivery ratio (PDR)}. We repeat the same set of experiments for 7 days at 9 AM, 2 PM, and 6 PM.
Figure~\ref{fig:chap4-nf_pdr} shows that the average PDR increases at each distance with the increase in the transmission power. Figure~\ref{fig:chap4-nf_time} depicts the result for 7-day experiments (only at a distance of 200m) and shows that the average PDR changes at different time of the day. Overall, Figure~\ref{fig:chap4-nf_pdr} and~\ref{fig:chap4-nf_time} confirms that the average PDR increases with the increase in the transmission power. Thus, the near-far power problem needs to be addressed in SNOW. To this extent, we propose an adaptive transmission power control for SNOW design. 


\subsubsection{Adaptive Transmission Power Control}\label{sec:chap4-atpc}
Our design objective for the adaptive Tx power control is to correlate the subcarrier-level Tx power and link quality (i.e., PDR) between each node and the BS. We thus formulate a predictive model to provide each node with a proper Tx power to make a successful transmission to the BS using its assigned subcarrier. Note that our work differs from the work in~\cite{lin2016atpc} in fundamental concepts of the network design and architecture. In~\cite{lin2016atpc}, the authors have considered a multi-hop wireless sensor network based on IEEE 802.15.4~\cite{ieee154} with no concurrency between a set of transmitters and a receiver. Additionally, our model is much more simpler since we deal with single hop communications. As such, the overheads (i.e., energy consumption and latency at each node) associated with our model are fundamentally lesser than that in~\cite{lin2016atpc}. In the following, we describe our model.

Whenever a node is assigned a new subcarrier, changes location (inside the SNOW network), or observes a lower PDR, e.g., PDR below quality of service (QoS) requirements, it runs a lightweight predictive model to determine the convenient Tx power to make successful transmissions to the BS.
Our predictive model uses an approximation function to estimate the PDR distribution at different Tx power levels. Over time, that function is modified to adapt to the node's changes. The function is built from the sample pairs of the Tx power levels and PDRs between the node and the BS via a curve-fitting approach. A node collects these samples by sending groups of packets to the BS at different Tx power levels. Thus, our predictive model uses two vectors: $TP$ and $L$, where $TP = \{ tp_1, tp_2, \cdots, tp_m \}$ contains $m$ different Tx power levels that the node uses to send $m$ groups of packets to the BS and $L = \{ l_1, l_2, \cdots, l_m \}$ contains the corresponding PDR values at different Tx power levels. Thus, $l_i$ represents the PDR value at Tx power level $tp_i$. We use the following linear function to correlate between Tx power and PDR.
\begin{equation}
	l(tp_i) = a~.~tp_i + b \label{eqn:chap4-linear_model}
\end{equation}

To lessen the computational overhead in the node, we adopt the {\em least square approximation} technique to determine the unknown coefficients $a$ and $b$ in Equation (\ref{eqn:chap4-linear_model}). Thus, we find the minimum of the function $S(a, b)$, where
\begin{equation}
\nonumber	S(a, b) = \sum |l_i - l(tp_i)|^2.
\end{equation}
The minimum of $S(a, b)$ is determined by taking the partial derivatives of $S(a, b)$ with respect to $a$ and $b$, respectively, and setting them to zero. Thus, $ \frac{\partial S}{\partial a} = 0$ and $\frac{\partial S}{\partial b} = 0$ give us as follows.
\begin{align}
	\nonumber a~\sum (tp_i)^2 + b~\sum tp_i &= \sum l_i.tp_i \\
  \nonumber a~\sum tp_i + b~m &= \sum l_i 
\end{align}
Simplifying the above two equations, we find the estimated values of $a$ and $b$ as follows.
\begin{equation}\nonumber
\begin{split}
	\begin{bmatrix}
		\hat{a}\\
        \hat{b}
	\end{bmatrix}
    = \frac{1}{m \sum (tp_i)^2 - (\sum tp_i)^2} \times \\
    \begin{bmatrix}
    	m \sum l_i.tp_i - \sum l_i \sum tp_i\\
    	\sum l_i \sum (tp_i)^2 - \sum l_i.tp_i \sum tp_i
    \end{bmatrix}
\end{split}
\end{equation}
Using the estimated values of $a$ and $b$, the node can calculate the appropriate Tx power as follows.
\begin{equation}\label{eqn:chap4-estimated}
tp = \big[\frac{PDR_{\text{threshold}} - \hat{b}}{\hat{a}}\big] \in TP
\end{equation}
Here, $PDR_{\text{threshold}}$ is the threshold set empirically or according to QoS requirements, and $[.]$ denotes the function that rounds the value to the nearest integer in the vector $TP$.

Now that the initial model has been established in Equation (\ref{eqn:chap4-estimated}), this needs to be updated continuously with the node's changes over time. In Equation (\ref{eqn:chap4-linear_model}), both $a$ and $b$ are functions of time that allow the node to use the latest samples to adjust the curve-fitting model dynamically. 
It is empirically found that (Figure~\ref{fig:chap4-nf_pdr}) the slope of the curve does not change much over time; hence $a$ is assumed time-invariant in the predictive model. On the other hand, the value of $b$ changes drastically over time (Figure~\ref{fig:chap4-nf_time}). Thus, Equation (\ref{eqn:chap4-linear_model}) is rewritten as follows that characterizes the actual relationship between Tx power and PDR.
\begin{equation}
	\nonumber l(tp(t)) = a.tp(t) + b(t)
\end{equation}
Now, $b(t)$ is determined by the latest Tx power and PDR pairs using the following feedback-based control equation~\cite{lin2016atpc}.
\begin{align}
	\nonumber \Delta \hat{b}(t) &= \hat{b}(t) - \hat{b}(t+1) \\
    			\nonumber	  &= \frac{\sum^K_{k=1} [PDR_{\text{threshold}} - l_k(t - 1)]}{K} \\ 
                      &= PDR_{\text{threshold}} - l(t-1) \label{eqn:chap4-control}
\end{align}
Here, $l(t-1)$ is the average value of $K$ readings denoted as follows. 
\begin{equation}
	\nonumber l(t-1) = \frac{\sum^K_{k=1} l_k(t - 1)}{K}
\end{equation}
Here, $l_k(t-1)$, for $k = \{1, 2, \cdots, K\}$, is one reading of PDR during the time period $t-1$ and $K$ is the number of feedback responses at time period $t-1$. Now, the error in Equation (\ref{eqn:chap4-control}) is deducted from the previous estimation; hence the new estimation of $b(t)$ can be written as: $\hat{b}(t) = \hat{b}(t-1) - \Delta \hat{b}(t)$.
Given the newly estimated $\hat{b}(t)$, the node now can set the Tx power at time $t$ as:
\begin{equation}
	\nonumber tp(t) = \big[\frac{PDR_{\text{threshold}} - \hat{b}(t)}{\hat{a}}\big].
\end{equation}

%% file: chapter4/deployment.tex
\subsection{Network Architecture and Deployment Cost}\label{sec:chap4-deploy}
\begin{figure}[!htb]
\centering
\includegraphics[width=0.85\textwidth]{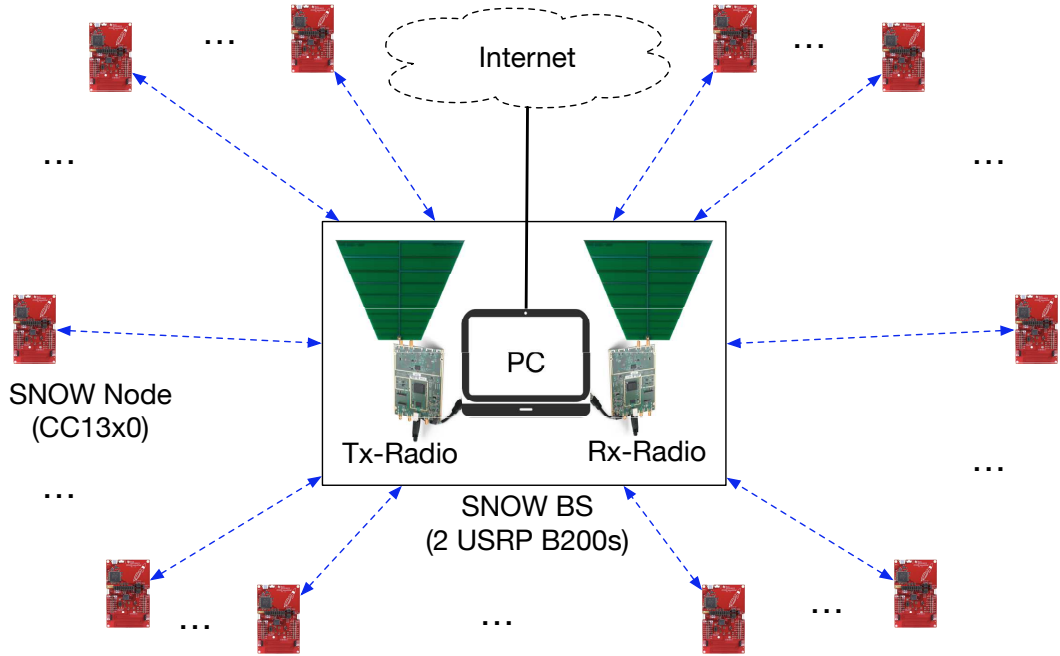}
\caption{The SNOW architecture for practical deployment (The PC may be replaced by a Raspberry Pi device. The two USRP B200 devices can be replaced by a USRP B2100 device that has two half-duplex radios.)}
\label{fig:chap4-deployment}
\end{figure}
In this section, we discuss the practical applicability of our implementation. Figure~\ref{fig:chap4-deployment} shows our network view. The SNOW BS is a PC that connects two USRP B200 devices (Tx-Radio and Rx-Radio). The BS is also connected to the Internet. In the BS, a USRP B210 device may be used which has two half-duplex radios. Also, a Raspberry Pi~\cite{raspberry} device may be used instead of the PC. All the CC13x0 nodes are battery-powered and directly connected to the BS.

\begin{figure}[!htb]
\centering
\includegraphics[width=0.5\textwidth]{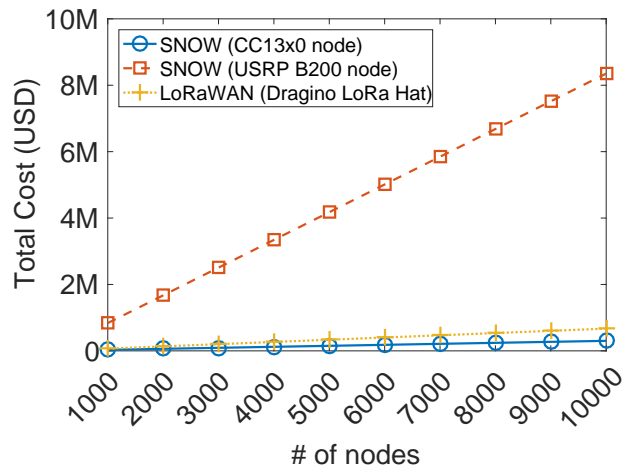}
\caption{Practical deployment cost with numerous nodes.}
\label{fig:chap4-cost}
\end{figure}

We now analyze the deployment cost of our CC13x0-based SNOW implementation and compare with the original USRP-based SNOW implementation in~\cite{snow_ton}. 
Figure~\ref{fig:chap4-cost} shows the total deployment cost of our CC13x0-based SNOW implementation for different numbers of nodes between 1000 and 10,000. A CC1310 or CC1350 device costs approximately \$30 USD (retail price). The price for the BS is approximately \$1600 USD (two USRP B200 devices \$750 USD each, and two antennas \$50 USD each). In this comparison, the cost of the PC is not considered since it is common for both implementations. For SNOW implementation in~\cite{snow_ton}, a node is a  USRP B200 device that has an antenna and runs on a Raspberry Pi. A Raspberry Pi device costs approximately \$35 USD. 
To provide an insight into the deployment cost of a LoRaWAN network, we consider the Dragino LoRa/GPS-Hat SX1276 IoT devices that run on Raspberry Pi and cost approximately \$32 USD per device (retail price)~\cite{dragino}. This LoRaWAN device has computational and RF capabilities that are almost similar to a TI CC13x0 device (e.g., both have Cortex-M MCU, similar energy profiles, sensors, software support, etc.). In addition, we consider a LoRaWAN gateway that costs approximately \$299 USD and can receive packets on multiple channels simultaneously~\cite{loracost}. Note that we rule out cheaper LoRaWAN devices (costs $\approx$\$10 USD) from the calculation since they do not have a similar profile as CC13x0 and do not provide any software support.
As shown in Figure~\ref{fig:chap4-cost}, to deploy an LPWAN with 1000 nodes, the CC13x0-based SNOW implementation may cost approximately \$31.6K USD, compared to \$836.6K USD for the USRP-based SNOW implementation proposed in~\cite{snow_ton}, and \$67.3K USD for the Dragino LoRa-Hat-based LoRaWAN. 
For a deployment of 10,000 nodes, the costs are \$301.6K, \$8.3M, and \$670.3K USD
for CC13x0-based SNOW implementation, USRP-based SNOW implementation, and Dragino LoRa-Hat-based LoRaWAN, 
respectively. As shown in Figure~\ref{fig:chap4-cost}, the cost of each LPWAN increases linearly with the increase in the number of nodes. However, the cost of our CC13x0-based SNOW implementation in unnoticeable.
Our new implementation of SNOW on the CC13x0 devices thus becomes highly scalable in terms of cost, making SNOW deployable for practical applications.

%% file: chapter4/evaluation.tex
\subsection{Evaluation}\label{sec:chap4-eval}
In this section, we provide an extensive evaluation of our CC13x0-based SNOW implementation. We evaluate both uplink and downlink performances with both stationary and mobile CC13x0 nodes.

\subsubsection{Setup}\label{sec:chap4-expsetup}
\begin{figure}[!htb]
\centering
\includegraphics[width=0.7\textwidth]{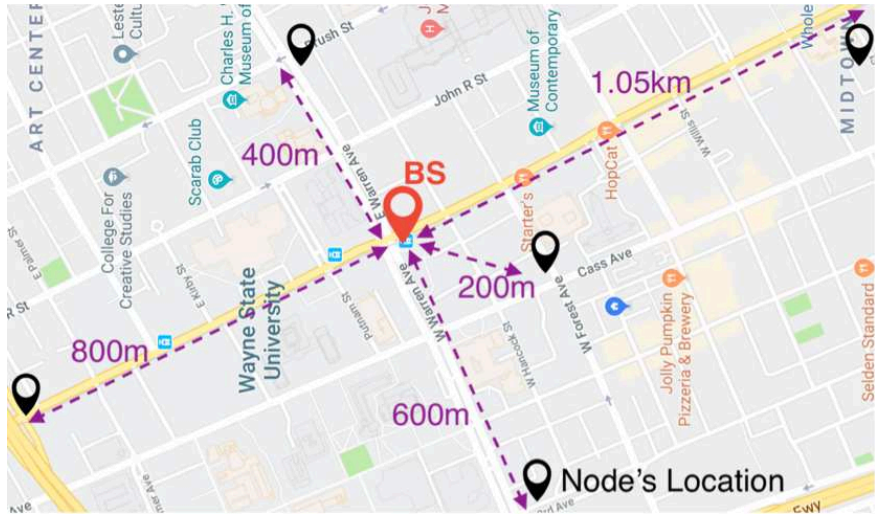}
\caption{SNOW deployment in Detroit, Michigan.}
\label{fig:chap4-testbed}
\end{figure}

Figure~\ref{fig:chap4-testbed} shows our deployment in the city of Detroit, Michigan. We deploy 22 CC1310 devices and 3 CC1350 devices (25 CC13x0 devices in total) at different distances between 200 and 1000m. 
We use the TV white space channel with frequency band 500--506MHz and split into 29 (numbered 1 to 29) overlapping (50\%) orthogonal subcarriers, each 400kHz wide. Note that the USRP-based SNOW also uses a similar subcarrier bandwidth~\cite{snow_ton}.
\begin{figure}[!htbp]
    \centering
      \subfigure[Packet reception rate at a node at different distances\label{fig:chap4-prr_dist}]{
    \includegraphics[width=0.50\textwidth]{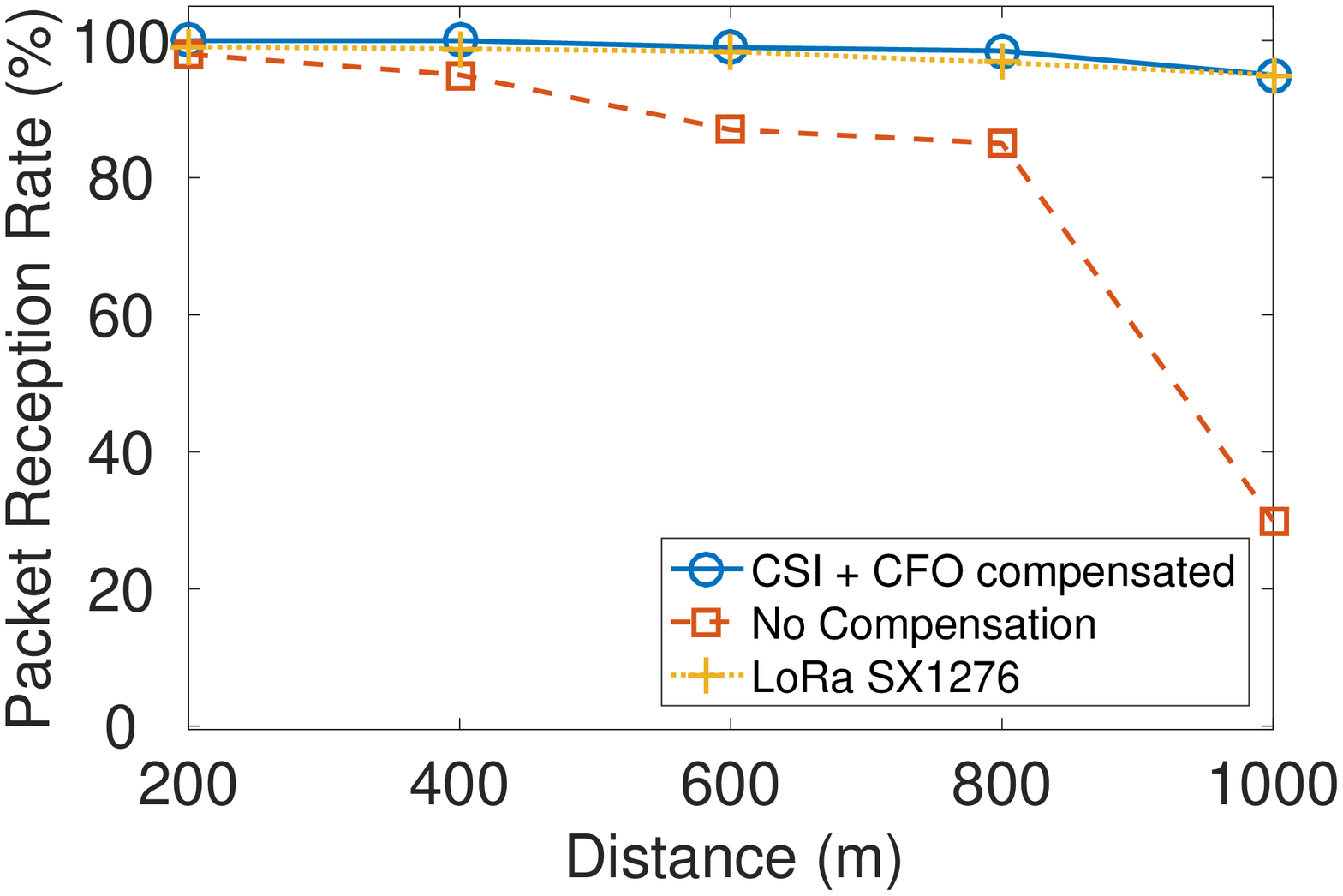}
      }\vfill
      \subfigure[Packet reception rate in uplink communication\label{fig:chap4-prr_txs}]{
        \includegraphics[width=.50\textwidth]{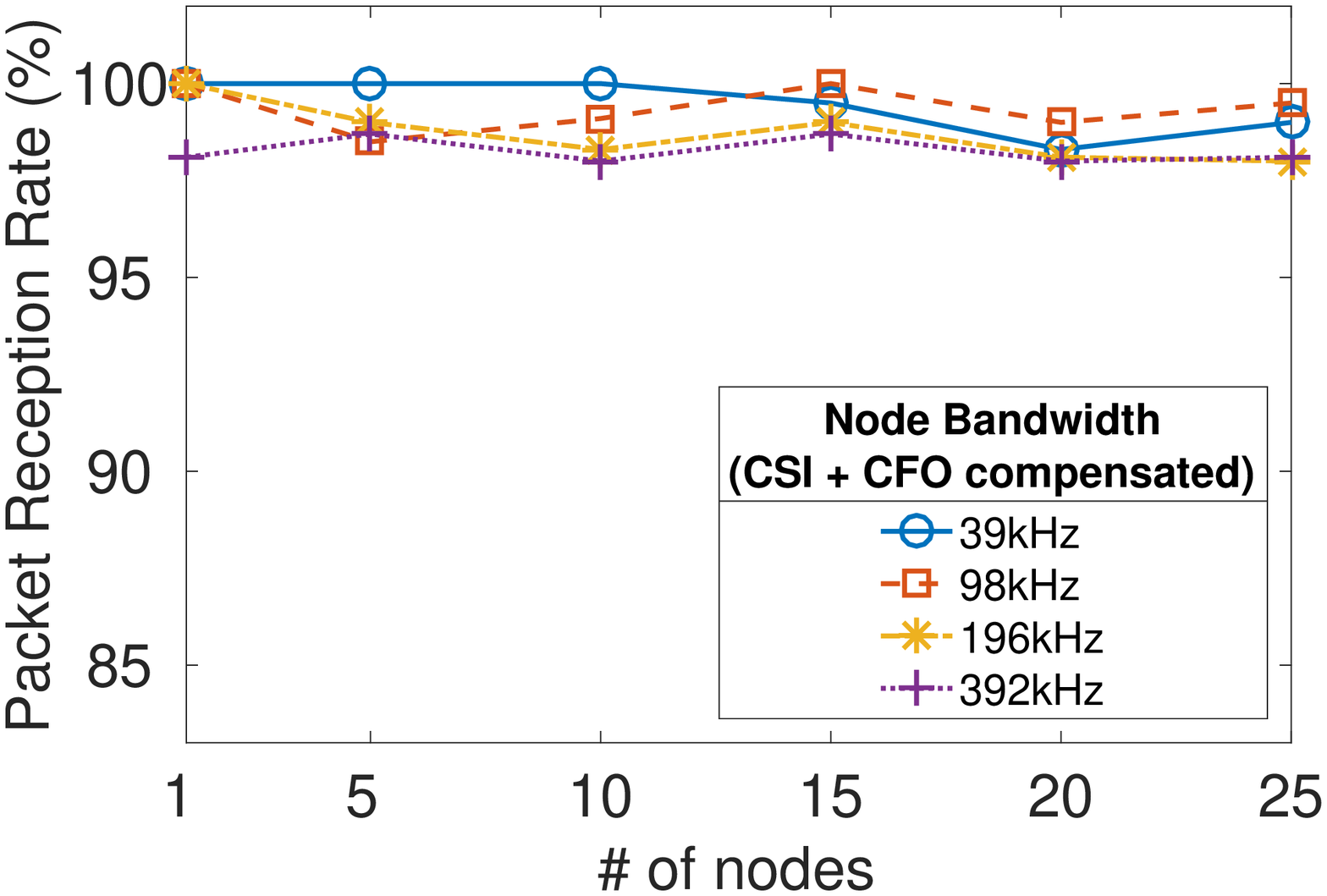}
      }\vfill
      \subfigure[Packet reception rate in downlink communication\label{fig:chap4-down_prr}]{
        \includegraphics[width=0.50\textwidth]{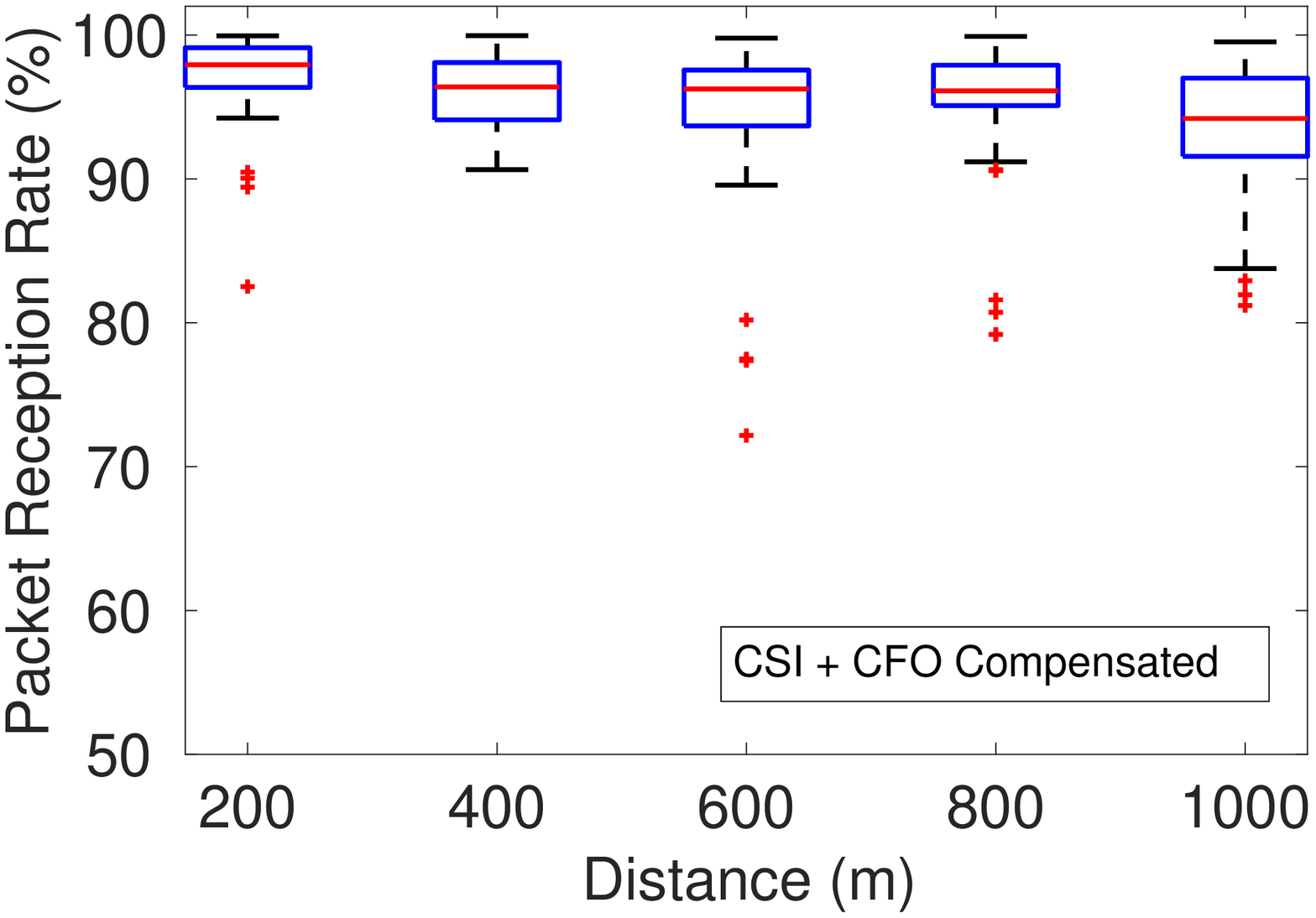}
      }
    \caption{Reliability in long distance communication.}
    \label{fig:chap4-prr}
\end{figure}
We use the 28th subcarrier as the join subcarrier and the 26th subcarrier as the downlink subcarrier. We do not use the 29th and the 27th subcarriers so that the join subcarrier may remain ICI-free (Section~\ref{sec:chap4-cfo}).
The remaining 25 subcarriers are assigned to different nodes.
We use the packet structure of the CC13x0 devices (\code{preamble}: (32 bits), \code{sync word}: (32 bits), \code{paylod length}:, \code{payload}: variable length, and \code{CRC} (16 bits)). Our default payload length is 30 bytes and contains random data. Our default bandwidth at the CC13x0 nodes is 39kHz. We use OOK modulation supported by the CC13x0 devices. Unlike the USRP-based SNOW, we do not use any spreading factor. Since the subcarrier bandwidths at the BS and the CC13x0 nodes are 400kHz and 39kHz, respectively, the oversampling at each subcarrier in the BS compensates for the spreading factor. Our default transmission power at the BS and the nodes is 15dBm. However, a CC13x0 device may choose to operate with any transmission power between 0 and 15dBm, as needed by our ATPC model (Section~\ref{sec:chap4-atpc}). The receive sensitivity at the BS is set to -114dBm, as per the white space regulations~\cite{whitespaceSurvey}. Unless stated otherwise, these are the default parameter settings.

\subsubsection{Reliability over Long Distance}\label{sec:chap4-prr_dist}
{\bf Achievable Distance.}
We first test the achievable communication range of our CC13x0-based SNOW implementation. We take one CC13x0 device and transmit to the BS from different distances between 200 and 1000m. We keep our antenna height at 3 meters above the ground for both the BS and the node. At each distance, the CC13x0 transmits 1000 packets with a random interval between 0 and 500ms. The transmission power is set to 15dBm. To show comparison, we repeat the same experiments without compensating for CSI and CFO as well. Additionally, we test the achievable distance between two LoRa SX1276 devices (bandwidth: 125kHz, spreading factor: 7, coding rate: 4/5) with the above settings. 
Figure~\ref{fig:chap4-prr_dist} shows that the packet reception rate (PRR) at the SNOW BS when packets are sent with and without compensating for CSI and CFO, comparing with LoRa. 
As shown in this figure, with CSI and CFO compensation, the BS achieves 95\% of PRR from a distance of 1km . 
Without CSI and CFO compensation, the PRR at the BS is as low as 30\% from 1km distance. This figure also shows that a LoRa SX1276 device can deliver packets to another over 1km with a PRR of 95\%, which is similar to the CC13x0-based SNOW node (CSI and CFO compensated). The results thus demonstrate that SNOW on the new platform is highly competitive against LoRa, an LPWAN leader that operates in the ISM band. Additionally,
we find that beyond approximately 1km, PRR stars decreasing in our implementation. Our best guess is that if we can place the BS or the node at a higher altitude (FCC allows up to 30 meters), we may achieve high reliability over much longer communication range.

{\bf Uplink Reliability.}
To show the uplink reliability under concurrent transmissions from different nodes (CFO and CSI compensated), we transmit from 1 to 25 nodes (using their assigned subcarriers) to the BS. In this experiment, all the nodes are distributed within 200 and 1000m of the BS. Each node uses different subcarrier bandwidths between 39 and 392kHz. For each bandwidth starting from 39kHz, a node sends consecutive 1000 packets. Between each bandwidth, a node sleeps for 500ms. Thus, the BS knows the change in the bandwidth. 
Note that in practical deployment scenarios, a node can let know the BS of its bandwidth during the joining process.
In this experiment, we show the performance of a node for different bandwidths. Figure~\ref{fig:chap4-prr_txs} shows that we can achieve up to 99\% reliability when 25 nodes transmit concurrently using 39kHz, and up to 98\% using 392kHz. Thus, ensuring high uplink reliability of our CC13x0-based implementation over long distances.

{\bf Downlink Reliability.}
In downlink, we test the reliability by sending 100 consecutive 30-byte (payload length) packets to each of the 25 nodes that are distributed within 200 and 1000m of the BS. We repeat the same experiment 50 times with an interval between 0 and 500ms. In this experiment, we compensate for both CSI and CFO.
Figure~\ref{fig:chap4-down_prr} shows our downlink reliability at different distances observed by different nodes. For better representation, we cluster the nodes that are located approximately at the same distance and plot the PRR against distance.
As shown in this figure, the PRR in downlink is as high as 99\% for 75\% of the nodes that are approximately 200m away from the BS. Also, 75\% of the nodes that are approximately 1km away from the BS achieve a PRR of 95\%. Thus, this experiment confirms high downlink reliability of our CC13x0-based implementation over long distances.

\subsubsection{Performance in Uplink Communication}\label{sec:chap4-tputuplink}
 \begin{figure}[!htbp]
    \centering
      \subfigure[Throughput under varying \# of nodes\label{fig:chap4-tput}]{
    \includegraphics[width=0.50\textwidth]{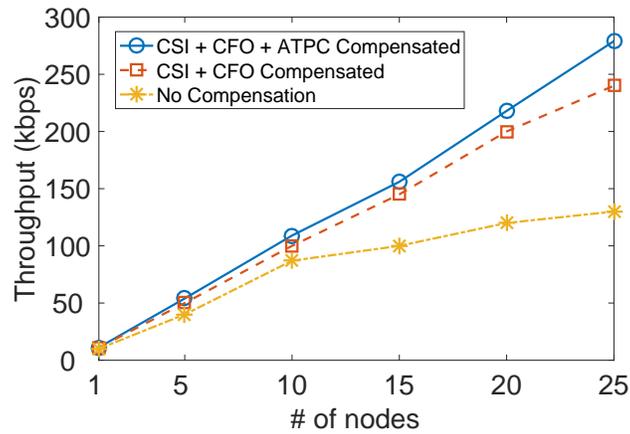}
      }\vfill
      \subfigure[End-to-end delay under varying \# of nodes\label{fig:chap4-delay}]{
        \includegraphics[width=.50\textwidth]{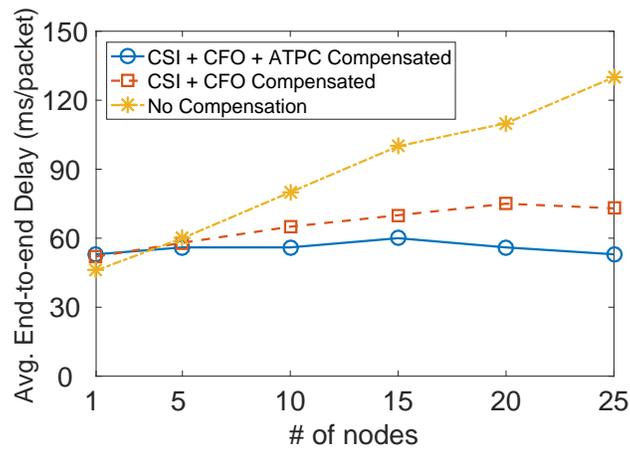}
      }\vfill
      \subfigure[Energy consumption under varying \# of nodes\label{fig:chap4-energy}]{
        \includegraphics[width=.535\textwidth]{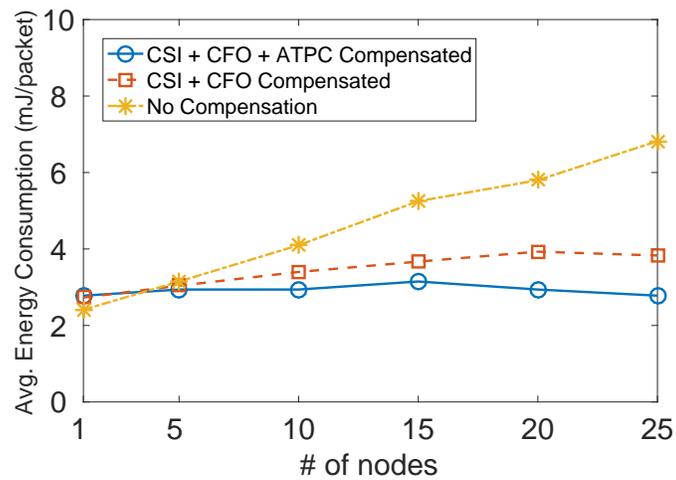}
      }
    \caption{Network performance in uplink communication.}
    \label{fig:chap4-performance}
 \end{figure}
In this section, we evaluate the uplink network performance in terms of throughput, end-to-end delay, and energy consumption. We calculate the throughput at the BS as the total achieved bitrate (kbps). We estimate the end-to-end delay at the nodes as the time (ms) elapsed between a packet transmit and its ACK receive. We also measure the energy consumption (mJoule) at the nodes. We allow from 1 to 25 nodes to transmit concurrently to the BS. We distribute the nodes between 200 and 1000m in our testbed. Each node transmits 1000 30-byte (payload length) packets with a random packet interval between 0 and 100ms. 
Such packet interval confirms that the node's transmissions are indeed asynchronous to the BS.
Each node uses a bandwidth of 39kHz. We evaluate the uplink network performance for three different cases: (1) nodes or/and BS {\bf compensate} for CSI, CFO, and ATPC; (2) nodes or/and BS {\bf compensate only} for CSI and CFO, but not ATPC; (3) nodes or/and BS {\bf do not compensate} for CSI, CFO, and ATPC. Note that ATPC applies to the nodes only, and hence we use "or/and" in the above three cases.
For each case, we run the experiments as long as at least 90\% of the packets are delivered to the BS. Thus, a node may try several times to deliver a packet to the BS.

{\bf Throughput.}
Figure~\ref{fig:chap4-tput} shows that the BS achieves up to 279kbps of throughput when 25 nodes transmit concurrently (case 1), yielding 11.16kbps per node. Additionally, the overall throughput at the BS increases linearly with an increase in the number of nodes. When only CSI and CFO are compensated for, the overall throughput at the BS also increases with an increase in the number of concurrent transmissions, however, it depends on the nodes' distribution (physical) across the network. If there is no near-far power problem, the overall throughput may be the same as observed in case 1. With no compensation, the achieved throughput per node is approximately 5kbps, thus 2x lesser than case 1. Note that a CC13x0 device can generate a baseband signal with a symbol rate of 11.2kbaud (OOK modulated). Thus, using a node bandwidth of 39kHz or 392kHz will not affect the per node throughput. However, a lower node bandwidth gives higher PRR (Section~\ref{sec:chap4-prr_dist}) due to longer symbol duration, combating the ICI to some extent.
Note that if we use any other COTS device that can generate a higher symbol rate for OOK at higher node bandwidth, the per node throughput may also increase with an increase in the node bandwidth.
Overall, CC13x0-based SNOW implementation shows high potential for practical deployments.

{\bf End-to-end Delay.}
Figure~\ref{fig:chap4-delay} shows the average end-to-end delay per packet at the nodes. When CSI, CFO, and ATPC are compensated for, the average end-to-end delay per packet in the network is 55ms with 25 concurrent transmissions. Also, for case 1, the average end-to-end delay per packet almost remains constant for any number of concurrent transmissions. For case 2, where only CSI and CFO are compensated for, the average end-to-end delay per packet increases a little bit with an increase in the number of concurrent transmissions. With no compensation, the average end-to-end delay per packet increases almost linearly with an increase in the number of concurrent transmissions. The reason is that a node retransmits several packets several times.  
Overall, our CC13x0-based SNOW implementation shows great promise for low-latency Industry 4.0 applications~\cite{modekurthy2018utilization}.

{\bf Energy Consumption.}
Figure~\ref{fig:chap4-energy} shows the average energy consumption per packet at the nodes. We use the CC13x0 energy profile to calculate the energy consumption during Tx, Rx, and idle time~\cite{cc1310, cc1350}. 
For case 1, where the CSI, CFO, and ATPC are compensated for, the average energy consumption per packet in the network is approximately 2.78mJoule with 25 concurrent transmissions. Also, the average energy consumption per packet almost remains constant for any number of concurrent transmissions. For case 2, where only CSI and CFO are compensated for, the average energy consumption per packet increases to 3.83mJoule for 25 concurrent transmissions. Also, when nothing is compensated for, the average energy consumption per packet increases almost linearly with an increase in the number of concurrent transmissions. The reason is that a node retransmits several packets several times. Overall, small energy consumption in case 1 confirms that the CC13x0-based SNOW may host long-lasting IoT applications.

\subsubsection{Performance Comparison with LoRaWAN in Uplink Communication}

In this section, we compare the performance of our CC13x0-based SNOW implementation with a LoRaWAN network. We have 8 Dragino LoRa/GPS-Hat Sx1276 transceivers that can transmit or receive on a single channel. We create a LoRaWAN gateway capable of receiving on 3 channels simultaneously using 3 of our LoRa-Hats, while the remaining 5 devices act as LoRaWAN nodes. To provide a fair comparison, we also allow 5 SNOW nodes (3 CC1350 devices and 2 CC1310 devices) to transmit to the SNOW BS, allowing only 3 subcarriers for data Rx/Tx. In LoRaWAN, the nodes transmit on 500kHz channels using a spreading factor of 7  and a coding rate of $\frac{4}{5}$. In SNOW, the nodes use a subcarrier bandwidth of 392kHz with no bit spreading factor. While choosing 500kHz or 392kHz has no differentiable impact in our CC13x0-based SNOW implementation (as discussed in Section~\ref{sec:chap4-tputuplink}), we choose the latter due to the configurable Tx bandwidth limitation of the devices. The LoRaWAN gateway uses 3 adjacent 500kHz channels in the 915MHz frequency band (in the US), while the SNOW BS, in this setup, uses 3 adjacent overlapping subcarriers, numbered 10, 11, and 12 (refer to Section~\ref{sec:chap4-expsetup} for subcarrier allocation) in the white spaces. Each node (for both LoRaWAN and SNOW) transmits 1000 thirty-byte (payload size) packets from a distance of approximately 1km to the gateway/BS with a random inter-packet interval between 500 and 1000ms and a Tx power of 15dBm. Each node randomly hops to a different channel/subcarrier after sending 200 packets. In LoRaWAN, the nodes use the pure ALOHA MAC protocol (Class-A operation~\cite{lorawan}). In SNOW, the nodes use the lightweight CSMA/CA MAC protocol (as discussed in Section~\ref{sec:chap4-snow_overview}). In the following, we compare LoRaWAN and SNOW in terms of reliability, throughput, and energy consumption with the above settings.
\begin{figure}[!htbp]
    \centering
      \subfigure[Packet reception rate at the gateway/BS\label{fig:chap4-lora_prr}]{
    \includegraphics[width=0.50\textwidth]{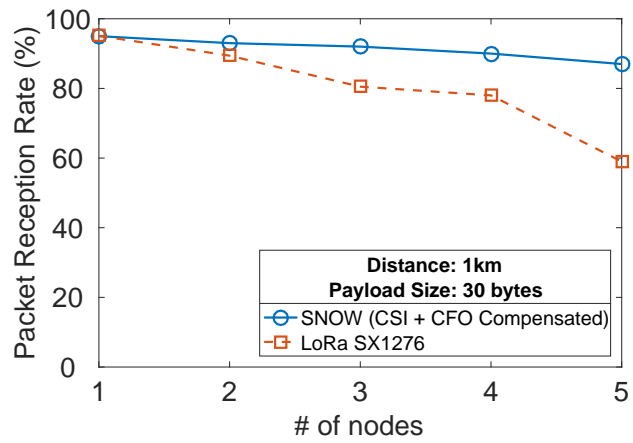}
      }\vfill
      \subfigure[Throughput at the gateway/BS\label{fig:chap4-lora_tput}]{
        \includegraphics[width=.50\textwidth]{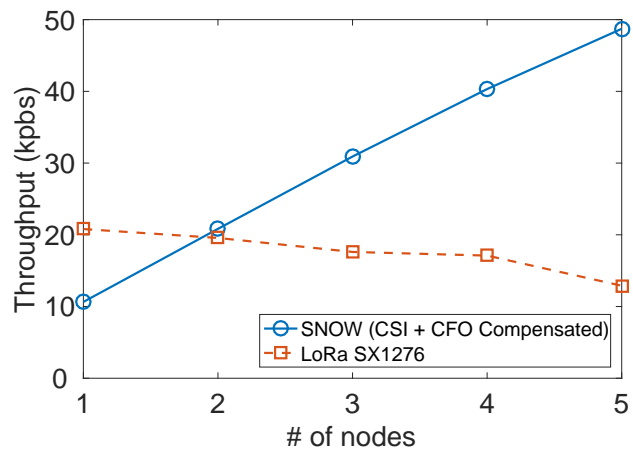}
      }
      \vfill 
      \subfigure[Energy consumption at the nodes\label{fig:chap4-lora_en}]{
        \includegraphics[width=.50\textwidth]{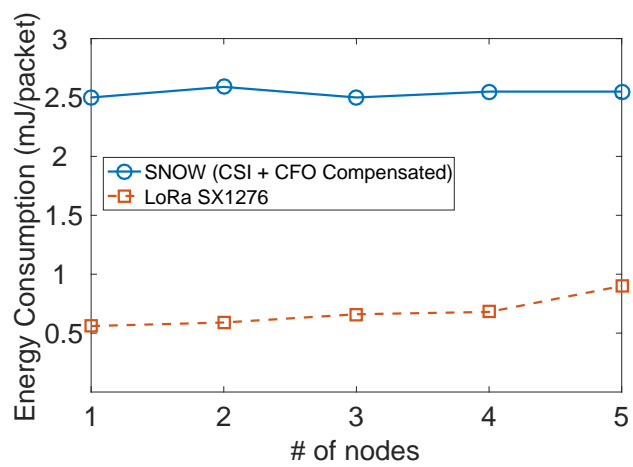}
      }
    \caption{Uplink performance comparison between SNOW and LoRaWAN.}
    \label{fig:chap4-lora-comp}
 \end{figure}

{\bf Reliability Comparison with Parallel Tx/Rx.}
Figure~\ref{fig:chap4-lora_prr} shows the packet reception rate at the gateway/BS for LoRaWAN and CC13x0-based SNOW implementation under varying number of nodes that transmit concurrently. As shown in this figure, when only one node transmits, the packet reception rate is approximately 95\% in both LoRaWAN and SNOW. Also, the packet reception rate of LoRaWAN decreases with the increase in the number of parallel transmissions. For SNOW, it remains almost similar with the increase in the number of parallel transmissions. For example, when 5 nodes transmit in parallel, LoRaWAN achieves a packet reception rate of 59\%, compared to 87\% in SNOW. Such a performance degradation in LoRaWAN happens as it uses an ALOHA-based MAC protocol without any collision avoidance. The packet reception rate of LoRaWAN may increase if we increase the inter-packet interval and will remain the same for SNOW even if we decrease the inter-packet interval. 

{\bf Throughput Comparison.}
Figure~\ref{fig:chap4-lora_tput} shows the overall throughput (in kbps based on the correctly received packets) comparison at the gateway/BS between LoRaWAN and SNOW. As shown in this figure, the throughput at the LoRaWAN gateway is approximately 20.8kbps, compared to 10.64kbps at the SNOW BS when only one node transmits. However, the throughput at the SNOW BS surpasses that at the LoRaWAN gateway when 2 or more nodes transmit concurrently. As shown in Figure~\ref{fig:chap4-lora_tput}, the throughput at the SNOW BS is $\frac{48.12}{12.9} \approx 3.7$x higher compared to LoRaWAN when 5 nodes transmit concurrently. Such a performance degradation in LoRaWAN happens as it uses an ALOHA-based MAC protocol without any collision avoidance and the inter-packet interval. Compared to LoRaWAN, CC13x0-based SNOW implementation thus shows better promise for high data-rate and low-latency IoT applications.

{\bf Energy Consumption Comparison.}
Figure~\ref{fig:chap4-lora_en} shows the per packet energy consumption (in mJ based on the correctly received packets) at the nodes of LoRaWAN and our CC13x0-based SNOW implementation. As shown in this figure, when 5 nodes transmit in parallel, a LoRaWAN node spends approximately 0.9mJ/packet, compared to 2.5mJ/packet in SNOW.
Even though here the per packet energy consumption in SNOW is slightly higher than that in LoRaWAN, the result shows that it increases in LoRaWAN and remains steady in SNOW with the increase in the number of nodes that transmit concurrently. SNOW is designed to enable a large number of concurrent transmissions to the BS and such a tendency in energy consumption shows its energy efficiency under that scenario. On the other hand, the number of retransmissions to deliver a packet increases with the increase in the number of nodes in LoRaWAN, thereby increasing the per packet energy consumption. Due to a limited number of devices, we are unable to demonstrate this in real experiment. However, we demonstrated the energy efficiency of SNOW over LoRaWAN under a large number of nodes (2000 nodes) through realistic simulations in our earlier work in~\cite{snow_ton}. As reported in~~\cite{snow_ton}, a LoRaWAN node consumes on average 450.56mJ of energy to send 100 forty-byte packets compared to 22.22mJ in a SNOW node when 2000 nodes transmit concurrently.


\subsubsection{Performance in Downlink Communication}
\begin{figure}[!htb]
\centering
\includegraphics[width=0.50\textwidth]{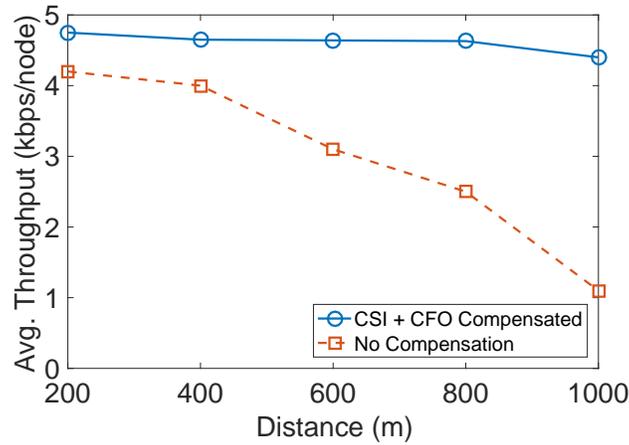}
\caption{Throughput in downlink communication.}
\label{fig:chap4-down_tput}
\end{figure}
In this section, we evaluate the downlink network performance in terms of throughput. The BS sends 1000 consecutive 30-byte (payload length) packets to each of the 25 nodes. Also, the BS and the nodes compensate for both CSI and CFO. In downlink, the BS uses a Tx bandwidth of 39kHz. We repeat the above experiment without compensating for CSI and CFO as well. Figure~\ref{fig:chap4-down_tput} shows the average throughput per node at different distances. For better representation, we cluster the nodes that are located approximately at the same distance and plot average throughput against the distance. As shown in this figure, a node that is approximately 200m away from the BS can achieve an average downlink throughput of 4.8kpbs, while both the BS and the node compensate for CSI and CFO. The average throughput remains almost the same as those observed at other distances, up to 1km as well. In contrast, the average throughput drops sharply with an increase in the distance when CSI and CFO are not compensated for. Note that a CC13x0 device can successfully receive an OOK-modulated signal with 4.8kbaud symbol rate and 39kHz bandwidth~\cite{cc1310}.
Overall, our CC13x0-based SNOW implementation holds the potentials for low-rate IoT applications.
\begin{figure}[t]
    \centering
      \subfigure[Throughput under varying payload size\label{fig:chap4-m_tput}]{
    \includegraphics[width=0.50\textwidth]{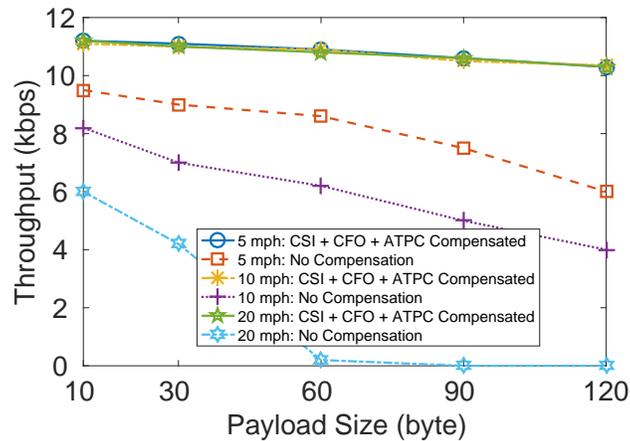} 
      }\vfill
      \subfigure[Energy consumption under varying payload size\label{fig:chap4-menergy}]{
        \includegraphics[width=.50\textwidth]{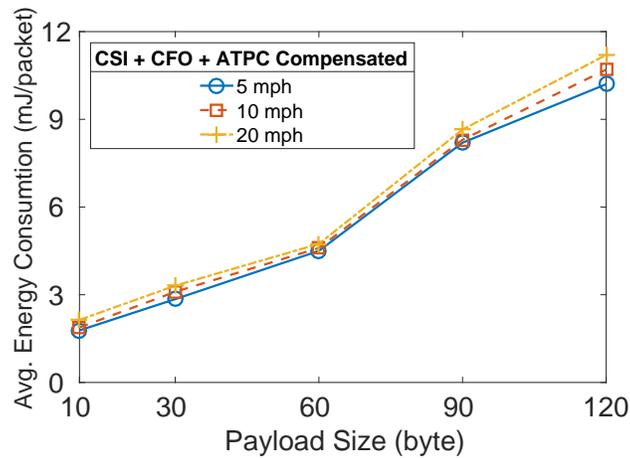}
      }
    \caption{Throughput and energy consumption under mobility.}
    \label{fig:chap4-mobility}
\end{figure}

\begin{figure}[!htbp]
    \centering
      \subfigure[End-to-end delay under varying payload size\label{fig:chap4-mdelay}]{
        \includegraphics[width=.50\textwidth]{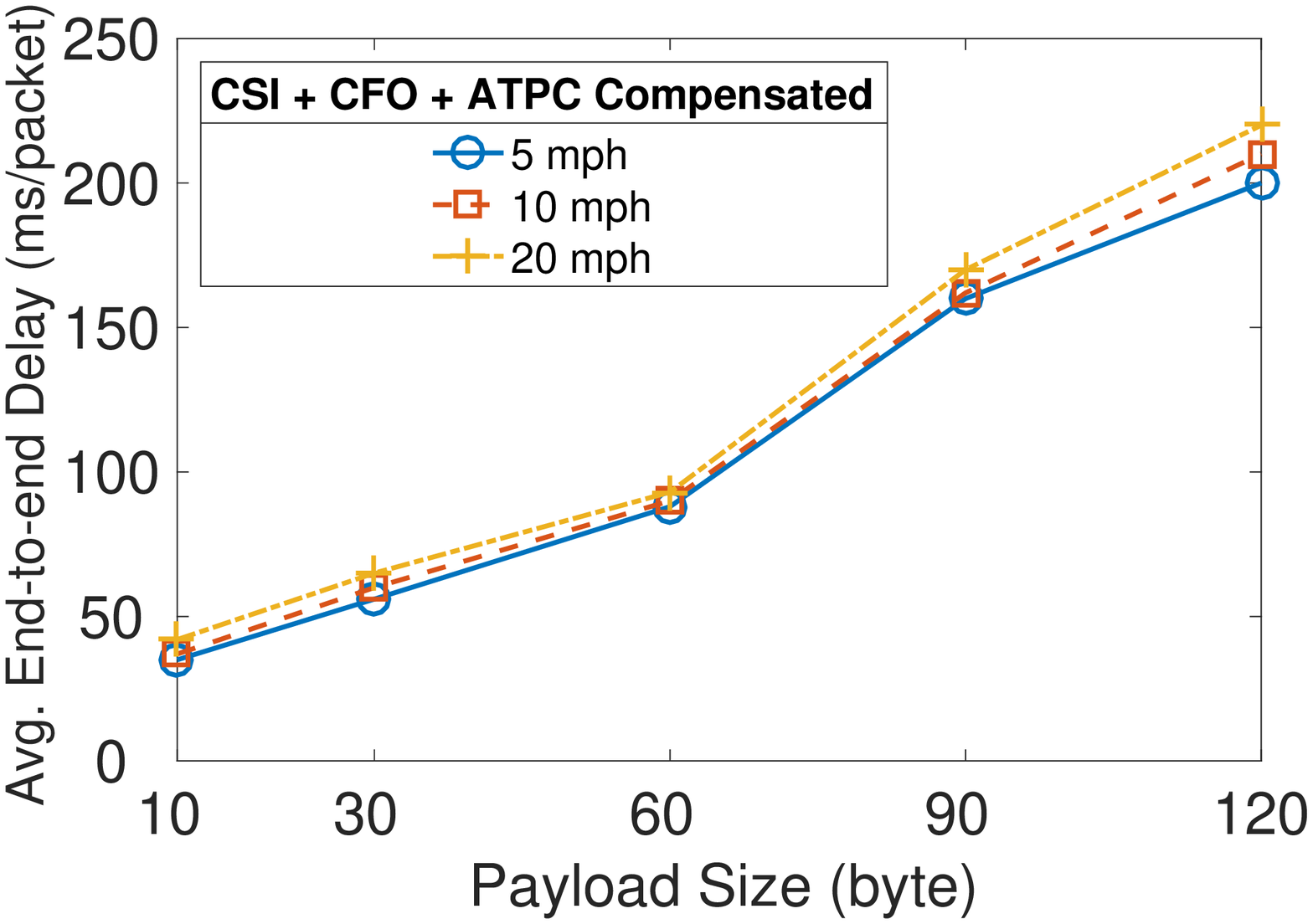}
      }\vfill 
      \subfigure[CDF of end-to-end delay at different payload sizes\label{fig:chap4-m_cdfpayload}]{
        \includegraphics[width=0.50\textwidth]{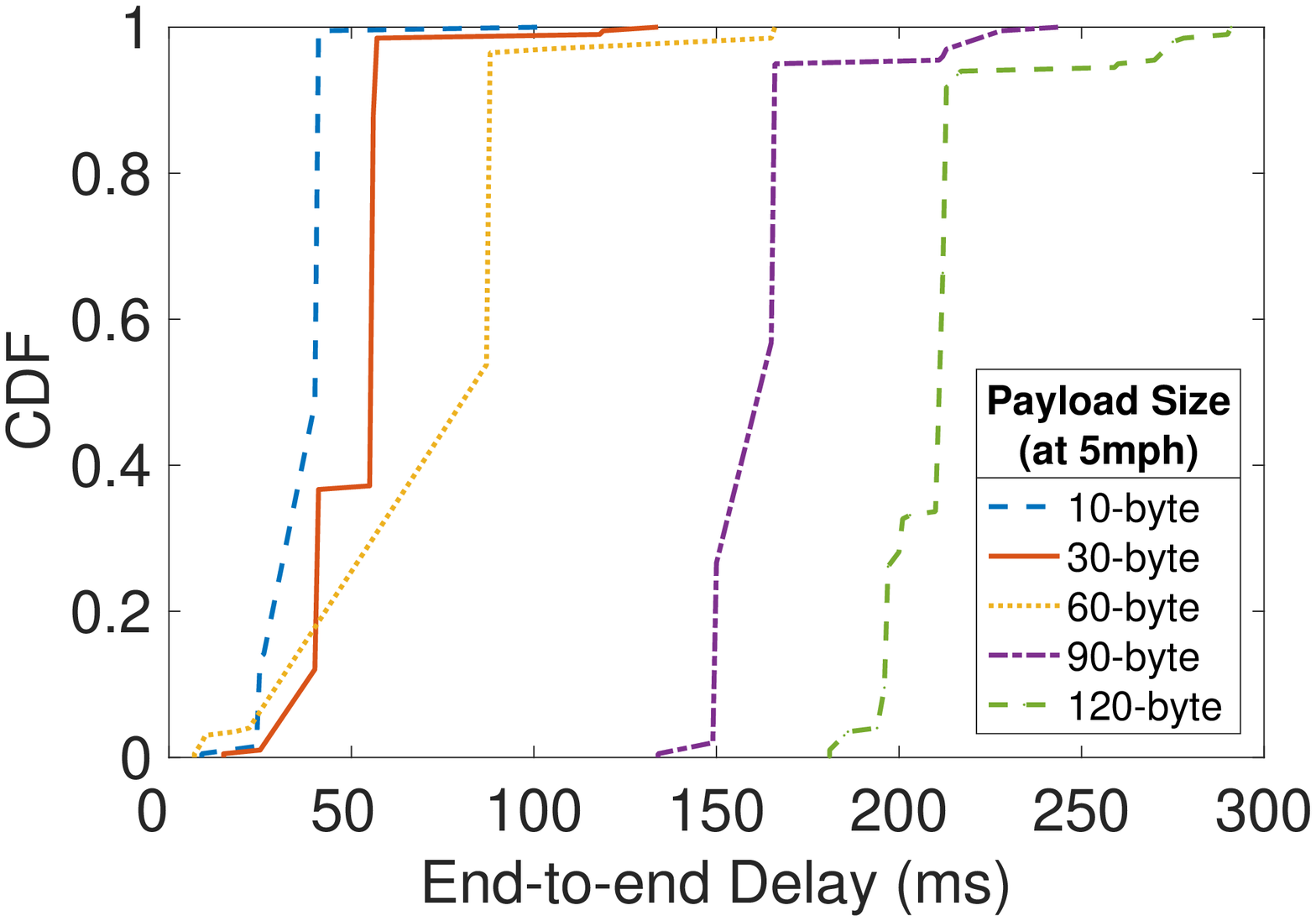}
      }\vfill 
      \subfigure[CDF of end-to-end delay at different speeds\label{fig:chap4-m_cdfmph}]{
        \includegraphics[width=.50\textwidth]{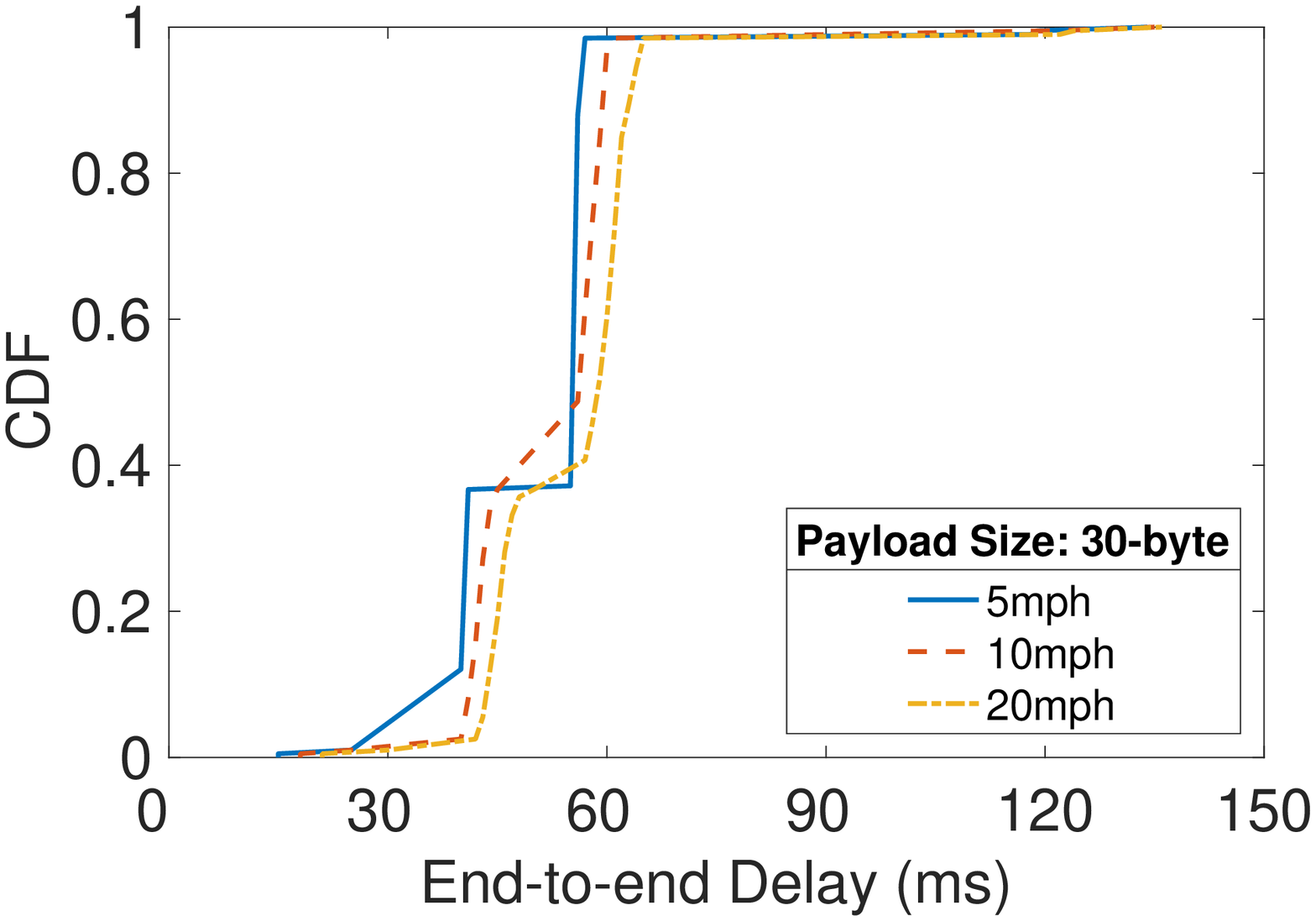}
      }
    \caption{End-to-end delay under node's mobility.}
    \label{fig:chap4-m_eted}
 \end{figure}

\subsubsection{Performance under Mobility}
In this section, we evaluate the network performance under CC13x0 node's mobility in terms of throughput, energy consumption, and end-to-end delay. 
We allow all 25 nodes to transmit concurrently to the BS. However, due to our limited resources, we enable mobility in only one node that is approximately 600m far from the BS and calculate its performance. 
All nodes except the mobile node continuously transmit to the BS 30-byte (payload size) packets with a random interval between 0 and 50ms, using their assigned subcarriers, each 39kHz wide. 
We vary the speed of the mobile node approximately to 5mph, 10mph, and 20mph in any arbitrary direction within our network range. At each speed, we change the payload size of the mobile node between 10 and 120bytes. For each payload size, the mobile node transmits to the BS 1000 packets with a random interval between 0 and 50ms. We run experiments with the above settings for two cases: (1) the mobile node or/and the BS {\bf compensate} for CSI, CFO, and ATPC; (2) the mobile node or/and the BS {\bf do not compensate} for CSI, CFO, and ATPC.

{\bf Throughput.}
Figure~\ref{fig:chap4-m_tput} shows the throughput at the BS (of the mobile node) for different speeds and payload sizes. As this figure suggests, the throughput decreases slightly from 11.18kbps to 10.3kbps at 5mph, 10.35kbps at 10mph, and 10.3kbps at 20mph for an increase in the payload size between 10 and 120bytes, as CSI, CFO, and ATPC are compensated for. When the mobile node or/and the BS do not compensate for CSI, CFO, and ATPC, the throughput decreases sharply with an increase in speed and packet size. For example, at 20mph, the throughput drops to approximately 0 for payload size of 60bytes. In general, the packet size is susceptible to node's mobility. In fact, if CSI and CFO are not compensated for, the effects of unknown channel conditions and frequency offset ripple through a longer packet and increase the BER. Thus, our SNOW implementation is resilient and robust under node's mobility.

{\bf Energy Consumption.}
Figure~\ref{fig:chap4-menergy} shows that the average energy consumption per packet increases slightly higher than linear with an increase in the payload size, when CSI, CFO, or/and ATPC are compensated for. For example, at 5mph, it takes on average 1.78mJoule, 2.85mJoule, 4.5mJoule, 8.2mJoule, and 10.2mJoule to transmit a payload of size 10, 30, 60, 90 and 120bytes, respectively. Also, the average energy consumption per packet increases with an increase in the speed. As shown in this figure, the average energy consumption per packet is approximately 1.78mJoule at 5mph and 2.14mJoule at 20mph, for a payload of size 10bytes. Our best guess is that at higher speeds the mobile node retransmits several packets multiple times due to ACK loss, high BER at BS, or/and ATPC. Overall, Figure~\ref{fig:chap4-menergy} confirms that our CC13x0-based SNOW implementation is energy efficient under node's mobility.

{\bf End-to-end Delay.}
Figure~\ref{fig:chap4-mdelay} shows that the average end-to-end delay per packet at the mobile node increases with an increase in speed and payload size. For example, at 5mph, the average end-to-end delay per packet with a payload of size 10, 30, 60, 90, and 120bytes are 35, 56, 88, 160, 200ms, respectively; at 10mph, the average end-to-end delays are 37, 60, 90, 162, 210ms, respectively; at 20mph, the average end-to-end delays are 42, 65, 93, 170, 220ms, respectively. Moreover, Figure~\ref{fig:chap4-m_cdfpayload} shows the cumulative distribution function (CDF) of the end-to-end delay at a constant speed of 5mph with varying payload sizes. This figure shows that 60\% of the 10-byte (payload length) packets observe an end-to-end delay more than 35ms, 65\% of the 30-byte (payload length) packets observe an end-to-end delay more than 55ms, 50\% of the 60-byte (payload length) packets observe an end-to-end delay more than 90ms, 98\% of the 90-byte (payload length) packets observe an end-to-end delay more than 150ms, and 95\% of the 120-byte (payload length) packets observe an end-to-end delay more than 195ms. Furthermore, Figure~\ref{fig:chap4-m_cdfmph} shows the CDF of end-to-end delays for a fixed payload length of 30bytes at varying speed. As this figure shows, 98\% of the packets at 5mph observe an end-to-end delay up to 55ms, 99.99\% of the packets at 10mph observe an end-to-end delay up to 60ms, and 98\% of the packets at 20mph observe an end-to-end delay up to 65ms. Overall, Figure~\ref{fig:chap4-m_eted} confirms that our CC13x0-based SNOW implementation may provide very low latency under node's mobility.

%% file: chapter4/related.tex
\subsection{Related Work}\label{sec:chap4-related}
Recently, a number of LPWAN technologies have been developed that operate in the licensed (LTE Cat M1~\cite{cat, saxena2016achievable}, NB-IoT~\cite{nbiot, chen2017narrowband}, EC-GSM-IoT~\cite{ecgsmiot, gozalvez2016new}, 5G~\cite{3gpp, akpakwu2017survey}) or unlicensed (LoRa~\cite{lorawan}, SigFox~\cite{sigfox}, RPMA (INGENU)~\cite{rpma}, IQRF \cite{iqrf}, Telensa~\cite{telensa},  DASH7~\cite{dash7}, WEIGHTLESS-N~\cite{weightless}, WEIGHTLESS-P~\cite{weightless-p}, IEEE 802.11ah~\cite{IEE80211_ah}, IEEE 802.15.4k \cite{IEE802154_k}, IEEE 802.15.4g~\cite{IEE802154_g}) spectrum. 
Operating in the licensed band is costly due to high service fee and costly infrastructure. On the contrary, most non-cellular LPWANs, except SNOW and WEIGHTLESS-W, operate in the ISM band. While the ISM band is unlicensed, it is getting heavily crowded due to the proliferation of LPWANs as well as other wireless technologies in this band. 
To avoid the high cost of licensed band and the crowd of the ISM band, SNOW was designed to exploit the widely available, less crowded, and wide spectrum of the TV white spaces. 
Existing work on white space focused on exploiting the white spaces for broadband access~\cite{whitespaceSurvey, zhang2015design, hasan2014gsm, kumar, harrison2015whitespace, ws_sigcomm09, WATCH, videostreaming, linkasymmetry, vehiclebased, ws_mobicom13, ws_nsdi10} and spectrum determination through spectrum sensing~\cite{saeed2017local, ws_dyspan08_kim, ws_mobicom08, ws_dyspan11, FIWEX} or/and geo-location approach~\cite{dbreq, database1, database2, database3, vehiclebased, hysim}. Alongside,  
various standards bodies (IEEE 802.11af~\cite{IEE802_af}, IEEE 802.15.4m~\cite{ieee154}, IEEE 802.19.1~\cite{IEE802_19}, IEEE 802.22~\cite{IEEE802_22}, IEEE 1900.4a~\cite{IEE19004_a}, IEEE 1900.7~\cite{IEE19007}, ECMA-392~\cite{ECMA392, kocks2012spectrum}) and industry leaders (Microsoft~\cite{4Africa, MSRAfrica}, Google~\cite{GoogleAfrica}) have also targeted the white spaces for unlicensed personal or commercial use. In contrast, SNOW exploits white spaces for highly scalable LPWAN.
With the rapid growth of IoT,  LPWANs will suffer from crowded spectrum due to long range. It is hence critical to exploit white spaces for IoT. 
Our paper focuses on implementing SNOW using the cheap and widely available COTS devices for practical and scalable deployment.

%% file: chapter5.tex
\section{Integrating Multiple LPWANs for Enhanced Scalability and Extended Coverage}   \label{chap:snow4}
Due to their capability of communicating over long distances at very low transmission power, Low-Power Wide-Area Networks (LPWANs) are evolving as an enabling  technology for Internet of Things (IoT). 
Despite their promise,  existing LPWAN technologies still face limitations in meeting scalability and covering very wide areas which make their adoption challenging for future IoT applications, specially in  infrastructure-limited rural areas. 
To address this limitation, in this paper, we consider achieving scalability and extended coverage by integrating multiple LPWANs. 
{\bf\slshape SNOW (Sensor Network Over White Spaces)}, a recently proposed LPWAN architecture over the TV white spaces, has demonstrated its advantages over  existing LPWANs in performance and  energy-efficiency.  
In this paper, we propose to scale up LPWANs through a seamless  integration of multiple SNOWs which enables concurrent inter-SNOW and intra-SNOW communications.  
We then formulate the tradeoff between scalability and inter-SNOW interference as a constrained optimization problem whose objective is to  maximize scalability by managing white space spectrum sharing across multiple SNOWs. We also prove the NP-hardness of this problem. 
We then propose an intuitive polynomial time heuristic algorithm for solving the scalability optimization problem which is highly efficient in practice. For the sake of theoretical bound, we also propose a simple polynomial-time $\frac{1}{2}$-approximation algorithm for the scalability optimization problem. 
Hardware experiments through deployment in an area of (25x15)km$^2$ as well as large scale simulations demonstrate the effectiveness of our algorithms and feasibility of achieving scalability through seamless integration of SNOWs with high reliability, low latency, and energy efficiency.

\input{chapter5/introduction}

\input{chapter5/snow_overview}

\input{chapter5/system_model}

\input{chapter5/inter_snow_comm}

\input{chapter5/technical}

\input{chapter5/experiment}

\input{chapter5/simulation}

\input{chapter5/related}

\subsection{Summary}\label{sec:chap5-conclusions}
LPWANs represent a key enabling  technology for Internet of Things (IoT) that offer long communication range at low power. While many competing LPWAN technologies have been developed recently, they still face limitations in meeting scalability and covering much wider area, thus making their adoption challenging for future IoT applications, specially in  infrastructure-limited rural areas. In this paper, we have addressed this challenge by integrating multiple LPWANs for scalability and extended coverage. Recently proposed {\slshape SNOW}, an LPWAN that operates over the TV white spaces has demonstrated its advantages over  existing LPWANs in performance and  energy-efficiency.  We have proposed to scale up LPWANs through a seamless  integration of multiple SNOWs that enables concurrent inter-SNOW and intra-SNOW communications.  We have then formulated the tradeoff between scalability and inter-SNOW interference as a scalability optimization problem, and have proved its NP-hardness.  We have proposed a polynomial time heuristic that is highly effective in experiments as well as a polynomial-time $1/2$-approximation algorithm. Testbed experiments as well as large scale simulations demonstrate the feasibility of achieving scalability through our proposed integration of SNOWs with high reliability, low latency, and energy efficiency.


%% file: chapter5/introduction.tex
\subsection{Introduction}\label{sec:chap5-introduction}
To overcome the range limit and scalability challenges in traditional wireless sensor networks (WSNs), Low-Power Wide-Area Networks (LPWANs) are emerging as an enabling  technology for Internet of Things (IoT).   Due to their escalating demand, LPWANs are gaining momentum, with multiple competing technologies being developed including  LoRaWAN, SigFox, IQRF, RPMA (Ingenu), DASH7, 
Weightless-N/P in the ISM band; and 
EC-GSM-IoT, NB-IoT,  LTE Cat M1  (LTE-Advanced Pro), and 5G in the licensed cellular band (see survey~\cite{ismail2018low}).
In parallel,
to avoid the crowd of the limited ISM band and the cost of the licensed band,
we developed {\bf\slshape  SNOW (Sensor Network Over White Spaces)}, an LPWAN architecture  to support wide-area WSN by exploiting the TV white spaces~\cite{snow_ton, snow, snow2}.  \revise{{\slshape White spaces} refer to the allocated but locally unused TV channels, and can be used by unlicensed devices as secondary users. Unlicensed devices need to either sense the medium or consult with a cloud-hosted geo-location database before transmitting~\cite{fcc_second_order}.}
Thanks to their lower frequencies (54--862MHz in the US), white spaces have excellent propagation characteristics over long distance and obstacles.
While their potentials have been explored mostly for broadband access (see survey~\cite{whitespaceSurvey}), our 
design and experimentation demonstrated the potential of SNOW to enable asynchronous, low power, bidirectional, and massively concurrent communications between numerous sensors and a base station (BS) directly over long distances~\cite{snow, snow2, snow_ton}. 

Despite their promise, existing LPWANs face challenge in very large-area (e.g., city-wide) deployment~\cite{charm, choir}. Without line of sight, communication range of LoRaWAN, a leading LPWAN technology that is commercially available~\cite{LoRaleader1, LoRaleader2, loraindoor, marcelis2017dare, linklab, semtech}, is  short~\cite{loraindoor}, specially indoors ($<$100m while its specified urban range is 2--5km~\cite{thailand}).  Its performance  drops sharply as the number of nodes grows~\cite{bor2016lora, LPWANchallenges, voigt2016mitigating, lorascale, lorascale2, loralimitation}, supporting only 120 nodes per 3.8 hectares~\cite{bor2016lora} which is not sufficient to meet the future IoT demand.
Apart from these scenarios, applications like 
agricultural IoT,  oil-field monitoring,  smart and connected rural communities would require much wider area coverage~\cite{ismail2018low, whitespaceSurvey}.  In this paper, we address this {\bf challenge} and propose LPWAN scalability by integrating multiple LPWANs.

Most LPWANs are limited to star topology, and rely on wired infrastructure (e.g., cellular LPWANs) or Internet (e.g., LoRaWAN) to integrate multiple networks to cover large areas. Lack of infrastructure (also raised in a hearing before the US Senate~\cite{monsanto}) hinders their adoption to enable rural and remote area applications such as {\bf agricultural IoT}  and {\bf industrial IoT}  (e.g., for oil/gas field~\cite{oilffield}) that may cover hundreds of square kms. According to the Department of Agriculture, $<20\%$ farmers can afford the cost of manual sensor data collection for smart farming \cite{doa}. Smart farming powered by IoT can double the produce at low cost by better measuring soil nutrients, moisture, fertilizer, seeds, and storage temperature through dense sensor deployment~\cite{chemistry, nature}. Industries like Microsoft~\cite{vasisht17farmbeats, farmbeat},  Monsanto~\cite{monsanto},  and many \cite{climate, deere, claas, cnh, att}  are now promoting agricultural IoT.  Monitoring a large oil-field (e.g., 74x8km$^2$ East TX Oil-field~\cite{texasof})  needs to connect tens of thousands of sensors ~\cite{WirelessHART_Emerson, WirelessHART_MOL}.  Such agricultural IoT and industrial IoT can be enabled by integrating multiple LPWANs specially SNOWs due to abundant white spaces.
Similar integration may also be needed in a smart city deployment for extended coverage or for running different applications on different LPWANs.

In this paper, we address the above scalability challenge  by integrating multiple  SNOWs that are under the same management/control. Such an integration raises several concerns. First, we have to design a  protocol to enable inter-SNOW communication, specially peer-to-peer communication (when a node in one SNOW wants to communicate with a node in a different SNOW). Second, since multiple coexisting SNOWs can interfere each other, thus affecting the scalability,   it is critical to handle the tradeoffs between scalability and inter-SNOW interference.  Specifically, we make the following  novel contributions.

\begin{itemize}

\item We propose to scale up LPWAN through seamless  integration of multiple SNOWs that enables concurrent inter- and intra-SNOW communications.  This is done by exploiting the characteristics of the SNOW physical layer.

\item We then formulate the tradeoff between scalability and inter-SNOW interference as a constrained optimization problem whose objective is to  maximize scalability by managing white space spectrum sharing across multiple SNOWs, and prove its NP-hardness.

\item We propose an intuitive polynomial time heuristic for solving the scalability optimization problem which is highly efficient in practice.

\item For the sake of analytical performance bound,  we also propose a simple polynomial-time approximation algorithm with an approximation ratio of $\frac{1}{2}$.

\item We implement the proposed SNOW technologies  in GNU radio~\cite{gnuradio} on Universal Software Radio Peripheral (USRP) devices~\cite{usrp}. We perform experiments by deploying 9 USRP devices in an area of (25x15)km$^2$ in Detroit, Michigan.  We also perform large scale simulations in NS-3~\cite{ns3}. Both experiments and simulations demonstrate the feasibility of achieving scalability through seamless integration of SNOWs allowing  concurrent intra- and inter-SNOW communications with high reliability, low latency, and energy efficiency while using our heuristic and approximation algorithms. Also, simulations show that SNOW cluster network can connect thousands of sensors over tens of kilometers of geographic area.
\end{itemize}


In the rest of the paper,  Section~\ref{sec:chap5-snow_overview} gives an overview of SNOW. Section~\ref{sec:chap5-system_model} explains the system model.  
Section~\ref{sec:chap5-p2p} describes our inter-SNOW communication technique. Section~\ref{sec:chap5-technical} formulates the scalability optimization problem for integration, proves its NP-hardness, and presents the heuristic and the approximation algorithm. Section~\ref{sec:chap5-implementation} explains the implementation of our network model. Section~\ref{sec:chap5-eval} presents our experimental and simulation results. Section~\ref{sec:chap5-related} presents related work. Finally, Section~\ref{sec:chap5-conclusions} concludes the paper.

%% file: chapter5/snow_overview.tex
\subsection{An Overview of SNOW}\label{sec:chap5-snow_overview}
\begin{figure}[!htb]
\centering
\includegraphics[width=0.7\textwidth]{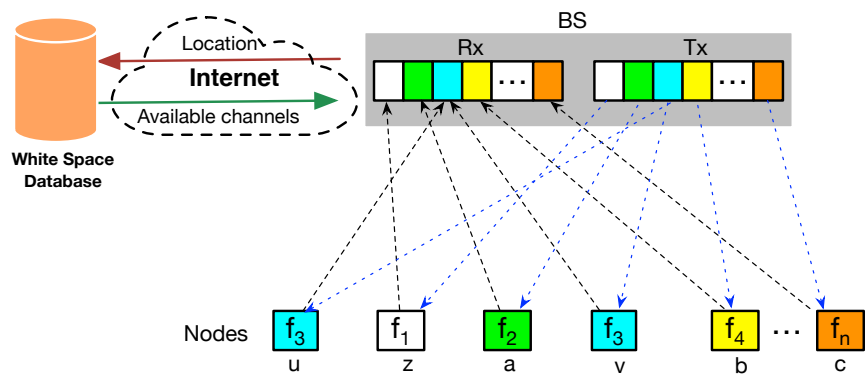}
\caption{SNOW architecture with dual radio BS and subcarriers.}
\label{fig:chap5-dualradio}
\end{figure}
Here we provide a brief overview of the design and architecture of a {\bf single} SNOW that we developed in~\cite{snow, snow2, snow_ton}.
SNOW is an asynchronous,  long range, low power WSN platform to operate over TV white spaces. A SNOW node has a single half-duplex narrowband radio.  
Due to long transmission (Tx) range, the nodes are directly connected to the BS and vice versa (Figure~\ref{fig:chap5-dualradio}). SNOW thus forms a star topology.    
\revise{The BS determines white spaces in the area by accessing a cloud-hosted database  through the Internet. 
Hence, it does not check on the incumbents or evaluate cross-technology interference.
The nodes  are power constrained and not directly connected to the Internet.  They do not do spectrum sensing or cloud access.}
The BS uses a wide channel split into orthogonal subcarriers.   As shown in Figure~\ref{fig:chap5-dualradio}, the BS uses two radios, both operating on the same spectrum --  one for only transmission (called {\bf\slshape Tx radio}),  and the other for only reception (called {\bf\slshape Rx radio}). Such a dual-radio of the BS allows concurrent
bidirectional communications in SNOW. 
We implemented SNOW on USRP (universal software radio peripheral) devices~\cite{usrp} using GNU Radio~\cite{gnuradio}.  
The implementation has been made {\bf open-source}~\cite{snow_bs, snow_cots}. A short video demonstrating how SNOW works is also available in YouTube~\cite{snow_demo}. In the following, 
we provide a brief overview of the SNOW physical layer
(PHY) and the Media Access Control (MAC) layer. A full description of this design is available in~\cite{snow_ton}.

\subsubsection{SNOW PHY Layer}\label{sec:chap5-phy}

A key design goal of SNOW is to achieve high scalability by exploiting wide spectrum of white spaces. Hence, its PHY is designed based on a 
{\bf D}istributed implementation of {\bf OFDM} for multi-user access, called {\bf D-OFDM}. D-OFDM  splits a wide spectrum into numerous narrowband orthogonal subcarriers enabling parallel data streams to/from numerous distributed nodes from/to the BS.  A subcarrier bandwidth is in kHz (e.g., 50kHz, 100kHz, 200kHz, or so depending on packet size and needed bit rate). Narrower bands have lower bit rate but longer range, and consume less power~\cite{channelwidth}. The nodes transmit/receive on orthogonal subcarriers, each using one. A subcarrier is modulated using Binary Phase Shift Keying (BPSK) or Amplitude Shift Keying (ASK).  If the BS spectrum is split into  $n$  subcarriers,  it can receive from $n$ nodes simultaneously using a single antenna. Similarly, it can transmit different data on different subcarriers through a single transmission. The BS can also use fragmented spectrum. This design is different from MIMO radio adopted in various wireless domains including IEEE 802.11n~\cite{mimo} as they rely on multiple antennas to enable the same. 


While OFDM has been adopted  for multi-access in the forms of OFDMA  and SC-FDMA in various  broadband (e.g., WiMAX~\cite{wimax}) and cellular  (e.g., LTE~\cite{lte_whitepaper}) technologies~\cite{3gpp, scfdma, OFDMAWiMAX}, they rely on strong time synchronization which is very costly for low-power nodes. We adopted OFDM for the first time in WSN design and without requiring time synchronization. D-OFDM enables multiple packet receptions that are transmitted asynchronously from different nodes which was possible as WSN needs low data rate and short packets. Time synchronization is avoided by extending the symbol duration (repeating a symbol multiple times) and sacrificing bit rate. The effect is similar to extending cyclic prefix (CP) beyond what is required to control  inter-symbol interference (ISI). CPs of adequate lengths have the effect of rendering asynchronous signals to appear orthogonal at the receiver, increasing guard-interval. As it reduces data rate, D-OFDM is suitable for LPWAN. Carrier frequency offset (CFO) is estimated using training symbols when a node joins the network on a subcarrier (right most) whose overlapping subcarriers are not used. Using this CFO, it is  determined on its assigned subcarrier and compensated for using traditional method to mitigate ICI.

\subsubsection{SNOW MAC Layer}

\revise{The BS spectrum  is split into  $n$ overlapping orthogonal subcarriers -- $f_1, f_2, \cdots, f_n$ -- each of equal width. Each node is assigned one subcarrier. When the number of nodes is no greater than the number of subcarriers, every node is assigned a unique subcarrier.  Otherwise, a subcarrier is shared by more than one node. 
The nodes that share the same subcarrier will contend for and access it using a CSMA/CA (Carrier Sense Multiple Access with Collision Avoidance) policy. The subcarrier assignment by the BS minimizes the interference and contention between the nodes. As long as there is an option, the BS thus tries to assign different subcarriers to the nodes that are hidden to each other.}

The subcarrier allocation is done by the BS.
The nodes in SNOW use a lightweight CSMA/CA protocol for transmission that uses a static interval for random back-off like the one used in TinyOS~\cite{tinyos} . 
Specifically, when a node has data to send, it wakes up by turning its radio on. Then it performs a random back-off in a fixed {\slshape initial back-off window}.  When the back-off timer expires, it runs CCA (Clear Channel Assessment) and if the subcarrier is clear, it transmits the data. If the subcarrier is occupied, then the node makes a random back-off in a fixed {\slshape congestion back-off window}. After this back-off expires, if the subcarrier is clean the node transmits immediately. This process is repeated until it makes the transmission. The node then can go to sleep again.

The nodes can autonomously transmit, remain in receive (Rx) mode, or sleep. 
Since D-OFDM allows handling asynchronous Tx and 
Rx, the link layer can send acknowledgment (ACK) for any transmission in either direction.   As shown in Figure~\ref{fig:chap5-dualradio}, both radios of the BS use the same spectrum and subcarriers - the subcarriers in the Rx radio are for receiving while those in the Tx radio are for transmitting. Since each node (non BS) has just a single half-duplex radio, it can be either receiving or transmitting, but not doing both at the same time.  
Both experiments and large-scale simulations show high efficiency  of SNOW in latency and energy with a linear increase in throughput with the number of nodes, demonstrating its superiority  over existing designs~\cite{snow, snow2}.

%% file: chapter5/system_model.tex
\subsection{System Model}\label{sec:chap5-system_model}

\begin{figure}[!htb]
\centering
\includegraphics[width=0.85\textwidth]{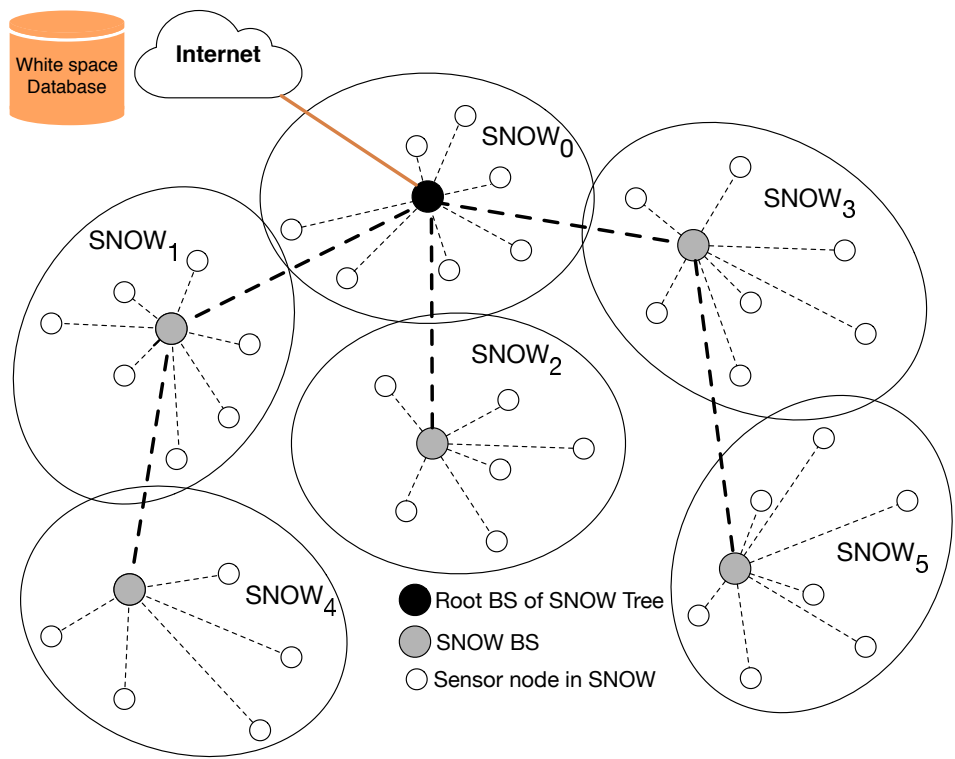}
\caption{A SNOW-tree.}
\label{fig:chap5-mesh}
\end{figure}
We consider  many coexisting SNOWs that are under the same management/control  and  need to coordinate among themselves for extended coverage in a  wide area or  to host different applications.   
As such, we consider an inter-SNOW network  as a SNOW-tree in the spirit of a  cluster tree  used in the new IEEE 802.15.4m standard~\cite{802154m}, each  cluster representing a personal area network under a coordinator. The {\bf root} of the tree is  connected to the white space database. In the similar spirit, our inter-SNOW network of the coordinated SNOWs is shown in Figure~\ref{fig:chap5-mesh} as a SNOW-tree. \revise{Each cluster is a star topology SNOW. All BSs form a tree that are connected through white space. Each BS is powerful or there can be multiple backup BSs for each cluster. So the chances of a BS failure is quite low in practice. Even if a BS fails, the root BS may reconstruct the tree.}

Let there be a total of $N$ BSs (and hence $N$ SNOWs) in the SNOW-tree, denoted by BS$_0$, BS$_1$, $\cdots$, BS$_{N-1}$, where BS$_i$ is the base station of SNOW$_i$. BS$_0$ is the {\bf root BS} and is connected to the white space database through the Internet. The remaining BSs are in remote places where Internet connection many not be available. Those BSs thus depend on BS$_0$ for white space information.
Every BS is assumed to know the location of its operating area (its location and the locations of its nodes).  Localization is not the focus of our work and can be achieved through  manual configuration or some  existing WSN localization technique such as those based on ultrasonic sensors or other sensing modalities~\cite{wsnlocalizationsurvey}. 
BS$_0$  gets the location information of all BSs and  finds the white space channels for all SNOWs. It also knows the topology of the tree and allocates the spectrum among all SNOWs. Each BS  splits its assigned spectrum and assigns subcarriers to its nodes. \revise{For simplicity, we consider that all nodes in the tree transmit with the same transmission power and receive with the same receive sensitivity.}

\revise{In an agricultural IoT, Internet connection is not available everywhere in the wide agricultural field. The farmer's home usually has the Internet connection and the root BS can be placed there. Microsoft's Farmbeats~\cite{vasisht17farmbeats} project for agricultural IoT also exhibits such a scenario. Similarly, in a large oil field, the root BS can be in the office or control room. The considered SNOW-tree thus represents practical scenarios of wide area deployments in rural fields. The IEEE 802.15.4m standard also aims to utilize the white spaces under the exact same tree network model. We shall consider the scalability through a seamless integration and communication protocol among such coexisting SNOWs.}

%% file: chapter5/inter_snow_comm.tex
\subsection{Enabling Concurrent Inter-SNOW and intra-SNOW Communications}\label{sec:chap5-p2p}

Here we describe our inter-SNOW communication technique to enable seamless integration of the SNOWs for scalability. Specifically, we explain how we can enable concurrent inter-SNOW and intra-SNOW communications by exploiting the PHY design of SNOW.    To explain this we consider  \emph{peer-to-peer} inter-cluster communication in the SNOW-tree. That is, one node in a SNOW wants or needs to communicate with a node in another SNOW.
\begin{figure}[!htb]
\centering
\includegraphics[width=0.85\textwidth]{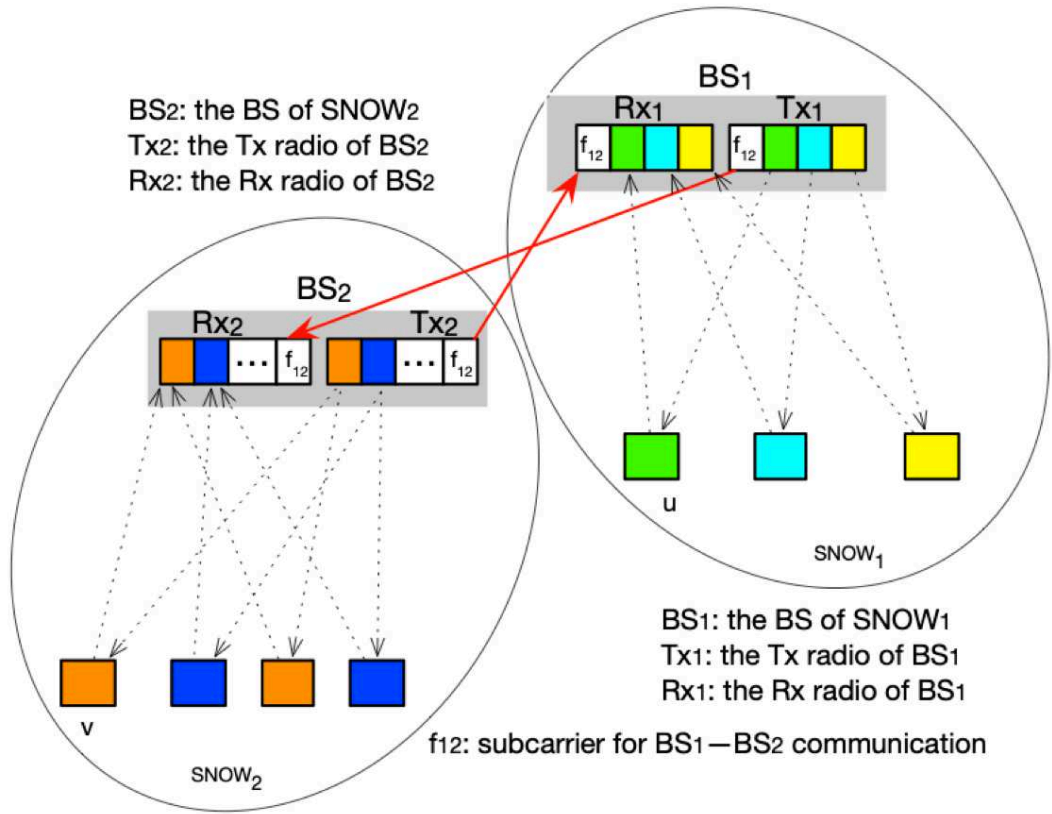}
\caption{Inter-SNOW communication.}
\label{fig:chap5-intercluster}
\end{figure}
  
For peer-to-peer communication across SNOWs, a node first sends its packet to its BS.
Note that two nodes may not communicate directly even if they are in communication range of each other as they may operate on different subcarriers.
The BS will then route to the destination SNOW's BS along the path given by the tree which in turn  will  forward to the destination node.  Hence, the first {\bf question} is {\slshape ``How do two neighboring BSs exchange packets without interrupting their communication with their own nodes?"} Let us consider SNOW$_1$ and SNOW$_2$ as  two neighboring SNOWs in Figure~\ref{fig:chap5-intercluster} which will communicate with each other. We allocate a special subcarrier from both of their spectrum (i.e., a common subcarrier among the two BSs) that will be used for communication between these two BSs. 
For a tree link BS$_i\rightarrow$ BS$_j$, this subcarrier is denoted by $f_{i,j}$. To each tree link BS$_i\rightarrow$ BS$_j$, we assign a distinct $f_{i,j}$, eliminating interference among the BS transmissions made along the tree links. This is always feasible because the number ($N$) of SNOWs, and hence the number of tree links ($N-1$), is very small compared to the total number of subcarriers.
\revise{Additionally, if the connecting subcarrier that forms a tree link for BS-BS communication fails, another subcarrier is assigned since usually there is much overlap between two neighboring BSs.}

As shown in Figure~\ref{fig:chap5-intercluster}, $f_{1,2}$ is a special subcarrier that enables BS$_1$-BS$_2$ communication as described above. D-OFDM allows us to encode any data on any subcarrier while the radio is transmitting. Thus the SNOW PHY will allow us to encode any time on any number of subcarriers and transmit. Exploiting this important feature of the SNOW PHY, Tx$_1$ radio will encode the packet on the subcarrier $f_{1,2}$ which is used for BS$_1$--BS$_2$ communication in Figure~\ref{fig:chap5-intercluster}.
If there are pending ACKs for its own nodes, they can also be encoded in their respective subcarriers. Then Tx$_1$ radio makes a single transmission. Rx$_2$ will receive it on subcarrier $f_{1,2}$ while the nodes of SNOW$_1$ will receive on their designated subcarriers. BS$_2$ can receive from BS$_1$ in the same way. 
They can similarly forward to next neighboring SNOWs. Thus both inter-SNOW and intra-SNOW communications can happen in parallel. 
Following are the several issues and our techniques to address those to enable such communication.

\subsubsection{Handling Collision in BS-BS Communication}
Using one subcarrier for  BS$_1$--BS$_2$ communication,   BS$_1$ and BS$_2$ cannot simultaneously transmit to each other. When Tx$_1$ transmits on  $f_{1,2}$, there is high energy on $f_{1,2}$ at Rx$_1$. The similar is the case  when Tx$_2$ transmits. If they start transmitting simultaneously, both packets will be lost. A straightforward solution is to use two different subcarriers for  Tx$_1\rightarrow$ Rx$_2$ and Tx$_2\rightarrow$ Rx$_1$ transmission. However, using two subcarriers dedicated for this may result in their underutilization and hinder  scalability. Hence, we  use a single subcarrier for BS$_1$--BS$_2$ communication and adopt random back-off within a fixed interval rule for this special subcarrier. That is, if BS-BS communication collides, they make random back-off after which they retry transmission.

\subsubsection{Dealing with Sleep/Wake up} 
When a node $u$ from SNOW$_1$ wants to send a packet to a node $v$ in SNOW$_2$, it first makes the transmission to BS$_1$ which then sends to BS$_2$ (Figure~\ref{fig:chap5-intercluster}). 
When BS$_2$ attempts to transmit to $v$, it can be sleeping which  BS$_2$ has no knowledge of. To handle this, we adopt a periodic beacon that the BS of each SNOW sends to its nodes. The nodes are aware of the period of beacon. All nodes in a BS that are participating in peer-to-peer communication wake up for beacon. Thus, $v$ will wake up for beacon as it participates in peer-to-peer communication. BS$_2$ will 
encode $v$'s message on the subcarrier used by $v$ in the beacon. Thus, $v$ can receive the message from the beacon of BS$_2$.

%% file: chapter5/technical.tex

\subsection{Handling Tradeoffs between Scalability and Inter-SNOW Interference}\label{sec:chap5-technical}
\revise{Our objective of integrating multiple SNOWs is scalability which can be achieved if every SNOW can support a large number of nodes. The number of nodes supported by a SNOW increases if the number of subcarriers used in that SNOW increases. However, if each SNOW uses the entire  spectrum available at its location,  there will be much spectrum overlap with the neighboring SNOWs. This will ultimately increase inter-SNOW interference, resulting in a lot of back-offs by the nodes during packet transmission. Like any other LPWAN, SNOW nodes are energy-constrained and cannot afford any sophisticated MAC protocol to avoid such interference, thereby wasting energy.
On the other end, if all neighboring SNOWs use non-overlapping spectrum, inter-SNOW interference will be minimized, but each SNOW in this way can support only a handful of nodes, thus degrading  the scalability.} This tradeoff between scalability and inter-SNOW interference due to integration raises a spectrum allocation which cannot be solved using traditional spectrum allocation approach in wireless networks. We propose to accomplish such an allocation by formulating a {\bf\slshape Scalability Optimization Problem (SOP)} where our objective is to optimize scalability while limiting the interference.
To our knowledge, this problem is unique and never arose in other wireless domains. We now formulate SOP, prove its NP-hardness, and provide polynomial-time near-optimal solutions.

\subsubsection{SOP Formulation}\label{sec:chap5-prob_formulation}
The root BS knows the topology of the BS connections, accesses the white space database for each BS, and allocates the spectrum among the BSs. The spectrum allocation has to balance between scalability and inter-SNOW interference as described above. For SOP, we consider a uniform bandwidth $\omega$ of a subcarrier across all SNOWs. Let $Z_i$ be the set of  orthogonal subcarriers available at BS$_i$ considering $\alpha$ as the {\slshape fraction of overlap} between two neighboring subcarriers, where $0\le \alpha\le 0.5$ (as we found in our experiments~\cite{snow, snow2} that two orthogonal subcarriers can overlap at most up to half). Thus, if $W_i$ is the total available bandwidth at BS$_i$, then its total number of orthogonal subcarriers is given by
\begin{equation*}\label{eqn:chap5-subcarrier}
 |Z_i| = \frac{W_i}{\omega\alpha} - 1.
 \end{equation*}

We consider that the values of $\omega$ and $\alpha$ are uniform across all BSs. Let the set of subcarriers to be assigned to BS$_i$ be $X_i\subseteq Z_i$,   with $|X_i|$ being the number of subcarriers in $X_i$.  We can consider the {\slshape total number of subcarriers}, $\sum_{i = 0}^{N-1} |X_i|$, assigned to all SNOWs as the  {\slshape scalability} metric. We will maximize this metric. Every BS$_i$ (i.e., SNOW$_i$) requires a {\slshape minimum number of subcarriers} $\sigma_i$ to support its  nodes. Hence, we define {\bf Constraint } (\ref{c:chap5-const1})   to indicate the minimum and maximum number of subcarriers for each BS. If some communication in SNOW$_i$ is interfered by another communication in SNOW$_j$, then SNOW$_j$ is its {\bf \slshape interferer.} Since the root BS knows the locations of all BSs (all SNOWs) in the SNOW-tree, it can determine all interference relationships (which SNOW is an interferer of which SNOWs) among the SNOWs based on the nodes' communication range. 
\begin{figure*}[!htb]
\centering
\includegraphics[width=0.99\textwidth]{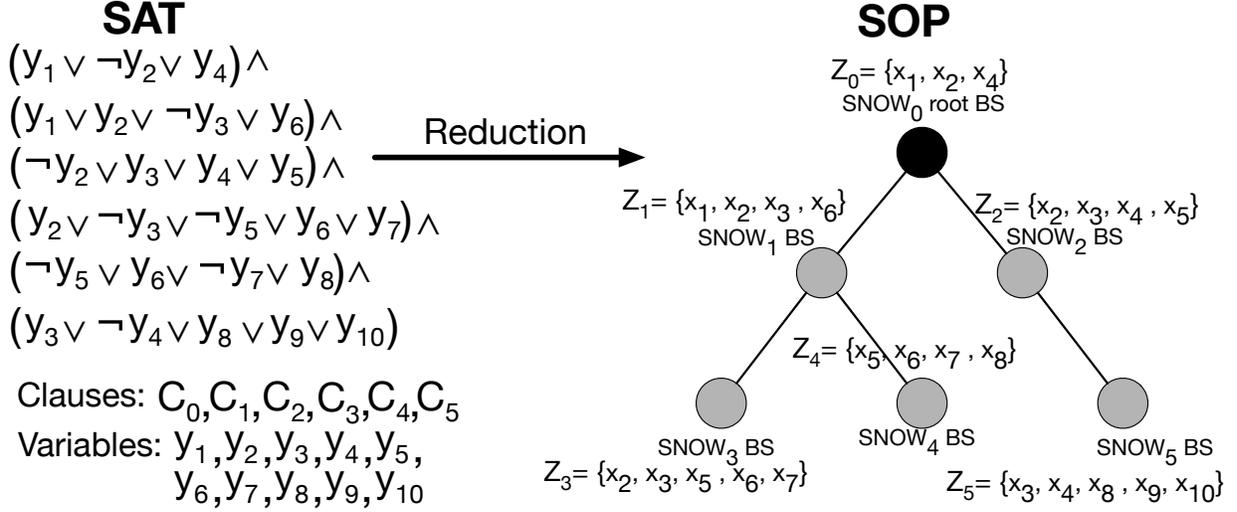}
\caption{Reduction from SAT.}
\label{fig:chap5-reduction}
\end{figure*}

Let $I_i\subset \{0, 1, \cdots, N-1\}$ be such that each SNOW$_j$ with $j\in I_i$ is an interferer of SNOW$_i$ (i.e., BS$_i$). In the SNOW-tree, let $p(i) \in \{0, 1, \cdots, N-1\}$  be such that BS$_{p(i)}$ is the parent of BS$_i$ and $Ch_j\subset \{1, 2, \cdots, N-1\}$ be such that each BS$_j$ with $j\in Ch_i$ is a child of BS$_i$. The SNOWs associated with a BS's parent and children are its interferer already, i.e., $(\{p(i)\} \cup Ch_i)\subseteq I_i$. To  limit inter-SNOW interference,  let  $\phi_{i, j}$ be the {\slshape maximum allowable number of subcarriers} that can overlap between two interfering SNOWS,  SNOW$_i$ and SNOW$_j$. As explained in Section~\ref{sec:chap5-p2p}, there must be at least one subcarrier common between a BS and its parent which is defined in {\bf Constraint} (\ref{c:chap5-const2}).
\revise{ Note that we can also use Constraint (\ref{c:chap5-const2}) to set $\phi_{i, p(i)}$ to indicate the number of on demand subcarriers between BSs BS$_i$ and BS$_{p(i)}$ in a SNOW-tree. Sometimes the demand can change and the root BS will re-run the SOP algorithm to take it into account.}
{\bf Constraint} (\ref{c:chap5-const3}) indicates the minimum and maximum number of overlapping subcarriers between other interfering pairs. Thus, SOP is formulated as follows where the root BS allocates the spectrum among all BSs (i.e., assigns subcarriers $X_i\subseteq Z_i$ to SNOW$_i$) in order to

\begin{align}
\centering
\text {Maximize~~~}  & \sum_{i = 0}^{N-1} |X_i| \nonumber \\
 \text{subject to~~~}& \sigma_i \leq |X_i| \leq |Z_i|, \; X_i \subseteq Z_i \label{c:chap5-const1} \\
                     & 1 \leq |X_i \cap X_{p(i)}| \leq \phi_{i, p(i)},  1 \le i < N  \label{c:chap5-const2}\\
                     & 0 \leq |X_i \cap X_j| \leq \phi_{i,j}, 0 \le i < N  \nonumber \\
                     &~~~~~~~~~~~~~~~~~~~~\forall j \in I_i  -  (\{p(i)\} \cup Ch_i) \label{c:chap5-const3}
\end{align}

SOP is a unique problem that we have observed first in integrating SNOWs. It is quite different from spectrum allocation in cellular network where towers are connected through a wired network and spectrum availability/dynamics~\cite{channel_cellular} do not change. Due to technology-specific features and unique communication primitive of SNOW, traditional channel allocation techniques for wireless networks (see survey~\cite{channel_wireless}), WSN (see survey~\cite{channel_wsn}), or cognitive radio networks (see survey~\cite{cognitive_channel_survey, cognitive_channel_survey2})  are also not applicable as SOP involves assigning a large number of subcarriers to each BS allowing some degree of overlaps among interfering BSs for enhanced scalability. In the following, we will first characterize SOP and then propose its solution strategy.

\subsubsection{NP-Hardness of SOP}

We now prove that SOP is NP-hard which can be proved through a reduction from the {\slshape SAT (Boolean Satisfiability)} problem. The SAT problem asks whether there exists a truth assignment that makes all clauses true~\cite{cook1971complexity}. Theorem~\ref{thm:chap5-np_hard_proof} formally proves the NP-hardness of SOP by proving that its decision version is NP-complete.

\begin{theorem}\label{thm:chap5-np_hard_proof}
Given a SOP for SNOW-tree, it is NP-complete to decide whether it is feasible or not.
\end{theorem}

\begin{proof} Given an instance of SOP in SNOW-tree with overlapping spectrum assignment for $N$ BSs and $m$ subcarriers, where BS$_i$, $0 \le i < N$ gets $m_i$ number of subcarriers. It is verifiable in $O(Nm)$ time whether the subcarrier assignment is feasible or not. Hence, the problem is in NP. To prove NP-hardness, we reduce an arbitrary instance $\mathcal{I}(SAT)$ of SAT to an instance $\mathcal{I}(SOP)$ of the SOP in SNOW-tree and show that $\mathcal{I}(SAT)$ has an \emph{interpretation} that satisfies a \emph{boolean formula} if and only if $\mathcal{I}(SOP)$ is feasible.

Let $\mathcal{I}(SAT)$ have $m$ boolean variables $y_0, y_1, y_2, ..., y_{m-1}$ and $N$ clauses $C_0, C_1, \cdots, C_{N-1}$ in conjunctive normal form. Now, for the set of variables in $\mathcal{I}(SAT)$ we create a set of subcarriers $Z = \{x_0, x_1, \cdots, x_{m-1}\}$ in $\mathcal{I}(SOP)$ that are available in SNOW-tree. Then, we create one SNOW BS$_i$ in $\mathcal{I}(SOP)$ for each clause $C_i$ in $\mathcal{I}(SAT)$. Also, we create one subset $Z_i \in Z$ for each BS$_i$ that corresponds to subset of boolean variables in clause $C_i$. 
 As an example, consider a boolean formula $(y_1 \lor \neg y_2 \lor y_4) \land (y_1 \lor y_2 \lor \neg y_3 \lor y_6) \land (\neg y_2 \lor y_3 \lor y_4 \lor y_5) \land (y_2 \lor \neg y_3 \lor \neg y_5 \lor y_6 \lor y_7) \land (\neg y_5 \lor y_6 \lor \neg y_7 \lor y_8) \land (y_3 \lor \neg y_4 \lor y_8 \lor y_9 \lor y_{10})$ of 10 variables and 6 clauses in $\mathcal{I}(SAT)$, thus in $\mathcal{I}(SOP)$, $Z_0 = \{x_1, x_2, x_4\},\; Z_1 = \{x_1, x_2, x_3, x_6 \},\; Z_2 = \{x_2, x_3, x_4, x_5 \}, \cdots,$ and $Z_{5} = \{x_3, x_4, x_8, x_9, x_{10}\}$. If a boolean variable $y_k$ exists as a positive literal in clause $C_i$ and negative literal in $C_j$, then corresponding BS$_i$ (i.e. SNOW$_i$) and BS$_j$ (i.e. SNOW$_j$) interfere each other and $x_k \in \{Z_i \cap Z_j\}$ is the interfering subcarrier between them. Thus, setting $y_k$ to $true$ in $\mathcal{I}(SAT)$ will yield assigning subcarrier $x_k$ to BS$_i$ or BS$_j$, and vice versa. In the previous example, if $y_2$ is set to $true$, then BS$_1$ and BS$_3$ get subcarrier $x_2$ and not BS$_0$ and BS$_2$.

To build the SNOW-tree, we consider BS$_0$ as the root BS that corresponds to clause $C_0$. 
We draw an edge between BS$_i$ and BS$_j$ if corresponding clauses $C_i$ and $C_j$ have at least one common positive or negative literal. The number of such literals in $\mathcal{I}(SAT)$ represents $\phi^{}_{ij}$ in $\mathcal{I}(SOP)$. While creating the SNOW-tree, we \emph{do not} draw an edge between BS$_j$ and BS$_k$ if BS$_j$ $\in (\{p(i)\} \cup Ch_i)$ and BS$_k$ $\in (\{p(i)\} \cup Ch_i)$, where, $i \ne j \ne k$. Thus, no loops are created and the number of edges in SNOW-tree become $N-1$, as shown in Figure~\ref{fig:chap5-reduction}. The whole reduction process runs in $O(m^2\lg N)$ time.

Suppose that $\mathcal{I}(SAT)$ has an interpretation that satisfies the boolean formula. Thus, each clause $C_i$ is also $true$. Also, a subset of variables in each clause $C_i$ is $true$ that corresponds to the subset of subcarriers $X_i$ that is assigned to BS$_i$ in $\mathcal{I}(SOP)$. The number of variables in clause $C_i$ that are set to $true$ represents the minimum number of subcarriers $\phi_i$ in $\mathcal{I}(SOP)$. Also, no two interfering BS$_i$ and BS$_j$ get more than $\phi^{}_{ij}$ number of common subcarriers between them. We also include a common subcarrier between neighboring BS$_i$ and BS$_j$ if there is none already, thus considering corresponding literal in $\mathcal{I}(SAT)$ as $true$ which does not change the satisfiability of boolean formula. Such inclusion also does not violate right hand side condition of Constraints (\ref{c:chap5-const2}) and (\ref{c:chap5-const3}). Thus, $\mathcal{I}(SOP)$ has a feasible subcarrier assignment where the root BS assigns at least $N$ subcarriers in total to all the BSs SNOW-tree, each having at least one.

Now, let $\mathcal{I}(SOP)$ have a feasible subcarrier assignment in SNOW-tree. Thus the root BS assigns at least $N$ subcarriers to $N$ BSs, each having at least one. Since each BS$_i$ in $\mathcal{I}(SOP)$ represents a clause $C_i$ in $\mathcal{I}(SAT)$ and two neighboring BS$_i$ and BS$_j$ in $\mathcal{I}(SOP)$ have at least one common subcarrier and SNOW-tree is connected, each clause $C_i$ has at least one literal that is set to be $true$. Thus, we have an interpretation in $\mathcal{I}(SAT)$ that satisfies the boolean formula.
\end{proof}

\subsubsection{Efficient Greedy Heuristic for SOP}\label{sec:chap5-proposed_solution}

Since an optimal solution of SOP cannot be achieved in polynomial time unless P=NP, we first propose an intuitive, highly efficient, polynomial time greedy heuristic. 
In the beginning, the greedy heuristic will assign to every BS the entire spectrum available in its location.  It will then keep removing subcarriers from their assignments until the constraints are satisfied. The target will be to remove as less number of subcarriers as possible. 

\begin{algorithm}
\DontPrintSemicolon
\KwData{ $Z_i$ for BS$_i$, $0 \le i < N$ in a SOP Instance.}
\KwResult{Subcarriers $X_i$ for BS$_i$, $0\le i <N$.}
\BlankLine

\For{$\text{each } BS_i \text{ in } \text{the SNOW-tree}$}{
$X_i = Z_i$.
}

\For{$\text{each } BS_i \text{ in } \text{inter-SNOW Tree}$}{
   \For{$\text{each } BS_j\; \in\; I_i$}{
        Let, $Z_{i,j} = Z_i \cap Z_j$.
        
        \For{$\text{each subcarrier}\; x_l\; \in\; Z_{i,j}$}{
            \If{$|X_i \cap X_j| > \phi_{i,j}$}{
                \If{$|X_i| \ge |X_j| \; and\; |X_i| > \sigma_i$ }{
                
                    Delete $x_l$ from $X_i$.
                }
                \ElseIf {$|X_j| > \sigma_j$}{
                
                    Delete $x_l$ from $X_j$.
                }
                \Else (\tcc*[f]{Infeasible solution}){
                
                    Don't delete $x_l$ from $X_i$ or $X_j$. 
                }
            }
            \Else{
                Break.
            }
        }
    }
}

\caption{Greedy Heuristic Algorithm}
\label{algo:chap5-greedy}
\end{algorithm}

The greedy heuristic is described as follows. In the beginning, the root BS greedily assigns to every BS$_i$ all the subcarriers that are available at the location of BS$_i$ (i.e.,  the entire spectrum available in BS$_i$'s  location).  Note that such an assignment maximizes the  {\slshape scalability} metric $\sum_{i = 0}^{N-1} |X_i|$, but violates the constrains of SOP. Specifically, it satisfies Constraint (\ref{c:chap5-const1}), but may violate Constraints (\ref{c:chap5-const2}) and (\ref{c:chap5-const3}) that are defined to keep the BSs connected as a tree and to limit interference between neighboring or interfering BSs by limiting their common usable subcarriers.  Now, with a view to satisfying those two constrains, the heuristic greedily removes some subcarriers  that are common between interfering BSs. Such removal of subcarriers is done to make the least decrease in the scalability and to ensure that Constraint (\ref{c:chap5-const1}) is not violated.  In other words, it tries to keep the subcarrier assignment balanced between BSs. Specifically, for every interfering BS pair, BS$_i$ and BS$_j$, we do the following until they satisfy Constraints (\ref{c:chap5-const2}) and (\ref{c:chap5-const3}):  Find the next common subcarrier between them and remove it from BS$_i$ if $| X_i | > | X_j |$ and  $| X_i | > \sigma_i$; otherwise remove it from BS$_j$ if $| X_j | > \sigma_j$.

The pseudocode of our greedy heuristic is shown as Algorithm~\ref{algo:chap5-greedy}. As shown in the pseudo code, the heuristic may not find feasible solution in some rare cases where some BS pairs, BS$_i$ and BS$_j$,  cannot satisfy the condition $|X_i \cap X_j| \le  \phi_{i,j}$.  In such cases, we can either use the infeasible solution and use  the found subcarrier allocation or relax the constraints for those BSs (violating the constraints) by changing their values of $\sigma_i$ or $\phi_{i,j}$ in Constraints (\ref{c:chap5-const1}),  (\ref{c:chap5-const2}), and (\ref{c:chap5-const3}) of the SOP. 

{\bf Time Complexity of the Greedy Heuristic.}  
Since the SNOW-tree has $N$ base stations (or  $N$ SNOWs),  Algorithm~\ref{algo:chap5-greedy} will find intersection of the subcarriers for each of  $O(N^2)$ pairs of BSs (line 5 of Algorithm~\ref{algo:chap5-greedy}). 
Finding intersection of the subcarriers for a pair of BSs takes   $O(M\lg M)$ time, where $M= \max \{   | Z_i|  \mid  0\le i <N  \} $. Thus the time complexity of Algorithm~\ref{algo:chap5-greedy} is  $O(N^2M\lg M)$.

\subsubsection{Approximation Algorithm for SOP}\label{sec:chap5-bounded_algorithm}

While the heuristic (Algorithm~\ref{algo:chap5-greedy}) can be highly efficient
in practice, we also propose an algorithm for which we can derive an analytical performance bound. 
Our reduction used in Theorem~\ref{thm:chap5-np_hard_proof} provides the key insights for developing such an approximation algorithm. 
Our key observation from the reduction is that a solution approach for SOP can be developed by extending a solution for 
the MAX-SAT (Maximum Satisfiability) problem and by incorporating the constraints of the former. MAX-SAT, a generalized version of SAT,
asks to determine the maximum number of clauses, of a given
Boolean formula in conjunctive normal form, that can be made
true by an assignment of truth values to the variables of the
formula~\cite{maxsat}. The observation allows us to leverage the well-established results for MAX-SAT. 
Specifically, we leverage a very simple but  analytically efficient approach adopted for MAX-SAT solution, and incorporate the SOP constraints 
to develop a constant approximation algorithm for SOP.

\begin{algorithm}
\DontPrintSemicolon
\KwData{ $Z_i$ for BS$_i$, $0 \le i < N$ in a SOP Instance.}
\KwResult{Subcarriers $X_i$ for BS$_i$, $0\le i <N$.}
\BlankLine

\For{$\text{each } BS_i \text{ in } \text{the SNOW-tree}$}{
$X'_i = X''_i =\emptyset;$ 
}

Let, $Z = Z_0 \cup Z_2 \cup \cdots \cup Z_{N-1}$

\For (\tcc*[f]{step 1}) {$\text{each subcarrier} \;x_l \in Z$}{
        Uniformly and independently add $x_l$ with a probability of $\frac{1}{2}$ to $X'_i$, $\forall i: x_l \in Z_i$.
}

  \If (\tcc*[f]{violates Constraint \ref{c:chap5-const1}}) {$\exists i $ such that $ |X'_i| <  \sigma_i $}{ 
Let, $Z^{\prime} = (Z_0 - X'_0) \cup (Z_1 - X'_1) \cup \cdots \cup (Z_{N-1} - X'_{N-1})$

\For (\tcc*[f]{step 2})  {$\text{each subcarrier} \;x_k \in Z^{\prime}$}{
    Uniformly and independently add $x_k$ with a probability of $\frac{1}{2}$ to $X''_i$, $\forall i: x_k \in (Z_i - X'_i)$.
}

}     

\For{$\text{each } BS_i \text{ in } \text{the SNOW-tree}$}{
$X_i = X'_i  \cup X''_i;$.
}

\caption{Probabilistic 1/2-Approximation Algorithm}
\label{algo:chap5-approx}
\end{algorithm}

Considering $\Omega$ as the total weight of all clauses, a simple approximation algorithm for MAX-SAT sets each variable to true with probability $\frac{1}{2}$. 
By linearity of expectation, the expected weight of the satisfied clauses is at least $\frac{1}{2}\Omega$, thus making the approach a randomized $\frac{1}{2}$-approximation algorithm. 
In solving the SOP  in a similar spirit, we shall consider assigning a {\slshape subcarrier} to a {\slshape SNOW} in place of a {\slshape variable} to a {\slshape clause}. \revise{Choosing a probability other than $\frac{1}{2}$ would require us to calculate different probabilities for different subcarriers based on the level of interference they contribute to different BSs which involves a costly approach. Therefore, it is very difficult and impractical for us to develop a faster approximation algorithm based on our proposed approach. Since MAX-SAT  does not have  Constraints (\ref{c:chap5-const1}), (\ref{c:chap5-const2}), and (\ref{c:chap5-const3}), we modify such a probabilistic assignment whose  pseudocode is shown  as Algorithm~\ref{algo:chap5-approx} to take into account these constraints.}

Algorithm~\ref{algo:chap5-approx} assigns subcarriers to the BSs in two steps. In step 1, it assigns each distinct subcarrier $x_l$ in the SNOW-tree uniformly and independently with probability of $\frac{1}{2}$ to each BS$_i$ such that $x_l \in Z_i$ (i.e., the BS where the subcarrier is available). The set of subcarriers that BS$_i$ gets after this step is $X'_i$. Thus, the expected number of subcarriers assigned to BS$_i$ in this step is $E[ |X'_i|]= \frac{|Z_i|}{2}$. Similarly, the expected number of common subcarriers between two interfering BSs, BS$_i$ and BS$_j$, after step 1 is 
$E[ |X'_i  \cap  X'_j   |]= \frac{|Z_i  \cap  Z_j |}{4}$. Our experiments (Sec. \ref{sec:chap5-experiments}, \ref{sec:chap5-simulations}) show that two interfering BSs can use even up to 60\% of their total available common subcarriers. That is, the values $\phi_{i,j}$ in Constraints  (\ref{c:chap5-const2}) and (\ref{c:chap5-const3}) can be up to 60\%  of  $|Z_i  \cap  Z_j |$. Thus after step 1, the probability of satisfying Constraints  (\ref{c:chap5-const2}) and (\ref{c:chap5-const3})  is very high. 
Hence, if some BS$_i$ violates Constraint  (\ref{c:chap5-const1}), i.e., if $|X'_i|  < \sigma_i$, we repeat subcarrier assignment in the same way in step 2. 
Specifically, step 2 assigns each distinct subcarrier $x_k$ uniformly and independently with probability of $\frac{1}{2}$ to each BS$_i$ such that $x_k \in Z_i  -  X'_i $. If $X''_i$  is the set of subcarriers assigned to BS$_i$ in step 2, then BS$_i$ finally gets subcarriers $X_i = X'_i  \cup X''_i$. While step 2  increases the probability of satisfying Constraints  (\ref{c:chap5-const1}), it decreases that of satisfying the other two constraints which was very high before this step. Hence, we do not adopt any further subcarrier addition.


{\bf Performance Analysis.}
As described above, Algorithm~\ref{algo:chap5-approx} sometimes can end up with an infeasible solution for SOP.  However, as we describe below, such chances are quite low, and the probability of finding a feasible solution is quite high ($\approx 1$). Then Theorem~\ref{thm:chap5-bound} proves that the Algorithm has an approximation ratio of $\frac{1}{2}$ for any solution it provides (feasible or infeasible).

As described before, after step 1  of Algorithm~\ref{algo:chap5-approx},   $E[|X_i^{\prime}|] = \frac{|Z_i|}{2}$ for each BS$_i$; and 
 $E[|X_i^{\prime} \cap X_j^{\prime}|] = \frac{|Z_i \cap Z_j|}{4}$, for each interfering BS pairs,  BS$_i$ and BS$_j$.  Step 2 runs  only if Constraint (\ref{c:chap5-const1}) remains violated after step 1. 
 Thus, if step 2 does not run, the probability of satisfying  Constraint (\ref{c:chap5-const1}) is 1. 
 Similarly, if step 2 runs, 
  $E[|X_i^{\prime\prime}|] = \frac{|Z_i - X_i^{\prime}|}{2}$ and $E[|X_i^{\prime\prime} \cap X_j^{\prime\prime}|] = \frac{|(Z_i - X_i^{\prime}) \cap (Z_j - X_j^{\prime})|}{4}$. 
Now, if both steps run, the expected number of subcarriers assigned to BS$_i$ is 

\begin{align} 
\centering
    E[|X_i|] &= E[|X'_i|] +  E[|X''_i|]\nonumber \\ 
    &=  \frac{|Z_i|}{2} + \frac{|Z_i - X'_i|}{2} \nonumber \\
    &=    \frac{|Z_i|}{2} + \frac{|Z_i|} {2}  -  \frac{|X'_i|} {2} \nonumber \\   
    &=     \frac{|Z_i|}{2} + \frac{|Z_i|} {2}  -  \frac{|Z_i|} {4} \nonumber \\  
    &=    \frac{3}{4} |Z_i|. \nonumber
\end{align}
Note that the value of $\sigma_i$ is set usually much smaller than the above value as a BS does not want to use all available subcarriers, allowing other SNOWs to use those. 
Thus, the probability of satisfying Constraint (\ref{c:chap5-const1}) is $\approx 1$.

If step 2 does not run, then the expected number of common subcarriers between each interfering BS pairs, BS$_i$ and BS$_j$, is $E[|X_i \cap X_j|]  = \frac{|Z_i \cap Z_j|}{4}$. 
As we have discussed before, the value of $\phi_{i,j}$ in Constraints  (\ref{c:chap5-const2}) and (\ref{c:chap5-const3})  is usually above $\frac{|Z_i  \cap  Z_j |}{2}$, which is twice the value of  $E[|X_i \cap X_j|]$. 
Thus, the probability of  satisfying Constraints (\ref{c:chap5-const2}) and (\ref{c:chap5-const3}) is also $\approx 1$.  If step 2 runs, then
\begin{align}
\centering
    E[|X_i \cap X_j|] &= E[|X_i^{\prime} \cap X_j^{\prime}|] +  E[|X_i^{\prime\prime} \cap X_j^{\prime\prime}|] \nonumber \\
    &= \frac{|Z_i \cap Z_j|}{4} +  \frac{|(Z_i - X_i^{\prime}) \cap (Z_j - X_j^{\prime})|}{4} \nonumber \\
    &= \frac{|Z_i \cap Z_j|}{4} + \frac{| (Z_i \cap Z_j )  - (X'_i \cap X'_j )  |}{4} \nonumber \\
    &= \frac{|Z_i \cap Z_j|}{4} + \frac{| Z_i \cap Z_j |}{4}  -  \frac{| X'_i \cap X'_j |}{4} \nonumber \\
     &= \frac{|Z_i \cap Z_j|}{4} + \frac{| Z_i \cap Z_j |}{4}  -  \frac{| Z_i \cap Z_j |}{16} \nonumber \\
     &= \frac{ 7}{16} |Z_i \cap Z_j|     <     \frac{|Z_i  \cap  Z_j |}{2}   \nonumber
\end{align}
which means that  the probability of satisfying Constraints (\ref{c:chap5-const2}) and (\ref{c:chap5-const3}) is $\approx 1$ even if step 2 runs. Thus the probability of satisfying all constraints is $\approx 1$.  

\begin{theorem}\label{thm:chap5-bound}
 Algorithm~\ref{algo:chap5-approx} has an approximation ratio of $\frac{1}{2}$. 
\end{theorem}

\begin{proof}
Since an optimal value of the objective (scalability metric) is unknown, a conservative upper-bound is given by $OPT = \sum_{i=0}^{N-1} |Z_i| $. 
If step 2 of the algorithm does not run, according to probabilistic assignments of subcarriers in step 1 of Algorithm~\ref{algo:chap5-approx}, we have
in step 1,

\begin{align}
\centering
E[\text{total }X_i] &=\sum_{i=0}^{N-1}\sum_{l = 0}^{|Z_i|-1}|x_l|\Pr \{ x_l\text{  assigned to }BS_i\} \nonumber \\
&= \sum_{i=0}^{N-1}|Z_i|.\frac{1}{2} \nonumber \\
&= \frac{1}{2} \sum_{i=0}^{N-1} |Z_i|    \ge \frac{1}{2}OPT \nonumber 
\end{align}

If step 2 of Algorithm~\ref{algo:chap5-approx} runs, then
\begin{align}
\centering
E[\text{total }X''_i] &= \frac{1}{2} \sum_{i=0}^{N-1} |Z_i  -      X'_i| \nonumber \\
&=\frac{1}{2} \sum_{i=0}^{N-1}|Z_i|    -   \frac{1}{2} \sum_{i=0}^{N-1}|X'_i| \nonumber \\
&=\frac{1}{2} \sum_{i=0}^{N-1}|Z_i|    -   \frac{1}{4} \sum_{i=0}^{N-1}|Z_i| \nonumber \\
&=  \frac{1}{4} \sum_{i=0}^{N-1}|Z_i|. \nonumber
\end{align}
Now using linearity of expectation, if step 2 runs, then
\begin{align*}
\centering
E[\text{total }X_i] &=  \frac{1}{2} \sum_{i=0}^{N-1} |Z_i|   +   \frac{1}{4} \sum_{i=0}^{N-1} |Z_i|  \\
                           & =   \frac{3}{4} \sum_{i=0}^{N-1} |Z_i|    \ge \frac{3}{4}OPT
\end{align*}
Thus the approximation bound follows. 
\end{proof}

{\bf Time Complexity of the Approximation Algorithm.}
Since the SNOW-tree has $N$ BSs, Algorithm~\ref{algo:chap5-approx} will do  {\slshape union} operations of all the $O(N)$ BSs' subcarriers (in line 3) in the step 1 run. If step 2 runs, {\slshape subtraction} and {\slshape union} operations will be done.  {\slshape Union} and {\slshape subtraction} operations will run in $O(M\lg M)$ time, where $M= \max \{   | Z_i|  \mid  0\le i <N  \} $. Thus, the time complexity of Algorithm~\ref{algo:chap5-approx} is $O(NM\lg M)$.

As we shall describe in Sections~\ref{sec:chap5-experiments} and \ref{sec:chap5-simulations} through evaluations, our heuristic (Algorithm~\ref{algo:chap5-greedy}) performs better in terms of scalability, energy consumption, and  latency while Algorithm~\ref{algo:chap5-approx} provides theoretical performance guarantee.

%% file: chapter5/experiment.tex
\subsection{Implementation}\label{sec:chap5-implementation}

We implement our proposed SNOW technologies on the GNU Radio~\cite{gnuradio} platform using USRP devices that can operate between 70MHz - 6GHz of spectrum~\cite{usrp}. We have 9 USRP (2 B210, 4 B200, and 3 USRP1) devices. 
To demonstrate the effectiveness of SOP in intra-SNOW communication we use 2x2 devices in 2 different SNOW BSs (each having one Tx-Radio and one Rx-Radio), where one BS is assigned 3 nodes (3 USRPs) and the other BS is assigned 2 nodes (2 USRPs). On the other hand, to demonstrate the inter-SNOW communication, we use 2x3 devices in 3 different SNOW BSs (each having one Tx-Radio and one Rx-Radio). In this case, each BS is assigned one USRP device as node.
\begin{figure}[!htb]
\centering
\includegraphics[width=0.85\textwidth]{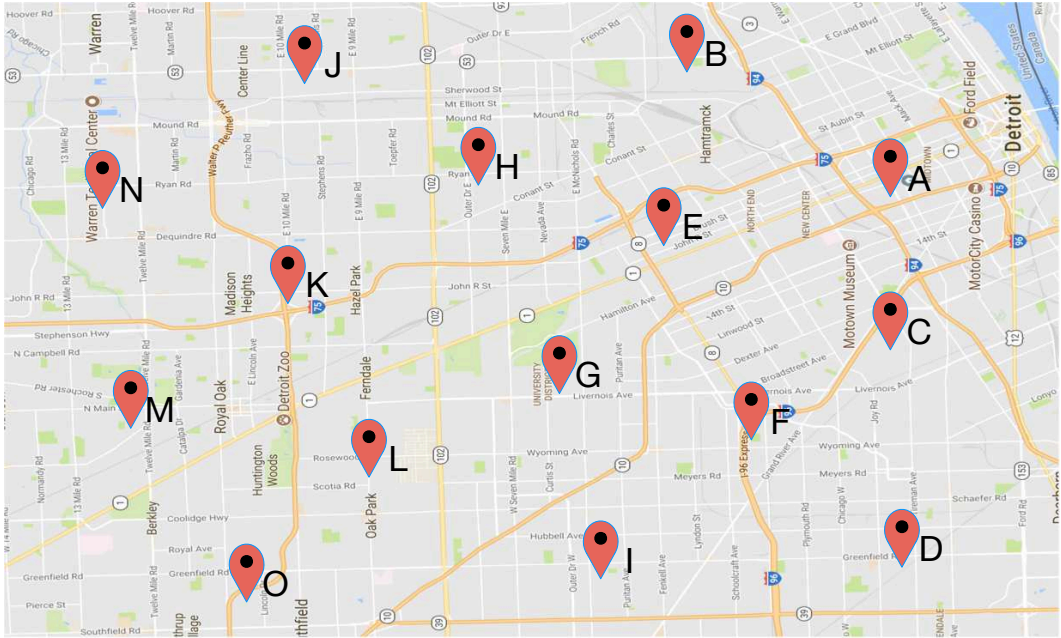}
\caption{SNOW BS positions used in experiments (and simulations).}
\label{fig:chap5-testbed}
\end{figure}

We evaluate the performance of our design by experimenting at 15 different candidate locations covering approximately (25x15)km$^2$ of a large metropolitan area in the city of Detroit, Michigan (Figure~\ref{fig:chap5-testbed}). 
Due to our limited number of USRP devices (3 BSs each having one node to demonstrate inter-SNOW communication) in real experiments, we create 5 different SNOW-trees at different candidate locations and do the experiments separately. 
In experiments, we choose to create 3 SNOWs to demonstrate the integration of as many SNOWs as we can with our limited number of devices, and most importantly to cover more area using a SNOW-tree. 
In~\cite{snow, snow2}, we have already performed extensive experiments considering multiple nodes in a single SNOW. Hence, here we will show the intra-SNOW communication using 2 SNOW BSs one having 3 nodes and the other having 2 nodes.
However, later in simulations, we create a single SNOW-tree of 15  SNOWs each having 1000 nodes. We perform experiments on white space availability at different locations and determine the values of $\phi_{i, p(i)}$ and $\phi_{i, j}$ in Constraints (\ref{c:chap5-const2}) and (\ref{c:chap5-const3}), respectively. We compare the performance of our greedy heuristic and our approximation algorithm for SOP with a direct allocation scheme. A \emph{direct allocation scheme} is unaware of scalability and inter-SNOW interference and hence will assign each BS all the subcarriers that are available at its location. Moreover, we perform exhaustive experiments on both intra- and inter-SNOW communications.

\subsection{Evaluation}\label{sec:chap5-eval}
In this section, we evaluate the performance of our SOP algorithms in inter- and intra-SNOW communications through experiments and simulations.
\subsubsection{Experiments}\label{sec:chap5-experiments}
{\bf Experimental Setup.}\label{sec:chap5-exp_setup}
Our testbed location has white spaces ranging between 518 and 686MHz (TV channels 21--51) for different BSs. We set each subcarrier bandwidth to 400kHz which is the default subcarrier bandwidth in SNOW~\cite{snow, snow2}. We use 40-byte (including header, random payload, and CRC) packets with a spreading factor of 8, modulated or demodulated as BPSK (Binary Phase-Shift Keying). With the similar spirit of IEEE 802.15.4, we set the Tx power to 0dBm in the SNOW nodes for energy efficiency. Receive sensitivity is set to -94dBm both in SNOW BSs and the nodes. Meanwhile, BSs transmit with a Tx power of 15dBm ($\approx$40mW) to their nodes and neighboring BSs that is the maximum allowable Tx-power limit in most of the white space channels at our testbed location. For energy calculations at the nodes, we use the energy profile of TI CC1070 RF unit by Texas Instruments that can operate in white spaces~\cite{cc1070}. Unless stated otherwise, these are our default parameter settings.

{\bf Finding Allowable Overlap of Spectrum.}\label{sec:chap5-exp_whitespaces}
\begin{figure}[!htb]
    \centering
      \subfigure[Available white spaces (presented as the TV channel indices used in the US) at different BS locations. A dot in the figure means that the TV channel in x-axis is white space at the location in y-axis.\label{fig:chap5-channels}]{
    \includegraphics[width=0.5\textwidth]{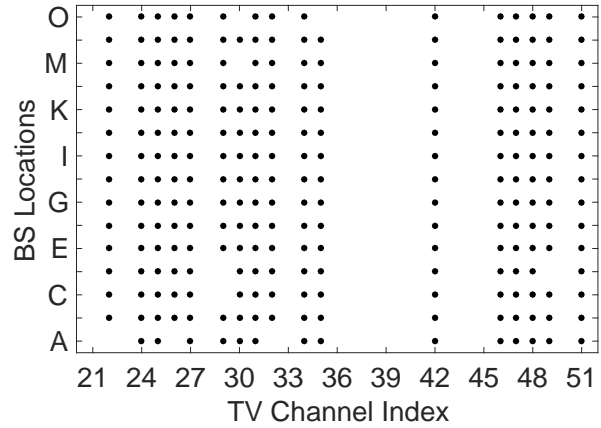}
      }
    \hfill
      \subfigure[Reliability in communication with different magnitude of overlaps in white spaces between BSs in different SNOW-trees.\label{fig:chap5-overlaps}]{
        \includegraphics[width=.5\textwidth]{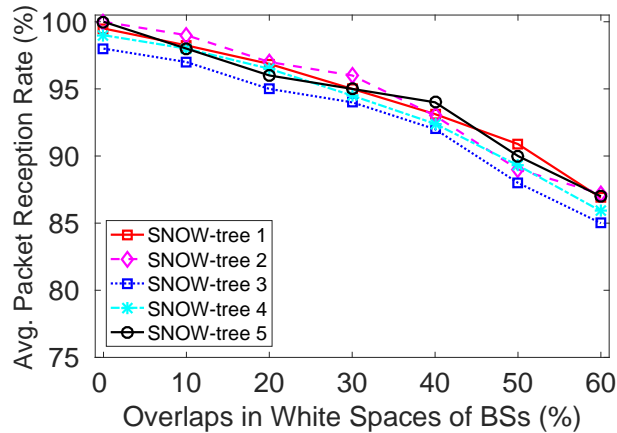}
      }
    \caption{White spaces availability and reliability in different SNOW-trees.}
    \label{fig:chap5-primitives}
\end{figure}
We first determine how many subcarriers can be common  between two interfering SNOWs without degrading their performance.  
We determine white spaces at 15 different locations from a cloud-hosted database~\cite{wdb}.
Figure~\ref{fig:chap5-channels} shows the available white spaces at different locations confirmed by both database and sensing. Also, we conduct experiments on 5 different SNOW-trees to determine the maximum allowable number of common subcarriers between interfering BSs. Locations of BSs in 5 trees are (1) B, A, E; (2) D, C, F; (3) G, I, L; (4) J, H, K; (5) N, M, O; respectively, where the BS in the middle location in each SNOW-tree is the root BS. In this paper, we also identify the SNOW BSs by their location indices. In each tree, we allow BSs to operate with different magnitudes of white space overlap between them. To determine the maximum allowable number of common subcarriers between interfering BSs in a tree, each node hops randomly to all the subcarriers that are available in its BS location and sends consecutive 100 packets to its BS. Each node repeats this procedure 1000 times.

\revise{As shown in Figure~\ref{fig:chap5-overlaps}, the BSs in each tree can overlap 60\% of their white spaces to yield an average Packet Reception Rate (PRR) of 85\%. We consider that an 85\% PRR is an acceptable rate. This figure also shows that the average PRR decrease with the increase in the magnitude of overlap. Finding the maximum allowable overlap needs to be done only once in the beginning of the network operation and may be recomputed if there is a significant change (e.g., some BS or a large number of nodes leave or join) in the network. A network deployment may choose its magnitude of overlap based on the target application’s quality of service (QoS) requirements. We thus set the values of $\phi_{i,p(i)}$ and $\phi_{i,j}$ in Constraints (\ref{c:chap5-const2}) and (\ref{c:chap5-const3}), respectively, based on this experiment. Finding the optimal values of these variables is out of the scope of this paper.}

{\bf Evaluating the Scalability Metric.}\label{sec:chap5-exp_sop}
\begin{figure}[!htb]
    \centering
      \subfigure[Scalability metric values achieved in different SNOW-trees.\label{fig:chap5-subs}]{
        \includegraphics[width=.5\textwidth]{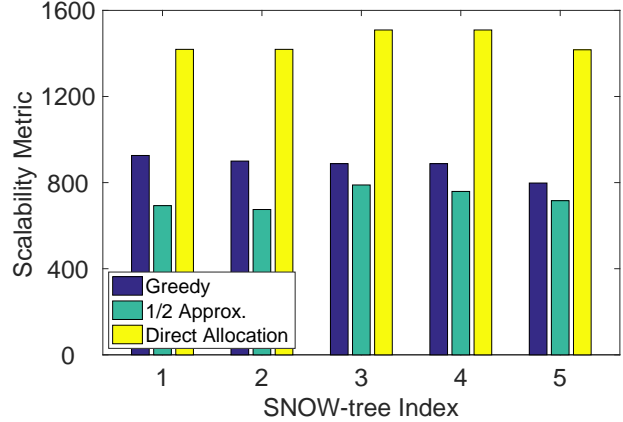}
      }
      \hfill
      \subfigure[Execution time of different SOP algorithms.\label{fig:chap5-latency_algo}]{
        \includegraphics[width=.5\textwidth]{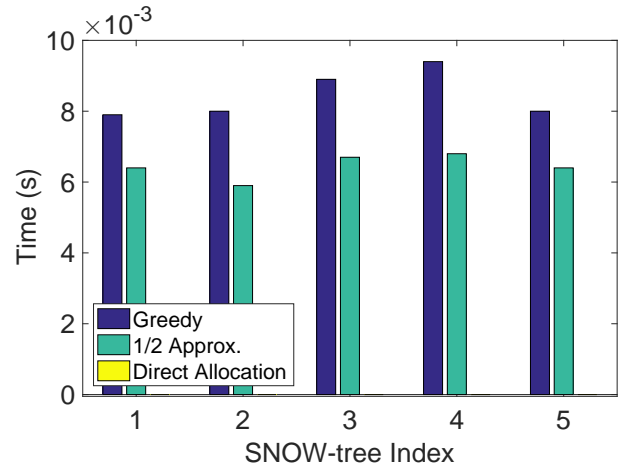}
      }
    \caption{Scalability metric and execution time of SOP algorithms by different root BSs in different SNOW-trees.}
    \label{fig:chap5-sop}
\end{figure}
To demonstrate the performances in maximizing the scalability metric under our approaches and the baseline approach, we set the value of $\sigma_i$ in Constraint (\ref{c:chap5-const1}) to 100 for all the BSs. 
We choose the same value for each BS since most (13 out of 15) of the BS locations have the same set of white space channels. 
Figure~\ref{fig:chap5-subs} shows the values of the scalability metric achieved in 5 different SNOW-trees using our greedy heuristic,  approximation algorithm, and the direct allocation approach. 
This figure shows that the direct allocation scheme assigns more subcarriers to all BSs.  Our later experiments will show that such an assignment suffers in terms of reliability, latency, and energy consumptions compared to our greedy heuristic and approximation algorithm due to its violation of Constraints (\ref{c:chap5-const2}) and (\ref{c:chap5-const3}) of SOP. 
Also, our greedy heuristic can offer higher scalability than our approximation approach, while the latter can be preferred when analytical performance bound is a concern. Thus, our greedy heuristic can be more effective in practice (even though its performance bound was not derived).

Figure~\ref{fig:chap5-latency_algo} shows the time taken by our greedy heuristic, our approximation algorithm, and direct allocation scheme to assign subcarriers to BSs. Our greedy heuristic observes 0.094ms compared to 0.068ms for our approximation algorithm in worst case in SNOW-tree 4. In the figure, time taken by the direct allocation scheme is not visible as it is approximately 0ms (since it does not employ any intelligent technique). However, time taken by our greedy heuristic and our approximation algorithm are very low and practical.

\begin{figure}[!htb]
    \centering
      \subfigure[Reliability in different SNOW BSs\label{fig:chap5-snow_reliability}]{
    \includegraphics[width=0.5\textwidth]{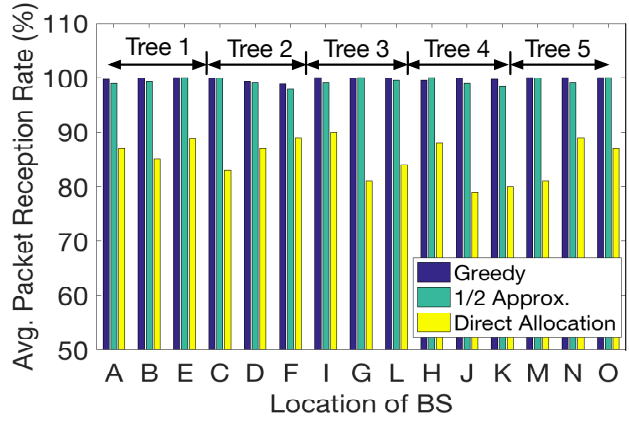}
      }
    \hfill
      \subfigure[Average Latency in intra-SNOW comm.\label{fig:chap5-snow_latency}]{
        \includegraphics[width=.5\textwidth]{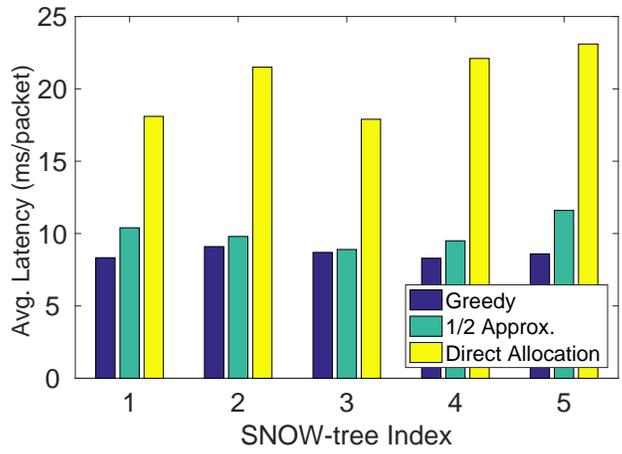}
      }
      \hfill
      \subfigure[Energy consumption in intra-SNOW comm.\label{fig:chap5-snow_energy}]{
        \includegraphics[width=.5\textwidth]{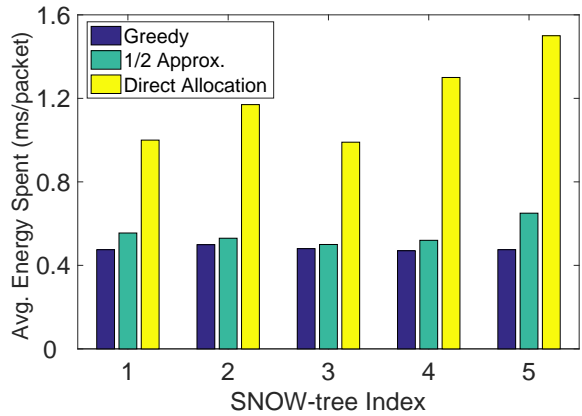}
      }
    \caption{Performance of intra-SNOW communications in different SNOW-trees.}
    \label{fig:chap5-snow_comm}
\end{figure}

{\bf Experiments on Intra-SNOW Communication.}
In this section, we demonstrate intra-SNOW communication performance when multiple interfering SNOWs are integrated together to coexist. Due to our limited number of USRP devices, we choose two interfering SNOWs in a SNOW-tree, to run intra-SNOW communications independently at the same time. For example in SNOW-tree 1, SNOWs at locations A and B perform intra-SNOW communications. Here, SNOWs at locations A and B are assigned 3 and 2 nodes, respectively (as explained in Section~\ref{sec:chap5-implementation}). 
Similarly, we allow SNOWs at locations B and C; C and A to do the same, respectively. In experiments, each node under a SNOW hops randomly on different subcarriers assigned by our greedy heuristic algorithm and sends 100 consecutive packets to the BS. We repeat the same set of experiments when subcarrier assignment is done by our approximation algorithm and the directly allocation scheme.  
We allow the nodes in a SNOW to hop across different subcarriers to emulate that as if all the subcarriers of that SNOW were assigned to different nodes. Figure~\ref{fig:chap5-snow_comm} shows the reliability, latency, and energy consumption in intra-SNOW communication under different SOP algorithms.

Figure~\ref{fig:chap5-snow_reliability} shows the average PRR in different SNOW BSs. In each SNOW-tree, the average PRR at each SNOW BS is calculated from all 3 pairs of intra-SNOW communication experiments. The highest average PRR is approximately 100\% in SNOW BSs located at E, I, M, N, and O, while the lowest average PRR is approximately 98.9\% in SNOW BS located at F when the subcarriers assigned by our greedy heuristic algorithm is used. For our approximation algorithm, the highest and lowest average PRR values are approximately 100\% and 97.9\%, respectively. For the direct allocation scheme, these values are 89\% and 79\%, respectively.
Figure~\ref{fig:chap5-snow_latency} shows that the average latency to successfully deliver an intra-SNOW packet to a SNOW BS is lower in all SNOW-trees while the subcarriers assigned by our greedy heuristic algorithm are used. For example, the average latency per packet is as low as 8.3ms in SNOW-tree 4 compared to 9.5ms and 22.1ms for our approximation algorithm and direct allocation subcarrier assignments, respectively.
Figure~\ref{fig:chap5-snow_energy} shows that the average energy consumption for each packet is also lower in all SNOW-trees when our greedy subcarrier assignment is used. In SNOW-tree 4, the average energy consumption per packet is as low as 0.47mJ compared to 0.52mJ and 1.31mJ for approximation and direct allocation subcarrier assignments, respectively. 
Thus, all the experiments in Figure~\ref{fig:chap5-snow_comm} confirm that both our greedy heuristic and approximation algorithm are practical choices for SOP.

\begin{figure}[!htb]
    \centering
      \subfigure[Reliability in inter-SNOW communication\label{fig:chap5-p2p_reliability}]{
    \includegraphics[width=0.50\textwidth]{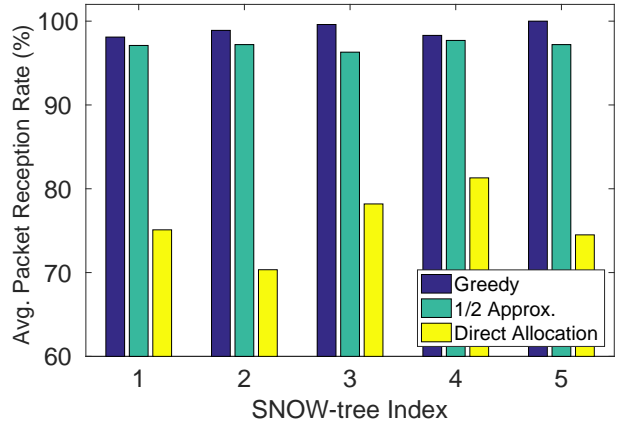}
      }
    \hfill
      \subfigure[Latency in inter-SNOW communication\label{fig:chap5-p2p_latency}]{
        \includegraphics[width=.50\textwidth]{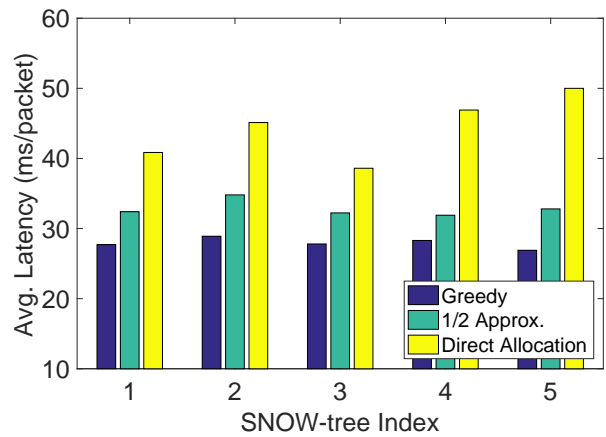}
      }
      \hfill
      \subfigure[Energy consumption in inter-SNOW comm.\label{fig:chap5-p2p_energy}]{
        \includegraphics[width=.50\textwidth]{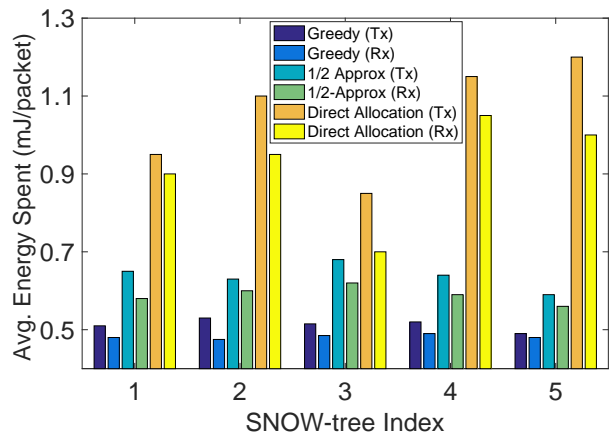}
      }
    \caption{Performance of inter-SNOW communications in different SNOW-trees.}
    \label{fig:chap5-p2p_comm}
\end{figure}

{\bf Experiments on Inter-SNOW Communications.}
To demonstrate inter-SNOW communication performance, we perform parallel communications between two nodes under two sibling BSs in each SNOW-tree, using the sets of subcarriers assigned to BSs by different SOP algorithms in our previous experiments. 
Since, we have only one node under each BS in a tree (as explained in Section~\ref{sec:chap5-implementation}), we allow those nodes to use all the subcarriers of their respective BSs.
Considering SNOW-tree 1, the node in BS located at B (and E) will send inter-SNOW packets to the node in BS located at E (and B) via root BS located at A. Thus, this is level three inter-SNOW communication.
In experiments, the node in BS at B (and E) randomly hops into different subcarriers of its BS and sends consecutive 100 packets destined for the node in BS at E (and B). BS at B (and E) first receives the packets (intra-SNOW) and then relays to its parent BS at A (inter-SNOW). Root BS at A then relays (inter-SNOW) the packets to BS at E (and B). Finally, BS at E (and B) sends (intra-SNOW) the packets to its node. Considering a single inter-SNOW packet, since the node is randomly hopping to different subcarriers, the BS sends (intra-SNOW) the same packet via all subcarriers, so that the node may receive it instantly. The whole process is repeated 1000 times in every SNOW-tree.
Figure~\ref{fig:chap5-p2p_comm} shows the average PRR, latency, and energy consumption in inter-SNOW communications, while the set of subcarriers used are given by our greedy heuristic, our approximation algorithm, and the direct allocation scheme in our previous experiments in Section~\ref{sec:chap5-exp_sop}.
Figure~\ref{fig:chap5-p2p_reliability} shows that the average PRR values are high in all SNOW-trees when the subcarriers are assigned using our greedy heuristic. For example, PRR is as high as 99.99\% in SNOW-tree 5 compared to 97.2\% and 74\% by our approximation algorithm and the direct allocation scheme, respectively.
Figure~\ref{fig:chap5-p2p_latency} shows that the per inter-SNOW packet latency is lower in all SNOW-trees in case of our greedy subcarrier assignments. In SNOW-tree 5, it is 26.2ms on average compared to 32.8ms and 50ms in cases of our approximation algorithm and the direct allocation scheme assignments, respectively.
Figure~\ref{fig:chap5-p2p_energy} shows average energy consumed per inter-SNOW packet at Tx and Rx nodes are lower in all SNOW-trees for our greedy assignments. In SNOW-tree 5, Tx and Rx nodes consume on average 0.49mJ and 0.48mJ energy, respectively. For our approximation algorithm, these values are 0.59mJ and 0.56mJ, while the direct allocation yields 1.2mJ and 1mJ. 
These experiments thus confirm that our greedy heuristic and approximation algorithms are practical choices for SOP.

%% file: chapter5/simulation.tex
\subsubsection{Simulation}\label{sec:chap5-simulations}
For evaluation under large-scale network, we perform simulations through NS-3~\cite{ns3}.

\begin{figure}[!htb]
    \centering 
      \subfigure[Locations of SNOW BSs in SNOW-tree.\label{fig:chap5-sim_tree}]{
    \includegraphics[width=0.45\textwidth]{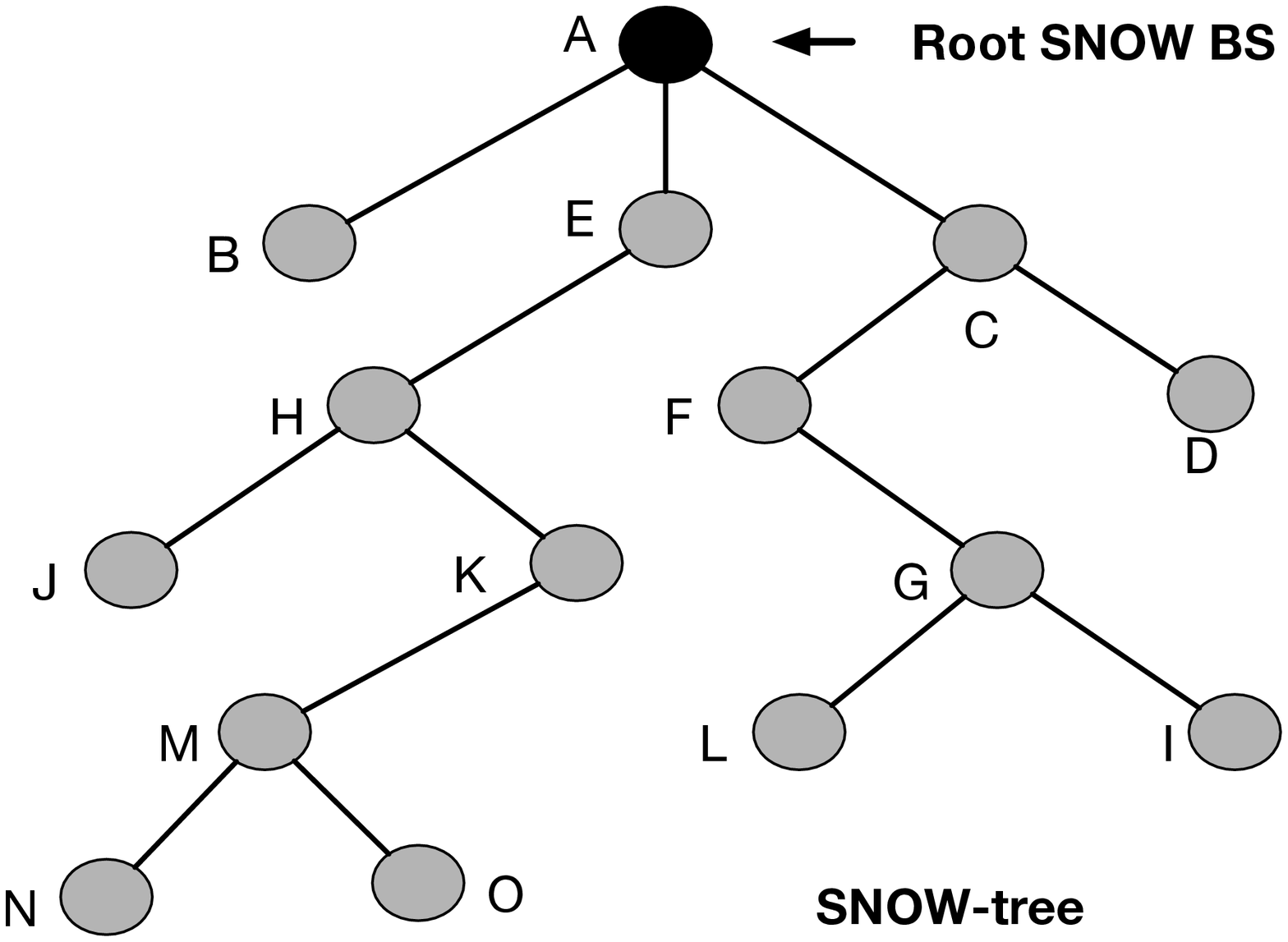}
      }
    \hfill
      \subfigure[Max. allowed number of common subcarriers between interfering SNOW pairs.\label{fig:chap5-interf_mat}]{
        \includegraphics[width=.45\textwidth]{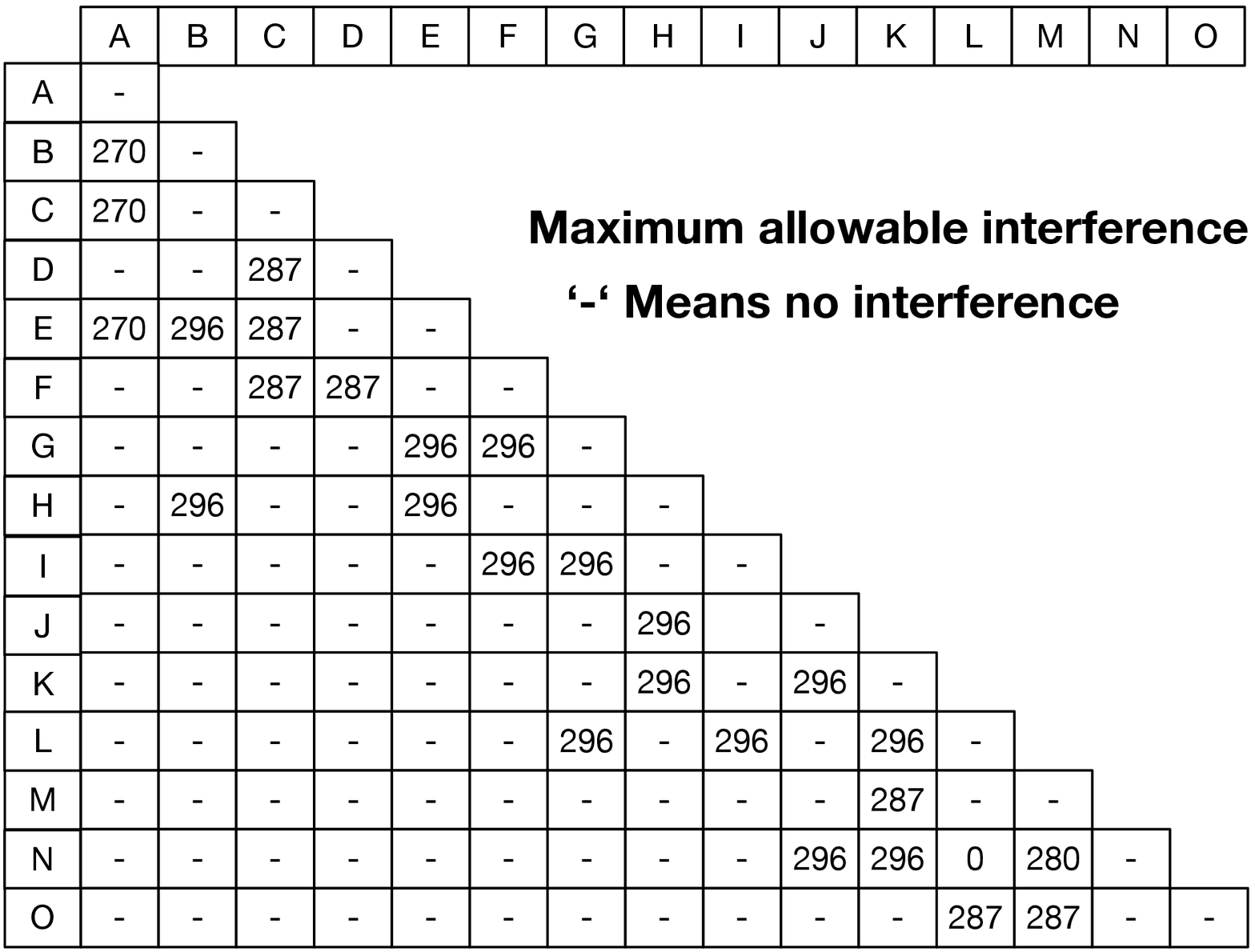}
      }
      \hfill
      \subfigure[Subcarrier assignment by root BS at location A in SNOW-tree.\label{fig:chap5-sop_sim}]{
        \includegraphics[width=0.45\textwidth]{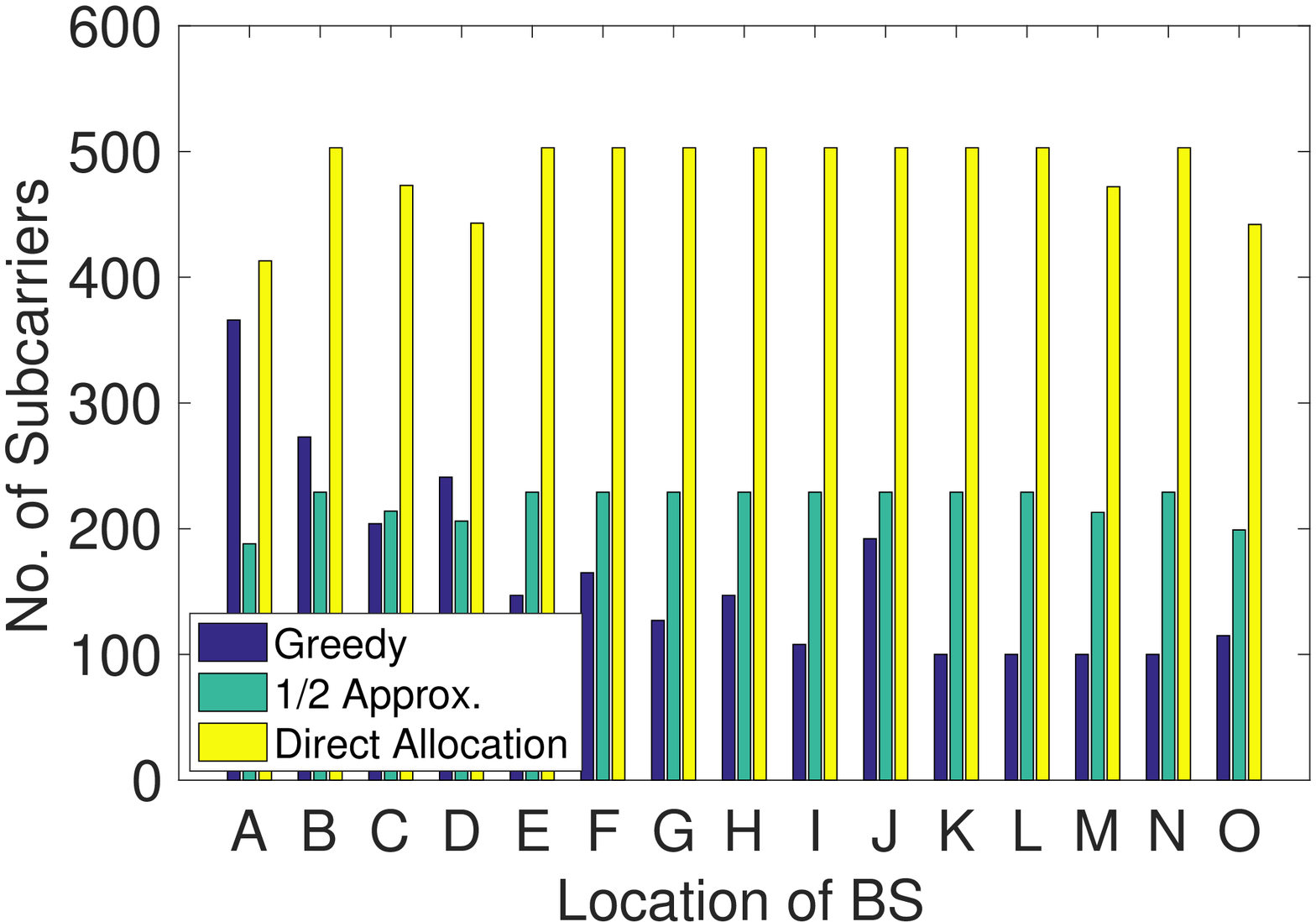}
      }
    \caption{SNOW-tree topology, allowable interference between BSs, and subcarrier allocation for BSs in simulation.}
    \label{fig:chap5-sim_setup}
\end{figure}

{\bf Simulation Setup.}\label{sec:chap5-sim_setup}
We create a SNOW-tree of 15 SNOWs (BSs) as shown in Figure~\ref{fig:chap5-sim_tree} and simulate the (25x15)km$^2$ area as shown in Figure~\ref{fig:chap5-testbed}. BS at location A is the root BS. Each SNOW has 1000 nodes, totaling 15000 thousand nodes in the SNOW-tree. We limit the maximum allowable number of common subcarriers between interfering BSs based on the white space availability at different BS locations (Figure~\ref{fig:chap5-channels}) and our experimental findings, which is shown in Figure~\ref{fig:chap5-interf_mat}. $\sigma_i$ in Constraint (\ref{c:chap5-const1}) is chosen to be 100 for all the BS. Thus, a subcarrier will be used by at most 10 nodes in worst case in intra-SNOW communication. Figure~\ref{fig:chap5-sop_sim} shows the subcarrier assignments for all BSs by the root BS at location A, while using our greedy heuristic algorithm, approximation algorithm, and the direct allocation scheme. Here, both greedy heuristic and approximation algorithms do not violate any of the Constraints of SOP. However, the direct allocation scheme violates Constraints (\ref{c:chap5-const2}) and (\ref{c:chap5-const3}) of SOP. The values for various parameters such as packet size, spreading factor, modulation, and Tx power are set the same as described in our real experiments (Section~\ref{sec:chap5-exp_setup}).

\begin{figure}[!htb]
    \centering
      \subfigure[Reliability in multi-level inter-SNOW comm.\label{fig:chap5-reliability_sim}]{
        \includegraphics[width=.45\textwidth]{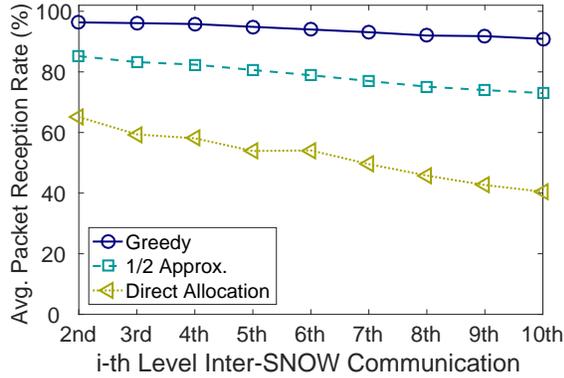}
      }
      \hfill
      \subfigure[Latency in multi-level inter-SNOW comm.\label{fig:chap5-latency_sim}]{
        \includegraphics[width=.45\textwidth]{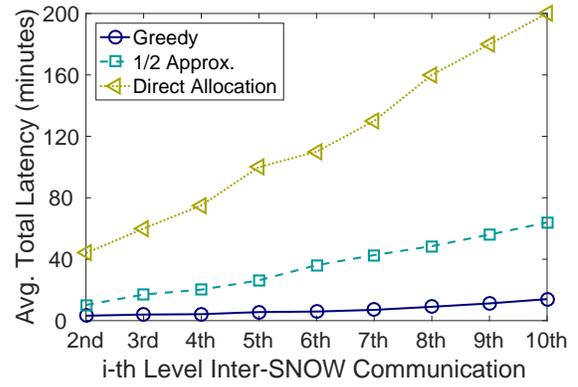}
      }
      \hfill
      \subfigure[Energy in multi-level inter-SNOW comm.\label{fig:chap5-energy_sim}]{
        \includegraphics[width=.45\textwidth]{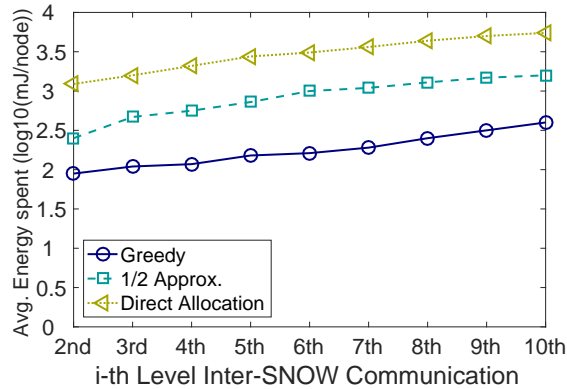}
      }
    \caption{Multi-level parallel inter-SNOW communications in SNOW-tree}
    \label{fig:chap5-p2p_comm_sim}
\end{figure}

{\bf Simulation Results.}
We evaluate the performance of our design using thousands of nodes by generating thousands of parallel multi-level inter-SNOW communications. In simulation, each node in each SNOW sends 100 packets with a random sleep interval of 0-50 ms, destined for another node in second level (adjacent SNOWs) and up to its maximum reachable level inside the SNOW-tree. In each SNOW, we identify nodes from 1 to 1000. In our simulation, a node with ID $i$ will send inter-SNOW packets to the nodes with ID $i$ in all other SNOWs.
Figure~\ref{fig:chap5-p2p_comm_sim} demonstrates the performances in terms of reliability, latency, and energy consumption when subcarriers assigned by our greedy heuristic, approximation algorithm, and direct allocation scheme are used.

Figure~\ref{fig:chap5-reliability_sim} shows that by using the subcarriers assigned by our greedy heuristic algorithm, we can achieve on average PRR of 93\% even in 10th level inter-SNOW communications. On the other hand, our approximation algorithm and direct allocation scheme can provide approximately 73\% and 40\% of average PRR, respectively.
Figure~\ref{fig:chap5-latency_sim} shows that by using the subcarriers assigned by our greedy heuristic algorithm, we observe on average total latency of 14 minutes to send all successful inter-SNOW packets to the second levels and up to the maximum achievable levels by all 15000 nodes. Using subcarriers given by our approximation algorithm and direct allocation scheme, these values are approximately 60 minutes and 200 minutes, respectively.
Figure~\ref{fig:chap5-energy_sim} shows that by using the subcarriers assigned by our greedy heuristic algorithm, the per node energy consumption to send all successful inter-SNOW packets to all possible levels is 389mJ. While in cases of our approximation algorithm and direct allocation scheme, these values are 1728mJ and 5580mJ, respectively. Thus, the simulation results demonstrate that the greedy heuristic or the approximation algorithm can be chosen to scale up LPWANs for future IoT applications.

\subsubsection{Discussion}
\revise{In Section~\ref{sec:chap5-proposed_solution}, we have justified that our greedy heuristic approach is an intuitive and highly scalable polynomial-time solution. Additionally, we have discussed that deriving an analytical bound (in terms of scalability) of our greedy heuristic is not immediate. Hence, for the cases when an analytical performance bound is needed, we have proposed a probabilistic optimization approach and derived its theoretical performance bound (Section~\ref{sec:chap5-bounded_algorithm}). Specifically, our probabilistic optimization approach is a $\frac{1}{2}$-approximation algorithm. In terms of performance, both experiments and simulations demonstrate that our greedy heuristic algorithm provides higher reliability, lower latency, and lower energy consumption in both intra- and inter-SNOW communications compared to our approximation algorithm, which is due to its interference-aware subcarrier assignments to different SNOW BSs. As our approximation algorithm assigns more subcarriers to most of the BSs (both in experiments and simulations), it also assigns a greater number of interfering subcarriers between neighboring BSs, resulting in frequent back-offs in transmissions by the nodes, an increase in latency and energy consumption in both intra- and inter-SNOW communications.}

\revise{As described in Algorithms~\ref{algo:chap5-greedy} and~\ref{algo:chap5-approx}, our greedy heuristic or/and approximation algorithms may fail to provide a feasible subcarrier assignment for few SOP problem instances.
In practice, either our greedy heuristic or our approximation algorithm may be adopted to handle the subcarrier assignment failure of each other. In cases when both fail, the target application's requirement will dictate which solution should be adopted. For example, if the application requires bounded performance and high spectrum utilization, our approximation algorithm may be adopted. On the other hand, greedy heuristic may be chosen in case higher reliability is expected. In experiments, we were unable to demonstrate such cases based on the available TV white spaces and environments at our testbed location. Our realistic simulations, where parameters are chosen based on our experiments, do not also showcase any infeasible cases of our greedy heuristic algorithm. In general, our experiments and simulations demonstrate that both greedy heuristic and approximation algorithms may be practically chosen to scale up LPWANs for future IoT applications.}

%% file: chapter5/related.tex
\subsection{Related Work}\label{sec:chap5-related}

\revise{The LPWAN technologies are still in their infancy with some still being developed (e.g., 5G, NB-IoT, LTE Cat M1, Weightless-P), some having only uplink capability (e.g., SigFox, Weightless-N), while, for some, there is still no publicly available documentation (e.g., SigFox)~\cite{ismail2018low, whitespaceSurvey}. Thus, developing generalized techniques to address integration is not our focus. Instead, we propose an integration of multiple SNOWs in the white spaces for scaling up, the insights of which may also be extended to other LPWANs in the future.
To cover a wide area, LoRaWAN integrates multiple gateways through the Internet~\cite{lorawan}. Cellular networks do the same relying on wired infrastructure~\cite{channel_cellular}.  Rural and remote areas lack such infrastructure.  Wireless integration that we have considered in this paper can be a solution for both urban and rural areas.}


\revise{While the proposed integration may look like channel allocation in traditional tiered or clustered multi-channel networks~\cite{si2010overview, alicherry2005joint, xing2007superimposed, ramachandran2006interference, ko2007distributed, raniwala2005architecture, kodialam2005characterizing, marina2010topology, 80211mesh, rad2006joint, gurewitz2007cooperative, aryafar2008distance}, it is a {\bf\slshape conceptually different} problem with new challenges.
{\bf First,} in traditional networks, the links operate on predefined fixed-bandwidth channels. In contrast, in integrating multiple SNOW networks we have to find proper bandwidths for all links and they are inter-dependent and can be different.  
{\bf Second,} SNOW integration involves assigning a large number of subcarriers to each BS allowing some degree
of overlaps among interfering BSs for enhanced scalability.
{\bf Finally,} through integration, we have to retain massive parallel communication (between a SNOW BS and its numerous nodes) and concurrent inter- and intra-SNOW communications.
%
%
Hence, traditional channel allocation for wireless networks~\cite{channel_wireless}, WSN~\cite{channel_wsn, leach}, or cognitive radio networks~\cite{cognitive_channel_survey} cannot be used in SNOW integration.
%
%
In regard to the white space networking, the closest work to ours is~\cite{gameapproach} that considers multiple WiFi-like 
networks in white spaces, where all users have access to white
space database, and every access point (AP) chooses a single
channel. Thus the problem is different from SOP.}

%% file: chapter7.tex
\section{Future Research Directions} \label{chap:future}

This thesis may lead the future work for solving the problems and challenges raised by the emerging Internet of Things (IoT) and Cyber-Physical Systems (CPS) applications. Our primary focus in this chapter is to envision the design and development of the scalable, energy-efficient, and intelligent systems for IoT/CPS. In the following, we outline the future research directions.

\subsection{Super-Massive Scalability in SNOW}\label{sec:chap6-massive_snow}
To enable the next generation IoT/CPS applications where hundreds of thousands of requests from thousands of sensors (e.g., real-time wide-area monitoring, wireless data centers) will need coordination, LPWANs have to perform at a super-massive scale. 
The following design choices in SNOW may help achieving such super-massive scalability.

\begin{enumerate}
	\item In the current D-OFDM-based SNOW PHY, the BS can receive from one node in one subcarrier. Thus, the concurrent transmissions from multiple nodes are received via multiple subcarriers. 
	The SNOW PHY may be augmented such that the BS will receive concurrently from multiple nodes in a single subcarrier as well, making it super scalable. 
	This may be done through an entanglement between D-OFDM and CDMA (Code Division Multiple Access), which has never been done for any other system. Such entanglement, however, will require to rethink the ISI (Inter-symbol-interference) for both D-OFDM and CDMA as well as the ICI in SNOW.

	\item The BS in SNOW currently assigns different subcarriers to different nodes randomly. A subcarrier assignment protocol in SNOW may be proposed considering the following aspects. A mathematical formulation of a constrained optimization problem where the objective may include minimizing the interference between the hidden nodes in SNOW and the constraints may include node's positions, number of nodes per subcarrier, and their transmission power. This protocol, however, will need to incorporate autonomous join/leave of the nodes to provide seamless operation in SNOW, which will be very challenging.

	\item The Current SNOW PHY allows a uniform data modulation technique (ASK (amplitude-shift keying)/BPSK (binary phase-shift keying)) in its D-OFDM subcarriers, which makes it difficult to host IoT/CPS applications where different nodes have different bitrate requirements. So, enabling different modulation techniques in different subcarriers may be explored. This, however, will require to address the ISI and ICI of the D-OFDM design.

	\item The current SNOW integration considers a tree structured system model, i.e., SNOW-tree~\cite{snow3, snow_integration_ton}. A more generalized SNOW integration may be achieved by considering a general graph model with the formulation of a constrained multi-objective optimization problem that will maximize the scalability and minimize the latency in data aggregation. For this, the interference between the SNOWs and PAPR of each SNOW BS has to be taken as constraints. To solve this multi-objective optimization problem, an {\em evolutionary algorithm} may be proposed.
\end{enumerate}

\subsection{Integration and Coexistence of Heterogeneous LPWANs}\label{sec:chap6-lpwan_coexist}

With the growth of multiple LPWAN technologies and heterogeneous devices, the existing IoT platforms are on the verge of the {\em Tower of Babble Effect}. As such, these technologies with heterogeneous requirements will face severe inter-technology interference, e.g., LoRaWAN, SigFox, IQRF, RPMA, etc. in the ISM band, or NB-IoT, LTE, 5G, etc. in the licensed band. Depending on the application requirements, these LPWANs will need an integration between themselves or/and to coexist together by avoiding each other's interference. This thesis may lead the research in both of these aspects. With the knowledge the seminal work on the SNOW integration~\cite{snow3, snow_integration_ton}, a multi-variable constrained optimization problem may be formulated, where the objective may include maximizing/minimizing different quality of service (QoS) parameters and the constraints may include the fair share of the wireless mediums and interference between different LPWAN technologies.

\subsection{Internet of Intelligent Things (IoIT)}\label{sec:chap6-iiot}
The IoIT has been evolving in the past few years to provide predictive, accurate, and faster data analytics in the IoT platforms. Involvement of AI in the IoT platforms is twofold: (1) {\em machine learning} algorithms are being used to make sense of the huge amount of data generated by the sensors and (2) the coordination between these sensors is governed by numerous variations of the {\em swarm intelligence}. The latter, however, has been tested in the computer simulations mostly~\cite{sworm} due to the lack of scalable and energy-efficient communication protocols between heterogeneous and resource-constrained (e.g., limited energy budget and computation power) IoT end devices (e.g., sensors/actuators), which may be explored involving the practical hardware platforms, as described below.

\begin{enumerate}
	\item A scalable and energy-efficient IoIT platform may be developed where heterogeneous nodes will coordinate towards several optimization goals such as collaborative sensing/actuation and data fusion. Depending on the nature of the optimization, the core of such a coordination will be based on a {\em genetic algorithm} (where the nodes try to avoid each other) or a {\em swarm intelligence} algorithm (where the nodes help each other). To bootstrap such coordination with proactive maintenance, it will require the introduction of {\em fuzziness} in the platform which may be governed by the gateway.

	\item To realize the IoIT platform envisioned above, the first set of challenges will be to build a robust, lightweight, low-latency, and energy-efficient communication protocol between the heterogeneous IoT devices, facilitating synchronization between themselves, and selecting a set of devices to perform a specific task based on their {\em fuzziness score}. These challenges may be explored in a greater detail and propose solution techniques.

	\item Once the design and development of the basic building blocks of the above IoIT platform is realized, the focus may be on developing protocols that will be needed for the adopting applications. For example, dynamic sensor/actuator placement protocols (can be used in UAV platforms), schedule or/and route planning protocols (can be used in smart waste management), etc. may be designed, while considering the resource limitations of the corresponding platform.
\end{enumerate}

%% file: chapter8.tex
\section{Conclusions}   \label{chap:conclusion}

Today, WSNs face scalability challenges in wide-area wireless monitoring and control applications (e.g., smart city) that require thousands of sensors to be connected over long distances. Existing WSN technologies (e.g., IEEE 802.15.4 in 2.4GHz) facilitate this by forming multi-hop mesh networks, complicating the protocol design and network deployment. To address this, we have designed SNOW -- a novel LPWAN
technology by exploiting the TV white spaces. SNOW achieves scalability and energy efficiency by enabling concurrent packets reception at a base station (BS) from numerous sensors and concurrent packets transmission to numerous sensors from the BS, simultaneously, over several kilometers. We have demonstrated the feasibility of SNOW by implementing it on a prototype hardware. 

To make the SNOW implementation widely available and practically deployable, we have implemented SNOW using the low-cost and small form-factored COTS IoT devices. The COTS devices, however, face a variety of practical challenges that are very difficult to handle with their cheap radios. Specifically, we have addressed the high peak-to-average power ratio (PARP) problem, calculated the channel state information, and carrier frequency offsets. We have also proposed an adaptive transmission power protocol for the nodes. We have demonstrated COTS SNOW implementation on the
TI CC1310 and TI CC1350 devices, reducing the cost and form-factor of a SNOW node by 30x and 10x, respectively. Overall, COTS SNOW implementation is practical for many IoT/CPS applications due to its low-cost, low-form-factored, and low-energy consumption features.

As the LPWANs still face limitations in meeting the scalability and coverage demand of very wide-area IoT/CPS deployments (e.g., (74x8) km$^2$ East Texas oilfield with ten thousand nodes). To enable this, we have proposed a network architecture called SNOW-tree through a seamless integration of multiple SNOWs where they form a tree structure and are under the same management/control at the tree root, addressing the inter-SNOW interference
by formulating a constrained optimization problem (which is NP-hard) whose objective is to maximize
scalability by managing the spectrum sharing across the SNOWs. We have also proposed two highly effective
polynomial-time methods to solve it: a greedy heuristic algorithm and a $\frac{1}{2}$-approximation algorithm. We have finally demonstrated the feasibility of this work by deploying 15 SNOWs, covering (25x15)km$^2$.

To enable the next generation IoT/CPS applications with hundreds of thousands of requests from numerous sensors (e.g., wireless data centers), we will make SNOW super scalable by realizing the following design choices. (1) We will enrich the SNOW physical layer such that the BS will receive concurrent packets from multiple nodes using a single subcarrier, in parallel to the concurrent packets reception in all other subcarriers. (2) We will design a hidden terminal-aware subcarrier assignment to the nodes, thus minimizing inter-node interference. (3) We will enable multi-modulation across the subcarriers to enable different bitrates at different nodes. (4) We will propose a
network model where multiple SNOWs will seamlessly coordinate by forming a general graph structure. Overall, our vision is to make SNOW super scalable, widely available, and cheaply deployable for wide-area IoT/CPS applications.

%% file: bibliography.tex
\section*{REFERENCES}

\addcontentsline{toc}{section}{References}

\bibliography{bib/mybib}

\bibliographystyle{abbrv}